\documentclass[reqno]{amsart}

\usepackage{latexsym,pifont}

\usepackage{epsfig}

\usepackage{ifpdf}
\ifpdf
\usepackage{epstopdf}
\usepackage{hyperref}
\else
\usepackage[hypertex]{hyperref}
\fi

\usepackage{amsfonts,amsmath,amssymb,mathrsfs,extarrows,MnSymbol}
\usepackage[all]{xy}

\newtheorem{Thm}{Theorem}
\newtheorem{Prop}[Thm]{Proposition}
\newtheorem{Lem}[Thm]{Lemma}
\newtheorem{Cor}[Thm]{Corollary}
\theoremstyle{definition}
\newtheorem{Rem}[Thm]{Remark}
\newtheorem{Def}[Thm]{Definition}
\newtheorem{Eg}[Thm]{Example}
\newtheorem{Conv}[Thm]{Convention}
\newtheorem{Quest}[Thm]{Question}
\newtheorem{Prob}[Thm]{Problem}

\newcommand{\alxydim}[2]{\begin{aligned}\xymatrix#1{#2}\end{aligned}}

\newcommand{\brem}{\begin{Rem}}
\newcommand{\erem}{\end{Rem}\medskip}
\newcommand{\beg}{\begin{Eg}}
\newcommand{\eeg}{\end{Eg}}
\newcommand{\bedef}{\begin{Def}}
\newcommand{\exdef}{
\end{Def}\vskip0.1cm}
\newcommand{\berop}{\begin{Prop}}
\newcommand{\eerop}{\end{Prop}}
\newcommand{\belem}{\begin{Lem}}
\newcommand{\elem}{\end{Lem}}
\newcommand{\bethe}{\begin{Thm}}
\newcommand{\ethe}{\end{Thm}}
\newcommand{\becor}{\begin{Cor}}
\newcommand{\ecor}{\end{Cor}}
\newcommand{\beroof}{\noindent\begin{proof}}
\newcommand{\eroof}{\qed\end{proof}}
\newcommand{\becon}{\begin{Conv}}
\newcommand{\econ}{\begin{flushright}$\checkmark$\end{flushright}\end{Conv}}
\newcommand{\bequest}{\begin{Quest}}
\newcommand{\equest}{\end{Quest}}
\newcommand{\brob}{\begin{Prob}}
\newcommand{\erob}{\end{Prob}}

\newcommand{\barr}{\begin{array}}
\newcommand{\earr}{\end{array}}
\newcommand{\ben}{\begin{enumerate}}
\newcommand{\een}{\end{enumerate}}
\newcommand{\bit}{\begin{itemize}}
\newcommand{\eit}{\end{itemize}}

\newcommand{\qq}{\begin{eqnarray}}
\newcommand{\qqq}{\end{eqnarray}}

\newcommand{\nn}{\nonumber}

\newcommand{\ovl}[1]{\overline{#1}}
\newcommand{\unl}[1]{\underline{#1}}

\newcommand{\Reqref}[1]{Eq.\,\eqref{#1}}
\newcommand{\Rcite}[1]{Ref.\,\cite{#1}}
\newcommand{\Rxcite}[2]{Ref.\,\cite[#1]{#2}}

\newcommand{\tx}[1]{\textrm{#1}} 

\newcommand{\gt}[1]{\mathfrak{#1}}

\def\cA{\mathcal{A}}
\def\cB{\mathcal{B}}

\def\cD{\mathcal{D}}
\def\cE{\mathcal{E}}

\def\cG{\mathcal{G}}
\def\ceH{\mathcal{H}}
\def\cI{\mathcal{I}}
\def\cJ{\mathcal{J}}
\def\cK{\mathcal{K}}

\def\cM{\mathcal{M}}

\def\cO{\mathcal{O}}
\def\cP{\mathcal{P}}

\def\cS{\mathcal{S}}
\def\cT{\mathcal{T}}
\def\cU{\mathcal{U}}

\def\cZ{\mathcal{Z}}

\newcommand{\CK}{\mathcal{K}}

\newcommand{\CP}{\mathcal{P}}

\def\xcA{\mathscr{A}}
\def\xcB{\mathscr{B}}
\def\xcC{\mathscr{C}}
\def\xcD{\mathscr{D}}

\def\xcF{\mathscr{F}}
\def\xcG{\mathscr{G}}
\def\xcH{\mathscr{H}}
\def\xcI{\mathscr{I}}
\def\xcJ{\mathscr{J}}
\def\xcK{\mathscr{K}}
\def\xcL{\mathscr{L}}
\def\xcM{\mathscr{M}}
\def\xcN{\mathscr{N}}

\def\xcR{\mathscr{R}}
\def\xcS{\mathscr{S}}

\def\xcV{\mathscr{V}}
\def\xcW{\mathscr{W}}
\def\xcX{\mathscr{X}}

\def\bH{{\mathbb{H}}}

\def\bR{{\mathbb{R}}}
\def\bS{{\mathbb{S}}}

\def\bZ{{\mathbb{Z}}}

\def\a{\alpha}
\def\b{\beta}
\def\g{\gamma}
\def\G{\Gamma}
\def\d{\delta}
\def\D{\Delta}

\def\vep{\varepsilon}

\def\tht{\theta}

\def\la{\lambda}
\def\La{\Lambda}
\def\om{\omega}
\def\Om{\Omega}

\def\si{\sigma}
\def\Si{\Sigma}

\def\z{\zeta}

\def\bgt{\gt{b}}
\def\Bgt{\gt{B}}

\def\Egt{\gt{E}}

\def\ggt{\gt{g}}

\def\Kgt{\gt{K}}

\def\Mgt{\gt{M}}

\def\Pgt{\gt{P}}

\def\sgt{\gt{s}}
\def\Sgt{\gt{S}}
\def\tgt{\gt{t}}

\def\Vgt{\gt{V}}

\def\Wgt{\gt{W}}

\def\Xgt{\gt{X}}

\newcommand{\sfd}{{\mathsf d}}

\newcommand{\sfE}{{\mathsf E}}

\newcommand{\sfi}{{\mathsf i}}

\newcommand{\sfk}{{\mathsf k}}

\newcommand{\sfL}{{\mathsf L}}

\newcommand{\sfP}{{\mathsf P}}

\newcommand{\sfT}{{\mathsf T}}

\newcommand{\txA}{{\rm A}}

\newcommand{\txB}{{\rm B}}

\newcommand{\txc}{{\rm c}}

\newcommand{\ee}{{\rm e}}
\newcommand{\txE}{{\rm E}}
\newcommand{\txf}{{\rm f}}
\newcommand{\txF}{{\rm F}}
\newcommand{\txg}{{\rm g}}
\newcommand{\txG}{{\rm G}}
\newcommand{\txh}{{\rm h}}
\newcommand{\txH}{{\rm H}}

\newcommand{\txK}{{\rm K}}

\newcommand{\txm}{{\rm m}}
\newcommand{\txM}{{\rm M}}
\newcommand{\txN}{{\rm N}}

\newcommand{\txP}{{\rm P}}

\newcommand{\txS}{{\rm S}}

\def\vC{\check{C}}
\def\Cv{\v{C}}

\def\vd{\check{\d}}

\def\exp{{\rm exp}}
\def\id{{\rm id}}
\newcommand{\pr}{{\rm pr}}

\def\obj{{\rm Ob}}
\def\mor{{\rm Hom}}
\def\morf{{\rm Mor}}
\def\dim{{\rm dim}}
\def\im{{\rm im}}
\def\ker{{\rm ker}}

\newcommand{\Id}{{\rm Id}}
\def\Inv{{\rm Inv}}

\def\bgrb{\gt{BGrb}}

\newcommand{\Gr}{{\rm Gr}}
\newcommand{\gtGr}{\mathfrak{Gr}}
\newcommand{\gtgr}{\mathfrak{gr}}

\newcommand{\Set}{{\rm Set}}
\newcommand{\Man}{{\rm Man}}
\newcommand{\GMan}{\textrm{$\txG$-$\Man$}}

\newcommand{\ic}{\imath}
\newcommand{\pLie}[1]{\,{-\hspace{-8pt}\xcL}_{#1}}
\def\p{\partial}

\def\con{\righthalfcup}

\def\emb{\hookrightarrow}

\def\Hol{{\rm Hol}}

\newcommand{\cGk}{\cG_\sfk}
\newcommand{\DQ}{\D\Qup}
\newcommand{\DT}{\D\Tup}
\newcommand{\DTn}{\D\Tnup}

\def\bd1{{\boldsymbol{1}}}
\def\brd0{{\boldsymbol{0}}}

\def\tr{{\rm tr}}

\def\Ad{{\rm Ad}}

\newcommand{\faff}[1]{P^{\sfk}_{+}(#1)}

\newcommand{\uj}{{\rm U}(1)}
\newcommand{\sug}{{\rm SU}(2)}

\def\x{\times}
\def\ox{\otimes}

\def\lx{{\hspace{-0.04cm}\ltimes\hspace{-0.05cm}}}

\newcommand{\Vcon}[2]{\left(\,#1\,,\,#2\,\right)_{\con}}

\newcommand{\GBra}[2]{\lsem\,#1\,,\,#2\,\rsem}

\newcommand{\Mup}{{}^{\tx{\tiny $M$}}\hspace{-1pt}}
\newcommand{\tMup}{{}^{\tx{\tiny $\tilde M$}}\hspace{-1pt}}
\newcommand{\xcMup}{{}^{\tx{\tiny $\xcM$}}\hspace{-1pt}}
\newcommand{\txcMup}{{}^{\tx{\tiny $\tilde\xcM$}}\hspace{-2pt}}
\newcommand{\xcFup}{{}^{\tx{\tiny $\xcF$}}\hspace{-2pt}}
\newcommand{\txcFup}{{}^{\tx{\tiny $\tilde\xcF$}}\hspace{-2pt}}
\newcommand{\Qup}{{}^{\tx{\tiny $Q$}}\hspace{-1pt}}
\newcommand{\tQup}{{}^{\tx{\tiny $\tilde Q$}}\hspace{-1pt}}
\newcommand{\Tup}{{}^{\tx{\tiny $T$}}\hspace{-1pt}}
\newcommand{\Tnup}{{}^{\tx{\tiny $T_n$}}\hspace{-1pt}}

\newcommand{\tTnup}{{}^{\tx{\tiny $\tilde T_n$}}\hspace{-1pt}}

\newcommand{\sfPup}{{}^{\tx{\tiny $\sfP$}}\hspace{-2pt}}

\numberwithin{equation}{section} \numberwithin{Thm}{section}

\begin{document}

\title{The gauging of two-dimensional bosonic sigma models\\
on world-sheets with defects}

\author{Krzysztof Gaw\c{e}dzki}
\address{K.G.:\ Laboratoire de Physique, C.N.R.S., ENS-Lyon,
Universit\'e de Lyon, 46 All\'ee d'Italie, 69364 Lyon, France}
\email{kgawedzk@ens-lyon.fr}
\author{Rafa\l ~R.~Suszek}
\address{R.R.S.:\ Katedra Metod Matematycznych Fizyki, Wydzia\l
~Fizyki Uniwersytetu Warszawskiego, ul.\ Ho\.za 74, PL-00-682
Warszawa, Poland} \email{suszek@fuw.edu.pl}
\author{Konrad Waldorf}
\address{K.W.:\ Fakult\"at f\"ur Mathematik, Universit\"at
Regensburg, Universit\"atsstra\ss e 31, 93053 Regensburg, Germany}
\email{konrad.waldorf@mathematik.uni-regensburg.de}
\medskip

\begin{abstract}
We extend our analysis of the gauging of rigid symmetries in bosonic
two-dimensional sigma models with Wess--Zumino terms in the action
to the case of world-sheets with defects. A structure that permits a
non-anomalous coupling of such sigma models to world-sheet gauge
fields of arbitrary topology is analysed, together with obstructions
to its existence, and the classification of its inequivalent
choices.
\end{abstract}


\maketitle

\tableofcontents

\section{Introduction}\label{sec:Intro}

\noindent In this paper, we study a non-anomalous gauging of rigid
symmetries in two-dimensional bosonic non-linear $\si$-models with
Wess--Zumino terms in the action functional. Classical fields of
such models take values in target manifolds equipped with
appropriate differential-geometric structures. We shall admit the
presence of defects at which the values of fields jump or change the
target space. Rigid symmetries of such $\si$-models are induced by
transformations of the target spaces that may be consistently lifted
to the differential-geometric structures required for the definition
of the action functional, e.g. a Riemannian metric, a gerbe, or a
gerbe bi-module. In this context, we present a detailed study of the
circumstances under which rigid symmetries may be promoted to local
ones by a gauging procedure that couples the original fields of the
model to world-sheet gauge fields. Our study leads to a formulation
of necessary conditions for the consistent gauging and to an
exhaustive cohomological classification of the resulting gauged
$\si$-models. In particular, we cover in a unified way the gauging
of continuous symmetries and, for discrete symmetries, the known
description of orbifold $\si$-models.

An analogous program was realized in \Rcite{Gawedzki:2010rn} in the
simpler case of the mono-phase $\si$-model. Completing the earlier
studies of the subject
\cite{Jack:1989ne,Hull:1990ms,Figueroa:1994ns,Figueroa:1994dj}, that
work also covered the case of gauge fields in nontrivial principal
bundles of the symmetry group $\,\txG\,$ and provided an exhaustive
study of global gauge anomalies that may obstruct invariance under
'large' gauge transformations non-homotopic to the identity. The
interested reader may consult \Rcite{Gawedzki:2010rn} for a
description of the physical background and for a more detailed
account of the history of the topic. The main point of the present
work is to extend the analysis of \Rcite{Gawedzki:2010rn} to the
case of multi-phase $\si$-models in which the different phases of
the two-dimensional theory are separated by one-dimensional domain
walls, called defect lines that may intersect at defect junctions.
Our analysis will lead to a natural completion of the analysis of
\Rcite{Gawedzki:2010rn} that accommodates the full-blown
2-categorial background of a general $\si$-model characterised in
Refs.\,\cite{Fuchs:2007fw,Runkel:2008gr,Suszek:2011hg}. In
application to boundary defects, it extends the results of
\Rcite{Figueroa:2005} by treating the case of gauge fields in
nontrivial principal bundles and by analyzing the global gauge
anomalies in boundary $\,\si$-models with Wess--Zumino terms.

The paper is organized as follows. In Section \ref{sec:prel}, we
briefly recall the mathematical structures involved in the
definition of Feynman amplitudes of the $\,\si$-models with defects
and in the description of their rigid symmetries. Section
\ref{sec:gauge-min} describes the coupling of the $\,\si$-model to
gauge fields in a trivial principal bundle of the symmetry group,
extending the approach of
\Rcite{Jack:1989ne,Hull:1990ms,Figueroa:2005} to the case of
$\,\si$-models with defects. The resulting coupling assures the
invariance of the gauged amplitudes under infinitesimal gauge
transformations, i.e. the absence of local gauge anomalies. In
Section \ref{sec:large}, we discuss the behaviour of the gauged
amplitudes under large gauge transformations, completing the
analysis of global gauge anomalies for topologically trivial gauge
fields performed in \Rcite{Gawedzki:2010rn} for $\,\si$-models
without defects. In Section \ref{sec:eg-backgrnd}, we illustrate the
preceding discussions on the example of the WZW model with defects.
Sections \ref{sec:groupoid} and \ref{sec:infinit-equiv} are an
interlude providing an interpretation of the conditions assuring the
absence of local gauge anomalies in terms of (Lie) algebroids
related to generalised geometry and in terms of mixed Deligne and
Lie-algebraic cohomology. Section \ref{sec:Gequiv} is devoted to an
extension of the notion of a $\,\txG$-equivariant structure on
gerbes, introduced in \Rcite{Gawedzki:2010rn}, to the rest of the
geometric structure needed to define Feynman amplitudes for a
$\,\si$-model with defects. In Section \ref{sec:coset}, the
definition of the full-fledged $\,\txG$-equivariant structure is
motivated by showing how it permits to push down the target-space
structure to the quotient of the target by the symmetry group
$\,\txG\,$ whenever the quotient space is smooth, and in Section
\ref{sec:nontriv}, it is shown how a $\,\txG$-equivariant structure
permits to couple the $\,\si$-model with defects to world-sheet
gauge fields in an arbitrary principal $\,\txG$-bundle. Finally,
Section \ref{sec:class-equiv-back} provides a detailed study of
obstructions to the existence and the classification of
$\,\txG$-equivariant structures, extending the results of
\Rcite{Gawedzki:2010rn} to the case with defects. Nine Appendices
contain a few more technical proofs, as well as some additional
material.

\bigskip
\noindent\textbf{Acknowledgements:} R.R.S. was partially funded by
the Collaborative Research Centre 676 ``Particles, Strings and the
Early Universe - the Structure of Matter and Space-Time'' at the
beginning of the work, and subsequently from the Polish Ministry of
Science and Higher Education grant No.\,N N201 372736. R.R.S. and
K.W. are grateful to the Laboratoire de Physique at ENS-Lyon for
hospitality generously extended to them at various stages of the
work on the project reported herein.
\bigskip

\section{Preliminaries}\label{sec:prel}

\subsection{The geometric structure of the $\si$-model}

\noindent Let us first recall some basic notions on which the
subsequent analysis of the $\,\si$-models with Wess--Zumino terms
and defects is founded. We start by identifying, after
\Rcite{Fuchs:2007fw,Runkel:2008gr}, the differential-geometric
structure on the target space underlying the definition of the
Feynman amplitudes of the field theory of interest.
\bedef\label{def:bckgrnd}
A \textbf{string background} is a triple $\,\Bgt=(\cM,\cB,\cJ)\,$
composed of the following  structure:
\begin{enumerate}
\item the \textbf{target} $\,\cM=(M,\txg,\cG)\,$ consisting of
a smooth manifold $\,M$ called the \textbf{target space}, a metric
$\,\txg$,\ and a bundle gerbe $\,\cG$;\medskip
\item the \textbf{$\cG$-bi-brane}
$\,\cB=(Q,\iota_1,\iota_2,\om,\Phi)\,$ consisting of a smooth
manifold $\,Q$ called the \textbf{$\cG$-bi-brane world-volume}, a
2-form $\,\om$ called the \textbf{$\cG$-bi-brane curvature}, two
smooth maps $\,\iota_1,\iota_2:Q \to M$,\ and a gerbe 1-isomorphism
\qq\nn
\Phi\ :\ \iota_1^*\cG\xrightarrow{\cong}\iota_{2}^*\cG\ox I_\om\,,
\qqq
written in terms of the trivial gerbe $\,I_\om\,$ with curving
$\,\om$;\medskip
\item
a collection $\mathcal{J} =
(\mathcal{J}_3,\mathcal{J}_{4},\mathcal{J}_5,...)$ in which
$\mathcal{J}_n$ is the \textbf{$n$-valent
$(\cG,\cB)$-inter-bi-brane}
\begin{equation*}
\mathcal{J}_n=\bigl(T_n;
\vep_n^{1,2},\vep_{n}^{2,3},...,\vep_n^{n-1,n},\vep_n^{n,1};\pi_n^{1,2},\pi_n^{2,3},...,\pi_n^{n-1,n},\pi_n^{n,1};\varphi_n
\bigr)\text{,}
\end{equation*}
consisting of
\begin{enumerate}

\item
a smooth manifold
$T_n$.

\item
maps $\,\vep^{*,*}_n:T_n\to\{-1,1\}\,$ called the
\textbf{orientation maps};
\item
smooth maps
$\,\pi^{*,*}_n:T_n\to Q\,$ subject to the constraints
\qq\label{eq:iopi-iopi}
\iota_2^{\vep_n^{k-1,k}}\circ\pi_n^{k-1,k}=\iota_1^{\vep_n^{k,k+1}}
\circ\pi_n^{k,k+1}=:\pi_n^k\,,\qquad k\in\mathbb{Z}/n\mathbb{Z}\,,
\qqq
where $\iota^{+1}_1 := \iota_1$, $\iota^{+1}_2 := \iota_2$,
$\iota_1^{-1}:=\iota_2$ and $\iota_2^{-1}:= \iota_1$;

\item
a  gerbe 2-isomorphism
\qq\label{diag:2iso}
\xy (50,0)*{\bullet}="G3"+(5,4)*{\cG_n^3\ox I_{\om_n^{1,2}+\om_n^{2,
3}}}; (25,-20)*{\bullet}="G2"+(-10,0)*{\cG_n^2\ox I_{\om_n^{1,2}}};
(75,-20)*{\ \vdots}="dots";
(35,-40)*{\bullet}="G1"+(0,-4)*{\cG_n^1};
(65,-40)*{\bullet}="G1add"+(10.5,-4)*{\cG_n^1\ox I_{\om_n^{1,2}+
\om_n^{2,3}+\ldots+\om_n^{n,1}}}; (50,-40)*{}="id";
\ar@{->}|*+{\Phi_n^{2,3}\ox\id} "G2";"G3"
\ar@{->}|*+{\Phi_n^{3,4}\ox\id} "G3";"dots"
\ar@{->}|*+{\Phi_n^{1,2}} "G1";"G2"
\ar@{->}|*+{\Phi_n^{n,n+1}\ox\id} "dots";"G1add" \ar@{=}|*+{\id}
"G1"+(2,0);"G1add"+(-2,0) \ar@{=>}|*+{\varphi_n}
"G3"+(0,-3);"id"+(0,+3)
\endxy
\qqq
written in terms of 1-isomorphisms $\,\Phi_n^{k,k+1}=(\pi_n^{k,k+1})^{*}\Phi^{\vep_n^{k,k+1}}\,$ between gerbes $\,\cG_n^k=
(\pi_n^{k})^{*}\cG$,\ and the trivial gerbes with curvings
$\,\om_n^{k,k+1}=\vep_n^{k,k+1}\,(\pi_n^{k,k+1})^{*}\om$.
\end{enumerate}
\end{enumerate}
The manifold $\,T=\bigsqcup_{n\geq 3}\,T_n$, is called the \textbf{$(\cG,
\cB)$-inter-bi-brane world-volume},
and the manifold
\qq\nn
\xcF:=M \sqcup Q\sqcup T
\qqq
will be called the \textbf{target space of background $\,\Bgt$}.\
Also, in what follows, we shall often write the inter-bi-brane data
in the form
$\,\cJ\equiv(T_n,(\vep_n^{k,k+1},\pi_n^{k,k+1});\varphi_n)\,$ for
brevity. \exdef \noindent We give an important example of a complete
string background in Section \ref{sec:eg-backgrnd}. \brem The above
definition invokes the structure of the 2-category
$\,\bgrb^\nabla(\xcF)\,$ of bundle gerbes with connection over the
target space. The abstract 2-category was first introduced in
\Rcite{Stevenson:2000wj} and further developed in
\Rcite{Waldorf:2007mm}. Its 1-morphisms can be composed in the usual
manner,
\qq\nn
\Phi_{2,3}\circ\Phi_{1,2}\ =\ \cG_1\xrightarrow{\Phi_{1,
2}}\cG_2\xrightarrow{\Phi_{2,3}}\cG_3\,,
\qqq
whereas 2-isomorphisms can be composed in two different ways, namely
horizontally
\qq\nn
\varphi_2\circ\varphi_1\ =\ \alxydim{}{\cG_1
\ar@/^1.6pc/[rrr]^{\Phi_{2,3}\circ\Phi_{1,2}}="5"
\ar@/_1.6pc/[rrr]_{\Phi_{2,3}'\circ\Phi_{1,2}'}="6"
\ar@{=>}"5";"6"|{\varphi_2\circ\varphi_1} &&& \cG_3}\;\;\equiv\;\;
\alxydim{@C=1cm}{\cG_1 \ar@/^1.6pc/[rr]^{\Phi_{1,2}}="1"
\ar@/_1.6pc/[rr]_{\Phi_{1,2}'}="2" \ar@{=>}"1";"2"|{\varphi_1} &&
\cG_2 \ar@/^1.6pc/[rr]^{\Phi_{2,3}}="3"
\ar@/_1.6pc/[rr]_{\Phi_{2,3}'}="4" \ar@{=>}"3";"4"|{\varphi_2} &&
\cG_3}\,,
\qqq
and vertically
\qq\nn
\varphi'\bullet\varphi\ =\ \alxydim{@C=2cm}{\cG_1
\ar@/^2.5pc/[r]^{\Phi_{1,2}}="1" \ar[r]^<<<<<<<{\Phi_{1,2}'}
\ar@/_2.5pc/[r]_{\Phi_{1,2}''}="3" \ar@{}[r]^{}="2" & \cG_2
\ar@{=>}"1";"2"|{\varphi} \ar@{=>}"2";"3"|{\varphi'}}\,.
\qqq
The two modes of composition are related by a coherence condition
\qq\nn
(\varphi_2'\bullet\varphi_2)\circ(\varphi_1'\bullet\varphi_1)\ =\
\alxydim{@C=2cm}{\cG_1 \ar@/^2.5pc/[r]^{\Phi_{1,2}}="1"
\ar[r]^<<<<<<<{\Phi_{1,2}'} \ar@/_2.5pc/[r]_{\Phi_{1,2}''}="3"
\ar@{}[r]^{}="2" & \cG_2 \ar@{=>}"1";"2"|{\varphi_1}
\ar@{=>}"2";"3"|{\varphi_1'} \ar@/^2.5pc/[r]^{\Phi_{2,3}}="1"
\ar[r]^<<<<<<<{\Phi_{2,3}'} \ar@/_2.5pc/[r]_{\Phi_{2,3}''}="3"
\ar@{}[r]^{}="2" & \cG_3 \ar@{=>}"1";"2"|{\varphi_2}
\ar@{=>}"2";"3"|{\varphi_2'}}\ =\
(\varphi_2'\circ\varphi_1')\bullet(\varphi_2\circ\varphi_1)\,.
\qqq
Furthermore, gerbes, 1-morphisms and 2-morphisms can be tensored and
pulled back along smooth maps, both in a way consistent with the
various compositions. \erem

\begin{figure}[h]
\begin{center}
\vskip 0.2cm
\leavevmode
\includegraphics[width=9cm,height=7cm]{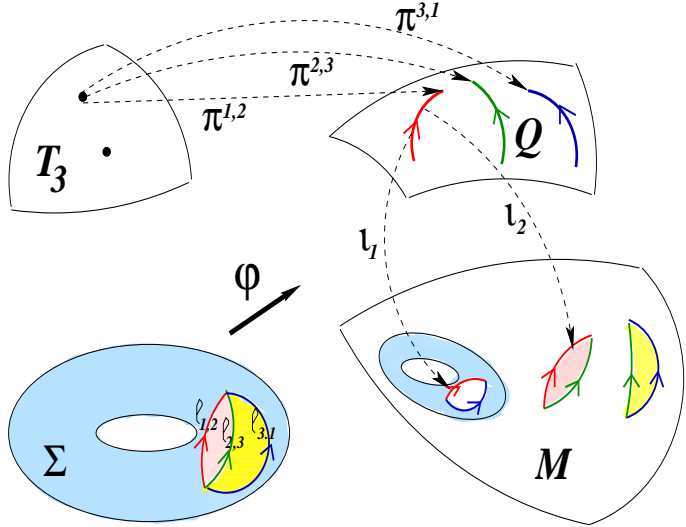}\\
\vskip 0.1cm \caption{A Riemann surface with a defect quiver, and a
network-field configuration.} \label{fig:1}
\end{center}
\end{figure}

\brem The bi-brane curvature satisfies the relations
\qq
&\sfd\om=-\D_Q\txH\,,& \label{eq:triv-restr}\\\cr
&\D_{T_n}\om=0\,,&\label{eq:triv-restr2}
\qqq
where we have introduced the shorthand notation
\qq\nn
\D_Q=\iota_2^*-\iota_1^*\,,\qquad\qquad\D_{T_n}=\sum_{k=1}^n\,
\vep_n^{k,k+1}\,\pi_n^{k,k+1\,*}
\qqq
for the linear combinations of pullbacks that are going to appear
frequently in the present paper. The two pullback operators satisfy
the identity
\qq\label{eq:DelTnoDelQ}
\D_{T_n}\circ\D_Q=0\,.
\qqq
\erem \noindent The target space of the string background is the
codomain of the dynamical fields of the theory that we want to
study, hence its name. This statement is rendered precise by
\bedef\label{def:net-field}
Let $\,(\Si,\g)\,$ be a closed oriented Riemann surface, termed the
\textbf{world-sheet}, and let $\,\G\,$ be an oriented graph
(possibly empty) embedded in $\,\Sigma$. We shall call it the
\textbf{defect quiver}. We shall assume that, if not empty, $\,\G\,$
is composed of oriented \textbf{defect lines}, closed or ending at
points $\,\jmath\,$\ called \textbf{defect junctions}  of valence
$\,\geq3$, \,see Fig.\,\ref{fig:1}. \,The connected components
$\,\wp\,$ of $\,\Sigma\setminus\G\,$ will be called patches. The set
of patches will be denoted by $\,\Pgt_\Si$,\ the set of defect lines
by $\,\Egt_\G$,\ and the set of defect junctions by $\,\Vgt_\G$.
\,Furthermore, let $\,\Bgt=(\cM,\cB,\cJ)\,$ be a string background
with target space $\,\xcF=M\sqcup Q\sqcup T$,\ as introduced in
Definition \ref{def:bckgrnd}. A \textbf{network-field configuration
$\,(\varphi\,\vert\,\G)\,$ in string background $\,\Bgt\,$ on
world-sheet $\,(\Si,\g)\,$ with defect quiver $\,\G\,$} is a pair
composed of the defect quiver $\,\G\,$ embedded in the world-sheet
$\,\Si$,\ together with a map
\qq\nn
\varphi\ :\ \Si\to\xcF
\qqq
such that
\bit
\item $\varphi\,$ restricts to a smooth map $\,\Si
\setminus\G\to M$,\ a smooth map $\,\G\setminus\Vgt_\G
\to Q$,\ and it sends $\,\Vgt_\G\,$ into $\,T\,$ in such a manner that a
defect junction $\,\jmath\,$ of valence $\,n_\jmath\,$ is mapped to
$\,T_{n_\jmath}$;
\item for each (oriented) defect line $\,\ell$, $\,\varphi\,$ extends
from the patch to the left (resp.\ right) of it (in the direction
defined by the orientation of the line) to a smooth map
$\,\varphi_{|1}({\rm resp.\ }
\varphi_{|2}):\ell\setminus\Vgt_\G\rightarrow M\,$ with the property
\qq\label{eq:phial}
\varphi_{|\a}=\iota_\a\circ \varphi|_{\ell\setminus\Vgt_\G}\,;
\qqq
\item for $\,\jmath\in\Vgt_\G\,$ an $n_\jmath$-valent defect
junction and $\,\ell_{k,k+1},\ k=1,...,n$,\ the defect lines meeting
at $\,\jmath\,$ and ordered in a cyclic way,\,\ the maps
$\,\varphi\vert_{\ell_{k,k+1}\setminus\Vgt_\G}\,$ into $\,Q\,$ admit
smooth extensions $\,\varphi_{k,k+1}\,$ to $\,\jmath\,$ such that
\qq\label{eq:phipi}
\varphi_{k,k+1}(\jmath)=\pi_{n_\jmath}^{k,k+1}\circ\varphi(\jmath)\,;
\qqq
\item for $\,(\jmath,\ell_{k,k+1})\,$ as above, the orientation map
takes the value
$\,\vep_{n_\jmath}^{k,k+1}\bigl(\varphi(\jmath)\bigr)=+ 1\,$ if
$\,\ell_{k,k+1}\,$ is oriented towards $\,\jmath\,$ and
$\,\vep_{n_\jmath}^{k,k+1}\bigl(\varphi(\jmath)\bigr)=- 1\,$ if the
line is oriented away from  $\,\jmath\,$ for vertices with defect line
ordered counter-clockwise and the opposite values for vertices with
defect lines ordered clockwise.
\eit

\vskip -0.2cm

\exdef \brem\label{rem:bdry-defcts} It is to be stressed that the
definition of a bi-brane encompasses the more familiar concept of a
(D-)brane, assigned to the boundary of the world-sheet, cf.\
\Rcite{Runkel:2008gr}. Indeed, upon taking the target space in the
special form $\,M':=M\sqcup\{ \bullet\}\,$ for an arbitrary
singleton $\,\{\bullet\}$,\ alongside any manifold $\,Q\,$ equipped
with a pair of smooth maps $\,\iota_1\ :\ Q\to M\subset M'\,$ and
$\,\iota_2\ :\ Q\to\{\bullet\}\subset M'\,$ (the constant map) as a
bi-brane world-volume, we readily recover the standard notion in a
new guise: The patch on one side of the defect line carrying the
data of the (boundary) bi-brane is mapped to a single point, thereby
mimicking an interface between a nontrivial phase of the $\si$-model
and the trivial (void) $\si$-model. \erem \noindent We are now fully
equipped to introduce the probability amplitudes of a
two-dimensional non-linear $\,\si$-model whose symmetries will be
studied in the remainder of the paper.
\bedef\label{def:sigma}
Let $\,(\varphi\,\vert\,\G)\,$ be a network-field configuration in
string background $\,\Bgt=(\cM,\cB,\cJ)\,$ on world-sheet
$\,(\Si,\g)\,$ with defect quiver $\,\G$,\ as introduced in
Definition \ref{def:net-field}. The euclidean Feynman amplitude of
$\,(\varphi\,\vert\,\G)\,$ is
\qq\label{eq:sigma}
\xcA[(\varphi\,\vert\,\G);\gamma]\ :=\
\exp\Big[-\tfrac{1}{2}\,\int_\Si\,\txg(\sfd\varphi\overset{\wedge}{,}
\star_\g\sfd \varphi)\Big]\ \Hol_{\cG,\Phi,
(\varphi_n)}(\varphi\,\vert\,\G)\,,
\qqq
in which
\bit
\item $\sfd\varphi(\si)=\p_u\varphi^\mu(\si)\,\sfd\si^u
\ox\p_\mu|_{\varphi(\si)}$\,\ in local coordinates
$\,\{\si^u\}\,$ on $\,\Si\,$ and
$\{\varphi^\mu\}\,$ on $\,M$,\ and the
target-space metric is assumed to act on the second factor of the
tensor product;
\item $\star_\g\,$ is the world-sheet Hodge operator determined
by the world-sheet metric $\,\g$;
\item the so-called `topological', or Wess--Zumino contribution
to the Feynman amplitude
\qq\nn
\xcA_{\rm WZ}[(\varphi\,\vert\,\G)]:=\Hol_{\cG,\Phi,
(\varphi_n)}(\varphi\,\vert\,\G)
\qqq
is given by the decorated-surface holonomy $\,\Hol_{\cG,\Phi,
(\varphi_n)}(\varphi\,\vert\,\G)\,$ for the network-field
configuration $\,(\varphi\,\vert\,\G)$, \,described in detail in
\Rcite{Runkel:2008gr} and generalising the holonomy  of
\Rcite{Fuchs:2007fw} to the case with defect networks.
\eit
\exdef

\brem\label{rem:align} The quantum euclidean non-linear
$\,\si$-model with defects is obtained by functional integration of
probability amplitudes \eqref{eq:sigma} over the network-field
configurations $\,(\varphi\,|\, \Gamma)$.\erem

In order to understand the fundamental property of the
decorated-surface holonomy which is its invariance under
transformations of the background realised by 1- and 2-isomorphisms
and accompanied by compensating transformations of the bi-brane
1-isomorphism and of the inter-bi-brane 2-isomorphism, respectively,
we shall invoke several basic properties of the 2-category
$\,\bgrb^\nabla(\xcM)\,$ after \Rcite{Waldorf:2007mm}.\ Thus, first
of all, recall that each gerbe $\,\cG\,$ has a dual, to be denoted
as $\,\cG^\vee$,\ with the opposite curvature. Furthermore, every
1-isomorphism $\,\Psi: \cG_1\xrightarrow{\cong}\cG_2\,$ between
gerbes $\,\cG_\a,\ \a=1,2,$ \,admits an inverse,
$\,\Psi^{-1}:\cG_2\xrightarrow{\cong} \cG_1$,\ defined in such a
manner that
\qq\nn
(\Psi_1\circ\Psi_2)^{-1}=\Psi_2^{-1}\circ\Psi_1^{-1}\,,
\qqq
and there exist 2-isomorphisms
\qq\label{eq:birth-and-death}
d_\Psi\ :\ \Psi^{-1}\circ\Psi\xLongrightarrow{\cong}\id_{\cG_1}\,,
\qquad\qquad b_\Psi\ :\ \id_{\cG_2}\xLongrightarrow{\cong}\Psi\circ
\Psi^{-1}\,,
\qqq
called the `death' and `birth' 2-isomorphism, respectively.
These satisfy the identities
\qq\nn
\la_{\Psi^{-1}}=\rho_{\Psi^{-1}}\bullet\bigl(d_\Psi\circ
\id_{\Psi^{-1}}\bigr)\bullet\bigl(\id_\Psi\circ b_\Psi\bigr)\,,
\qquad\qquad\rho_\Psi=\la_\Psi\bullet\bigl(\id_\Psi\circ d_\Psi
\bigr)\bullet\bigl(b_\Psi\circ\id_\Psi\bigr)\,,
\qqq
written in terms of the natural 2-isomorphisms
\qq\nn
\la_\Psi\ :\ \Psi\circ\id_{\cG_1}\xLongrightarrow{\cong}\Psi\,,
\qquad\qquad\rho_\Psi\ :\ \id_{\cG_2}\circ\Psi
\xLongrightarrow{\cong}\Psi\,.
\qqq
They can be used to explicitly construct, for any 2-isomorphism
$\,\psi:\Psi_1\xLongrightarrow{\cong}\Psi_2\,$ between
1-isomorphisms $\,\Psi_\a,\ \a=1,2$,\ \,the associated 2-isomorphism
$\,\psi^\sharp:\Psi_2^{-1}\xLongrightarrow{\cong} \Psi_1^{-1}$.\
Furthermore, there exists a canonical 2-isomorphism
$\,i_\Psi:\bigl(\Psi^{-1}\bigr)^{-1}\xLongrightarrow{\cong} \Psi$.\
So far, we have pedantically kept the labels on the identity
morphisms. For the sake of brevity, we shall drop these labels from
now onwards, leaving it to the reader to reconstruct them from the
context. We complete the summary of useful facts concerning the
2-category $\,\bgrb^\nabla(\xcM)\,$ by giving the following
\berop\cite{Gawedzki:1987ak,Carey:2002,Waldorf:2007phd}
\label{prop:torsors} Let $\,\xcM\,$ be a smooth manifold.
\begin{enumerate}
\item
Given two
gerbes $\,\cG_\a,\ \a=1,2$,\ over $\,\xcM\,$ of the same curvature,
there exists a flat gerbe $\,\cD\,$ and a canonical 1-isomorphism
\qq\nn
\cG_2\xrightarrow{\cong}\cD\ox\cG_1\,.
\qqq
The two gerbes are 1-isomorphic iff $\,\cD\cong I_0$.\

\item
Given two 1-isomorphisms $\,\Psi_\a:\cG\xrightarrow{\cong}\ceH,\
\a=1,2$,\ between two gerbes $\,\cG\,$ and $\,\ceH\,$ over
$\,\xcM$,\ there exists a flat line bundle $\,D\to\xcM\,$ and a
canonical 2-isomorphism
\qq\nn
\Psi_2\xLongrightarrow{\cong}D\ox\Psi_1\,,
\qqq
where on the right-hand side $\,D\,$ is viewed as a 1-isomorphism of
the trivial gerbe $\,I_0\,$ using the canonical equivalence of
categories $\,{\rm
Bun}:\gt{End}(I_0)\xrightarrow{\equiv}\gt{Bun}_0^\nabla(\xcM)\,$
(for $\,\gt{End}(I_0)\,$ the $\mor$-category of endomorphisms of the
trivial gerbe $\,I_0\,$ and $\,\gt{Bun}_0^\nabla(\xcM)\,$ the
category of vector bundles with a traceless curvature). The two
1-isomorphisms are 2-isomorphic iff $\,D\cong J_0\,$ for $\,J_0\,$
the trivial line bundle with the trivial connection.

\item
Given two 2-isomorphisms $\,\psi_\a:
\Psi\xLongrightarrow{\cong}\Xi,\ \a=1,2$,\ of two 1-isomorphisms
$\,\Psi\,$ and $\,\Xi\,$ between a (common) pair of gerbes over
$\,\xcM$,\ there exists a locally constant map $\,d\in C^\infty
\left(\pi_0(\xcM),\uj\right)\,$ (for $\,\pi_0(\xcM)\,$ the set of
connected components of $\,\xcM$) such that
\qq\nn
\psi_2=d\ox\psi_1\,.
\qqq
\end{enumerate}
\eerop

We are now ready to formulate
\berop
\cite[Sec.\,2.7]{Runkel:2008gr}\label{prop:hol-inv} Let
$\,\Bgt_\a=(\cM_\a,\cB_\a,\cJ_\a),\ \a=1,2$,\ be a pair of string
backgrounds as in Definition \ref{def:bckgrnd}, differing in the
choice of the structure $\,(\cG_\a,\Phi_\a, \varphi_{\a\,n}\ \vert\
n\geq 3)\,$ exclusively. \,Assume the existence of a 1-isomorphism
\qq\nn
\Psi\ :\ \cG_1\xrightarrow{\cong}\cG_2\,,
\qqq
and of a 2-isomorphism
\qq\nn
\psi\ :\ \Phi_1\xLongrightarrow{\cong}\bigl(\iota_2^*\Psi^{-1}\ox
\id\bigr)\circ\Phi_2\circ\iota_1^*\Psi
\qqq
satisfying the equality
\qq\nn
\varphi_{1\,n}&=&\bigl(d_\Psi\bigr)_n^1\bullet\bigl(\id\circ\bigl(
i_\Psi^{\vep_n^{1,2}}\bigr)_n^1\bigr)\cr\cr
&&\bullet(\id\circ\la_{\Psi^1_n})\bullet\bigl(\id\circ\varphi_{2\,
n}\circ\id\bigr)\bullet\bigl(\id\circ\la_{\Phi_{2\,n}^{n,1}\ox\id}
\circ\la_{\Phi_{2\,n}^{n-1,n}\ox\id}\circ\cdots\circ\la_{\Phi_{2\,
n}^{1,2}\ox\id}\circ\id\bigr)\cr\cr
&&\bullet\bigl(\id\circ\bigl(b_\Psi^{-1}\bigr)_n^n\circ\id\circ
\bigl(b_\Psi^{-1}\bigr)_n^{n-1}\circ\id\circ\cdots\circ\id\circ
\bigl(b_\Psi^{-1}\bigr)_n^2\circ\id\bigr)\cr\cr
&&\bullet\bigl(\id\circ\bigl(i_\Psi^{\vep_n^{n,1}}\bigr)_n^n\circ
\id\circ\bigl(i_\Psi^{\vep_n^{n-1,n}}\bigr)_n^{n-1}\circ\id\circ
\cdots\circ\id\circ\bigl(i_\Psi^{\vep_n^{2,3}}\bigr)_n^2\circ\id
\bigr)\cr\cr
&&\bullet\bigl(\bigl(\psi_n^{n,1}\ox\id\bigr)\circ\bigl(\psi_n^{n-1
,n}\ox\id\bigr)\circ\cdots\circ\psi_n^{1,2}\bigr),
\qqq
the latter being written in terms of the 1-isomorphism
\qq\nn
\Psi^1_n:=\pi_n^{1\,*}\Psi\,,
\qqq
and of the 2-isomorphisms
\qq\nn
\bigl(b_\Psi^{-1}\bigr)_n^k:=\pi_n^{k\,*}b_\Psi^{-1}\,,\qquad
\qquad\bigl(d_\Psi\bigr)_n^1:=\pi_n^{1\,*}d_\Psi\,,
\qqq
as well as
\qq\nn
\bigl(i_\Psi^{\vep_n^{k,k+1}}\bigr)_n^k:=\pi_n^{k\,*}
i_\Psi^{\vep_n^{k,k+1}}\,,\qquad\qquad i_\Psi^{\vep_n^{k,k+1}}:=
\left\{ \barr{lcl} \id_\Psi\ & \tx{if} & \vep_n^{k,k+1}=1\,, \\
i_\Psi\ & \tx{otherwise} & \earr \right.
\qqq
and
\qq\nn
\psi_n^{k,k+1}:=\pi_n^{k,k+1\,*}\psi^{\vep_n^{k,k+1}}\,,\qquad
\qquad\psi^{\vep_n^{k,k+1}}:=\left\{ \barr{lcl} \psi\ & \tx{if} &
\vep_n^{k,k+1}=1\,, \\ \psi^{\sharp\,-1}\ & \tx{otherwise.} & \earr
\right.
\qqq
Then,
\qq\nn
\Hol_{\cG_2,\Phi_2,(\varphi_{2\,n})}(\varphi\,\vert\,\G)=\Hol_{\cG_1,
\Phi_1,(\varphi_{1\,n})}(\varphi\,\vert\,\G)\,.
\qqq
\eerop
\noindent The last proposition will play a central r\^ole in our
discussion of the global gauge anomalies in the background of
topologically trivial gauge fields.\bigskip

\subsection{$\txG$-actions via simplicial $\txG$-spaces}
\label{sub:simplicial}

\noindent Equivariance properties of the differential-geo\-met\-ric
structures on $\,\xcF\,$ with respect to the group action may be
conveniently described using categorial concepts such as the action
groupoid $\,\txG\lx\xcF\,$ and the nerve of the small category
$\,\txG\lx\xcF$.\ We begin by recalling a few basic concepts.
\bedef\label{def:Gspace}
Let $\,\xcM\,$ be a smooth manifold equipped with a smooth (left)
action of a Lie group $\,\txG\,$
\qq\label{eq:Gact-M}
\xcMup\ell\ :\ G\x\xcM\to\xcM\ :\ (g,m)\mapsto g.m\equiv\xcMup\ell_g
(m)\,.
\qqq
Such a manifold will be referred to as a \textbf{$\txG$-space} from
now onwards. Let $\ggt$ be the Lie algebra of $\,\txG$. To every
element $\,X\in\ggt$,\ we associate a vector field $\,\ovl X\,$ on
$\,\xcM\,$ acting on smooth functions according to the formula
\qq\label{eq:X-gen-fun}
(\ovl X f)(m)=\tfrac{\sfd\ }{\sfd t}\big\vert_{t=0}f\bigl(\ee^{-t\,
X}.m\bigr)\,.
\qqq

\exdef \noindent In order to keep subsequent equations more
transparent, we set up
\becon\label{con:fund-vec-f}
Let $\,\xcM\,$ be a $\txG$-space and let $\,\{\xcMup\xcK_a\}_{a=1,2,
\ldots,\dim\,\ggt}\,$ be the fundamental vector fields on $\,\xcM\,$
associated, through \Reqref{eq:X-gen-fun}, to the generators
\qq\label{eq:defta}
t_a,\ a=1,2,\ldots,\dim\,\ggt\,,\qquad\qquad [t_a,t_b]=f_{abc}\,t_c
\qqq
of the Lie algebra $\,\ggt$,\ with the corresponding Lie bracket
\qq\label{eq:Kill-alg}
[\xcMup\xcK_a,\xcMup\xcK_b]=f_{abc}\,\xcMup\xcK_c\,.
\qqq
Contraction with $\,\xcMup\xcK_a\,$ will be denoted as
\qq\nn
\ic_a:=\ic_{\xcMup\xcK_a}\,,
\qqq
and the corresponding Lie derivative as
\qq\nn
\pLie{a}:=\pLie{\xcMup\xcK_a}\,.
\qqq
Furthermore, given a $\ggt$-valued 1-form $\,\eta=\eta^a\ox t_a\in
\Om^1(\xcM)\ox \ggt\,$ on $\,\xcM$,\ and a $p$-form $\,\xi\in\Om^p(
\xcM)$,\ we write
\qq\nn
\ovl\eta:=\eta^a\ox\xcMup\xcK_a
\qqq
and define
\qq\nn
\ovl\eta\cdot\xi:=\eta^a\wedge\bigl(\ic_a\xi\bigr)\,.
\qqq
This formula gives an obvious meaning to symbols such as
$\,\ee^{\ovl\eta}\cdot\xi$. \econ \noindent The following definition
fixes the standard terminology used throughout this paper.
\bedef\label{def:Gbas-form}
A smooth $p$-form $\,\eta\in\Om^p(\xcM)\,$ will be called
\textbf{$\ggt$-horizontal} iff it is annihilated by contraction with
all the vector fields $\,\xcMup \xcK_a$,
\qq\nn
\ic_a\eta=0\,.
\qqq
It will be termed \textbf{$\ggt$-invariant} resp.\
\textbf{$\txG$-invariant} iff it satisfies the conditions
\qq\nn
\pLie{a}\eta=0
\qqq
resp.
\qq\nn
\xcMup\ell_g^*\eta=\eta\,,\qquad g\in\txG\,.
\qqq
A $p$-form which is both $\ggt$-horizontal and $\ggt$-invariant
resp.\ $\txG$-invariant will be called \textbf{$\ggt$-basic} resp.\
\textbf{$\txG$-basic}.
\smallskip

Next, let $\,\z=\z_a\ox\tau^a\in\Om^p(\xcM)\ox\ggt^*\,$ be a smooth
$\ggt^*$-valued $p$-form on $\,\xcM$,\ written in terms of the basis
$\,\{\tau^a\}\,$ of $\,\ggt^*\,$ dual to $\,\{t_a\}$.\ For
$\,X\in\ggt$,\ we shall write
\qq\nn
\z(X):=\tau^a(X)\,\z_a\,.
\qqq
A $p$-form $\eta\,$ will be called \textbf{$\ggt$-equivariant}
resp.\ \textbf{$\txG$-equivariant} iff it satisfies the condition
\qq\nn
\pLie{a}\z_b=f_{abc}\,\z_c
\qqq
resp.
\qq\nn
\left(\xcMup\ell_g^*\z\right)(X)=\z\left(\Ad_{g^{-1}}X\right)
\qqq
for all $\,X\in\ggt$. \exdef \noindent We may now introduce
\bedef
Let $\,\xcM\,$ be a $\txG$-space, with the (left) action of the
group $\,\txG\,$ given by \Reqref{eq:Gact-M}. The \textbf{action
groupoid}
\qq\nn
\txG\lx\xcM\ :\ \alxydim{}{\txG\x\xcM \ar@<.5ex>[r]^{\quad s}
\ar@<-.5ex>[r]_{\quad t} & \xcM}
\qqq
is a (small) category with the object and morphism sets
\qq\nn
\obj\,(\txG\lx\xcM)=\xcM\,,\qquad\qquad\morf\,(\txG\lx\xcM)=\txG\x
\xcM\,,
\qqq
with the source and target maps
\qq\nn
s(g,x)=x\,,\qquad\qquad t(g,x)=g.x\,,
\qqq
with the identity morphisms
\qq\nn
\id_x=(e,x)
\qqq
($e\,$ is the group unit), and, finally, with the composition of
morphisms
\qq\nn
(g,h.x)\circ(h,x)=(g\cdot h,x)\,.
\qqq
\exdef \noindent Both $\,\txG\x\xcM\,$ and $\,\xcM\,$ carry the
structure of a $\txG$-space, the two being related by
\berop
Let $\,\txG\lx\xcM\,$ be the action groupoid over a $\txG$-space
$\,\xcM\,$ equipped with a (left) $\txG$-action \eqref{eq:Gact-M},
and let $\,\xcMup \xcK_a,\ a=1,2,\ldots,\dim\,\ggt\,$ be the
corresponding fundamental vector fields on $\,\xcM$,\ introduced in
Definition \ref{eq:Gact-M}. Denote by $\,L_a\,$ (resp.\ $\,R_a$) the
left-invariant (resp.\ right-invariant) vector fields on $\,\txG\,$
dual to the left-invariant (resp.\ right-invariant) Maurer-Cartan
1-forms $\,\theta_L^a$,\ with $\,\theta_L(g)=g^{-1}\,\sfd
g=\theta_L^a(g)\ox t_a\,$ (resp.\ $\,\theta_R^a$,\ with
$\,\theta_R(g) =\sfd g\,g^{-1}=\theta_R^a(g)\ox t_a$). The source
and target maps of $\,\txG\lx\xcM\,$ intertwine the left action
\eqref{eq:Gact-M} of $\,\txG\,$ on $\,\xcM\,$ with the combined
adjoint and left action of $\,\txG\,$ on $\,\txG\x\xcM$,
\qq\nn
s\bigl(\Ad_h(g),h.x\bigr)=h.s(g,x)\,,\qquad\qquad t\bigl(\Ad_h(g),h.
x\bigr)=h.t(g,x)\,.
\qqq
In particular, the vector field $\,\xcMup\xcK^{(0)}_a\equiv\xcMup
\xcK_a\,$ on $\,M\,$ can be regarded as the pushforward of the
vector field
\qq\label{eq:pushf-Kill}
\xcMup \xcK_a^{(1)}(g,x)=(L_a-R_a)(g)+\xcMup\xcK_a(x)\,.
\qqq
\eerop
\beroof
Obvious, through inspection. \eroof\bigskip

In the next
step, we invoke some elementary notions of simplicial analysis that
will prove useful in what follows. The interested reader is urged to
consult, e.g., \Rcite{Weibel:1994} for further details. The point of
departure is
\bedef
The \textbf{category $\,\D\,$} has subsets
$\,[m]:=\{0,1,\ldots,m\}\,$ of the set of non-negative integers as
its objects, and order-preserving maps
\qq\nn
\theta\ :\ [l]\to[m]\ :\ i\mapsto\theta(i)\,,\qquad\qquad i<j\
\Rightarrow\ \theta(i)<\theta(j)
\qqq
as its morphisms. We shall use the notation $\,\mor_\D([l],[m])=:\D(
l,m)$. \exdef \noindent We have
\berop
The $\mor$-sets of the category $\,\D\,$ have the following
properties:
\bit
\item[(i)] $\,l>m\ \Rightarrow\ \D(l,m)=\emptyset$;
\item[(ii)] $\,\D(m,m)=\{\id_{[m]}\}$;
\item[(iii)] $\,\D(m-1,m)=\{\ \theta^{(m)}_k\ \vert \ k=0,1,\ldots,m
\ \}$,\ with the \textbf{universal coface maps}\linebreak
$\,\theta^{(m)}_k(i) =\left\{ \barr{lcl} i\ & \tx{if} & i<k \\ i+1\
& \tx{otherwise} & \earr \right.$,\ satisfying the
\textbf{cosimplicial identities}
\qq\label{eq:cosimp-id}
\theta^{(m+1)}_k\circ\theta^{(m)}_l=\theta^{(m+1)}_l\circ\theta^{(m
)}_{k-1}
\qqq
for $\,0\leq l<k\,$ and $\,k=0,1,\ldots,m+1$.
\eit
\eerop
\beroof Obvious, through direct inspection. \eroof
\bigskip

\noindent Next, we introduce
\bedef
An \textbf{incomplete simplicial set} is a family
$\,\{S^m\}_{m=0,1,\ldots} \,$ of sets, indexed by non-negative
integers, together with a collection of maps
$\,\si^{(m+1)}_k:S^{m+1}\to S^m,\ k=0,1,\ldots,m+ 1\,$ between them,
termed the \textbf{face maps}, that satisfy the \textbf{simplicial
identities}
\qq\label{eq:simp-id}
\si^{(m)}_l\circ\si^{(m+1)}_k=\si^{(m)}_{k-1}\circ\si^{(m+1)}_l
\qqq
for all $\,0\leq l<k\,$ and $\,k=0,1,\ldots,m+1$.\exdef Upon
employing the straightforward one-to-one correspondence between
incomplete simplicial manifolds, i.e. incomplete simplicial sets
internal to the subcategory $\,\Man\,$ of smooth manifolds, and
contravariant functors $\,M:\D\to\Man\,$ (cf.\
\cite{Gawedzki:2009jj}), we arrive at
\bedef\label{def:nerves}
The \textbf{nerve}
\qq\label{eq:nerve-cover}\qquad\qquad
\alxydim{}{ \cdots \ar@<.75ex>[r] \ar@<.25ex>[r] \ar@<-.25ex>[r]
\ar@<-.75ex>[r] & \txG^2\times\xcM \ar@<.5ex>[r] \ar@<0.ex>[r]
\ar@<-.5ex>[r] & \txG\times\xcM \ar@<.5ex>[r] \ar@<-.5ex>[r] & \xcM}
\qqq
\textbf{of the action groupoid} $\,\txG\lx\xcM\,$ is a contravariant
functor $\,\txG\xcM:\D\to\GMan\,$ from $\,\D\,$ to the category
$\,\GMan\,$ of $\txG$-spaces, with the object component
\qq\nn
\txG\xcM\bigl([0]\bigr):=\xcM\,,\qquad\qquad
\txG\xcM\bigl([m]\bigr):= \txG^m\x\xcM\,,\qquad m\geq 1\,,
\qqq
and the morphism component determined by its restriction to the
universal coface maps,
\qq\nn
\txG\xcM\bigl(\theta^{(m)}_k\bigr):=\xcMup d_k^{(m)}\,,
\qqq
with the resultant \textbf{face maps of the nerve}, $\,\xcMup d_k^{(
m)}:\txG^m\times\xcM\to\txG^{m-1}\times\xcM\,$ (represented by the
arrows in the diagram), given explicitly in the form
\qq\nn
\xcMup d_0^{(1)}(g,x)&=&x\,,\cr\cr \xcMup d_1^{(1)}(g,x)&=&g.x\,,\cr
\cr\cr \xcMup d_0^{(m)}(g_{m},g_{m-1},\ldots,g_1,x)&=&(g_{m-1},g_{m-2},
\ldots,g_1,x)\,,\cr\cr \xcMup d_{m}^{(m)}(g_{m},g_{m-1},\ldots,g_1
,x)&=&(g_{m},g_{m-1},\ldots,g_2 ,g_1.x)\,,\cr\cr \xcMup
d_i^{(m)}(g_{m},g_{m-1},\ldots,g_1,x)&=&(g_{m},g_{m-1},\ldots,g_{m+2-i}
,g_{m+1-i}\cdot g_{m-i},g_{m-1-i},\ldots,g_1,x)\,.
\qqq
The nerve $\,\txG\xcM\,$ is additionally endowed with the
\textbf{degeneracy maps} $\,\xcMup s_k^{(m)}:
\txG^m\times\xcM\to\txG^{m+1}\times\xcM\,$ defined as
\qq\nn
\xcMup s_0^{(0)}(x)&=&(e,x)\,,\cr\cr \xcMup s_i^{(m)}(g_{m},g_{m-1},
\ldots,g_1,x)&=&(g_{m},g_{m-1},\ldots,g_{m+1-i},e,g_{m-i},g_{m-1-i},
\ldots,g_1,x)\,.
\qqq
\exdef \brem The nerve $\,\txG\xcM\,$ of the action groupoid
$\,\txG\lx\xcM\,$ is the standard nerve of the small category
$\,\txG\lx\xcM$,\ understood in the sense of \Rcite{Segal:1968}, in
which the set of $n$-tuples of composable morphisms in
$\,\txG\lx\xcM\,$ has been identified with $\,\txG^m\x\xcM\,$ in a
natural manner,
\qq\nn
\bigl(\bigl(g_{m},(g_{m-1}\cdot g_{m-2}\cdots
g_1).x\bigr),\bigl(g_{m-1}, (g_{m-2}\cdot g_{m-3}\cdots
g_1).x\bigr),\ldots,(g_1,x)\bigr)\,\,
\cr\cr\equiv
(g_{m},g_{m-1},\ldots,g_1,x)\,.
\qqq
Note also that the face maps of $\,\txG\xcM\,$ satisfy the
simplicial identities, including, in particular, the quadratic
relations
\qq\label{eq:simpl-rel-dddd}
\xcMup d^{(m)}_i\circ\xcMup d^{(m+1)}_j=\xcMup d^{(m)}_{j-1}
\circ\xcMup d^{(m+1)}_i\qquad\tx{for } i<j\,.
\qqq
\erem \noindent The face maps of $\,\txG\xcM\,$ are
$\txG$-equivariant. The corresponding $\txG$-action on
$\,\txG^m\times\xcM\,$ is
\qq
\xcMup\ell^{(m)}\ &:&\ \txG\x(\txG^m\x\xcM)\to\txG^m\x\xcM \nonumber\\
\label{eq:Gact-on-nerve}\\
&:&\ \bigl(h,(g_{m},g_{m-1},\ldots,g_1,x)\bigr)\mapsto\left(\Ad_h(g_{m}),
\Ad_h(g_{m-1}),\ldots,\Ad_h(g_1),h.x\right)\,,\nonumber
\qqq
with $\,\xcMup\ell^{(0)}\equiv\xcMup\ell$.\ The action gives rise to
vector fields on $\,\txG^n\x\xcM\,$ of the form
\qq\label{eq:fund-vect-on-nerve}
\xcMup\xcK^{(m)}_a(g_{m},g_{m-1},\ldots,g_1,x)=\sum_{k=1}^{m}\,(
L_a-R_a)(g_k)+\xcMup\xcK_a(x)\,.
\qqq
The face maps can also be used to associate with $\,\txG\xcM\,$ (and
for a choice of a differential sheaf over the nerve) a pullback
cohomology in the sense of \Rcite{Murray:1994db}.
\berop\label{prop:pull-ops}
The pullback operators
\qq
\xcMup\d_\txG^{(m)}=\sum_{i=0}^{m+1}\,(-1)^{m+1-i}\,\xcMup
d^{(m+1)*}_i\,, \label{eq:pullback-cob}
\qqq
square to zero,
\qq\nn
\xcMup\d_\txG^{(m)}\circ\xcMup\d_\txG^{(m-1)}=0\,.
\qqq
\eerop
\beroof
A straightforward check. \eroof\bigskip \noindent The cohomology
will be seen to encode the $\txG$-equivariance relations satisfied
by elements of the description of the $\si$-model in the presence of
defects.
\bigskip

Next, specialising our discussion to the $\si$-model context, we
assume that the components $\,M$, $\,Q\,$ and $\,T_n\,$ of the target
space are $\txG$-spaces,
and that the various maps between them involved in the string
background are $\txG$-maps, i.e.\ that they commute with the actions
of $\,\txG$.\ We have the following straightforward
\berop\label{prop:Gnat-IBB-maps}
Let $\,\Bgt=(\cM,\cB,\cJ)\,$ be a string background as in Definition
\ref{def:bckgrnd} such that the components $\,M$, $\,Q\,$ and
$\,T_n\,$ of the target space $\,\xcF=M\sqcup Q\sqcup T\,$ are
$\txG$-spaces. If the associated maps
$\,\iota_\a\equiv\iota_\a^{(0)},\ \a=1,2\,$ and
$\,\pi_n^{k,k+1}\equiv\pi_n^{k, k+1\,(0)},\ k=1,2,\ldots,n$,\ are
$\txG$-maps, then they extend to natural transformations
\qq\nn
\txG\iota_\a\ :\ \txG Q\to\txG M\,,\qquad\qquad\txG\pi_n^{k,k+1}\ :
\ \txG T_n\to\txG Q
\qqq
between the respective nerve functors of Definition \ref{def:nerves}
by setting
\qq
\iota_\a^{(m)}&\equiv&\txG\iota_\a([m]):=\id_{\txG^m}\times\iota_\a\ :\
\txG^m\x Q\to\txG^m\x M\,,\cr &&\label{eq:funct-bibb-maps}\\
\pi_n^{k,k+1\,(m)}&\equiv&\txG\pi_n^{k,k+1}([m])
:=\id_{\txG^m}\times\pi_n^{k,k+1}\ :\ \txG^m\x T_n\to\txG^m\x Q\,.
\nonumber
\qqq
\eerop

\noindent The last proposition will prove useful in a cohomological
classification of $\txG$-equivariant string backgrounds in
Section \ref{sec:class-equiv-back}. From now on, we suppose that the
assumptions of Proposition \ref{prop:Gnat-IBB-maps} are satisfied.

The $\txG$-equivariant maps $\,\iota^{(m)}_\a\,$ give rise to
operators
\qq\nn
\D_Q^{(m)}:=\iota^{(m)*}_2-\iota^{(m)*}_1\,,\qquad\qquad\D_{T_n}^{(
m)}:=\sum_{k=1}^n\,\vep_n^{k,k+1}\,\pi_n^{k,k+1\,(m)\,*}
\qqq
extending $\,\D_Q\equiv\D^{(0)}_Q\,$ and $\,\D_{T_n}\equiv\D^{(0
)}_{T_n}$,\ respectively, and satisfying the simple relations
\qq\label{eq:tildelDQ-DQdel}
\Qup\d^{(m)}_\txG\circ\D_Q^{(m)}=\D_Q^{(m+1)}\circ\Mup\d^{(m
)}_\txG
\qqq
and
\qq\nn
\Tnup\d^{(m)}_\txG\circ\D_{T_n}^{(m)}=\D_{T_n}^{(m+1)}\circ\Qup
\d^{(m)}_\txG\,,
\qqq
to be exploited presently.
\becon
In order to unclutter the notation, and, at the same time, avoid
confusion with the various indexing schemes employed in the paper,
we fix once and for all a convention for pullbacks of differential
forms along canonical projection maps. Thus, for any $p$-form
$\,\eta\,$ on a smooth space $\,\xcM:=\xcM_1\x\xcM_2\x\cdots\x
\xcM_N\,$ equipped with canonical projections $\,\pr_{i_1,i_2,\ldots
,i_n}:\xcM\to\xcM_{i_1}\x\xcM_{i_2}\x\cdots\x\xcM_{i_n}\,$ given for
$\,1\leq i_1<i_2<\ldots<i_n\leq N$,\ we denote
\qq\nn
\eta_{[i_1,i_2,\ldots,i_n]^* }:=\pr_{i_1,i_2,\ldots,i_n}^*\eta\,.
\qqq
In particular,
\qq\nn
\eta_{i^*}\equiv\eta_{[i]^*}=\pr_i^*\eta
\qqq
for any $\,1\leq i\leq N$.\ Analogous convention will be used for
geometric objects such as bundles, gerbes etc. \econ

\subsection{Rigid symmetries of the $\si$-model}\label{sub:Vino}

\noindent These are the symmetries induced by a group action on the target
space $\,\xcF$. \,They transform fields of the model by composing
them with the target-space transformations in a way that leaves the
probability amplitudes unchanged. The study of continuous rigid
symmetries entails, on the infinitesimal level, the analysis of
variations of the Feynman amplitudes of \Reqref{eq:sigma} generated
by the action of smooth vector fields on the target space. The
following result will, therefore, prove instrumental.
\berop\cite[App.\,A.2]{Runkel:2008gr}
Let $\,\xcA\,$ be the functional \eqref{eq:sigma}, and let
$\,\xcV\,$ be a vector field on $\,\xcF$,\ with restrictions $\,\xcV
\vert_M=\Mup\xcV,\ \xcV\vert_Q=\Qup\xcV\,$ and $\,\xcV\vert_{T_n}=
\Tnup\xcV$,\ subject to the \textbf{alignment conditions}
\qq\label{eq:MQT-align}
\iota_{\a\,*}\Qup\xcV=\Mup\xcV\vert_{\iota_\a(Q)}\,,\qquad\qquad
\pi_{n\,*}^{k,k+1}\hspace{0.03cm}\Tnup\xcV=\Qup\xcV\vert_{\pi_n^{k,k+1}(T_n)}\,.
\qqq
Denote by $\,\xi_t:\xcF\to\xcF\,$ the (local) flow of $\,\xcV\,$
(assumed to exist). \,The variation of $\,\xcA\,$ along $\,\xi\,$ is
then given by
\qq\label{eq:var-act-gen}
\tfrac{\sfd\ }{\sfd t}\big\vert_{t=0}\xcA[(\varphi\,
\vert\,\G);\g]\hspace{7cm}\qqq \vskip -0.4cm
\qq
=\Big(-\tfrac{1}{2}\,\int_\Si\,\bigl(\pLie{\xcV}\txg
\bigr)_\varphi(\sfd\varphi\overset{\wedge}{,}\star_\g\sfd\varphi)+
\sfi\,\int_\Si\,\varphi^*(\ic_\xcV\txH)+\sfi\,\int_\G\,
\varphi^*(\ic_\xcV\om)\Big)\,\xcA[(\xi_t\circ\varphi\,
\vert\,\G);\g].\nonumber
\qqq
\eerop

\noindent Properties of vector fields whose flows preserve the
Feynman amplitudes may be read off from \Reqref{eq:var-act-gen}.
\berop\label{prop:rigid-inf-inv}
Let $\,\xcFup\xcK_a\,$ be the fundamental vector fields on
$\,\xcF$,\ introduced in Definition \ref{eq:Gact-M} (their
restrictions $\,\xcMup\xcK_a,\ \xcM=M,Q,T_n\,$ satisfy the alignment
conditions \eqref{eq:MQT-align}). Then, the $\si$-model Feynman
amplitudes of \Reqref{eq:sigma} are invariant under
rigid translations of the $\si$-model field $\,\varphi:\Si\to\xcF\,$
along the flows of the vector fields $\,\xcFup\xcK_a\,$ on
$\,\xcF\,$ iff the components $\,\Mup\xcK_a\,$ are Killing
for $\,\txg$,
\qq\label{eq:Kill}
\pLie{a}\txg=0\,,
\qqq
and there exist a collection of 1-forms $\,\kappa_a\,$ on $\,M\,$
and a collection of functions $\,k_a\,$ on $\,Q\,$ such that
\qq
\ic_a\txH&=&-\sfd\kappa_a\,,\label{eq:HS-exact}\\\cr
\ic_a\om+\D_Q\kappa_a&=&-\sfd k_a\,,\label{eq:bdry-exact}
\\\cr
\D_{T_n}k_a&=&0\,.\label{eq:junct-exact}
\qqq
\eerop
\beroof Using integration by parts,
it is straightforward to see that conditions
\eqref{eq:Kill}-\eqref{eq:junct-exact} are sufficient for the
variation of the amplitudes in the direction of $\,\xcFup\xcK_a\,$
to vanish. A proof of the claim that they are necessary is given in
Appendix \ref{app:rigid-inf-inv}. \eroof\bigskip \noindent Taken together,
the fields $\,\xcK_a,\ \kappa_a\,$ and $\,k_a\,$ compose a section
of a generalised tangent bundle over (the respective components of)
the target space. On smooth sections of the latter, we may define a
bracket operation generalising the Lie bracket of sections of the
tangent bundle and closing on the linear span of sections generating
rigid symmetries of the $\si$-model. It will be seen to play an
important and natural r\^ole in our considerations of the
constraints for a consistent gauging later on. In the meantime, let
us be more specific about the algebraic structure on the generalised
tangent bundle of the target space that emerges from our analysis.
\bedef\label{def:tw-bra-str}
Consider a smooth manifold $\,\xcF=M\sqcup Q\sqcup T\,$ whose
components $\,M,\ Q\,$ and $\,T=\bigsqcup_{n\geq 3}\,T_n\,$ support
smooth maps $\,\iota_\a:Q\to M,\ \a=1,2\,$ and $\,\pi_n^{k,k+1}:T_n
\to Q,\ k=1,2,\ldots,n$,\ satisfying \Reqref{eq:iopi-iopi}, as well
as smooth forms: a closed $\,\txH\in \Om^3(M)\,$ and
$\,\om\in\Om^2(Q)\,$ related to $\,\txH\,$ as in
\Reqref{eq:triv-restr} and satisfying \Reqref{eq:triv-restr2}.
Denote by $\,\sfE^{(p,q)}\xcM:=\bigwedge^p\sfT\xcM\oplus\bigwedge^q
\sfT^*\xcM\,$ the \textbf{generalised tangent bundle of type
$\,(p,q)\,$} over $\,\xcM=M,Q,T_n$,\ and introduce the \textbf{total
generalised tangent bundle}
\qq\nn
\sfE\xcF:=\sfE^{(1,1)}M\sqcup\sfE^{(1,0)}Q\sqcup\sfT T\,.
\qqq
The \textbf{$(\txH,\om;\D_Q)$-twisted bracket structure} on $\,\sfE
\xcF\,$ is the quadruple
\qq\nn
\bigl(\sfE\xcF,\GBra{\cdot}{\cdot}^{(\txH,\om;\D_Q)},
\Vcon{\cdot}{\cdot},\a_{\sfT\xcF}\bigr)=:\Mgt^{(\txH,\om;\D_Q)}\xcF
\qqq
in which $\,\a_{\sfT\xcF}:\sfE\xcF\to\sfT\xcF\,$ is the canonical
(component-wise) projection, $\,\Vcon{\cdot}{\cdot}:\G(\sfE\xcF)^2
\to C^\infty(M,\bR)\,$ is the scalar product on the component
$\,\sfE^{(1,1)}M\,$ of $\,\sfE\xcF$,\ given by
\qq\nn
\Vcon{\xcV\oplus\upsilon}{\xcW\oplus\varpi}:=\tfrac{1}{2}\,(
\ic_\xcV\varpi+\ic_\xcW\upsilon)\,,\qquad\xcV\oplus\upsilon,\xcW\oplus
\varpi\in\G\bigl(\sfE^{(1,1)}M\bigr)\,,
\qqq
and $\,\GBra{\cdot}{\cdot}^{(\txH,\om;\D_Q)}\,$ is a bilinear
antisymmetric operation on smooth sections of $\,\sfE\xcF$,\ with
restrictions {\small
\qq\nn
\GBra{\Vgt_1}{\Vgt_2}^{(\txH,\om;\D_Q)}\vert_M\,&=&[\Mup\xcV_1,\Mup
\xcV_2]\oplus\bigl(\pLie{\xcV_1}\upsilon_2-\pLie{\xcV_2}
\upsilon_1-\tfrac{1}{2}\,\sfd(\ic_{\xcV_1}\upsilon_2-
\ic_{\xcV_2}\upsilon_1)+\ic_{\xcV_1}\ic_{\xcV_2}\txH\bigr)\,,\cr\cr
\GBra{\Vgt_1}{\Vgt_2}^{(\txH,\om;\D_Q)}\vert_Q\,\,&=&[\Qup\xcV_1,\Qup
\xcV_2]\oplus\bigl(\pLie{\xcV_1}f_2-\pLie{\xcV_2}f_1+
\ic_{\xcV_1}\ic_{\xcV_2}\om+\tfrac{1}{2}\,(\ic_{\xcV_1}\D_Q
\upsilon_2-\ic_{\xcV_2}\D_Q\upsilon_1)\bigr)\,,\cr\cr
\GBra{\Vgt_1}{\Vgt_2}^{(\txH,\om;\D_Q)}\vert_{T_n}&=&[\Tnup\xcV_1,
\Tnup\xcV_2]\,,
\qqq}
\noindent\hspace*{-0.1cm}expressed in terms of the restrictions
$\,\Vgt_i\vert_M=\Mup\xcV_i \oplus\upsilon_i,\
\Vgt_i\vert_Q=\Qup\xcV_i\oplus f_i\,$ and $\,\Vgt_i
\vert_{T_n}=\Tnup\xcV_i\,$ of $\,\Vgt_i\in\G(\sfE\xcF),\ i=1,2$.\
\exdef \brem It is to be noted that the restriction of the
$(\txH,\om;\D_Q)$-twisted bracket structure to $\,\sfE^{(1,1)}M\,$
defines the familiar Courant algebroid of
Refs.\,\cite{Courant:1990,Dorfman:1993,Liu:1997}, twisted by the
gerbe curvature $\,\txH\,$ in the manner first discussed in
\Rcite{Severa:2001qm}. Its r\^ole in the context of $\si$-model
symmetries was observed already in \Rcite{Alekseev:2004np}, and an
intrinsic gerbe-theoretic interpretation (in terms of generalised
tangent bundles twisted by local data of a gerbe) was put forward in
\Rcite{Hitchin:2004ut}. A full-fledged canonical interpretation of
the complete algebraic structure on the generalised tangent bundle
of the target space alongside a unified description in the language
of the 2-category of bundle gerbes with curving and connection over
$\,\xcF\,$ is given in \Rcite{Suszek:2012}. Below, we invoke some of
these results. \erem\medskip

\berop\cite[Prop.\,5.5]{Suszek:2012}\label{prop:sisym}
In the notation of Definition \ref{def:tw-bra-str}, denote by
$\,\G_\si(\sfE\xcF)\,$ the subspace of all smooth sections
$\,\Vgt=(\Mup\xcV\oplus\upsilon,\Qup\xcV\oplus f,\Tnup\xcV)\,$ of
$\,\sfE\xcF$,\ to be termed \textbf{$\si$-symmetric}, that satisfy
the alignment conditions \eqref{eq:MQT-align} and obey
\qq\nn
\pLie{\xcV}\txg=0\,,\qquad\qquad\left\{\barr{l}
\ic_\xcV\txH=-\sfd\upsilon\\\cr \ic_\xcV\om+\D_Q\upsilon=-\sfd f
\\\cr \D_{T_n}f=0\earr\right.\,.
\qqq
The $(\txH,\om;\D_Q)$-twisted bracket
$\,\GBra{\cdot}{\cdot}^{(\txH,\om;\D_Q)}\,$ closes on $\,\G_\si(\sfE
\xcF)$,\ that is
\qq\nn
\Vgt,\Wgt\in\G_\si(\sfE\xcF)\qquad\Longrightarrow\qquad
\GBra{\Vgt}{\Wgt}^{(\txH,\om;\D_Q)}\in\G_\si(\sfE\xcF)\,,
\qqq
and so it endows $\,\G_\si(\sfE\xcF)\,$ with the structure of an
algebra over $\,\bR$.
\eerop
\noindent It deserves to be emphasised that the symmetry algebra
$\,\bigl(\G_\si(\sfE\xcF),\GBra{\cdot}{\cdot}^{(\txH,\om;\D_Q)}
\bigr)\,$ is not, in general, a Lie algebra. Its field-theoretic
significance is reflected in
\berop\cite[Props.\,3.9 \& 5.5]{Suszek:2012}\label{prop:sympl-real}
Let $\,\sfP_\si\,$ be the phase space of the non-linear
two-dimensional $\si$-model for network-field configurations
$\,(\varphi\,\vert\,\G)\,$ in string background
$\,\Bgt=$\break$(\cM,\cB, \cJ)\,$ with target space $\,\xcF=M\sqcup
Q\sqcup T\,$ on a lorentzian world-sheet $\,(\Si,\g)\,$ with defect
quiver $\,\G$.\ To every $\si$-symmetric section
$\,\Vgt\in\G_\si(\sfE\xcF)\,$ of the total generalised tangent
bundle $\,\sfE\xcF\,$ over the target space of $\,\Bgt$,\ there is
associated a hamiltonian $\,h_\Vgt\in C^\infty(\sfP_\si,\bR)\,$
generating, through Poisson brackets, the action of the Lie algebra
$\,\ggt\,$ of the symmetry group $\,\txG\,$ on
$\,C^\infty(\sfP_\si,\bR)$,\ and the map $\,\Vgt\mapsto h_\Vgt\,$ is
an $\bR$-algebra homomorphism.
\eerop
\noindent The algebra $\,\bigl(\G_\si(\sfE\xcF),
\GBra{\cdot}{\cdot}^{(\txH,\om;\D_Q)}\bigr)\,$ of $\si$-model
symmetries, endowed with the scalar product $\,\Vcon{\cdot}{\cdot}$,
\ will be seen to play a very natural and important r\^ole in the
characterisation of the conditions in which the rigid symmetries
from $\,\txG\,$ can be gauged in topologically simple circumstances.
\bigskip

We conclude this introductory section by discussing, in the
simplicial framework, some $\txG$-equivariance properties of the
smooth forms
\qq\label{eq:rho-on-G}\qquad\qquad
\rho=\kappa_{a\,2^*}\wedge\theta_{L\,1^*}^a-\tfrac{1}{2}\,\txc_{ab
\,2^*}\,\theta_{L\,1^*}^a\wedge\theta_{L\,1^*}^b\in\Om^2(\txG\x M)
\,,
\qqq
and
\qq\label{eq:la-on-G}
\la=-k_{a\,2^*}\,\theta_{L\,1^*}^a\in\Om^1(\txG\x Q)
\qqq
on the nerve $\,\txG\xcF$,\ written in terms of the functions
\qq\label{eq:cAB}
\txc_{ab}:=\ic_a\kappa_b
\qqq
and of the respective canonical projections $\,\pr_1:\txG\x\xcF\to
\txG\,$ and $\,\pr_2:\txG\x\xcF\to\xcF$.\ These objects will be of
relevance in later sections, where they enter the analysis of the
$\txG$-equivariance of the string background. In our discussion, we
shall make extensive use of the following
\berop\label{prop:pulls-reex}
Let $\,\eta\in\Om^p(\xcM)\,$ be a smooth $p$-form on a $\txG$-space
$\,\xcM$,\ with the group $\,\txG\,$ acting on $\,\xcM\,$ as in
\Reqref{eq:Gact-M}, and with vector fields $\,\ovl X\,$
for $\,X\in\ggt\,$ as in Definition
\ref{def:Gspace}. The identities
\qq\label{eq:pull-thru-con}
\xcMup\ell_h^*(\ic_{\ovl X}\eta)=\ic_{\ovl{(\Ad_{h^{-1}}X)}}\xcMup
\ell_h^*\eta
\qqq
and
\qq\label{eq:pull-as-shift-act}
\xcMup\ell^*\eta(h,m)=\bigl(\ee^{-\ovl{\theta_L(h)}}.\xcMup\ell_h^*
\eta\bigr)(m)
\qqq
hold for any pair $\,(h,m)\in\txG\x\xcM$.
\eerop
\beroof
The first of the two identities is a consequence of the behaviour of
the fundamental vector fields
\qq
\xcFup\xcK^\mu_a(h.m)\equiv\tfrac{\sfd\ }{\sfd t}\big\vert_{t=0}
\varphi^\mu\bigl(\ee^{-tt_a}.h.m\bigr)=\tfrac{\sfd\ }{\sfd t}
\big\vert_{t=0}\varphi^\mu\bigl(h.\ee^{-t\Ad_{h^{-1}}t_a}.m\bigr)
=\ic_{\ovl{\Ad_{h^{-1}}t_a}}\ell_h^*\sfd\varphi^\mu(m)\cr
\label{eq:K-shift}
\qqq
under arbitrary shifts $\,h\in\txG\,$ of the argument $\,m\in\xcF$.\
Here, $\,\{\varphi^\mu\}\,$ is a local coordinate system on
$\,\xcM$.

The second identity becomes evident upon writing it out for the
basis 1-forms on $\,\xcM\,$ associated with the coordinates
$\,\varphi^\mu\,$ at $\,h.m\,$ and with those at $\,m$,\ to be
denoted $\,\psi^\nu,\ \nu=1,2,\ldots,\dim\,\xcM$.\ This also entails
introducing local coordinates $\,X^a,\ a=1,2,\ldots,\dim\,\txG$,\ at
$\,h\in\txG$.\ We then find the desired relation
\qq
\xcMup\ell^*\sfd\varphi^\mu(h,m)&=&\tfrac{\p\varphi^\mu(h.m)}{\p
\psi^\nu(m)}\,\sfd\psi^\nu(m)+\sfd X^a\,\p_a\varphi^\mu(h.m)\cr\cr
&\equiv&\xcMup\ell_h^*\sfd\varphi^\mu(m)+\theta_L^a(h)\,\tfrac{\sfd
\ }{\sfd t}\big\vert_{t=0}\varphi^\mu\bigl(\left(h\cdot\ee^{t t_a}
\right).m\bigr)\cr\cr
&=&\xcMup\ell_h^*\sfd\varphi^\mu(m)-\ovl{\theta_L(h)}.\xcMup
\ell_h^*\sfd\varphi^\mu(m)\equiv\ee^{-\ovl{\theta_L(h)}}.\ell_h^*
\sfd\varphi^\mu(m)\,.\label{eq:trafo-dphi}
\qqq
\eroof\bigskip \noindent We shall also assume that the various forms
on $\,M\,$ and $\,Q$,\ as well as their extensions, previously taken
to be $\ggt$-invariant resp.\ $\ggt$-equivariant, are in fact
$\txG$-invariant resp.\ $\txG$-equivariant,\ cf.\ Definition
\ref{def:Gbas-form}. Using the above identities for $\txG$-spaces
$\,\xcM\,$ of $\,\txG\xcF$,\ we can readily verify
\berop\cite[Lemmas 3.11 \& 3.13]{Gawedzki:2010rn}\label{prop:just-rho}
Let $\,\rho\,$ be the 2-form on $\,\txG\x M\,$ given in
\Reqref{eq:rho-on-G}, with $\,\kappa=\kappa_a\ox\tau^a\in\Om^1(M)
\ox\ggt^*$,\ defined in terms of the 1-forms $\,\kappa_a\,$ of
\Reqref{eq:HS-exact}, $\txG$-equivariant in the sense of Definition
\ref{def:Gbas-form} and subject to the additional constraints
\qq\nn
\txc_{ba}=-\txc_{ab}\,,
\qqq
cf.\ \Reqref{eq:cAB}. Then, in the notation of
\Reqref{eq:Gact-on-nerve} and Proposition \ref{prop:pull-ops}, the
following relations hold true:
\qq
\Mup\ell_h^{(1)*}\rho&=&\rho\,,\qquad h\in\txG\label{eq:rho-inv}\\
\cr
\Mup\d_\txG^{(1)}\rho&=&0\,,\label{eq:delro}\\\cr
\Mup\d_\txG^{(0)}\txH&=&\sfd\rho\,.\label{eq:delH-rho}
\qqq
\eerop\medskip
\noindent The proposition was proven in \Rcite{Gawedzki:2010rn}.
Here, on the other hand, we want to study properties of forms
$\,\om\,$ and $\,\la$.\
\berop
Under the assumptions of Proposition \ref{prop:Gnat-IBB-maps}, let
$\,\om\,$ be a $\txG$-invariant 2-form on $\,Q\,$ satisfying
\Reqref{eq:triv-restr} for $\,\txH\,$ a closed $\txG$-invariant
3-form on $\,M$.\ Finally, let $\,\rho\,$ and $\,\la\,$ be the
corresponding forms on $\,\txG\x M\,$ and $\,\txG\x Q$,\
respectively, given in Eqs.\,\eqref{eq:rho-on-G} and
\eqref{eq:la-on-G} and related to $\,\om\,$ as in
\Reqref{eq:bdry-exact}, with $\,\kappa=\kappa_a\ox\tau^a\in\Om^1(M)
\ox\ggt^*\,$ as in Proposition \ref{prop:just-rho} and with $\,k=k_a
\ox\tau^a\in C^\infty(Q,\bR)\ox\ggt^*$,\ defined in terms of functions
$\,k_a\,$ of \Reqref{eq:bdry-exact}, $\txG$-equivariant in the sense
of Definition \ref{def:Gbas-form}. Then, the following relations are
satisfied:
\qq
\Qup\ell_h^{(1)*}\la&=&\la\,,\label{eq:ellh-la}\\\cr
\Qup\d_\txG^{(0)}\om&=&-\D_Q^{(1)}\rho+\sfd\la\,,\label{eq:dla}\\
\cr
\Qup\d^{(1)}_\txG\la&=&0\,.\label{eq:della}
\qqq
\eerop
\beroof
Using \Reqref{eq:Gact-on-nerve} and the obvious identity
\qq\nn
\left(\Ad_h^*\ox\id_\ggt\right)\theta_L=\left(\id_{\Om^1(\txG)}\ox
\Ad_h\right)\theta_L\,,
\qqq
together with the $\txG$-equivariance property of $\,k$,
\qq\label{eq:Gequiv-xi}
\left(\Qup\ell_h^{(0)*}k\right)(X)=k\left(\Ad_{h^{-1}}X\right)\,,
\qqq
written for arbitrary $\,X\in\ggt$,\ we find, in the natural
shorthand notation $\,\la\equiv-k_{2^*}(\theta_{L\,1^*})$,
\qq\nn
\Qup\ell_h^{(1)*}\la=-\left(\Qup\ell_h^{(0)*}k\right)_{2^*}\left(
\left(\Ad_h^*\ox\id_\ggt\right)\theta_{L\,1^*}\right)=-k_{2^*}
\left(\left(\Ad_h^*\ox \Ad_{h^{-1}}\right)\theta_{L\,1^*}\right)=
\la\,,
\qqq
which gives relation \eqref{eq:ellh-la}. Next,
\Reqref{eq:bdry-exact} and the $\ggt$-equivariance property of
$\,k\,$ (implied by its $\txG$-equivariance), in conjunction with
the $\txG$-invariance property of $\,\om$,
\qq\nn
\Qup\ell^{(0)*}_h\om=\om\,,
\qqq
and the Maurer-Cartan equation
\qq\nn
\sfd\theta_L^a=-\tfrac{1}{2}\,f_{abc}\,\theta_L^b\wedge\theta_L^c
\,,
\qqq
yield
\qq\nn
\Qup d_1^{(1)*}\om(g,m)\hspace{-0.2cm}&\equiv&\hspace{-0.2cm}
\Qup\ell^{(0)\,*}\om(g,m)=\bigl(
\ee^{-\ovl{\theta_L(g)}}.\Qup\ell_g^{(0)*}\om\bigr)(m)=\ee^{-
\ovl{\theta_L(g)}}\om(m)\cr\cr
\hspace{-0.2cm}&=&\hspace{-0.2cm}\om(m)+\theta_L^a(g)\wedge\bigl(\sfd k_a+\D_Q^{(0)}\kappa_a
\bigr)(m)+\tfrac{1}{2}\,\theta^a_L\wedge\theta^b_L(g)\,
\ic_a\bigl(\sfd k_b+\D_Q^{(0)}\kappa_b\bigr)(m)\cr\cr
\hspace{-0.2cm}&=&\hspace{-0.2cm}\bigl(\Qup d_0^{(1)}\om-\D_Q^{(1)}\rho
+\sfd\la\bigr)(g,m)
\,,
\qqq
which is \Reqref{eq:dla}. Finally, taking, say, \Reqref{eq:dla}, we
note that it implies, by virtue of Eqs.\,\eqref{eq:tildelDQ-DQdel}
and \eqref{eq:delro}, the identity
\qq\nn
\sfd\Qup\d_\txG^{(1)}\la\equiv\Qup\d_\txG^{(1)}\circ\Qup\d_\txG^{(0
)}\om+\Qup\d_\txG^{(1)}\circ\D_Q^{(1)}\rho=\D_Q^{(2)}\circ\Mup
\d_\txG^{(1)}\rho=0\,.
\qqq
Using
\qq\nn
\theta_L(g\cdot h)=\theta_L(h)+\Ad_{h^{-1}}\theta_L(g)\,,
\qqq
we demonstrate that the de Rham cocycle $\,\Qup\d_\txG^{(1)}\la\,$
actually trivialises as
\qq\nn
\Qup d^{(2)*}_1\la(g_2,g_1,m)&\equiv&\la(g_2\cdot g_1,m)=-k\bigl(
\theta_L(g_1)+\Ad_{g_1^{-1}}\theta_L(g_2)\bigr)(m)\cr\cr
&=&-k\bigl(\theta_L(g_1)\bigr)(m)-\Qup\ell^{(0)*}_{g_1}k\bigl(
\theta_L(g_2)\bigr)(m)\cr\cr
&=&\bigl(\Qup d^{(2)*}_0\la+\Qup d^{(2)*}_2\la\bigr)(g_2,g_1,m)\,,
\qqq
which, upon dropping the point dependence, yields relation
\eqref{eq:della}. This completes the proof of the proposition.
\eroof

\section{The coupling to gauge fields: topologically trivial sector}
\label{sec:gauge-min}

\noindent Rigid symmetries of a field theory can sometimes be
gauged, i.e.\ promoted to the rank of local ones. In the $\si$-model
context, in the presence of extra differential-geometric structures
on the target space, it has to be ensured that the action of the
symmetry group on the target space lifts to those structures. Thus,
e.g., in the case in which the smooth target space $\,\xcF\,$
carries a metric and a gerbe, the action should be isometric and
such that the gerbe is equivariant with respect to it in an
appropriate sense. The gauging of group $\,\txG\,$ of rigid
symmetries involves a coupling to background gauge fields, given by
connections on a principal $\txG$-bundle $\,\sfP\,$ over the
space-time (in our case, the world-sheet) $\,\Si$. \,The
topologically trivial sector corresponds to the case with a trivial
$\txG$-bundle $\,\sfP=\Sigma\times G$. \,In this case, the gauge
fields are represented by global 1-forms on the world-sheet with
values in the Lie algebra $\,\ggt\,$ of the symmetry group $\,\txG$.
\,Somewhat misleadingly, we shall call such gauge fields
topologically trivial in what follows. \,It may turn out to be
physically appropriate, however, not to restrict attention to the
case of trivial (or trivializable) $\txG$-bundles and to consider
topologically nontrivial sectors with nontrivial $\txG$-bundles
$\,\sfP\,$ over $\Sigma$. \,For such bundles, the inclusion of
space-time gauge fields is accompanied by the replacement of the
original fields of the theory by (global) sections of the associated
bundle $\,\sfP\x_\txG\xcF$. In both topologically trivial and
topologically nontrivial sectors, the gauging has to take into
account topological issues associated with large gauge
transformations non-homotopic to the identity transformation. The
present section is dedicated to a detailed study of the gauging
procedure in the more straightforward case of the topologically
trivial sector. Already this analysis will lead to some interesting
compatibility conditions for a consistent lift of the geometric
symmetries on $\,\xcF\,$ to the geometric and cohomological
structures over the target space. We shall provide a precise
algebraic, cohomological and canonical characterisation of the
ensuing constraints for the consistent gauging.

\subsection{Insights from the study of trivial backgrounds}

\noindent In order to develop some intuition as to the possible nature of the
coupling between the world-sheet gauge fields and the target-space
structures, we further simplify the physical setting by taking all
target-space structures to be trivial, that is described by tensor
fields, the latter being further assumed $\txG$-invariant. Thus, we
set
\qq
&\cG=I_\txB\,,\qquad\qquad\ \,\txB\in\Om^2(M)\,,\qquad\qquad&\tx{with}\qquad
\Mup\ell_g^*\txB=
\txB\,,\label{eq:back-triv-gerb}\\\cr
&\Phi=J_\txP\,,\qquad\qquad\ \,\,\txP\in\Om^1(Q)\,,\qquad\qquad&\tx{with}
\qquad\Qup\ell_g^*\txP
=\txP\,,\\\cr
&\phi_n=K_{\txf_n}\,,\qquad\txf_n\in C^\infty(T_n,\uj)\,,\qquad&\tx{with}\qquad
\Tnup\ell_g^*\txf_n=\txf_n\,,\label{eq:back-triv-ibb}
\qqq
for arbitrary $\,g\in\txG$, \,with,\ as previously, $\,I_\txB\,$ the
trivial gerbe with (global) curving $\,\txB$, $\,J_\txP\,$ the
trivial bi-brane defined by the smooth 1-form $\,\txP$,\ and
$\,K_{\txf_n}\,$ the trivial inter-bi-branes determined by the
respective smooth $\uj$-valued functions $\,\txf_n\,$ on $\,T_n$.\
The tensors entering the above definitions are subject to the
relations
\qq\label{eq:back-cohom-triv}
\D_Q\txB=\sfd\txP-\om\,,\qquad\qquad\D_{T_n}\txP-\sfi\,\sfd\log
\txf_n=0\,.
\qqq
The Feynman amplitude \eqref{eq:sigma} now takes the simple form
\qq
\xcA[(\varphi\,\vert\,\G);\gamma]\ =\
\exp\Big[-\tfrac{1}{2}\,\int_\Si\,
\txg(\sfd\varphi\overset{\wedge}{,}\star_\g\sfd\varphi)+
\sfi\,\int_\Si\,\varphi^*\txB+\sfi\,\int_\G\,(\varphi\vert_\G)^*
\txP\Big]\prod_{\jmath\in\Vgt_\G}\varphi^*\txf_{n_\jmath}(\jmath
)\,,\cr\label{eq:si-act-triv}
\qqq
where the product is taken over the set $\,\Vgt_\G\,$ of defect
junctions. Note that we may take
\qq\nn
\kappa_a=\ic_a\txB\,,\qquad\qquad k_a=\ic_a\txP
\qqq
in this case, and this choice is consistent with Proposition
\ref{prop:rigid-inf-inv} as
\qq\nn
\D_Q\kappa_a&\equiv&\D_Q(\ic_a\txB)=\ic_a\D_Q\txB=\ic_a(\sfd\txP-
\om)=-\sfd(\ic_a\txP)-\ic_a\om\cr\cr
&\equiv&-\sfd k_a-\ic_a\om\,,\cr\cr
\D_{T_n}k_a&\equiv&\D_{T_n}(\ic_a\txP)=\ic_a\D_{T_n}\txP=\sfi\,
\ic_a\sfd\log\txf_n =0\,.
\qqq
Finally, we readily check the identities
\qq
\pLie{a}\kappa_b&=&f_{abc}\,\kappa_c\,, \label{eq:HS1-ids-triv}\\\cr
\pLie{a}k_b&=&f_{abc}\,k_c\label{eq:FFM-ids-triv}
\qqq
and
\qq\label{eq:HS2-ids-triv}
\txc_{ba}=-\txc_{ab}\,.
\qqq
Actually, in consequence of the assumed $\txG$-invariance of the
background, a little more is true: With the help of Proposition
\ref{prop:pulls-reex}, we readily establish the $\txG$-equivariance
of $\,\kappa=\kappa_a\ox\tau^a\,$ and $\,k=k_a\ox\tau^a$.

Let $\,\txA=A_u^a\,\sfd\si^u\ox t_a\in\Om^1(\Si)\ox \ggt\,$ be a
(topologically trivial) world-sheet gauge field, undergoing a
transformation
\qq\label{eq:trafo-A}
\txA\mapsto\Ad_\chi\txA-\sfd\chi\,\chi^{-1}\equiv{}^{\tx{\tiny $\chi$}}
\hspace{-2pt}\txA\label{eq:chiA}
\qqq
under a gauge transformation $\chi\in C^\infty(\Si,\txG)$ that maps
a $\si$-model field according to
\qq\label{eq:trafo-phi}
\varphi\mapsto\xcFup\ell(\chi,\varphi)\equiv\chi.\varphi\,.\label{eq:chiphi}
\qqq
The standard `minimal-coupling' recipe, familiar from models of
point-particle physics, is tantamount to the replacement of the
vector field-valued world-sheet forms $\,\sfd\varphi\,$ in the term
involving the world-sheet metric by
\qq\nn
\sfd\varphi(\si)\mapsto\ee^{-\ovl{\txA(\si)}}.\sfd\varphi(\si)=
\bigl(\p_u\varphi^\mu-A^a_u\,\xcK_a^\mu(\varphi)\bigr)(\si)\,\sfd
\si^u\ox\p_\mu=:D_\txA\varphi(\si)\,,
\qqq
and of the target-space forms $\,\txB\,$ and $\,\txP\,$ in the `topological'
term by the corresponding objects
\qq\nn
\txB\mapsto\ee^{-\ovl\txA_{1^*}}.\txB_{2^*}\,,\qquad\qquad\txP
\mapsto\ee^{-\ovl\txA_{1^*}}.\txP_{2^*}
\qqq
on the product space $\,\Si\x\xcF\,$ (the latter coming with the
canonical projections $\,\pr_1:\Si\x\xcF\to\Si\,$ and $\,\pr_2:\Si\x
\xcF\to\xcF$,\ indicated by the subscripts), to be subsequently
pulled back to $\,\Si\,$ along the map
\qq\label{eq:ext-phi}
\phi=(\id_\Si,\varphi)\ :\ \Si\to\Si\x\xcF\,.
\qqq
We have
\berop\label{prop:Gau-inv-triv}
In the notation of Definition \ref{def:sigma}, and for a
$\txG$-invariant topologically trivial string background $\,\Bgt\,$
defined by Eqs.\,\eqref{eq:back-triv-gerb}-\eqref{eq:back-triv-ibb},
let $\,\txA\in\Om^1(\Si)\ox\ggt\,$ be a topologically trivial gauge
field on $\,\Si$.\ The gauged non-linear $\si$-model amplitudes for
network-field configurations $\,(\varphi\,\vert\,\G)\,$ in
background $\,\Bgt$,\ given by the expression
\qq\label{ampltrivback}
\widehat{\xcA}[(\varphi\,\vert\,\G);\txA,\g]\hspace{7cm}
\qqq \vskip
-0.4cm
\qq\nn
=\exp\Big[-\tfrac{1}{2}\int_\Si\txg(D_\txA\varphi
\overset{\wedge}{,}\star_\g D_\txA\varphi)+\sfi\int_\Si\phi^*
\ee^{-\ovl\txA_{1^*}}.\txB_{2^*}+\sfi\int_\G(\phi\vert_\G)^*
\ee^{-\ovl\txA_{1*}}.\txP_{2^*}\Big]\prod_{\jmath\in\Vgt_\G}
\varphi^*\txf_{n_\jmath}(\jmath)\,,
\qqq
written in terms of the extended
world-sheet field $\,\phi\,$ of
\Reqref{eq:ext-phi}, are invariant under simultaneous gauge
transformations
$\,(\varphi,\txA)\mapsto\bigl(\chi.\varphi,{}^{\tx{\tiny
$\chi$}}\hspace{-2pt}\txA\bigr)\,$ for $\,\chi\in C^\infty(\Si,\txG)$.
\eerop
\beroof
The `metric' and `topological' terms are independently invariant
under the gauge transformations. We begin by demonstrating the
invariance of the former. To this end, we examine the transformed
\textbf{covariant derivative},
\qq\nn
D_\txA\varphi^\mu\mapsto D_{{}^{\tx{\tiny $\chi$}}\hspace{-2pt}
\txA}(\chi.\varphi)^\mu\,,
\qqq
where both $\,\varphi\,$ and $\,\chi\,$ in the last expression are
functions of the world-sheet coordinates $\,\si^u,\ u=1,2$.\ Taking
into account this $\si^u$-dependence of $\,\varphi\,$ and that of
$\,\chi$,\ respectively, we decompose the transformed world-sheet
differential $\,\sfd(\chi.\varphi)^\mu\,$ into two terms as in
\Reqref{eq:trafo-dphi}, whereby we obtain, using
\Reqref{eq:K-shift},
\qq
D_{{}^{\tx{\tiny $\chi$}}\hspace{-2pt}\txA}(\chi.\varphi)^\mu&=&
\bigl(\ell_\chi^*\sfd\varphi^\mu-\ic_{\ovl{\chi^*\theta_L}}
\ell_\chi^*\sfd\varphi^\mu\bigr)-\bigl(\Ad_\chi\txA-\sfd\chi\,
\chi^{-1}\bigr)^a\,\ic_{\ovl{\Ad_{\chi^{-1}}t_a}}\ell_\chi^*\sfd
\varphi^\mu\cr\cr\label{eq:cov-der-cov}
&=&\bigl(\ell_\chi^*\sfd\varphi^\mu-\ic_{\ovl{\chi^*\theta_L}}
\ell_\chi^*\sfd\varphi^\mu\bigr)-\ic_{\ovl{\txA-\chi^{-1}\,\sfd\chi}}
\ell_\chi^*\sfd\varphi^\mu\\\cr
&=&\bigl(\ell_\chi^*\sfd\varphi^\mu-\ic_{\ovl\txA}\ell_\chi^*\sfd
\varphi^\mu\bigr)=\tfrac{\p(\chi.\varphi)^\mu}{\p\varphi^\nu}\,
D_\txA\varphi^\nu\,.\nonumber
\qqq
This justifies the name given to the object $\,D_\txA\varphi^\mu\,$
and straightforwardly implies the gauge invariance of the `metric'
term by virtue of the $\txG$-invariance of the target-space metric.
\vskip 0.1cm

Next, we pass to the `topological' term. Its gauge invariance is a
simple corollary to the following observation.
\belem
Let $\,\eta\,$ be a $\txG$-invariant $p$-form on $\,\xcF$.\ Then, in
the above notation,
\qq\nn
(\chi.\phi)^*\ee^{-\ovl{{}^{\tx{\tiny $\chi$}}\hspace{-2pt}
\txA_{1^*}}}.\eta_{2^*}=\phi^*\ee^{-\ovl{\txA_{1^*}}}.\eta_{2^*}\,.
\qqq
\elem
\beroof
Our proof employs a technique that will turn out useful later on. It
associates to a gauge transformation
\qq\nn
\chi\ :\ \Si\to\txG
\qqq
a transformation on the product space $\,\Si\x\xcF\,$
\qq\label{eq:gauge-trans-ext}
L_\chi\ :\ \Si\x\xcF&\to&\Si\x\xcF\cr
(\si,m)&\mapsto&\bigl(\si,\chi(
\si).m\bigr)\,,
\qqq
and subsequently decomposes the lift as
\qq\nn
L_\chi\ :\ \Si\x\xcF&\xrightarrow{K_\chi}&\Si\x\txG\x\xcF \
\,\xrightarrow{\id_\Si\times\xcFup\ell}\ \,\Si\x\xcF\cr \
(\si,m)&\longmapsto&
\left(\si,\chi(\si),m\right)\quad\longmapsto\quad
\left(\si,\chi(\si).m\right)\,.
\qqq
Upon invoking Proposition \ref{prop:pulls-reex}, we find the
sought-after result
\qq\nn
\left((\chi.\phi)^*\ee^{-\ovl{{}^{\tx{\tiny $\chi$}}\hspace{-2pt}
\txA_{1^*}}}.\eta_{2^*}\right)(\si)&=&\left((L_\chi\circ\phi)^*
\ee^{-\ovl{{}^{\tx{\tiny $\chi$}}\hspace{-2pt}\txA_{1^*}}}.
\eta_{2^*}\right)(\si)=\phi^*\left(\ee^{-\ovl{\left(\Ad_{\chi^{-
1}}{}^{\tx{\tiny $\chi$}}\hspace{-2pt}\txA\right)_{1^*}}}.L_\chi^*
\eta_{2^*}\right)(\si)\cr\cr
&=&\ee^{-\ovl{\left(\Ad_{\chi^{-1}}{}^{\tx{\tiny $\chi$}}
\hspace{-2pt}\txA\right)(\si)}}.\ee^{-\ovl{\chi^*\theta_L(\si)}}.
\xcFup\ell_{\chi(\si)}^*\varphi^*\eta(\si)=\ee^{-\ovl{\txA(\si)}}.
\varphi^*\eta(\si)\cr\cr
&\equiv&\phi^*\left(\ee^{-\ovl{\txA_{1^*}}}.\eta_{2^*}\right)(\si
)\,.
\qqq
\eroof \bigskip \noindent It is now clear that the explicit form of the
`minimal coupling' used, in conjunction with the assumed
$\txG$-invariance of the string background imply the thesis of the
proposition. \eroof\bigskip

With view to a subsequent generalisation, let us rewrite the Feynman
amplitude \eqref{ampltrivback} in the more suggestive form
\qq\nn
\widehat{\xcA}[(\varphi\,\vert\,\G);\txA,\g]\hspace{7cm}
\qqq \vskip
-0.4cm
\qq\nn
=\,\exp\Big[-\tfrac{1}{2}\,\int_\Si\,\txg_\txA(\sfd\phi
\overset{\wedge}{,}\star_\g\sfd\phi)+\sfi\int_\Si\,\phi^*(
\txB_{2^*}+\rho_\txA)+\sfi\int_\G\,(\phi\vert_\G)^*
(\txP_{2^*}+\la_\txA)\Big]\prod_{\jmath\in\Vgt_\G}\varphi^*
\txf_{n_\jmath}(\jmath)
\qqq
upon introducing the forms
\qq
\rho_\txA&:=&\kappa_{a\,2^*}\wedge\txA_{1^*}^a-\tfrac{1}{2}\,
\txc_{ab\,2^*}\,\txA_{1^*}^a\wedge\txA_{1^*}^b\,\in\,\Om^2(\Si\x M)
\,,\label{eq:rhoA}\\ \cr
\la_\txA&:=&-k_{a\,2^*}\,\txA_{1^*}^a\,\in\,\Om^1(\Si\x Q)\,,
\label{eq:laA}
\qqq
and the (possibly degenerate) metric tensor
\qq\label{eq:gA-def}
\txg_\txA:=\txg_{2^*}-\txK_{a\,2^*}\ox\txA^a_{1^*}-\txA^a_{1^*}\ox
\txK_{a\,2^*}+\txh_{ab\,2^*}\,\left(\txA^a\ox\txA^b\right)_{1^*}
\qqq
on the `extended' target-space $\,\Si\x\xcF$. \,The latter tensor
is defined in terms of the 1-forms
\qq\label{eq:KA-def}
\txK_a:=\txg\bigl(\Mup\xcK_a,\cdot\bigr)\in\Om^1(M)
\qqq
and of the symmetric tensor
\qq\label{eq:hAB-def}
\txh_{ab}:=\txg\bigl(\Mup\xcK_a,\Mup\xcK_b\bigr)\,,
\qqq
and implicitly understood to act on the second tensor factor in
\qq\nn
\sfd\phi(\si)=\left(\sfd\si^u\ox\p_u,\p_u\varphi^\mu(\si)\,\sfd
\si^u\ox\p_\mu\vert_{\varphi(\si)}\right)\,.
\qqq
The lesson to be drawn from the above analysis can now be phrased as
follows: The coupling of a topologically trivial gauge field
$\,\txA\,$ to the $\si$-model fields boils down to replacing the
original quadruple $\,(\txg,\cG,\Phi,\varphi)\,$ over the target
space $\,\xcF\,$ in the amplitude by the corrected one $\,(\txg_\txA
,\cG_\txA,\Phi_\txA,\phi)\,$ over $\,\Si\x\xcF$,\ with
$\,\txg_\txA\,$ given by \Reqref{eq:gA-def},
\qq\nn
\cG_\txA:=\cG_{2^*}\ox I_{\rho_\txA}\,,\qquad\qquad\Phi_\txA:=
\Phi_{2*}\ox J_{\la_\txA}\,,
\qqq
and with $\,\phi\,$ defined by \Reqref{eq:ext-phi}.

\subsection{An Ansatz for nontrivial backgrounds, and
constraints for a consistent gauging}

\noindent Passing to general backgrounds, while keeping the
world-sheet gauge field topologically trivial, we imitate the simple
prescription established at the end of the previous section and
study the ensuing constraints. This is the route taken in
\Rcite{Gawedzki:2010rn} in the defect-free case. It was first
proposed by Jack \emph{et al.} in \Rcite{Jack:1989ne} and by Hull
and Spence in \Rcite{Hull:1989} for world-sheets without defects,
and by Figueroa-O'Farrill and Mohammedi in \Rcite{Figueroa:2005} for
those with boundary defects, where the circumstances were examined
in which the symmetry generated by the vector fields
$\,\Mup\xcK_a\,$ and $\,\Qup\xcK_a\,$ can be gauged. The constraints
for the possible coupling terms derived below agree, in the domain
of common applicability, with those deduced in the earlier papers,
however, the present method of derivation is slightly more
straightforward. \vskip 0.1cm

Let us recall, by way of a warm-up, the result of
\Rcite{Gawedzki:2010rn}, obtained by postulating a simple form of
the coupling of the standard Wess--Zumino amplitude in the absence
of defects,
\qq\nn
\xcA_{\rm WZ}(\varphi)=\Hol_\cG(\varphi)\,,
\qqq
to topologically trivial gauge fields, and by subsequently using the presumed
infinitesimal gauge-invariance
of the extended amplitude to fix its free parameters. More specifically,
we have
\berop\cite{Jack:1989ne,Hull:1989}\cite[Prop.\,3.1]{Gawedzki:2010rn}\label{prop:gequiv-ext-AWZ}
Consider a $\,\ggt^*$-valued 1-form and a
$\,\ggt^*\wedge\ggt^*$-valued function
\qq\nn
\a=\a_a\ox\tau^a\in\Om^1(M)\ox\ggt^*\,,\qquad\qquad\b=\b_{ab}\ox
\bigl(\tau^a\wedge\tau^b\bigr)\in\Om^0(M)\ox(\ggt^*\wedge\ggt^*)
\qqq
on $\,M$, \,written in the notation of Definition \ref{def:Gbas-form}.
\,An extension
\qq\nn
\widehat\xcA_{{\rm WZ}}(\varphi;\txA)=\ee^{\sfi\,\int_\Si\,
\phi^*\varsigma_\txA}\,\xcA_{{\rm WZ}}(\varphi)
\qqq
of $\,\xcA_{{\rm WZ}}(\varphi)$,\ expressed in terms of the map
$\,\phi =(\id_\Si,\varphi):\Si\to\Si\x M\,$ and of a 2-form
\qq\nn
\varsigma_\txA=\a_{a\,2^*}\wedge\txA_{1^*}^a-\tfrac{1}{2}\,
\b_{ab\,2^*}\,\txA_{1^*}^a\wedge\txA_{1^*}^b\,,
\qqq
on $\,\Si\x M\,$, \,is invariant under infinitesimal gauge
transformations
\qq\label{eq:inf-gtransfo}
\varphi\mapsto\varphi+\ic_{\ovl\La}\sfd\varphi\,,\qquad\qquad\txA
\mapsto\txA+\sfd\La-[\La,\txA]\,,
\qqq
with $\,\La=\La^at_a\,$ and $\,\ovl\La=\La^a\,\Mup\xcK_a\,$
for $\,\La^a\in
C^\infty(\Si,\bR)$,\, iff the following conditions:
\qq\nn
&\ic_{\ovl X}\txH=-\sfd\a(X)\,,\qquad\qquad\b(X\wedge Y)= \ic_{\ovl
X}\a(Y)\,,&\cr\cr
&\pLie{\ovl X}\a(Y)=\a\bigl([X,Y]\bigr)\,,\qquad\qquad\ic_{\ovl
X}\a(Y)=-\ic_{\ovl Y}\a(X)&
\qqq
are satisfied for all $\,X,Y\in\ggt$.
\eerop
\noindent It yields the all-important
\becor
In the notation of Proposition \ref{prop:gequiv-ext-AWZ}, the
formula
\qq\label{eq:WZamp-bulk}
\widehat\xcA_{{\rm
WZ}}(\varphi;\txA)=\ee^{\sfi\,\int_\Si\,\phi^*
\rho_\txA}\,\xcA_{{\rm WZ}}(\varphi)
\qqq
with $\,\rho_\txA\,$ given by \Reqref{eq:rhoA} defines an
infinitesimally gauge-invariant extension of $\,\xcA_{{\rm
WZ}}(\varphi)\,$ iff the 1-forms $\,\kappa_a\,$ satisfy conditions
\eqref{eq:HS-exact}, \eqref{eq:HS1-ids-triv} and
\eqref{eq:HS2-ids-triv}. \ecor Constraints \eqref{eq:HS-exact},
\eqref{eq:HS1-ids-triv} and \eqref{eq:HS2-ids-triv}, in conjunction
with the closedness of $\,\txH$,\ admit a number of interpretations.
Below, we point out two of them, a cohomological one and a
symplectic one. Two more, of a gerbe-theoretic and
generalised-geometric flavour, respectively, are postponed until
Sections \ref{sec:groupoid} and \ref{sec:infinit-equiv}, where we
come to discuss the infinitesimal equivariance of the string
background and the groupoid structure underlying rigid $\si$-model
symmetries. \brem Obstructions to gauging infinitesimal (rigid)
symmetries of the defect-free $\si$-model (or, equivalently, of
those of its symmetries that belong to the connected component of
the group unit in $\,\txG$) were first discussed in the framework of
$\ggt$-equivariant cohomology by Figueroa-O'Farrill and Stanciu in
Refs.\,\cite{Figueroa:1994ns,Figueroa:1994dj}, cf.\ also
Refs.\,\cite{Witten:1991mm,Wu:1993iia}. The upshot of their analysis
is the following
\berop\label{prop:gequiv-ext-H} Let $\,\txH\,$ be a smooth
closed $\txG$-invariant 3-form on a $\txG$-space $\,M$.\ Relations
\eqref{eq:HS-exact}, \eqref{eq:HS1-ids-triv} and
\eqref{eq:HS2-ids-triv} are satisfied iff $\,\txH\,$ admits a
$\ggt$-equivariantly closed extension
\qq\nn
\widehat\txH=\txH-\kappa\,,\qquad\qquad\kappa\in\Om^1(M)\ox\ggt^*
\qqq
given in terms of the $\ggt$-equivariant 1-form $\,\kappa=\kappa_a
\ox\tau^a\,$ in the Cartan model of the $\ggt$-equivariant
cohomology of $\,M$.
\eerop
\medskip
\noindent Above, the
Cartan model is taken with the $\ggt$-equivariant differential
defined on $\ggt$-equivariant $p$-forms $\,\eta$, polynomially dependent
on $\,X\in\ggt$, \,by the formula
\qq\nn
\widehat\sfd\eta(X)=\sfd\eta(X)-\ic_{\ovl X}\eta(X)\,.
\qqq
An equivalent description is provided
by the Weil model of the $\ggt$-equivariant cohomology.
\berop\cite[Lemma 3.9]{Gawedzki:2010rn}\label{prop:Weil-ext-H}
Let $\,M\,$ be a $\txG$-space with vector fields $\,\Mup\xcK_a\,$ as
in Definition \ref{eq:Gact-M}. Furthermore, let $\,\txA\,$ be a
$\ggt$-valued 1-form on an oriented two-dimensional manifold
$\,\Si$,\ and denote $\,\txF=\sfd
\txA+\txA\wedge\txA\in\Om^2(\Si)\ox\ggt$.\ The Weil transform
$\,\txH_\txA=\ee^{-\ovl{\txA_{1^*}}}.\widehat\txH_{2^*}(\txF_{1^*})\,$
of the $\ggt$-equivariantly closed extension $\,\widehat\txH\,$ of a
$\txG$-invariant 3-form $\,\txH\,$ on $\,M$,\ given in Proposition
\ref{prop:gequiv-ext-H}, reads
\qq\label{eq:txHA}
\txH_\txA=\txH_{2^*}+\sfd\rho_\txA\,,
\qqq
where $\,\rho_\txA\,$ is the 2-form on $\,\Si\x M\,$ defined in
\Reqref{eq:rhoA}. Hence, $\,\txH_\txA\,$ is the curvature 3-form of
the gerbe $\,\cG_\txA=\cG_{2^*}\ox I_{\rho_\txA}$.
\eerop
\erem\medskip

\brem The discussion of a symplectic realisation of the rigid
symmetries of the $\si$-model on its phase space was originally put
in the context of generalised geometry of
Refs.\,\cite{Hitchin:2004ut,Gualtieri:2003dx} by Alekseev and Strobl
in \Rcite{Alekseev:2004np}. The more complete picture,
incorporating, in particular, bi-brane data, is presented in
\Rcite{Suszek:2012}. Here, we merely cite
\berop\cite{Alekseev:2004np}\cite[Prop.\,3.10]{Suszek:2012}
The symplectic realisation, mentioned in Proposition
\ref{prop:sympl-real}, of the symmetry algebra $\,\ggt\,$ on the
phase space $\,\sfP_\si\,$ of the non-linear two-di\-men\-sion\-al
$\si$-model on a lorentzian world-sheet with an empty defect quiver
is hamiltonian iff condition \eqref{eq:HS1-ids-triv} is satisfied.
If, in addition, also \Reqref{eq:HS2-ids-triv} holds true, the
equitemporal Poisson bracket of the associated Noether currents is
anomaly-free and the resulting Poisson algebra of Noether currents
is isomorphic with $\,\ggt$.
\eerop
\erem\medskip

Prior to taking up the case of a non-empty defect quiver, we point
out that relations \eqref{eq:HS-exact}, \eqref{eq:HS1-ids-triv} and
\eqref{eq:HS2-ids-triv}, in conjunction with closedness of
$\,\txH$,\ imply further identities
\qq
\ic_a\ic_b\txH&=&-f_{abc}\,\kappa_c+\sfd\txc_{ab}
\,,\nonumber\\ \label{eq:impl-HS}\\
\ic_a\ic_b\ic_c\txH&=&-f_{bcd}\,\txc_{a
d}-f_{abd}\,\txc_{cd}+f_{acd}\,\txc_{bd}\,,\nonumber
\qqq
to be invoked later in our discussion.\bigskip

We may now reiterate the previous reasoning in the presence of a
defect quiver $\,\G$,\ and for a general background $\,\Bgt=(\cM,\cB
,\cJ)$,\ whereby the starting point is, this time, the
$\G$-corrected Wess--Zumino amplitude
\qq\label{eq:WZ-amp-def}
\xcA_{{\rm
WZ}}(\varphi\,\vert\,\G)=\Hol_{\cG,\Phi,(\varphi_n)}(\varphi\,
\vert\,\G)\,.
\qqq
Taking into account Proposition \ref{prop:gequiv-ext-AWZ}, we obtain
\berop\label{prop:gequiv-ext-AWZGam}
Let $\,\xcA_{{\rm WZ}}(\varphi\,\vert\,\G)\,$ be the Wess--Zumino
amplitude \eqref{eq:WZ-amp-def} defining the `topological' term of
the $\si$-model \eqref{eq:sigma} with the target as in Proposition
\ref{prop:gequiv-ext-AWZ}, and with a bi-brane $\,\cB=(Q,\iota_\a
,\om,\Phi )\,$ of a world-volume given by a $\txG$-space $\,Q\,$ and
a $\txG$-invariant curvature $\,\om$. \,Consider a $\,\ggt$-valued
function $\,\g=\g_a\ox\tau^a\in C^\infty(Q,\bR)\ox\ggt^*\,$ on
$\,Q$. \,In the notation of Proposition \ref{prop:gequiv-ext-AWZ},
an extension
\qq\label{eq:def-hol-gauge}
\widehat\xcA_{{\rm WZ}}\bigl[(\varphi\,\vert\,\G);\txA\bigr]=
\ee^{\sfi\,\int_\Si\,\phi^*\rho_\txA}\,\ee^{\sfi\,\int_\G\,(\phi
\vert_\G)^*\mu_\txA}\,\xcA_{{\rm WZ}}(\varphi\,\vert\,\G)
\qqq
of $\,\xcA_{{\rm WZ}}(\varphi\,\vert\,\G)$,\ expressed in terms of
the map $\,\phi=(\id_\Si,\varphi)$,\ of the 2-form $\,\rho_\txA\,$
given by \Reqref{eq:rhoA}, and of a 1-form
\qq\nn
\mu_\txA=-\g_{a\,2^*}\,\txA^a_{1^*}
\qqq
on $\,\Sigma\times Q$, \,is invariant under infinitesimal gauge transformations
\eqref{eq:inf-gtransfo}, written for
$\,\ovl\La=\La^a\,\xcFup\xcK_a\,$ with the fundamental vector fields
$\,\xcFup\xcK_a$, \,iff, in addition to Eqs.\,\eqref{eq:HS-exact},
\eqref{eq:HS1-ids-triv} and \eqref{eq:HS2-ids-triv},
the following conditions:
\qq\nn
&\ic_{\ovl X}\om+(\D_Q\kappa+\sfd\g)(X)=0\,,\qquad\qquad\D_{T_n}\g(X)
=0\,,&\cr\cr
&\pLie{\ovl X}\g(Y)=\g\bigl([X,Y]\bigr)&
\qqq
are satisfied for all $\,X,Y\in\ggt$.
\eerop
\beroof
Writing out the variation of the extension under transformation
\eqref{eq:inf-gtransfo}, we find, in addition to the bulk term
whose vanishing is ensured by the argument of Proposition
\ref{prop:gequiv-ext-AWZ}, a collection of terms localised on
$\,\G$,
\qq\nn
\int_\G\,(\phi\vert_\G)^*\bigl[\La_{1^*}^a\,(\ic_a\om
)_{2^*}-\La_{1^*}^a\,(\unl\ic_a\unl\D_Q\rho_\txA)-
\La_{1^*}^a\,(\ic_a\sfd\g_b)_{2^*}\wedge\txA^b_{1^*}\bigr]
\cr\cr
+\int_\G\,(\phi\vert_\G)^*\bigl[\La^a_{1^*}\,\unl\D_Q\bigl(
\kappa_{a\,2^*}+\txc_{ab\,2^*}\,\txA^b_{1^*}\bigr)+\La^a_{1^*}\,
\sfd\g_{a\,2^*}+\g_{2^*}\bigl([\La,\txA]_{1^*}\bigr)\bigr]\cr\cr
-\sum_{\jmath\in\Vgt_\G}\,\La^a_{1^*}\,\bigl(\D_{T_{n_\jmath}}\g_a
\bigr)_{2^*}(\jmath)\,,
\qqq
with $\,\unl\D_Q=(\id_\G\times\iota_2)^*-(\id_\G\times\iota_1)^*\,$
and $\,\Qup\unl\xcK_a(\si,q)=\Qup\xcK_a(q)$,\ and with
$\,\unl\ic_a=\ic_{\Qup\unl\xcK_a}$.\ In the above expression, the
first term is the defect contribution to \Reqref{eq:var-act-gen},
the second one comes from the exact term in the variation of the
bulk extension $\,\int_\Si\,\phi^*\rho_\txA$,\ the third one is the
variation of the target-space tensors in the defect extension
$\,\int_\G\,(\phi \vert_\G)^*\mu_\txA$,\ the fourth one is produced
- upon integration by parts - by the summand of the variation of the
bulk extension sourced by the gauge transformation of the connection
1-form $\,\txA\,$ that is proportional to $\,\sfd\La$,\ and the
penultimate one represents the infinitesimal transformation of the
gauge field in the defect extension. Finally, the contribution from
the defect junctions is the boundary term of the integration by
parts of the gauge-field variation in the defect extension. Using
the arbitrariness of the embedding map $\,\varphi\,$ and that of the
gauge potential $\,\txA$,\ we may next set to zero separately the
terms independent of and linear in $\,\txA$,\ whereupon we obtain
the following set of constraints:
\qq\nn
&\ic_a\om+\D_Q\kappa_a+\sfd\g_a=0\,,\qquad\qquad
\D_{T_{n_\jmath}}\g_a=0\,,&\cr\cr
&\pLie{a}\g_b=f_{abc}\,\g_c\,,&
\qqq
which proves the claim. \eroof\bigskip \noindent We are thus led to
the conclusion
\becor\label{cor:infinit-Ginv-act}
In the notation of Proposition \ref{prop:gequiv-ext-AWZGam}, the
formula
\qq\label{eq:ampl.with.defects}
\widehat\xcA_{{\rm WZ}}\bigl[(\varphi\,\vert\,\G);\txA\bigr]=
\ee^{\,\sfi\,\int_\Si\,\phi^*\rho_\txA}\,\ee^{\,\sfi\,\int_\G\,(\phi
\vert_\G)^*\la_\txA}\,\xcA_{{\rm WZ}}(\varphi\,\vert\,\G)
\qqq
with $\,\rho_\txA\,$ and $\,\la_\txA\,$ given by \Reqref{eq:rhoA}
and \Reqref{eq:laA}, respectively, defines an infinitesimally
gauge-invariant extension of $\,\xcA_{{\rm WZ}}(\varphi\,\vert\,\G
)\,$ in the presence of a defect quiver $\,\G\,$ embedded in
$\,\Si\,$ iff the 1-forms $\,\kappa_a$,\ introduced by
\Reqref{eq:HS-exact}, and the functions $\,k_a$,\ defined by
Eqs.\,\eqref{eq:bdry-exact}-\eqref{eq:junct-exact}, satisfy
conditions \eqref{eq:HS1-ids-triv}-\eqref{eq:HS2-ids-triv}.
In other words, the latter conditions assure the absence of
local gauge anomalies in the amplitudes \eqref{eq:ampl.with.defects}. \ecor

\noindent Similarly to the defect-free case, the new conditions for
a consistent gauging admit a simple cohomological, canonical,
gerbe-theoretic and generalised-geometric interpretation. As
previously, we focus on the first two, relegating the latter two to
Sections \ref{sec:groupoid} and \ref{sec:infinit-equiv}. \brem The
constraints arising in the presence of boundary defects (equivalent
to boundaries of an open world-sheet) were rephrased in the language
of the $\ggt$-equivariant cohomology of the world-volume $\,Q\,$ of
the corresponding boundary bi-brane (or D-brane) by
Figueroa-O'Farrill and Mohammedi in \Rcite{Figueroa:2005}. Here, we
extend their results to arbitrary defect quivers embedded in a
closed world-sheet.
\berop\label{prop:gequiv-ext-om} Let $\,\xcF=M\sqcup Q\sqcup T\,$ be
a $\txG$-space satisfying the assumptions of Proposition
\ref{prop:Gnat-IBB-maps}, and let $\,\om\,$ be a smooth
$\txG$-invariant 2-form on $\,Q\,$ obeying relations
\eqref{eq:triv-restr} and \eqref{eq:triv-restr2}, the former with
respect to a 3-form $\,\txH\,$ on $\,M\,$ with properties listed in
Proposition \ref{prop:gequiv-ext-H}. Relations
\eqref{eq:bdry-exact}-\eqref{eq:junct-exact} and
\eqref{eq:FFM-ids-triv} are satisfied iff $\,\om\,$ admits a
$\ggt$-equivariant extension
\qq\nn
\widehat\om=\om-k\,,\qquad\qquad k\in\Om^0(Q)\ox\ggt^*,
\qqq
given in terms of a $\ggt$-equivariant 0-form $\,k=k_a\ox\tau^a\,$
in the Cartan model of the $\ggt$-equivariant cohomology, that
satisfies the identities
\qq
&\widehat\sfd\widehat\om=-\D_Q\widehat\txH\,,&
\label{eq:hatdom-hatdelHa}\\\cr
&\D_{T_n}\widehat\om=0&\,.\label{eq:hatdelom-nil}
\qqq
\eerop
\beroof
Obvious, through inspection. \eroof \bigskip\noindent In the Weil
model of the $\ggt$-equivariant cohomology, we find
\berop\label{prop:Weil-ext-om}
In the notation of Proposition \ref{prop:Weil-ext-H} and under the
assumptions of Proposition \ref{prop:gequiv-ext-om}, the Weil
transform $\,\om_\txA=\ee^{-\ovl{\txA_{1^*}}}.\widehat\om_{2^*}(
\txF_{1^*})\,$ of $\,\widehat\om\,$ reads
\qq\label{eq:omA}
\om_\txA=\om_{2^*}-\unl\D_Q\rho_\txA+\sfd\la_\txA\,.
\qqq
It satisfies the identities
\qq\nn
\sfd\om_\txA=-\unl\D_Q\txH_\txA\,,\qquad\qquad\unl\D_{T_n}\om_\txA=0
\,,
\qqq
written in terms of $\,\unl\D_{T_n}=\sum_{k=1}^n\,\vep_n^{k,k+1}\,
\bigl(\id_{\Vgt_\G}\times\pi_n^{k,k+1}\bigr)^*$.
\eerop
\beroof
Obvious, through inspection. \eroof\erem\bigskip

\brem The canonical description of the $\si$-model in the presence
of conformal defects was set up in \Rcite{Suszek:2011hg}, and its
symmetries were subsequently studied at length in
\Rcite{Suszek:2012}. The analysis yields
\berop\cite[Props.\,3.10 \& 5.11]{Suszek:2012}
The symplectic realisation, mentioned in Proposition
\ref{prop:sympl-real}, of the symmetry algebra $\,\ggt\,$ on the
phase space $\,\sfP_\si\,$ of the non-linear two-dimensional
$\si$-model on a lorentzian world-sheet with an embedded defect
quiver is hamiltonian iff conditions \eqref{eq:HS1-ids-triv} and
\eqref{eq:FFM-ids-triv} are satisfied.
\eerop
\noindent One can also interpret relation \eqref{eq:junct-exact} as
an intertwiner condition for symmetry-generating hamiltonians on
twisted multi-string phase spaces, cf.\
\Rxcite{Thm.\,6.4}{Suszek:2012}.\erem\medskip

The hitherto findings can be phrased succinctly as
\becor\label{cor:triv-Ansatz}
The Feynman amplitudes of the non-linear $\si$-model on a closed
oriented world-sheet $\,(\Si,\g)$,\ with an embedded defect quiver
$\,\G$,\ coupled to the topologically trivial gauge field $\,\txA\,$
on $\,\Si\,$ take the form
\qq\qquad\qquad\label{eq:gauges-sigmod}
{\xcA}\bigl[(\varphi\,\vert\,\G);\txA,\g\bigr]=\exp\Big[-\tfrac{1}{2}
\,\int_\Si\,\txg_\txA\bigl(\sfd\phi\overset{\wedge}{,}\star_\g\sfd
\phi\bigr)\Big]\,\Hol_{\cG_\txA,\Phi_\txA,(\varphi_{n\,\txA})}
(\phi\,\vert\,\G)\,,
\qqq
where $\,\phi=(\id_\Si,\varphi)$,\ and the extended string
background $\,\Bgt_\txA=(\cM_\txA,\cB_\txA,\cJ_\txA)\,$ defining the
holonomy has components:
\bit
\item the extended target $\,\cM_\txA\,$ composed of the target
space $\,\Si\x M\,$ with metric $\,\txg_\txA\,$ of
\Reqref{eq:gA-def} and gerbe $\,\cG_\txA=\cG_{2^*}\ox
I_{\rho_\txA}$,\ with $\,\rho_\txA\,$ as in \Reqref{eq:rhoA}, of
curvature $\,\txH_\txA\,$ given by \Reqref{eq:txHA};
\item the extended bi-brane $\,\cB_\txA\,$ with world-volume
$\,\G\x Q$,\ bi-brane maps $\,\unl\iota_\a=\id_\G\x\iota_\a,\ \a=
1,2$,\ curvature $\,\om_\txA\,$ given by \Reqref{eq:omA}, and the
1-isomorphism $\,\Phi_\txA=\Phi_{2^*}\ox J_{\la_\txA}$,\ where
$\,\la_\txA\,$ is as in \Reqref{eq:laA};
\item the extended inter-bi-brane $\,\cJ_\txA\,$ with
component world-volumes $\,\Vgt_\G^{(n)}\x T_n,\ n\geq 3$,\ defined
in terms of the subsets $\,\Vgt_\G^{(n)}\subset\Vgt_\G\,$ composed
of vertices of valence $n$,\ with inter-bi-brane maps
$\,\unl\pi_n^{k,k+1}=\id_{\Vgt_\G^{(n)}}\x\pi_n^{k,k+1},\
k=1,2,\ldots,n$,\ and 2-isomorphisms
$\,\varphi_{n\,\txA}:=\varphi_{n\, 2^*}$.
\eit
Amplitudes \eqref{eq:gauges-sigmod} are invariant
under infinitesimal gauge transformations
\eqref{eq:inf-gtransfo}.
\ecor

\section{Large gauge transformations in the topologically trivial sector}
\label{sec:large}

\noindent In the preceding sections, we consistently gauged rigid
symmetries of the $\si$-model target in a way that renders the
$\si$-model invariant under infinitesimal gauge transformations.
Invariance under the latter ensures also invariance under the
so-called `small' gauge transformations $\,\chi:\Si\to\txG\,$ that
can be continuously deformed to the identity (and, thus, necessarily
take values in the connected component of the group unit of
$\,\txG$). For general $\,\txG\,$ (in particular, whenever the group
manifold is not connected or not simply connected, or both), we are
still left with the task of incorporating large gauge
transformations (not homotopic to the identity transformation) into
the formalism developed above. To accommodate non-connected groups,
we shall assume that the $\,\ggt^*$-valued 1-form $\,\kappa=
\kappa_a\ox\tau^a\,$ on $\,M\,$ and the $\,\ggt^*$-valued function
$\,k=k_a\ox\tau^a\,$ on $\,Q\,$ are not only $\,\ggt$-equivariant,
as required by Eqs.\,(\ref{eq:HS1-ids-triv}) and
(\ref{eq:FFM-ids-triv}), but $\,\txG$-equivariant. Let us carry out
a detailed analysis of invariance properties of the gauged
decorated-surface holonomy (the invariance of the minimally coupled
`metric' term has already been demonstrated in the proof of
Proposition \ref{prop:Gau-inv-triv}). We shall consider arbitrary
group-valued gauge transformations $\,\chi\in C^\infty(\Si,\txG)\,$
(for the trivial principal $\txG$-bundle) that lift to the product
space $\,\Si\x\xcF\,$ as in \Reqref{eq:gauge-trans-ext}. The gauge
transformations of the $\si$-model field and those of the gauge
field are given by Eqs.\,(\ref{eq:chiphi}) and (\ref{eq:chiA}),
respectively. The attendant gauge transformations of the gauge-field
strength read
\qq\label{eq:trafo-AF}
\txF\mapsto\Ad_\chi\txF=:{}^{\tx{\tiny $\chi$}}\hspace{-1pt}\txF\,.
\qqq
The following auxiliary result will prove useful.
\belem
The Weil transforms $\,\txH_\txA\in\Om^3(\Si\x M)\,$ and $\,\om_\txA
\in\Om^2(\G\x Q)$,\ defined in Eqs.\,\eqref{eq:txHA} and
\eqref{eq:omA}, respectively, transform under $\,L_\chi\,$ of
\Reqref{eq:gauge-trans-ext} as
\qq
L_\chi^*\txH_\txA&=&\txH_{{}^{\tx{\tiny $\chi^{-1}$}}\hspace{-2pt}
\txA}\,,\label{eq:trafo-HA}\\\cr
L_\chi^*\om_\txA&=&\om_{{}^{\tx{\tiny $\chi^{-1}$}}\hspace{-2pt}
\txA}\,.\label{eq:trafo-omA}
\qqq
\elem
\beroof
The first of the two relations, \Reqref{eq:trafo-HA}, was proven in
\Rxcite{Lemma 4.1}{Gawedzki:2010rn}. We shall therefore restrict
ourselves to proving the second one. The proof is conceptually
analogous to that of \Reqref{eq:trafo-HA}. Invoking Proposition
\ref{prop:pulls-reex}, in conjunction with Eqs.\,(\ref{eq:chiA}) and
(\ref{eq:trafo-AF}), we
find
\qq\nn
&&L_\chi^*\om_\txA(\si,q)\ =\
\om_\txA\bigl(\si,\chi(\si).q\bigr)=
\bigl(\ee^{-\ovl{\chi^*\theta_L(\si)}}.\ell_{\chi(\si)}^*\om_\txA(
\si,\cdot)\bigr)(q)\cr\cr
&&\equiv\
\bigl\{\ee^{-\ovl{\chi^*\theta_L(\si)}}.\ell_{\chi(\si)}^*
\ee^{-\ovl{\txA(\si)}}.\bigl[\om-k\bigl(\txF(\si)\bigr)\bigr]
\bigr\}(q)=\bigl\{\ee^{-\ovl{{}^{\tx{\tiny $\chi^{-1}$}}
\hspace{-2pt}\txA(\si)}}.\bigl[\om-k\bigl({}^{\tx{\tiny $\chi^{-
1}$}}\hspace{-2pt}\txF(\si)\bigr)\bigr]\bigr\}(q)\cr\cr
&&\equiv\ \om_{{}^{\tx{\tiny $\chi^{-1}$}}\hspace{-2pt}\txA}(\si,q)
\,.
\qqq
\eroof \bigskip \noindent The significance of the above
transformation laws stems from the following observations: In the
defect-free case, the gauge invariance of the `topological' term of
the $\si$-model is equivalent to the equality between the
transformed Wess--Zumino amplitude
\qq\label{eq:trafo-WZ-bulk}
\widehat\xcA_{{\rm WZ}}\bigl(\chi.\varphi;{}^{\tx{\tiny $\chi$}}
\hspace{-2pt}\txA\bigr)=\Hol_{\cG_{{}^{\tx{\tiny $\chi$}}
\hspace{-2pt}\txA}}(L_\chi\circ\phi)=\Hol_{L_\chi^*\cG_{{}^{\tx{\tiny
$\chi$}} \hspace{-2pt}\txA}}(\phi)
\qqq
and the one before the transformation,
\qq\nn
\widehat\xcA_{{\rm WZ}}(\varphi;\txA)=\Hol_{\cG_\txA}(\phi)\,.
\qqq
This leads us to compare the gerbes $\,L_\chi^*\cG_{{}^{\tx{\tiny
$\chi$}}\hspace{-2pt}\txA}\,$ and $\,\cG_\txA\,$ over $\,\Si\x M$,\
and \Reqref{eq:trafo-HA}, stating the equality of the respective
curvatures, indicates that the two gerbes are related.
\berop\cite[Prop.\,4.2]{Gawedzki:2010rn}\label{prop:flat-gerbe-D}
Let $\,\cD\,$ be the flat gerbe determined, alongside a canonical
1-isomorphism
\qq\nn
\Mup\ell^*\cG\xrightarrow{\cong}\cD\ox\cG_{2^*}\ox I_\rho\,,
\qqq
by \Reqref{eq:delH-rho} and Proposition \ref{prop:torsors}, and
let $\,\cG_\txA\,$ be the
gerbe defined in Corollary \ref{cor:triv-Ansatz}. Finally, let
$\,L_\chi\,$ be the map given by \Reqref{eq:gauge-trans-ext}. Then,
the gerbes $\,L_\chi^* \cG_{{}^{\tx{\tiny
$\chi$}}\hspace{-2pt}\txA}\,$ and $\,\cG_\txA\ox
(\chi\times\id_M)^*\cD\,$ are 1-isomorphic.
\eerop
\noindent The last proposition, taken together with
\Reqref{eq:trafo-WZ-bulk}, readily implies
\bethe\label{thm:glanbk}\cite[Thm.\,4.3]{Gawedzki:2010rn}
Let $\,\widehat\xcA_{{\rm WZ}}(\varphi;\txA)\,$ be the gauged
Wess--Zumino amplitude of the defect-free $\si$-model, introduced in
Corollary \ref{cor:triv-Ansatz}. Then, for any $\,\chi:\Si\to
\txG$,\ we have
\qq\label{eq:glanbk}
\widehat\xcA_{{\rm WZ}}\bigl(\chi.\varphi;{}^{\tx{\tiny $\chi$}}
\hspace{-2pt}\txA\bigr)=\widehat\xcA_{{\rm WZ}}(\varphi;\txA)\cdot
\Hol_\cD\bigl((\chi,\varphi)\bigr)\,.
\qqq
\ethe
\vskip 0.2cm

\noindent Taking, in identity \eqref{eq:glanbk}, the gauge field
$\,\txA=\chi^*\theta_L\,$ for which $\,{}^{\tx{\tiny $\chi$}}
\hspace{-2pt}\txA=0$,\ one infers
\becor \ The holonomy of the flat gerbe $\,\cD\,$ from Proposition
\ref{prop:flat-gerbe-D} satisfies the identity
\qq\label{eq:glanbk1}
\Hol_\cD\bigl((\chi,\varphi)\bigr)\ =\ \Hol_{\cG}(\chi.\varphi)\,\,
\Hol_{\cG}(\varphi)^{-1}\,\,\ee^{-\,\sfi\,\int_\Si\,\phi^*\rho_{\chi^*\theta_L}}\,.
\qqq
\ecor \noindent The above relation was used in
\Rcite{Gawedzki:2010rn} to identify the global gauge anomalies in
the examples of gauged $\si$-models without defects, cf.\ also
Section \ref{sec:eg-backgrnd} below. \vskip 0.2cm

\noindent Theorem \ref{thm:glanbk} has an important
\becor\cite[Cor.\,4.5]{Gawedzki:2010rn}\label{cor:equiv-1iso}
The gauged Wess--Zumino amplitude $\,\widehat\xcA_{{\rm WZ}}(
\varphi;\txA)\,$ of the defect-free $\si$-model, introduced in
Corollary \ref{cor:triv-Ansatz}, is gauge invariant iff there exists
a 1-isomorphism
\qq\nn
\Upsilon\ :\ \Mup\ell^*\cG\xrightarrow{\cong}\cG_{2^*}\ox I_\rho
\qqq
for $\,\rho\,$ as in \Reqref{eq:rho-on-G}. It yields, in particular,
the 1-isomorphism
\qq\label{Upsilon}
(\chi\times\id_M)^*\Upsilon\ :\ L_\chi^*\cG_{{}^{\tx{\tiny $\chi$}}
\hspace{-2pt}\txA}\xrightarrow{\cong}\cG_\txA\,.
\qqq\ecor
\bigskip

We may next follow a similar reasoning in the case
of a defect quiver with no junctions, $\,\Vgt_\G=\emptyset$.\ Here,
we compare
\qq\label{eq:trafo-WZ-dcircles}
\widehat\xcA_{{\rm WZ}}\bigl[(\chi.\varphi\,\vert\,\G);
{}^{\tx{\tiny $\chi$}}\hspace{-2pt}\txA\bigr]=
\Hol_{\cG_{{}^{\tx{\tiny $\chi$}}\hspace{-2pt}\txA},
\Phi_{{}^{\tx{\tiny $\chi$}}\hspace{-2pt}\txA}}(L_\chi\circ\phi\,
\vert\,\G)=\Hol_{L_\chi^*\cG_{{}^{\tx{\tiny $\chi$}}\hspace{-2pt}
\txA},L_\chi^*\Phi_{{}^{\tx{\tiny $\chi$}}\hspace{-2pt}\txA}}(\phi
\,\vert\,\G)
\qqq
with
\qq\nn
\widehat\xcA_{{\rm WZ}}\bigl[(\varphi\,\vert\,\G);\txA\bigr]=
\Hol_{\cG_\txA,\Phi_\txA}(\phi\,\vert\,\G)\,.
\qqq
The bulk analysis gives us the 1-isomorphism $\,(\chi\times\id_M)^*
\Upsilon$,\ and so, in the light of Proposition \ref{prop:hol-inv},
we are led to inspect the transformed bi-brane 1-isomorphism
$\,L_\chi^*\cB_{{}^{\tx{\tiny $\chi$}}\hspace{-2pt}\txA}$.\ For this
purpose, we decompose $\,L_\chi$,\ similarly as in
\Rcite{Gawedzki:2010rn}, into
\qq\nn
L_\chi\ :\ \G\x Q&\xrightarrow{K_\chi}&\G\x\txG\x Q \
\,\xrightarrow{\id_\G\times\Qup\ell}\ \,\G\x Q\cr \
(\si,q)&\longmapsto&
\left(\si,\chi(\si),q\right)\quad\longmapsto\quad
\left(\si,\chi(\si).q\right)
\qqq
and compute
\qq\label{eq:ellPhiA-decomp}
(\id_\G\times\Qup\ell)^*\Phi_\txA=\bigl(\Qup\ell^*\Phi\bigr)_{[2,3]^*}
\ox J_{\bigl(\Qup\ell^*\la\bigr)_\txA}\,,
\qqq
whereupon we take a closer look at the first factor of the tensor
product. Consider the pair of gerbes
\qq\label{eq:Qella-ends}
\Qup\ell^*\iota_1^*\cG\equiv\iota_1^{(1)\,*}\Mup\ell^*\cG\,,\qquad
\qquad\Qup\ell^*\iota_2^*\cG\ox I_{\Qup\ell^*\om}\equiv\iota_2^{(1)
\,*}\Mup\ell^*\cG\ox I_{\Qup\ell^*\om}\,,
\qqq
the equalities following from the assumed $\txG$-equivariance of the
bi-brane maps. Recalling \Reqref{eq:dla}, we readily convince
ourselves that both 1-isomorphisms $\,\Qup\ell^*\Phi\,$ and
$\,\bigl(\bigl(\iota_2^{(1)
\,*}\Upsilon^{-1}\ox\id_{I_{\Qup\ell^*\om-\sfd\la}}\bigr)\circ
\bigl(\Phi_{2^*}\ox\id_{I_{\iota_1^{(1)\,*}\rho}}\bigr)\circ
\iota_1^{(1)\,*}\Upsilon\bigr)\ox J_\la\,$ map the first of the two
gerbes in \Reqref{eq:Qella-ends} into the second one. Thus, in
virtue of Proposition \ref{prop:torsors}, there exist a flat line
bundle $\,D\xrightarrow{\pi_D}\txG\x Q\,$ and a 2-isomorphism
\qq\label{eq:triv-bund-corr}\qquad\qquad
\psi\ :\ \Qup\ell^*\Phi\xLongrightarrow{\cong}D\ox
\bigl(\bigl(\iota_2^{(1)\,*}\Upsilon^{-1}\ox\id\bigr)\circ\bigl(
\Phi_{2^*}\ox\id\bigr)\circ\iota_1^{(1)\,*}\Upsilon\bigr)\ox J_\la
\,.
\qqq
We are now ready to formulate the important
\berop\label{prop:flat-bundl-circle}
Let $\,D\,$ and $\,\psi\,$ be the flat line bundle and the
2-isomorphism over $\,\txG\x Q\,$ from \Reqref{eq:triv-bund-corr},
respectively, and let $\,\Phi_\txA\,$ be the 1-isomorphism on
$\,\G\x Q\,$ defined in Corollary \ref{cor:triv-Ansatz}. Finally,
let $\,L_\chi\,$ be the map given by \Reqref{eq:gauge-trans-ext}.
Then, we have the 2-isomorphism
\qq\nn
&&\hspace{-0.3cm}(\chi\times\id_Q)^*\psi\ox\id\ :\ L_\chi^*\Phi_{{}^{\tx{\tiny $\chi$}}
\txA}\xLongrightarrow{\cong}\cr\cr
&&\hspace{0.3cm}
(\chi\times\id_Q)^*D\ox\bigl[\bigl(\unl\iota_2^{(1)\,*}
(\chi\times\id_M)^*
\Upsilon^{-1}\ox\id\bigr)\circ\bigl(\Phi_\txA\ox\id\bigr)
\circ\unl\iota_1^{(1)\,*}(\chi\times\id_M)^*\Upsilon\bigr]\,,
\qqq
written in terms of the maps $\,\unl\iota_\a^{(1)}=\id_\G\times
\iota_\a,\ \a=1,2$.
\eerop
\beroof
Putting together Eqs.\,\eqref{eq:ellPhiA-decomp} and
\eqref{eq:triv-bund-corr}, as well as the identity
\qq\nn
\pr_{2,3}\circ K_\chi=\chi\times\id_Q\,,
\qqq
we obtain
\qq\nn
&&\hspace{-0.3cm}(\chi\times\id_Q)^*\psi\ox\id_{K_\chi^*J_{\bigl(\Qup\ell^*\la
\bigr)_{{}^{\tx{\tiny $\chi$}}\hspace{-2pt}\txA}}}=K_\chi^*\bigl(
\psi_{[2,3]^*}\ox\id_{J_{\bigl(\Qup\ell^*\la\bigr)_{{}^{\tx{\tiny
$\chi$}}\hspace{-2pt}\txA}}}\bigr)\ :\ L_\chi^*\Phi_{{}^{\tx{\tiny
$\chi$}}\txA}\xLongrightarrow{\cong}\cr\cr
&&\hspace{0.2cm}
(\chi\times\id_Q)^*D
\ox\bigl[\bigl((\chi\times\id_Q)^*\iota_2^{(1)\,*}\Upsilon^{-1}\ox\id
\bigr)\circ\bigl(\Phi_{2^*}\ox\id\bigr)\circ(\chi\times\id_Q)^*
\iota_1^{(1)\,*}\Upsilon\bigr]\cr\cr
&&\hspace{5.9cm}\ox J_{(\chi\times\id_Q)^*\la+K_\chi^*
\bigl(\Qup\ell^*\la\bigr)_{{}^{\tx{\tiny $\chi$}}\hspace{-2pt}\txA}}\cr\cr
&&\hspace{-0.15cm}=(\chi\times\id_Q)^*D\ox\bigl[\bigl(
\unl\iota_2^{(1)\,*}(\chi\times\id_M)^*\Upsilon^{-1}\ox\id\bigr)\circ
\bigl(\Phi_{2^*}\ox\id\bigr)\circ\unl\iota_1^{(1)\,*}(\chi\times\id_M)^*
\Upsilon\bigr]\cr\cr
&&\hspace{8.8cm}\ox J_{(\chi\times\id_Q)^*\la+L_\chi^*\la_{{}^{\tx{\tiny
$\chi$}}\hspace{-2pt}\txA}}\,.
\qqq
At this stage, it remains to verify the identity
\qq\nn
(\chi\times\id_Q)^*\la+L_\chi^*\la_{{}^{\tx{\tiny $\chi$}}\hspace{-2pt}
\txA}=\la_\txA\,.
\qqq
The latter follows upon summing up the right-hand sides of
\qq\nn
(\chi\times\id_Q)^*\la(\si,q)=\la\bigl(\chi(\si),q\bigr)=-k\bigl(\chi^*
\theta_L(\si)\bigr)(q)
\qqq
and
\qq\nn
&&\hspace{-0.2cm}L_\chi^*\la_{{}^{\tx{\tiny $\chi$}}\hspace{-2pt}\txA}(\si,q)\cr\cr
&&\hspace{-0.2cm}=\
\la_{{}^{\tx{\tiny $\chi$}}\hspace{-2pt}\txA}\bigl(\si,\chi(\si).q
\bigr)=\bigl(\ee^{-\ovl{\chi^*\theta_L(\si)}}.\Qup\ell_{\chi(\si
)}^*\la_{{}^{\tx{\tiny $\chi$}}\hspace{-2pt}\txA}(\si,\cdot)\bigr)(q
)\equiv-\bigl[\ee^{-\ovl{\chi^*\theta_L(\si)}}.\Qup\ell_{\chi(\si
)}^*k\bigl({}^{\tx{\tiny $\chi$}}\hspace{-2pt}\txA(\si)\bigr)
\bigr](q)\cr\cr
&&\hspace{-0.2cm}=\ -k\bigl(\Ad_{\chi(\si)^{-1}}{}^{\tx{\tiny$\chi$}}\hspace{-2pt}
\txA(\si)\bigr)(q)=\la_\txA(\si,q)+k\bigl(\chi^*\theta_L(\si)\bigr)
(q)\,,
\qqq
where in the last computation we used Proposition
\ref{prop:pulls-reex} and, subsequently, \Reqref{eq:chiA}.\eroof
\bigskip

\noindent It is now straightforward to demonstrate the validity of
\bethe\label{thm:glanbd}
Let $\,\widehat\xcA_{{\rm WZ}}[(\varphi\,\vert\,\G);\txA]\,$ be the
gauged Wess--Zumino amplitude, introduced in Corollary
\ref{cor:triv-Ansatz}, of the $\si$-model for network-field
configurations $\,(\varphi\,\vert\,\G)\,$ for a defect quiver
$\,\G\,$ with no junctions. Suppose that there exists a
1-isomorphism $\,\Upsilon\,$ defined in Corollary
\ref{cor:equiv-1iso}, and let $\,D\to\txG\x Q\,$ be the flat line
bundle given in \Reqref{eq:triv-bund-corr}. Then, for any
$\,\chi:\Si\to\txG$,\ we have
\qq\label{eq:glanbd}
\widehat\xcA_{{\rm WZ}}\bigl[(\chi.\varphi\,\vert\,\G);{}^{\tx{\tiny
$\chi$}} \hspace{-2pt}\txA\bigr]=\widehat\xcA_{{\rm WZ}}\bigl[(
\varphi\,\vert\,\G);\txA\bigr]\cdot\Hol_D\bigl((\chi,\varphi)
\vert_\G\bigr)\,.
\qqq \ethe
\beroof
Putting together Corollary \ref{cor:equiv-1iso}, Proposition
\ref{prop:flat-bundl-circle} and Proposition \ref{prop:hol-inv}, we
immediately conclude that the sole thing that remains to be proven
is the equality between the standard line-bundle holonomy $\,\Hol_D
\bigl((\chi,\varphi)\vert_\G\bigr)\,$ and the holonomy-like defect
contribution, as derived in \Rxcite{Sec.\,2.7}{Runkel:2008gr}, of
$\,(\chi\times\id_Q)^*D\,$ to the gauge-transformed Wess--Zumino
amplitude. To this end, we recall the explicit description of the
inverse functor $\,{\rm Bun}^{-1}\,$ given in
\Rxcite{Sec.\,2.5}{Waldorf:2007phd}. In the geometric language of
the thesis, the latter assigns to the (flat) bundle $\,D\,$ over
$\,\txG\x Q\,$ the 1-isomorphism $\,I_0\to I_0\,$ defined by the
trivial surjective submersion $\,\id_{\txG\x Q}$, \,the same bundle
$\,D\,$ over it, and the trivial isomorphism $\,\id_D\,$ as the
bundle isomorphism compatible with the (trivial) groupoid structure
on the fibres of the (trivial) bundle of $\,I_0$.\ The equality now
follows by construction. \eroof \vskip 0.2cm

\noindent Taking, in identity \eqref{eq:glanbd}, the gauge field
$\,\txA=\chi^*\theta_L\,$ for which $\,{}^{\tx{\tiny $\chi$}}
\hspace{-2pt}\txA=0$, \,one infers
\becor
The holonomy of the flat line bundle $\,D\,$ from Proposition
\ref{prop:flat-bundl-circle} satisfies the identity
\qq\label{eq:glanbd1}
\Hol_D\bigl((\chi,\varphi)|_\Gamma\bigr)\ =\
\Hol_{\cG,\Phi}\bigl(\chi.\varphi|\Gamma)\,\,
\Hol_{\cG,\Phi}\bigl(\varphi|\Gamma\bigr)^{-1}\,
\ee^{-\,\sfi\int_{\Sigma\setminus\Gamma}
\phi^*\rho_{\chi^*\theta_L}\,-\,
\sfi\int_\Gamma(\phi|_\Gamma)^*\lambda_{\chi^*\theta_L}}.
\qqq
\ecor
\vskip 0.2cm

\noindent We shall use the above relation to identify the
contribution to the global gauge anomaly from circular defects in
examples of gauged $\,\si$-models considered in Section
\ref{sec:eg-backgrnd}.

\medskip \noindent Theorem \ref{thm:glanbd} is at the basis of the
following result:
\berop\label{prop:equiv-2iso}
The gauged Wess--Zumino amplitude $\,\widehat\xcA_{{\rm
WZ}}(\varphi; \txA\,\vert\,\G)\,$ of the $\si$-model for
network-field configurations $\,(\varphi\, \vert\,\G)\,$ with a
defect quiver $\,\G\,$ without defect junctions is gauge invariant
iff there exist: a 1-isomorphism
\qq\nn
\Upsilon\ :\ \Mup\ell^*\cG\xrightarrow{\cong}\cG_{2^*}\ox I_\rho
\qqq
with $\,\rho\,$ as in \Reqref{eq:rho-on-G}, and a 2-isomorphism
\qq\label{Xi}
\Xi\ :\ \Qup\ell^*\Phi\xLongrightarrow{\cong}\bigl(\bigl(\iota_2^{(1
)\,*}\Upsilon^{-1}\ox\id\bigr)\circ\bigl(\Phi_{2^*}\ox\id\bigr)
\circ\iota_1^{(1)\,*}\Upsilon\bigr)\ox J_\la\,,
\qqq
with $\,\la\,$ as in \Reqref{eq:la-on-G}.\eerop
\beroof
The holonomy, along $\,(\chi,\varphi)$,\ of the flat bundle $\,D\,$
over $\,\txG\x Q\,$ that obstructs the desired equality of the
Wess--Zumino amplitudes is trivial if $\,D\,$ is isomorphic to the
trivial bundle $\,J_0$.\ The latter condition, by virtue of the
definition of $\,D$, \,is equivalent to the existence of $\,\Xi$.
\,This provides the proof of the ``if'' part of the proposition. A
proof of its ``only if'' part is given in Appendix
\ref{app:equiv-2isom}.\eroof
\bigskip

Finally, we may consider a
world-sheet with a general defect quiver. This time, we are to
compare
\qq\label{eq:trafo-WZ-dgen}
\widehat\xcA_{{\rm WZ}}\bigl[(\chi.\varphi\,\vert\,\G);
{}^{\tx{\tiny $\chi$}}\hspace{-2pt}\txA\bigr]=
\Hol_{\cG_{{}^{\tx{\tiny $\chi$}}\hspace{-2pt}\txA},
\Phi_{{}^{\tx{\tiny $\chi$}}\hspace{-2pt}\txA},(\varphi_{n\,
{}^{\tx{\tiny $\chi$}}\hspace{-2pt}\txA})}(L_\chi\circ\phi\,\vert\,
\G)=\Hol_{L_\chi^*\cG_{{}^{\tx{\tiny $\chi$}}\hspace{-2pt}\txA},
L_\chi^*\Phi_{{}^{\tx{\tiny $\chi$}}\hspace{-2pt}\txA},(L_\chi^*
\varphi_{n\,{}^{\tx{\tiny $\chi$}}\hspace{-2pt}\txA})}(\phi\,\vert\,
\G)\
\qqq
with
\qq\nn
\widehat\xcA_{{\rm WZ}}\bigl[(\varphi\,\vert\,\G);\txA\bigr]=
\Hol_{\cG_\txA,\Phi_\txA,(\varphi_{n\,\txA})}(\phi\,\vert\,\G)\,.
\qqq
Upon decomposing the defect-junction restriction of $\,L_\chi\,$ as
\qq\nn
L_\chi\ :\ \Vgt_\G^{(n)}\x T_n&\xrightarrow{K_\chi}&\Vgt_\G^{(n)}\x
\txG\x
T_n\xrightarrow{\id_{\Vgt_\G^{(n)}}\times\Tnup\ell}\Vgt_\G^{(n)} \x
T_n\cr \
(\jmath,t_n)&\longmapsto&\bigl(\jmath,\chi(\jmath),t_n\bigr)\quad
\longmapsto\quad\bigl(\jmath,\chi(\jmath).t_n\bigr)\,,
\qqq
we focus our attention on the 2-isomorphisms
\qq\nn
\bigl(\id_{\Vgt_\G^{(n)}}\x\Tnup\ell\bigr)^*\varphi_{n\,\txA}\equiv
\bigl(\id_{\Vgt_\G^{(n)}}\x\Tnup\ell\bigr)^*\varphi_{n\,2^*}=\bigl(
\Tnup\ell^*\varphi_n\bigr)_{[2,3]^*}\,.
\qqq
Reasoning along the same lines as in the case of non-intersecting
defect lines, we infer, with the help of the identity
\qq\nn
\D_{T_n}^{(1)}\la=0\,,
\qqq
implied by \Reqref{eq:junct-exact}, and of Proposition
\ref{prop:torsors}, the equality of the 2-isomorphisms
\qq
\Tnup\ell^*\varphi_n&=&d_n\ox\bigl[\bigl(d_\Upsilon\bigr)_n^{1\,(1
)}\bullet\bigl(\id\circ\bigl(i_\Upsilon^{\vep_n^{1,2}}\bigr)_n^{1\,
(1)}\bigr)\bullet(\id\circ\la_{\Upsilon^{1\,(1)}_n})\bullet\bigl(
\id\circ\varphi_{n\,2^*}\circ\id\bigr)\cr\cr
&&\bullet\bigl(\id\circ\la_{(\Phi_n^{n,1\,(1)})_{2^*}\ox\id}\circ
\la_{(\Phi_n^{n-1,n})_{2^*}\ox\id}\circ\cdots\circ\la_{(\Phi_n^{2,
3})_{2^*}\ox\id}\circ\id\bigr)\cr\cr
&&\bullet\bigl(\id\circ\bigl(b_\Upsilon^{-1}\bigr)_n^{n\,(1)}\circ
\id\circ\bigl(b_\Upsilon^{-1}\bigr)_n^{n-1\,(1)}\circ\id\circ\cdots
\circ\id\circ\bigl(b_\Upsilon^{-1}\bigr)_n^{2\,(1)}\circ\id\bigr)
\label{eq:loc-const-corr}\\\cr
&&\bullet\bigl(\id\circ\bigl(i_\Upsilon^{\vep_n^{n,1}}\bigr)_n^{n\,
(1)}\circ\id\circ\bigl(i_\Upsilon^{\vep_n^{n-1,n}}\bigr)_n^{n-1\,(1
)}\circ\id\circ\cdots\circ\id\circ\bigl(i_\Upsilon^{\vep_n^{2,3}}
\bigr)_n^{2\,(1)}\circ\id\bigr)\cr\cr
&&\bullet\bigl(\bigl(\Xi_n^{n,1\,(1)}\ox\id\bigr)\circ\bigl(
\Xi_n^{n-1,n\,(1)}\ox\id\bigr)\circ\cdots\circ\Xi_n^{1,2\,(1)}
\bigr)\bigr]\label{eq:locfctn}
\qqq
for some locally constant $\,\uj$-valued functions $\,d_n\,$ on
 $\,\txG\x T_n$.\ \,When writing the above, we used the maps
\qq\nn
\pi_n^{k\,(1)}:=\iota_1^{\vep_n^{k,k+1}\,(1)}\circ\pi_n^{k,k+1\,(1
)}\,,
\qqq
the 1-isomorphism
\qq\nn
\Upsilon_n^{1\,(1)}:=\pi_n^{1\,(1)\,*}\Upsilon\,,
\qqq
and the 2-isomorphisms
\qq\nn
\bigl(b_\Upsilon^{-1}\bigr)_n^{k\,(1)}:=\pi_n^{k\,(1)\,*}
b_\Upsilon^{-1}\,,\qquad\qquad\bigl(d_\Upsilon\bigr)_n^{k\,(1)}:=
\pi_n^{k\,(1)\,*}d_\Upsilon
\qqq
as well as
\qq
\bigl(i_\Upsilon^{\vep_n^{k,k+1}}\bigr)_n^{k\,(1)}:=\pi_n^{k\,(1)\,
*}i_\Upsilon^{\vep_n^{k,k+1}}\,,\qquad\qquad i_\Upsilon^{\vep_n^{k,
k+1}}:=\left\{ \barr{lcl} \id_\Upsilon\ & \tx{if} & \vep_n^{k,k+1}=
1\,, \\ i_\Upsilon\ & \tx{otherwise} & \earr \right.\cr
\label{eq:iUpsi-weight}
\qqq
and
\qq
\Xi_n^{k,k+1\,(1)}:=\pi_n^{k,k+1\,(1)\,*}\,\Xi^{\vep_n^{k,k+1}}\,,
\qquad\Xi^{\vep_n^{k,k+1}}:=\left\{ \barr{lcl} \Xi\ & \tx{if}
& \vep_n^{k,k+1}=1\,, \\ \Xi^{\sharp\,-1}\ & \tx{otherwise.} & \earr
\right.\cr\label{eq:gamma-weight}
\qqq
We thus obtain
\berop\label{prop:loc-const-junct}
Let $\,d_n\,$ be the locally constant maps on the $\,\txG\x T_n$
from \Reqref{eq:loc-const-corr}, and let $\,\varphi_{n\,\txA}\,$ be
the 2-isomorphism on $\,\Vgt_\G^{(n)}\x T_n\,$ defined in Corollary
\ref{cor:triv-Ansatz}. Finally, let $\,L_\chi\,$ be the map given by
\Reqref{eq:gauge-trans-ext}. Then, we have the equality
\qq\nn
&&\hspace{-0.7cm}L_\chi^*\varphi_{n\,{}^{\tx{\tiny $\chi$}}\hspace{-2pt}\txA}\cr\cr
&=&(\chi
,\id_{T_n})^*d_n\ox\bigl\{\bigl(d_{(\chi\x\id_M)^*\Upsilon}
\bigr)_n^{\unl 1}\bullet\bigl(\id\circ\bigl(i_\Upsilon^{\vep_n^{1,
2}}\bigr)_n^{\unl 1}\bigr)\bullet(\id\circ\la_{((\chi\x\id_M)^*
\Upsilon)^{\unl 1}_n})\bullet\bigl(\id\circ\varphi_{n\,\txA}\circ\id
\bigr)\cr\cr
&&\bullet\bigl(\id\circ\la_{(\Phi_n^{n,1})_{2^*}\ox\id}\circ\la_{(
\Phi_n^{n-1,n})_{2^*}\ox\id}\circ\cdots\circ\la_{(\Phi_n^{2,3}
)_{2^*}\ox\id}\circ\id\bigr)\cr\cr
&&\bullet\bigl(\id\circ\bigl(b_{(\chi\x\id_M)^*\Upsilon}^{-1}
\bigr)_n^{\unl n}\circ\id\circ\bigl(b_{(\chi\x\id_M)^*\Upsilon}^{-1}
\bigr)_n^{\unl{n-1}}\circ\id\circ\cdots\circ\id\circ\bigl(b_{(\chi\x
\id_M)^*\Upsilon}^{-1}\bigr)_n^{\unl 2}\circ\id\bigr)\cr\cr
&&\bullet\bigl(\id\circ\bigl(i_{(\chi\x\id_M)^*\Upsilon}^{\vep_n^{n,
1}}\bigr)_n^{\unl n}\circ\id\circ\bigl(i_{(\chi\x\id_M)^*
\Upsilon}^{\vep_n^{n-1,n}}\bigr)_n^{\unl{n-1}}\circ\id\circ\cdots
\circ\id\circ\bigl(i_{(\chi\x\id_M)^*\Upsilon}^{\vep_n^{2,3}}
\bigr)_n^{\unl 2}\circ\id\bigr)\cr\cr
&&\bullet\bigl[\bigl[\bigl((\chi\x\id_Q)^*\Xi\bigr)_n^{\unl{n,1}}\ox
\id\bigr]\circ\bigl[\bigl((\chi\x\id_Q)^*\Xi\bigr)_n^{\unl{n-1,n}}
\ox\id\bigr]\circ\cdots\circ\bigl((\chi\x\id_Q)^*\Xi\bigr)_n^{\unl{1
,2}}\bigr]\bigr\}\,,
\qqq
written in terms of the maps
\qq\nn
\unl\pi_n^k:=\unl\iota_1^{\vep_n^{k,k+1}}\circ\unl\pi_n^{k,k+1}\,,
\qqq
the 1-isomorphisms
\qq\nn
\Upsilon_n^{\unl k}:=\unl\pi_n^{k\,*}\Upsilon\,,
\qqq
and the 2-isomorphisms
\qq\nn
\bigl(b_\Upsilon^{-1}\bigr)_n^{\unl k}:=\unl\pi_n^{k\,*}
b_\Upsilon^{-1}\,,\qquad\qquad\bigl(d_\Upsilon\bigr)_n^{\unl k}:=
\unl\pi_n^{k\,*}d_\Upsilon
\qqq
and
\qq\nn
\bigl(i_\Upsilon^{\vep_n^{k,k+1}}\bigr)_n^{\unl k}:=\unl\pi_n^{k\,
*}i_\Upsilon^{\vep_n^{k,k+1}}\,,\qquad\qquad\Xi_n^{\unl{k,k+1}}:=
\unl\pi_n^{k,k+1\,*}\,\Xi^{\vep_n^{k,k+1}}\,,
\qqq
with the $\,i_\Upsilon^{\vep_n^{k,k+1}}\,$ and
$\,\Xi^{\vep_n^{k,k+1}}\,$ as in Eqs.\,\eqref{eq:iUpsi-weight} and
\eqref{eq:gamma-weight}, respectively.
\eerop
\beroof
The proof is a simple exercise using the assumed $\txG$-equivariance
of the (inter-)bi-brane maps. \eroof\bigskip  \noindent The above proposition
immediately yields
\bethe\label{thm:glanjt}
Let $\,\widehat\xcA_{{\rm WZ}}[(\varphi\,\vert\,\G);\txA]\,$ be the
gauged Wess--Zumino amplitude, introduced in Corollary
\ref{cor:infinit-Ginv-act}, of the $\si$-model for network-field
configurations $\,(\varphi\,\vert\,\G)\,$ for an arbitrary defect
quiver $\,\G$.\ Suppose that there exist a 1-isomorphism
$\,\Upsilon\,$ defined in Corollary \ref{cor:equiv-1iso} and a
2-isomorphism $\,\Xi\,$ defined in Proposition
\ref{prop:equiv-2iso}. Finally, let $\,d_n\,$ be the locally
constant $\,\uj$-valued functions on $\,\txG\times T_n\,$ introduced
in \Reqref{eq:loc-const-corr}. Then, for any $\,\chi:\Si\to\txG$,\
we have
\qq\label{eq:viol-inv}
\widehat\xcA_{{\rm WZ}}\bigl[(\chi.\varphi\,\vert\,\G);
{}^{\tx{\tiny $\chi$}}\hspace{-2pt}\txA\bigr]=\widehat\xcA_{{\rm
WZ}}\bigl[(\varphi\,\vert\,\G);\txA\bigr]\cdot\prod_{\jmath\in
\Vgt_\G}\,(\chi,\varphi)^*d_{n_\jmath}(\jmath)^{\pm 1}\,.
\qqq
where the exponent $\,+1\,$ is taken for the positive defect
junctions and $\,-1\,$ for the negative ones. \ethe
\beroof
An obvious consequence of Corollary \ref{cor:equiv-1iso} and
Proposition \ref{prop:equiv-2iso}, taken in conjunction with
Proposition \ref{prop:loc-const-junct}. \eroof
\vskip 0.3cm

\noindent Similarly as before, taking $\,\txA=\chi^*\theta_L$, \,we
infer from identity \eqref{eq:viol-inv} that, under the assumptions
of the above theorem,
\qq\label{eq:glanjt1}
\prod_{\jmath\in
\Vgt_\G}\,(\chi,\varphi)^*d_{n_\jmath}(\jmath)^{\pm 1}\,&=&\,
\Hol_{\cG,\Phi,(\varphi_n)}\bigl(\chi.(\varphi|\Gamma))\,\,
\Hol_{\cG,\Phi,(\varphi_n)}\bigl(\varphi|\Gamma\bigr)^{-1}\cr
&&\,\cdot\
\ee^{-\,\sfi\int_{\Sigma\setminus\Gamma}
\phi^*\rho_{\chi^*\theta_L}\,-\,
\sfi\int_\Gamma(\phi|_\Gamma)^*\lambda_{\chi^*\theta_L}}.
\qqq
Theorem \ref{thm:glanjt} permits to obtain the following result that
establishes necessary and sufficient conditions for gauge invariance
of the gauged Wess--Zumino amplitudes with general network-field
configurations:
\bethe\label{thm:equiv-12iso-constr}
The gauged Wess--Zumino amplitudes $\,\widehat\xcA_{{\rm
WZ}}(\varphi ;\txA\,\vert\,\G)$,\ introduced in Corollary
\ref{cor:infinit-Ginv-act}, are gauge invariant under all gauge
transformations iff there exist: a 1-isomorphism
\qq\nn
\Upsilon\ :\ \Mup\ell^*\cG\xrightarrow{\cong}\cG_{2^*}\ox I_\rho
\qqq
for $\,\rho\,$ as in \Reqref{eq:rho-on-G}, and a 2-isomorphism
\qq\nn
\Xi\ :\ \Qup\ell^*\Phi\xLongrightarrow{\cong}\bigl(\bigl(\iota_2^{(1
)\,*}\Upsilon^{-1}\ox\id\bigr)\circ\bigl(\Phi_{2^*}\ox\id\bigr)\circ
\iota_1^{(1)\,*}\Upsilon\bigr)\ox J_\la\,,
\qqq
for $\,\la\,$ as in \Reqref{eq:la-on-G}, such that the identities
\qq\nn
\Tnup\ell^*\varphi_n&=&\bigl(d_\Upsilon\bigr)_n^{1\,(1)}\bullet
\bigl(\id\circ\bigl(i_\Upsilon^{\vep_n^{1,2}}\bigr)_n^{1\,(1)}
\bigr)\bullet(\id\circ\la_{\Upsilon^{1\,(1)}_n})\bullet\bigl(\id
\circ\varphi_{n\, 2^*}\circ\id\bigr)\cr\cr
&&\bullet\bigl(\id\circ\la_{(\Phi_n^{n,1\,(1)})_{2^*}\ox\id}\circ
\la_{(\Phi_n^{n-1,n})_{2^*}\ox\id}\circ\cdots\circ\la_{(\Phi_n^{2,
3})_{2^*}\ox\id}\circ\id\bigr)\cr\cr
&&\bullet\bigl(\id\circ\bigl(b_\Upsilon^{-1}\bigr)_n^{n\,(1)}\circ
\id\circ\bigl(b_\Upsilon^{-1}\bigr)_n^{n-1\,(1)}\circ\id\circ\cdots
\circ\id\circ\bigl(b_\Upsilon^{-1}\bigr)_n^{2\,(1)}\circ\id\bigr)
\cr\cr
&&\bullet\bigl(\id\circ\bigl(i_\Upsilon^{\vep_n^{n,1}}\bigr)_n^{n\,
(1)}\circ\id\circ\bigl(i_\Upsilon^{\vep_n^{n-1,n}}\bigr)_n^{n-1\,(1
)}\circ\id\circ\cdots\circ\id\circ\bigl(i_\Upsilon^{\vep_n^{2,3}}
\bigr)_n^{2\,(1)}\circ\id\bigr)\cr\cr
&&\bullet\bigl(\bigl(\Xi_n^{n,1\,(1)}\ox\id\bigr)\circ\bigl(
\Xi_n^{n-1,n\,(1)}\ox\id\bigr)\circ\cdots\circ\Xi_n^{1,2\,(1)}
\bigr)
\qqq
hold true for all $\,n\geq 3$. \ethe
\beroof If the conditions listed above are satisfied then the
locally constant functions $\,d_n\,$ on $\,\txG\times T_n\,$
introduced by \eqref{eq:loc-const-corr} are equal to 1 and the gauge
invariance of the amplitudes follows from Theorem
\ref{thm:glanjt}. A proof of the ``only if'' part of the theorem
is given in Appendix \ref{app:equiv-2isom}. \eroof \brem In
particular, the conditions listed in Theorem
\ref{thm:equiv-12iso-constr} assure the absence of global gauge
anomalies in amplitudes \eqref{eq:ampl.with.defects}. \erem

\section{An example of a string background: the WZW model}
\label{sec:eg-backgrnd}

\subsection{The WZW target}\label{app:WZW-target}

\noindent We present here in some detail a distinguished class of
string backgrounds, with all components given by subspaces of the
group manifold of some connected compact semi-simple Lie
group\footnote{One could extend the discussion below to the setting
of not necessarily compact Lie groups.} $\,\xcG$, \,not necessarily
simply connected, endowed with additional cohomological structure.
These are the backgrounds in which the lagrangian fields of the
so-called Wess--Zumino--Witten (WZW) $\si$-model of
\Rcite{Witten:1983ar} take values. \,One has: $\,\xcG=\tilde\txG/Z$,
\,where $\,\tilde\txG=\mathop{\times}_{l}\tilde\txG_l\,$ is the
covering group of $\,\xcG\,$ that decomposes into the product of
simple factors, and $\,Z\,$ is a subgroup of the center $\,\tilde Z
=\mathop{\times}_{l}\tilde Z_{l}\,$ of $\,\tilde\txG$.\ The Lie
algebra $\,\ggt\,$ of $\,\tilde\txG\,$ decomposes as
$\,\oplus_{l}\ggt_{l}\,$ into the direct sum of simple factors. Let
$\,\txG=\tilde\txG/\tilde Z$. \,We shall consider $\,\xcG\,$ with
the adjoint action of $\,\txG$. \vskip 0.2cm

Let
$\,\sfk\,\tr_\ggt\,XY:=\sum\sfk_l\hspace{0.03cm}\tr_{\ggt_l}\,X^lY^l$
be an $\,ad$-invariant negative-definite bilinear form on the Lie
algebra $\,\ggt$, \,given by the sum of forms on $\,\ggt_l\,$ with
the same properties, where $\,\sfk=(\sfk_l)$,\ with $\sfk_l>0$,\ is
called the \textbf{level}. We shall consider the Cartan--Killing
metric on $\,\xcG\,$ given by
\qq\nn
\txg_\sfk=-\tfrac{1}{4\pi}\,\sfk\,\tr_\ggt(\tht_L\ox\tht_L)\,,
\qqq
and the closed 3-form
\qq\label{eq:Cart3}
\txH_\sfk=\tfrac{1}{12\pi}\,\sfk\,\tr_\ggt(\tht_L\wedge\tht_L\wedge
\tht_L)\,,
\qqq
where $\,\theta_L\,$ stands for the left-invariant Maurer-Cartan
form on $\,\xcG$. \,The bilinear forms $\,\tr_{\ggt_l}\,$ on the
simple Lie-algebra factors $\,\ggt_l\,$ are assumed to be normalized
so that if the group $\,\xcG\,$ is simply connected (i.e.
$\,\xcG=\tilde\txG$) \,then the periods of the 3-form
$\,\txH_\sfk\,$ take values in $\,2\pi\bZ\,$ iff all $\,\sfk_l\,$
are integers. For non-simply connected $\,\xcG$, \,the integrality
of the periods of $\,\frac{1}{2\pi}\txH_\sfk\,$ imposes more
stringent selection rules on the level $\,\sfk$, \,cf.\
Refs.\,\cite{Felder:1988sd,Gawedzki:2003pm}. Such integrality is
necessary and sufficient for the existence of a gerbe $\,\cGk\,$
over group $\,\xcG\,$ with curvature 3-form $\,\txH_\sfk\,$ whose
holonomy provides the Wess--Zumino part of Feynman amplitudes for
the defect-free WZW $\,\si$-models of conformal field theory.

\bedef\label{def:WZW-target}
\textbf{The level-$\sfk$ WZW target} is the triple $\,(\xcG,\txg_\sfk,
\cGk)$.
\exdef

We shall consider the defect-free WZW model defined for the above
target with rigid symmetries induced by the adjoint action of
$\,\txG\,$ on $\,\xcG\,$ (the other symmetries from the left-right
symmetry group $\,\xcG\x\xcG$, \,that extend to the loop-group
$\,L\xcG\x L\xcG\,$ symmetries of the defect-free WZW model, cannot
be gauged as they suffer from local gauge anomalies). Choose
generators $\,t_a,\ a=1,2,\ldots,\dim\,\ggt$, \,of the Lie algebra
$\,\ggt\,$ as in \Reqref{eq:defta}. \,They induce on $\,\xcG\,$ the
fundamental vector fields
\qq
{}^\xcG\hspace{-0.1cm}\xcK_a=L_a-R_a\,,
\qqq
where the $\,L_a\,$ (resp.\ $R_a$) \,are the left-invariant (resp.\
right-invariant) vector fields on $\,\xcG\,$ corresponding to the
generators $\,t_a\in\ggt$.
\berop\label{prop:rig-WZW-bulk} \ The collection
$\,\{v_a\}_{a=1,2,\ldots,\dim\,\ggt}\,$ of 1-forms on $\,\xcG\,$
given by
\qq\nn
v_a=-\tfrac{1}{4\pi}\,\sfk\,\tr_\ggt\bigl(t_a\,(\theta_L+\theta_R)
\bigr)
\qqq
defines a $\ggt$-equivariantly closed $\txG$-equivariant
(Cartan-model) extension of the Cartan 3-form $\,\txH_\sfk\,$ on
$\,\xcG$,\ as defined in \Reqref{eq:Cart3}, through
\qq\nn
\widehat\txH_\sfk=\txH_\sfk+v\,,\qquad v(t_a)=v_a\,.
\qqq
\eerop \beroof Obvious, through inspection.\eroof
\medskip
\noindent The last proposition lays out the basic setting for the
gauging of (continuous) rigid symmetries of the WZW
model\footnote{Another class of examples of constructions considered
in this paper is provided by the orbifold WZW models of
Refs.\,\cite{Gawedzki:2002se,Gawedzki:2003pm,Gawedzki:2004tu}, cf.\
also Refs.\,\cite{Gawedzki:2007uz,Gawedzki:2008um} for a
generalisation to WZW orientifolds.}.

\subsection{The boundary maximally symmetric WZW bi-branes}
\label{sub:maxym-bound}

\noindent In the next step, we consider $\cGk$-bi-branes for those
\emph{boundary} WZW defects that preserve a maximum amount of the
loop-group symmetry $\,\sfL\xcG\x\sfL\xcG\,$ of the WZW model in the
interior of the world-sheet, that is a single copy of $\,\sfL\xcG\,$
embedded `diagonally' in the double cartesian product. As argued in
Remark \ref{rem:bdry-defcts} (and originally in
\Rxcite{p.\,12}{Runkel:2008gr}), a consistent description of a
$\si$-model on a world-sheet with a boundary in the language of
defects and bi-branes assigned to them prerequires a special choice
of the target, which - in the case in hand - is simply the disjoint
union of the level-$\sfk$ WZW target group $\,\xcG\,$ with an arbitrary
singleton $\,\{\bullet\}\,$ with trivial geometric data over it. By a slight
abuse of the language, we shall refer also to this target as the
level-$\sfk$ WZW target in what follows.

\bedef\label{def:WZW-bdry-bib}
A \textbf{boundary maximally symmetric $\cGk$-bi-brane}, or a
\textbf{maximally symmetric $\cGk$-brane} for short, is a quintuple
$\,\cB_\la^\p:=(\xcC_\la,\iota_\la,\bullet, \om^\p_\la,\cT_\la)\,$
with the following components:
\bit
\item[(D.1)] the world-volume given by the conjugacy class
\qq\label{eq:conj-cl}
\xcC_\la=\left\{\ \Ad_x(t_\la) \quad\big\vert\quad x\in\xcG\
\right\}\,,
\qqq
of an element $\,t_\la=\ee^{\,2\pi\la}\in\xcG\,$ of
the Cartan subgroup of $\,\xcG$,\ with $\,\la=\oplus_l\la_l\,$ from the
fundamental affine Weyl alcove\footnote{In what follows, we shall
always identify $\,\tgt\,$ with its dual $\,\tgt^*\,$ using the
bilinear form $\,\tr_\ggt\,$ on $\,\ggt$.},
\qq\nn
\qquad\quad\xcA_W(\ggt)=\bigl\{\,\la\in\tgt \quad \big\vert\quad \tr_\ggt(\la
\,\vartheta)\leq 1\,,\quad\tr_\ggt(\la\,\a_i)\geq 0\,,\
i=1,2,\ldots,{\rm rank}\,\ggt \,\bigr\}\,,
\qqq
of the Cartan subalgebra $\,\tgt=\oplus_l\tgt_l\subset\ggt\,$ defined in terms of
the simple roots $\,\a_i\,$ of $\,\ggt\,$ and its highest root
$\,\vartheta$;
\item[(D.2)] the embedding $\,\iota_\la:\xcC_\la\emb\xcG\,$ of the
conjugacy class in the group manifold, alongside the constant map
$\,\bullet\ :\ \xcC_\la\to\{\bullet\}$;
\item[(D.3)] the curvature 2-form
\qq\label{eq:WZW-brane-curv}
\om^\p_\la=\tfrac{1}{8\pi}\,\sfk\,\tr_\ggt\bigg(\tht_L\wedge
\tfrac{\id_\ggt+\Ad_\bullet}{\id_\ggt-\Ad_\bullet}\,\tht_L\bigg)
\qqq
providing a global primitive for $\,\iota_\la^*\txH_\sfk$;
\item[(D.4)] the 1-isomorphism given by a trivialisation
$\,\cT_\la\ :\ \iota_\la^*\cGk\xrightarrow{\cong}I_{\om^\p_\la}\,$
of gerbe $\,\cGk\,$ restricted to $\,\xcC_\la$.
\eit
\exdef \brem It is vital to note that for an
arbitrary compact simple 1-connected Lie group $\,\xcG\,$ trivializations
$\,\cT_\la\,$ exist iff
\qq\nn
\sfk\la:=\oplus_l\sfk_l\lambda_l\,\in\,\sfk\,\xcA_W(\ggt)\cap
P(\ggt)=:\faff{\ggt}
\qqq
where $\,P(\ggt)\,$ denotes the weight lattice of $\,\ggt$,
\,whereas for non-simply connected $\,\txG\,$ additional selection
rules may be required for $\,\la$, \,cf.\ \Rcite{Gawedzki:2004tu}.
\,In the first case, the conjugacy classes $\,\xcC_\la$, \,each
corresponding to a unique $\,\la\in \xcA_W(\ggt)$, \,are
1-connected, whereas in the second case, they are connected but not
necessarily simply connected, and different $\,
\la\in\xcA_W(\ggt)\,$ may correspond to the same $\,\xcC_\la$. \erem

The world-volume of a maximally symmetric WZW $\cGk$-brane is
naturally endowed with the structure of a $\txG$-space coming from
the adjoint action of $\,\txG\,$ on $\,\xcG$.
\berop
Adopt the notation of Definition \ref{def:WZW-target} and
Proposition \ref{prop:rig-WZW-bulk}, and let $\,\xcC_\la\,$ be a
conjugacy class in $\,\xcG$,\ as defined in \Reqref{eq:conj-cl}. The
2-form $\,\om^\p_\la\,$ on $\,\xcC_\la\,$ given in
\Reqref{eq:WZW-brane-curv} is its own (trivial) $\xcG$-equivariant
(Cartan-model) extension satisfying \Reqref{eq:hatdom-hatdelHa} with
respect to the Cartan 3-form $\,\txH_\sfk\,$ on $\,\xcG$,\ and for
the pair of $\txG$-equivariant maps $\,(\iota_\la,\bullet)\,$ given
in Definition \ref{def:WZW-bdry-bib}.
\eerop \beroof Through inspection.\eroof
\bigskip
\bedef
A \textbf{general boundary maximally symmetric $\cGk$-bi-brane} is a
disjoint union of the elementary boundary maximally symmetric
$\cGk$-bi-branes of Definition \ref{def:WZW-bdry-bib}.\exdef

\subsection{The non-boundary maximally symmetric WZW bi-branes}
\label{sub:maxym-nonbound}

\noindent As the last example of WZW $\cGk$-bi-branes, we present
those associated with \emph{non-boundary} WZW defects that preserve
the full loop-group symmetry $\,\sfL\xcG\x\sfL\xcG\,$ of the WZW
model. They implement jumps by elements of the target Lie group in
the sense that the limiting values attained by the one-sided local
extensions $\,g_{|1}\,$ (from the left) and $\,g_{|2}\,$ (from the
right) of the field $\,g:\Si\setminus\G\to\xcG\,$ to the defect line
$\,\G\,$ (as given in Definition \ref{def:net-field}) are, in
general, different. A special class of such defects, the
central-jump defects for which $\,g_{|1}^{-1}g_{|2}\in Z$,\ where
$\,Z\,$ is the center of $\,\xcG$, \,were considered at length in
\Rcite{Runkel:2008gr}. The more general jump defects, with the jump
given by $\,g_{|1}^{-1}\cdot g_{|2}\in\xcC_\la$, \,were first
considered in \Rcite{Fuchs:2007fw}, where the notion of a bi-brane
was introduced. They will be discussed below, cf.\ also
\Rcite{Runkel:2010} for more details.
\bedef\label{def:WZW-non-bdry-bib}
A \textbf{non-boundary maximally symmetric $\cGk$-bi-brane} is a
quintuple $\,\cB_\la:=(Q_\la,\pr_1,\txm, \om_\la,\Phi_\la)\,$ with
the following components:
\bit
\item[(B.1)] the world-volume
\qq\nn
Q_\la=\xcG\x \xcC_\la\,,\qquad\la\in\faff{\ggt}\,,
\qqq
isomorphic with the bi-conjugacy class
\qq\nn
\xcB_{(t_\la,e)}=\bigl\{\ \bigl(x\cdot t_\la\cdot y^{-1},x\cdot y^{-
1}\bigr) \quad\big\vert\quad x,y\in\xcG \ \}\,,
\qqq
of \Rcite{Fuchs:2007fw} through
\qq\nn
D\ :\ \xcG\x\xcC_\la\xrightarrow{\cong}\xcB_{(t_\la,e)}\ :\
(g,h_\la) \mapsto(g\cdot h_\la,g)\,;
\qqq
\item[(B.2)] the canonical projection $\,\pr_1\ :\ Q_\la\to\xcG\ :\
(g,h_\la)\mapsto g\,$ and the multiplication map $\,\txm\ :\ Q_\la
\to\xcG\ :\ (g,h_\la)\mapsto g\cdot h_\la$;
\item[(B.3)] the curvature 2-form
\qq\label{eq:WZW-bibrane-curv}
\om_\la=-\om^\p_{\la\,2^*}+\varrho_\sfk\,,\qquad\qquad\varrho_\sfk=
\tfrac{1}{4\pi}\,\sfk\,\tr_\ggt\bigl(\tht_{L\,1^*}\wedge\tht_{R\,2^*}
\bigr)\,,
\qqq
providing a global primitive for $\,\txH_{\sfk\,1^*}-\txm^*
\txH_\sfk$;
\item[(B.4)] the 1-isomorphism
$\Phi_\la=\left(\cM_\sfk\ox\id_{I_{-\om^\p_\la}}\right)\circ\left(
\id_{\cG_{\sfk\,1^*}}\ox\cT_{\la\,2^*}^{-1}\ox\id_{I_{-\om^\p_\la}}
\right)\,$ (a gerbe bi-module) defined as
\qq\nn
\Phi_\la\ :\ \cG_{\sfk\,1^*}\equiv\cG_{\sfk\,1^*}\ox I_{\om^\p_\la}
\ox I_{-\om^\p_\la}\xrightarrow{\cong}\txm^*\cGk\ox I_{\om_\la}
\qqq
in terms of a 1-isomorphism
\qq\nn
\cM_\sfk\ :\ \cG_{\sfk\,1^*}\ox\cG_{\sfk\,2^*}\xrightarrow{\cong}
\txm^*\cGk\ox I_{\varrho_\sfk}
\qqq
of the multiplicative structure on $\,\cGk$,\ as introduced in
\Rcite{Carey:2004xt} and developed in
Refs.\,\cite{Waldorf:2008mult,Gawedzki:2009jj}.
\eit
\exdef \brem It ought to be emphasised that in the case of simple
1-connected groups $\,\xcG$, \,the uniqueness (up to a
2-isomorphism) of the 1-isomorphism $\,\cM_\sfk\,$ ensures that
non-boundary maximally symmetric WZW bi-branes are in a one-to-one
correspondence with their boundary analogs. \erem

We shall consider $\,Q_\la\,$ as a $\,\txG$-space with the diagonal
adjoint action of $\,\txG$.

\berop
Adopt the notation of Definition \ref{def:WZW-target} and
Proposition \ref{prop:rig-WZW-bulk}, and let $\,\xcC_\la\,$ be a
conjugacy class in $\,\xcG$,\ as defined in \Reqref{eq:conj-cl}. The
2-form $\,\om_\la\,$ on $\,\xcG\x\xcC_\la\,$ given in
\Reqref{eq:WZW-bibrane-curv} is its own (trivial) $\xcG$-equivariant
(Cartan-model) extension satisfying \Reqref{eq:hatdom-hatdelHa} with
respect to the Cartan 3-form $\,\txH_\sfk\,$ on $\,\xcG$,\ as
defined in \Reqref{eq:Cart3}, and for the pair of $\xcG$-equivariant
maps $\,\iota_1:=\pr_1\,$ and $\,\iota_2:=\txm$.
\eerop \beroof Through inspection.\eroof
\bigskip
\bedef
A \textbf{general non-boundary maximally symmetric $\cGk$-bi-brane}
is the disjoint union of the elementary non-boundary maximally
symmetric $\cGk$-bi-branes of Definition \ref{def:WZW-non-bdry-bib}.
\exdef

\subsection{The maximally symmetric WZW inter-bi-brane}
\label{sub:maxsyminter}

\noindent Inter-bi-branes describe the behaviour of $\si$-model
fields at junctions of defect lines. As such, they come in a
countably infinite variety that encodes the various relative
orientations (towards and away from the junction) of intersecting
defect lines and the arbitrary valence of junctions. In what
follows, we focus - for the sake of brevity - on the most elementary
defect junctions, namely those with two incoming defect lines (i.e.\
oriented towards the junction) and a single outgoing defect line
(i.e.\ oriented away from the junction). In the case of intersecting
WZW defects each carrying the data of a non-boundary maximally
symmetric WZW bi-brane, the world-sheet picture immediately suggests
two alternative descriptions, namely: We can represent a point in
the inter-bi-brane world-volume as a triple
$\,(g,h_\la,h_\mu)\in\xcG\x\xcC_\la\x \xcC_\mu\,$ in which $\,g\,$
gives a reference value of the $\si$-model field, $\,h_\la\,$ is the
jump at the first incoming defect line, $\,h_\mu\,$ is the jump at
the second incoming defect line, and the latter two are constrained
so that the total jump $\,h_\la\cdot h_\mu\,$ belongs to the second
factor $\,\xcC_\nu\,$ of the world-volume of the outgoing bi-brane.
The world-volume now arises as a disjoint union of (a subset of) all
those $\xcG\x\xcG$-orbits within $\,\xcG\x\xcC_\la\x\xcC_\mu\,$ with
respect to the action
\qq\nn
(\xcG\x\xcG)\x(\xcG\x\xcC_\la\x\xcC_\mu)&\longrightarrow&\xcG\x\xcC_\la\x\xcC_\mu\cr\cr
\left((x,y),(g,h_\la,h_\mu)\right)&\longmapsto&\left(x\cdot g\cdot y^{-
1},\Ad_y(h_\la),\Ad_y(h_\mu)\right)
\qqq
that are mapped to $\,\xcG\x\xcC_\nu\,$ by $\,\id_\xcG\x\txm$.
This is the representation introduced and used in the original
Refs.\,\cite{Runkel:2009sp,Suszek:2011hg}. Equivalently, we may
identify the world-volume of interest as a subspace in the fibred
product
\qq\nn
Q_\la{\,}_\txm\hspace{-3pt}\x_{\pr_1}\hspace{-2pt}Q_\mu=\{\ \left((g,
h_\la),(g',h_\mu)\right)\in Q_\la\x Q_\mu \quad\vert\quad g\cdot
h_\la=g' \ \}
\qqq
of the world-volumes of the two `incoming' bi-branes composed of (a
subset of) all orbits of the $\xcG\x\xcG$-action
\qq
(\xcG\x\xcG)\x\left(Q_\la{\,}_\txm\hspace{-3pt}\x_{\pr_1}
\hspace{-2pt}Q_\mu\right)&\longrightarrow& Q_\la{}_\txm\hspace{-3pt}\x_{\pr_1} \
\hspace{-2pt}Q_\mu\cr\cr
\left((x,y),\left((g,h_\la),(g\cdot
h_\la,h_\mu)\right)\right)&\longmapsto&\left(\left(x\cdot g\cdot y^{-1},
\Ad_y(h_\la)\right),\left(x\cdot g\cdot h_\la\cdot y^{-1},\Ad_y(
h_\mu)\right)\right)\qquad\label{eq:GG-act}
\qqq
from the preimage of $\,Q_\nu\,$ under the map
\qq
\pi_3^{1,3}\ :\ Q_\la{\,}_\txm\hspace{-3pt}\x_{\pr_1}\hspace{-2pt}
Q_\mu&\longrightarrow&\xcG\x\xcG\cr\cr
\left((g,h_\la),(g\cdot h_\la,h_\mu)\right)
&\longmapsto&\left(g,g^{-1}\cdot\txm(g\cdot h_\la,h_\mu)\right)\equiv(g,
h_\la\cdot h_\mu)\,. \label{eq:pi313}
\qqq
This is the representation that appears in \Rcite{Runkel:2010}. In
general, there are many orbits of the type described in the fibred
product $\,Q_\la{\,}_\txm\hspace{-3pt}\x_{\pr_1}\hspace{-2pt}
Q_\mu\,$ of a given pair of $\cGk$-bi-brane world-volumes. We shall
denote them as
$\,\left[Q_\la{\,}_\txm\hspace{-3pt}\x_{\pr_1}\hspace{-2pt}
Q_\mu\right]_{(\nu,b)}$,\ adding an extra degeneracy label
$\,b=0,1,\ldots,\,$ to distinguish them from one another.
\bedef
A \textbf{maximally symmetric $(\cGk,\cB_\sfk
)$-inter-bi-brane} is a quintuple $\,\cJ_{\la,\mu}^\nu:=\left(T_{\la
,\mu}^{\ \nu},\pr_1,\pr_2,\pi_3^{1,3},\varphi_{\la,\mu}^{\ \nu}
\right)\,$ with the following components:
\bit
\item[(B.1)] the world-volume
\qq\nn
T_{\la,\mu}^{\ \nu}:=\bigsqcup_{b=1}^{N_{\la,\mu}^{\ \nu}}\,\left[
Q_\la{\,}_\txm\hspace{-3pt}\x_{\pr_1}\hspace{-2pt}Q_\mu\right]_{(\nu,
b)}
\qqq
given as a disjoint sum of those ($N_{\la,\mu}^{\ \nu}$) orbits of
the $\xcG\x\xcG$-action of \Reqref{eq:GG-act} within the fibred
product $\,Q_\la{\,}_\txm\hspace{-3pt}\x_{\pr_1}\hspace{-2pt}Q_\mu\,$
that are mapped to $\,Q_\nu\,$ by $\,\pi_3^{1,3}\,$ of
\Reqref{eq:pi313} and that support a 2-isomorphism $\,\varphi_{\la,
\mu}^{\ \nu}\,$ of $\tx{(B.3)}$;
\item[(B.2)] the canonical projections $\,\pr_1\ :\ T_{\la,\mu}^{\
\nu}\to Q_\la\,$ and $\,\pr_2\ :\ T_{\la,\mu}^{\ \nu}\to Q_\mu$,\
alongside map $\,\pi_3^{1,3}\,$ of \Reqref{eq:pi313};
\item[(B.3)] a 2-isomorphism
\qq\nn
\xy (35,0)*{\bullet}="mGGI"+(-22,4)*{\tx{\scriptsize$\barr{c}
(\txm^*\cGk)_{[1, 2]^*}\ox\cG_{\sfk\,4^*}\ox I_{\om_{\la\,[1,2]^*}-
\om^\p_{\mu\,4^*}} \\ \equiv\cG_{\sfk\,3^*}\ox\cG_{\sfk\,4^*}\ox
I_{\om_{\la\,[1,2]^*}-\om^\p_{\mu\,4^*}}\earr$}};
(65,0)*{\bullet}="mGI"+(25,4)*{\tx{\scriptsize$\barr{c}\left(\left(
\txm\circ(\txm\x\id)\right)^*\cGk\right)_{[1,2,4]^*}\ox
I_{\om_{\la\,[1,2]^*}+\left((\txm\x\id)^*\om_\mu\right)_{[1,2,4]^*}}
\\ \equiv\left(\txm^*\cG\right)_{[3,4]^*}\ox I_{\om_{\la\,[1,2]^*}+
\om_{\mu\,[3,4]^*}}\earr$}};
(25,-20)*{\bullet}="GI"+(-23,0)*{\tx{\scriptsize$\cG_{\sfk\,1^*}\ox
\cG_{\sfk\,2^*}\ox\cG_{\sfk\,4^*}\ox I_{-\om^\p_{\la\,2^*}-
\om^\p_{\mu\,4^*}}$}};
(75,-20)*{\bullet}="GmGI"+(25,0)*{\tx{\scriptsize$\barr{c}\cG_{\sfk
\,1^*}\ox\left(\txm^*\cGk\right)_{[2,4]^*}\\ \ox
I_{\om_{\la\,[1,2]^*}+\om_{\mu\,[3,4]^*}-\left((\id\x\txm)^*
\varrho_\sfk\right)_{[1,2,4]^*}}\earr$}};
(35,-40)*{\bullet}="G1"+(-22,-4)*{\tx{\scriptsize$\cG_{\sfk\,1^*}
\equiv\cG_{\sfk\,1^*}\ox I_{\om^\p_{\la\,2^*}}\ox I_{\om^\p_{\mu\,
4^*}}\ox I_{-\om^\p_{\la\,2^*}-\om^\p_{\mu\,4^*}}$}};
(65,-40)*{\bullet}="G1I"+(22,-4)*{\tx{\scriptsize$\cG_{\sfk\,1^*}
\ox I_{\om_{\la\,[1,2]^*}+\om_{\mu\,[3,4]^*}-\left((\id\x\txm)^*
\om_\nu\right)_{[1,2,4]^*}}$}}; (50,0)*{}="hup";
(50,-40)*{}="hdown"; \ar@{->}|{\id\ox\cT_{\la\,
2^*}^{-1}\ox\cT_{\mu\,4^*}^{-1}\ox\id} "G1";"GI"
\ar@{->}|{\cM_{\sfk\,[1,2]^*}\ox\id} "GI";"mGGI"
\ar@{->}|{\cM_{\sfk\,[3,4]^*}\ox\id} "mGGI";"mGI"
\ar@{->}|{\left((\id\x\txm)^*\cM_\sfk^{-1}\right)_{[1,2,4]^*}\ox\id}
"mGI";"GmGI" \ar@{->}|{\id\ox\left(\txm^*\cT_\nu\right)_{[2,4]^*}\ox
\id} "GmGI";"G1I" \ar@{=}|{\ \id\ } "G1"+(2,0);"G1I"+(-2,0)
\ar@{=>}|{\ \varphi_{\la,\mu}^{\ \nu}\ } "hup"+(0,-3);"hdown"+(0,+3)
\endxy\,,
\qqq
where the various pullback labels refer to the cartesian components
in the decomposition $\,\xcG\x\xcC_\la\x\xcG\x\xcC_\mu\supset Q_\la
{\,}_\txm\hspace{-3pt}\x_{\pr_1}\hspace{-2pt}Q_\mu$.
\eit
\exdef \brem The precise form of the $\,T_{\la,\mu}^{\ \nu}\,$ is
known for $\,\xcG=\sug\,$ exclusively, cf.\ \Rcite{Runkel:2009sp}.
It is conjectured in that paper, as well as in \Rcite{Runkel:2010},
where partial evidence in favour of this conjecture is presented,
that numbers $\,N_{\la,\mu}^{\ \nu}\,$ coincide for simply connected
groups $\,\xcG\,$ with the structure
constants of the Verlinde fusion ring (also known as the Verlinde
dimensions) of the level-$\sfk$ WZW model. \erem
\vskip -0.2cm

\subsection{Global gauge anomalies of the WZW-model amplitudes}

\noindent The global gauge anomalies of the Feynman amplitudes of
defect-free WZW models were analysed in \Rcite{Gawedzki:2010rn}
basing on Eq.\,\eqref{eq:glanbk1} for pairs
$\,(\chi,\varphi):\Sigma\rightarrow\txG\x\xcG\,$ that generate the
integral homology group $\,H_2(\txG\x\xcG)$. \,It was shown there
that the relevant 2-cycles that probe the global gauge anomaly come
from  $\,\Sigma=\bS^1\times \bS^1\,$ and
$$(\chi,\varphi)(\ee^{i\sigma_1},\ee^{i\sigma_2})=(\exp[\sigma_1
\tilde p^\vee],\exp[\sigma_2 p^\vee])\,,$$ with $\,\tilde
p^\vee,\,p^\vee\in\tgt\,$ such that $\,\ee^{2\pi \tilde
p^\vee}\in\tilde Z\,$ and $\,e^{2\pi  p^\vee}\in Z$. \,For such
fields $\,\chi.\varphi=\varphi\,$ resulting in the relation
\qq\label{eq:bkanom}
\Hol_\cD\bigl((\chi,\varphi)\bigr)\ =\ \exp[2\pi\sfi\hspace{0.03cm}
\sfk\,{\rm tr}_{\ggt}\, \tilde p^\vee p^\vee]\,.
\qqq
The triviality of the right-hand side gives an easily verifiable
condition. It was used in \Rcite{Gawedzki:2010rn} to detect the
presence of global anomalies in multiple cases of defect-free WZW
$\,\si$-models with non-simply connected target groups $\,\xcG$.
\,This was further studied in \Rcite{DeFromont:2010mtrx}.
\medskip

Now add a circular defect $\,\Gamma\subset\Sigma\,$ with the field
$\,\varphi|_\Gamma\,$ mapping into the world-volume $\,Q_\la$ of a
non-boundary maximally symmetric $\,\cG_\sfk\,$ bi-brane. If the
bulk is without global gauge anomaly, then the source of remaining
global anomaly is the holonomy of the flat bundle $\,D$, \,defined
in \Reqref{eq:triv-bund-corr}, around the 1-cycle
$\,(\chi,\varphi)|_\Gamma\,$ in $\,\txG\x Q_\la$, \,cf.\
Eq.\,\eqref{eq:glanbd}. Relation \eqref{eq:glanbd1} now reduces to
the identity
\qq\label{eq:glanbd2}
\Hol_D\bigl((\chi,\varphi)|_\Gamma)\bigr)\ =\
\Hol_{\cG_\sfk,\Phi_\la}\bigl(\chi.\varphi|\Gamma\bigr)\,\,
\Hol_{\cG_\sfk,\Phi_\la}\bigl(\varphi|\Gamma\bigr)^{-1}\,
\ee^{-\,\sfi\int_{\Sigma\setminus\Gamma}
\phi^*\rho_{\chi^*\theta_L}}
\qqq
since the 1-form $\,\lambda_{\txA}=0\,$ in the case of maximally
symmetric $\,\cG_\sfk$-bi-branes. \,To test the contribution of
circular defects to the global gauge anomaly, one should take
$\,\chi\,$ and $\,(\varphi|\Gamma)\,$ such that
$\,(\chi,\varphi)|_\Gamma\,$ generate the kernel of the map
\qq\label{eq:homom}
(\iota^{(1)}_2)_*-(\iota^{(1)}_1)_*\,:\,\,H_1(\txG\x Q_\la)
\rightarrow H_1(\txG\x\xcG)\,,
\qqq
cf.\ Appendix \ref{app:equiv-2isom}. One has
\qq
H_1(\txG\x Q_\la)\,\cong\,H_1(\txG)\oplus H_1(\xcG)\oplus
H_1(\xcC_\la)\,,
\qqq
and it is easy to see that the kernel of \eqref{eq:homom}
corresponds to the subgroup $\,H_1(\txG)\oplus H_1(\xcG)\,\subset\,
H_1(\txG\x Q_\la)$. \,It is then enough to take
$\,\Sigma=\bS^1\times \bS^1$, $\,\Gamma=\{1\}\times \bS^1\,$ and
\qq
\chi(e^{i\sigma_1},e^{i\sigma_2})&=&\exp[\sigma_2\tilde
p^\vee]\,,\cr\cr
\varphi(e^{i\sigma_1},e^{i\sigma_2})&=&\exp[\sigma_1\lambda+\sigma_2p^\vee]
\ \ {\rm for}\ \ \sigma_1\not=0\,\,{\rm mod}\,2\pi\,,\cr\cr
\varphi(1,e^{i\sigma_2})&=&(\exp[\sigma_2p^\vee],\exp[2\pi\lambda])\in
Q_\la \nonumber
\qqq
with $\,\tilde p^\vee\,$ and $\,p^\vee\,$ as before. In this case,
$\,\chi.\varphi|\Gamma=\varphi|\Gamma\,$ and identity
\eqref{eq:glanbd2} gives
\qq\label{eq:glanbd3}
\Hol_D\bigl((\chi,\varphi)|_\Gamma\bigr)\,=\, \exp[2\pi
i\,\sfk\,\tr_\ggt\,\lambda\hspace{0.03cm}\tilde p^\vee].
\qqq
Hence, the absence of the global gauge anomaly requires that
$\,\sfk\,\tr_\ggt\lambda\hspace{0.03cm}\tilde p^\vee\in{\bZ}\,$ for
all $\,\tilde p^\vee\in\tgt\,$ such that
$\,\ee^{2\pi\tilde{p}^\vee}\in \tilde Z$, \,i.e. for
$\,\tilde{p}^\vee\,$ in the coweight lattice. This holds iff
$\,\sfk\lambda\,$ is in the root lattice of $\,\ggt$. For the other
non-boundary maximally symmetric $\,\cG_\sfk$-bi-branes, the global
gauge anomaly occurs. Note that the right-hand side of
Eq.\,\eqref{eq:glanbd3} does not depend on the winding of
$\,\varphi\,$ along the second $\,\bS^1\,$ factor in $\,\Sigma$.
\,This will find its explanation in the cohomological analysis of
Section \ref{sec:class-equiv-back}, cf.\ Corollary
\ref{cor:obstrXi}, where it will be shown that the isomorphism class
of the flat line bundle $\,D\,$ (given by its holonomy) is
represented by a cohomology class in $\,H^1(\txG,\uj)\subset
H^1(\txG\x Q_\la,\uj)$.
\medskip

For a single boundary maximally-symmetric $\cG_\sfk$-bi-brane with
world-volume $\,\xcC_\la$, \,a similar calculation gives
\qq
\Hol_D\bigl((\chi,\varphi)|_\Gamma)\ =\ 1
\qqq
since the kernel of the map
\qq\label{eq:homom1}
(\iota^{(1)}_2)_*-(\iota^{(1)}_1)_*\,:\,\,H_1(\txG\x\xcC_\la)
\rightarrow H_1(\txG\x(\xcG\sqcup\{\bullet\}))\,,
\qqq
to which the homology class generated by
$\,(\chi,\varphi)|_\Gamma\,$ belongs, is easily seen to vanish. This
is no more the case for general maximally symmetric
$\,\cG_\sfk$-branes. Taking, for example, the disjoint union of two
$\,\cG_\sfk$-branes with world-volume
$\,\xcC_{\la_1}\sqcup\xcC_{\la_2}$, \,one sees that the kernel of
the map
\qq
(\iota^{(1)}_2)_*-(\iota^{(1)}_1)_*\,:\,\,H_1\left(\txG\x(\xcC_{\la_1}\sqcup\xcC_{\la_2})\right)
\rightarrow H_1(\txG\x(\xcG\sqcup\{\bullet\}))
\qqq
is nontrivial. The 1-cycles that probe the global anomaly in that
case correspond to $\,\Sigma=\bS^1\x \bS^1$, $\,\Gamma=\{1\}\x
\bS^1\sqcup\{-1\}\x \bS^1$, and
\qq
\chi(e^{i\sigma_1},e^{i\sigma_2})&=&\exp[\sigma_2\tilde
p^\vee]\,,\cr\cr
\varphi(e^{i\sigma_1},e^{i\sigma_2})&=&\exp[2\pi\lambda_1+2\sigma_1(\lambda_2
-\lambda_1)] \ \ {\rm for}\ \ 0<\sigma_1<\pi\,\,{\rm
mod}\,2\pi\,,\cr\cr \varphi(e^{i\sigma_1},e^{i\sigma_2})&=&\bullet\
\ {\rm for}\ \ \pi<\sigma_1<2\pi\,\,{\rm mod}\,2\pi\,,\cr\cr
\varphi(1,e^{i\sigma_2})&=&\exp[2\pi\lambda_1]\in\xcC_{\la_1}\cr\cr
\varphi(-1,e^{i\sigma_2})&=&\exp[2\pi\lambda_2]\in\xcC_{\la_2}
\nonumber
\qqq
for which
\qq\label{eq:bdanom}
\Hol_\cD\bigl((\chi,\varphi)\bigr)\ =\ \exp[2\pi\sfi\hspace{0.03cm}
\sfk\,{\rm tr}_{\ggt}\, (\la_1-\la_2)\tilde p^\vee]\,.
\qqq
Hence, the absence of contributions to the global anomaly from
circular defects requires, in this case, that
$\,\sfk(\la_1-\la_2)\,$ belongs to the root lattice. As will follow
from the analysis of Section \ref{sec:class-equiv-back}, and in
particular from Corollary \ref{cor:obstrXi}, this is also the
necessary condition for the absence of such contributions. Similar
condition holding for all pair differences $\,\sfk(\la_i-\la_j)\,$
is required by the absence of boundary contributions to the global
gauge anomaly for general maximally symmetric $\,\cG_\sfk$-branes
with world-volumes that are disjoint unions of more than two
conjugacy classes.
\medskip

The analysis in Section \ref{sec:class-equiv-back} below, and in
particular Corollary \ref{cor:glancF}, shows also that the
$\,\uj$-valued functions $\,d_n\,$ that appear in the identity
\eqref{eq:viol-inv} are trivial in the case in question so that
introduction of maximally symmetric inter-bi-branes and intersecting
defects does not lead to further global gauge anomalies.

\section{A groupoidal interpretation of the
constraints}\label{sec:groupoid}

\noindent In this section, we reappraise the constraints for a consistent
gauging, Eqs.\,\eqref{eq:HS1-ids-triv}-\eqref{eq:HS2-ids-triv}, in
more geometric terms, and - in so doing - perform a more precise
identification of the structure underlying the rigid symmetries of
the $\si$-model. This goal will be achieved by establishing a
straightforward link between the $(\txH,\om;\D_Q)$-twisted bracket
structure $\,\Mgt^{(\txH,\om;\D_Q)}\xcF\,$ on the distinguished
$\si$-symmetric sections of the total generalised tangent bundle
$\,\sfE\xcF$,\ introduced in Section \ref{sub:Vino}, and the
(action-)groupoid structure. The relevance of the latter transpires
from our discussion of large gauge transformations in Section
\ref{sec:large} and will be demonstrated convincingly in Section
\ref{sec:nontriv}, where we extend the gauging procedure to nontrivial
gauge bundles.

In what follows, we assume knowledge of basic notions of the
theory of Lie groupoids and Lie algebroids. For the reader's
convenience, we recall the relevant facts in Appendix
\ref{app:oid}.\medskip

Let us, first, be more specific about the algebraic structure on the
$\si$-symmetric sections of $\,\sfE\xcF$.\ We have
\berop\cite[Prop.\,5.10]{Suszek:2012}\label{prop:obstr-Lie-alg}
Let $\,\G_\si(\sfE\xcF)\,$ be the subspace of the $\si$-symmetric
sections, defined in Proposition \ref{prop:sisym}, of the total
generalised tangent bundle $\,\sfE\xcF$,\ introduced in Definition
\ref{def:tw-bra-str}, over the target space $\,\xcF=M\sqcup Q\sqcup
T\,$ of the string background $\,\Bgt$,\ as characterised in
Definition \ref{def:bckgrnd}. Furthermore, let $\,\Mgt^{(\txH,\om;
\D_Q)}_\si\xcF\,$ be the restriction of the $(\txH,\om;\D_Q
)$-twisted bracket structure $\,\Mgt^{(\txH,\om;\D_Q)}\xcF\,$ on
smooth sections of $\,\sfE\xcF\,$ to $\,\G_\si(\sfE \xcF)$.\ The
subspace $\,\a_{\sfT\xcF}\bigl(\G_\si(\sfE\xcF)\bigr)\,$ is a Lie
subalgebra in the Lie algebra of vector fields on $\,\xcF\,$ and it
is isomorphic with the Lie algebra $\,\ggt\,$ of the Lie group
$\,\txG\,$ of the rigid $\si$-model symmetries.\ Fix a basis
$\,\{\xcFup\xcK_a \}\,$ in
$\,\a_{\sfT\xcF}\bigl(\G_\si(\sfE\xcF)\bigr)\,$ composed of the
fundamental vector fields on $\,\xcF\,$ corresponding to the generators
$\,t_a\,$ of $\,\ggt$. \,The associated
$\si$-symmetric sections $\,\Kgt_a\,$ of $\,\G_\si(\sfE\xcF)$,\ with
restrictions
\qq\nn
\Kgt_a\vert_M=\Mup\xcK_a\oplus\kappa_a\,,\qquad\qquad\Kgt_a\vert_Q=
\Qup\xcK_a\oplus k_a\,,\qquad\qquad\Kgt_a\vert_T=\Tnup\xcK_a
\qqq
in which the smooth 1-forms $\,\kappa_a\,$ on $\,M\,$ and the smooth
functions $\,k_a\,$ on $\,Q\,$ are related as in
Eqs.\,\eqref{eq:HS-exact}-\eqref{eq:junct-exact}, satisfy the
algebra
\qq\label{eq:Vinbra-Ka-tw}
\GBra{\Kgt_a}{\Kgt_b}^{(\txH,\om;\D_Q)}=f_{abc}\,\Kgt_c+0\oplus\a_{ab}
\qqq
with
\qq\nn
\a_{ab}\vert_\xcM=\left\{ \barr{ll}
\pLie{a}\kappa_b-f_{abc}\,\kappa_c-\sfd c_{(ab)} \quad & \tx{on}
\quad \xcM=M \cr\cr \pLie{a}k_b-f_{abc}\,k_c+\D_Q c_{(ab)} \quad &
\tx{on} \quad \xcM=Q \cr\cr 0 \quad & \tx{on} \quad \xcM=T_n\earr
\right.\,,\qquad c_{(ab)}=\Vcon{\Kgt_a}{\Kgt_b} \,.
\qqq
\eerop
\noindent It is well-known, cf.\,\Rcite{Gualtieri:2003dx}, that a
non-vanishing scalar product $\,\Vcon{\cdot}{\cdot}\,$ of sections
of the Courant algebroid on $\,\sfE^{(1,1)}M\,$ violates the Leibniz
identity and the Jacobi identity for the (Courant) bracket, and so
prevents the Courant algebroid from becoming a Lie algebroid. A
similar situation occurs in the present setup. In particular, we
find
\qq\label{eq:Leib-oid}\qquad\qquad
\GBra{\Vgt}{f\,\Wgt}^{(\txH,\om;\D_Q)}=f\,\GBra{\Vgt}{\Wgt}^{(\txH,
\om;\D_Q)}+\bigl(\ic_{\a_{\sfT\xcF}(\Vgt)}\sfd f\bigr)\,\Wgt-0\oplus
\tfrac{1}{2}\,\Vcon{\Vgt}{\Wgt}\,\sfd f
\qqq
for arbitrary $\,\Vgt,\Wgt\in\G(\sfE\xcF)\,$ and $\,f\in C^\infty(
\xcF,\bR)$,\ so that, once again, the Leibniz identity is obstructed
by a non-vanishing scalar product. Inspection of the previous
proposition yields
\berop\label{prop:sym-algebroid}
Let $\,\{\Kgt_a\}\,$ be the basis, introduced in Proposition
\ref{prop:obstr-Lie-alg}, of the subspace of $\si$-symmetric
sections of the total generalised tangent bundle $\,\sfE\xcF\,$ over
the target space $\,\xcF=M\sqcup Q\sqcup T\,$ of the string
background $\,\Bgt\,$ of Definition \ref{def:bckgrnd}. The triple
\qq\nn
\Sgt_\Bgt:=\bigl(\oplus_{a=1}^{\dim\,\ggt}\,C^\infty(\xcF,\bR)\,
\Kgt_a,\GBra{\cdot}{\cdot}^{(\txH,\om;\D_Q)},\a_{\sfT\xcF}\bigr)
\qqq
carries a canonical structure of a Lie algebroid iff the
$\,\Kgt_a\,$ satisfy constraints corresponding to
Eqs.\,\eqref{eq:HS1-ids-triv}-\eqref{eq:HS2-ids-triv}.
\eerop
\beroof
The structure of a Lie algebroid on $\,\Sgt_\Bgt\,$ means that the
Leibniz and Jacobi identities for $\,\GBra{\cdot}{\cdot}^{(\txH,\om;
\D_Q)}\,$ hold true on the subspace $\,\oplus_{a=1}^{\dim\,\ggt}\,
C^\infty(\xcF,\bR)\,\Kgt_a$,\ the latter being closed under the
bracket. Inspection of \Reqref{eq:Leib-oid} immediately shows that
the Leibniz identity is tantamount to the isotropy of the subspace,
\qq\label{eq:isotropy}
\txc_{(ab)}=0\,,
\qqq
that is to condition \eqref{eq:HS2-ids-triv}. The closure condition
then takes the form
\qq\nn
\a_{ab}=0\,,
\qqq
which, upon taking into account the earlier \Reqref{eq:isotropy},
becomes the conjunction of conditions \eqref{eq:HS1-ids-triv} and
\eqref{eq:FFM-ids-triv}. The Jacobi identity for the bracket
$\,\GBra{\cdot}{\cdot}^{(\txH,\om;\D_Q)}\,$ restricted to the thus
constrained subspace is simply the Jacobi identity for the structure
constants of $\,\ggt$.\eroof\bigskip  The above considerations lead us to
\bedef\label{def:gsym-algroup}
The Lie algebroid $\,\Sgt_\Bgt\,$ introduced in Proposition
\ref{prop:sym-algebroid} will be called the \textbf{gauge-symmetry
(Lie) algebroid of string background $\,\Bgt$}. Similarly, the
structure
\qq\nn
\txS_\Bgt:=\bigl(\txG\lx\xcF,(\iota_\a\ |\ \a=1,2),(\pi_n^{k,k+1}\
|\ k=1,2,\ldots,n,\ n\geq 3)\bigr)\,,
\qqq
consisting of the action groupoid $\,\txG\lx\xcF\,$ and of the
$\txG$-maps of Proposition \ref{prop:Gnat-IBB-maps}, will be termed
the \textbf{gauge-symmetry (Lie) groupoid of string background
$\,\Bgt$}. \exdef \noindent We conclude the present section with a
theorem that relates the two constructs, $\,\Sgt_\Bgt\,$ and
$\,\txS_\Bgt$,\ and thus provides us with a novel interpretation of
the constraints for a consistent gauging.
\bethe\label{thm:gtAlgebroid}
Let $\,\Bgt\,$ be a string background from Definition
\ref{def:bckgrnd}, and let $\,\ggt\lx\xcF\,$ be the tangent
algebroid, as characterised in Definition \ref{def:tan-alg}, of the
gauge-symmetry groupoid $\,\txS_\Bgt\,$ of Definition
\ref{def:gsym-algroup}. Finally, let $\,\sgt_\Bgt\,$ be the
gauge-symmetry algebroid of background $\,\Bgt\,$ introduced in the
same definition. Then, there exists a canonical isomorphism
\qq\nn
\sgt_\Bgt\cong\ggt\lx\xcF
\qqq
in the sense of Definition \ref{def:grpd} of Appendix
\ref{app:oid}.\ethe
\noindent A constructive proof of the theorem is given in
Appendix \ref{app:proof-algebroid}.

\section{The gauging vs.\ the $\ggt$-equivariance of string
backgrounds}\label{sec:infinit-equiv}

\noindent In the previous section, we established a purely geometric
interpretation of the (infinitesimal) conditions for a consistent
gauging, without any direct reference to the differential-geometric
structures present over the target space\footnote{In fact, there is
a natural relation between the $(\txH,\om;\D_Q)$-twisted bracket
structure $\,\Mgt^{( \txH,\om;\D_Q)}\xcF\,$ on $\si$-symmetric
sections of the total generalised tangent bundle $\,\sfE\xcF\,$ and
the 2-category of bundle gerbes with curving and connection over
$\,\xcF$.\ Over $\,M$,\ it boils down to the Hitchin morphism, first
reported in \Rcite{Hitchin:2005in}, between the $\txH$-twisted
Courant algebroid on $\,\sfE^{(1,1)}M\,$ and an untwisted Courant
algebroid on the generalised tangent bundle over $\,M\,$ twisted by
(local data of) the gerbe $\,\cG\,$ of curvature $\,\txH$.\ The
relation is worked out in all generality in \Rcite{Suszek:2012}. We
shall not pursue this aspect of $\si$-model symmetries in the
present paper.}. Here, we shall change the angle once more and
reinterpret the same conditions in terms of a $\ggt$-equivariant
extension of the Deligne hypercohomology that captures the local
description of the target-space structure $\,(\cG,\Phi,\varphi_n\
\vert\ n\geq3)$.\ This is to be viewed as a natural step towards the
full-fledged $\txG$-equivariant structure on $\,\Bgt\,$ that will be
developed in order to analyse topologically nontrivial sectors of
gauged $\si$-model.

\subsection{The local description of the background structure}
\label{subsec:Locdesc}

\noindent Below, we briefly review the sheaf-cohomological description of the
target-space gerbe $\,\cG$,\ the bi-brane 1-isomorphism $\,\Phi\,$
and the inter-bi-brane 2-isomorphisms $\,\varphi_n$.\ We confine
ourselves to the basic elements of the description, relegating the
full-blown discussion to Section \ref{sub:Deligne} in which we
employ the cohomological description in a classification of
$\txG$-equivariant structures on string backgrounds.

The point of departure is the Deligne complex
\qq\label{Delcom}
\cD(4)^\bullet\ :\ 0\to\unl{2\pi\bZ}_\xcF\to\unl\Om^0(\xcF)
\xrightarrow{\sfd}\unl\Om^1(\xcF)\xrightarrow{\sfd}\unl\Om^2(\xcF)
\xrightarrow{\sfd}\unl\Om^3(\xcF)
\qqq
of the following differential sheaves over the target space: the
sheaf $\,\unl{2\pi\bZ}_\xcF\,$ of locally constant $2\pi\bZ$-valued
functions on $\,\xcF\,$ (trivially injected in its successor), and
the sheaves $\,\unl\Om^p(\xcF),\ p=0,1,2\,$ of locally smooth (real)
$p$-forms on $\,\xcF$.\ Given a choice
$\,\cO^\xcF=\{O^\xcF_i\}_{i\in\xcI^\xcF} \,$ of a good open cover of
$\,\xcF\,$ (i.e.\ an open cover with contractible non-empty multiple
intersections of its elements, to be denoted as
$\,\cO^\xcF_{ij}=\cO^\xcF_i\cap\cO^\xcF_j,\ \cO^\xcF_{ijk}=
\cO^\xcF_i\cap\cO^\xcF_j\cap\cO^\xcF_k\,$ etc.), the Deligne complex
is extended in the direction of the associated \v Cech cohomology,
whereby the standard \v Cech-Deligne double complex $\,\vC^\bullet
\bigl(\cO^\xcF,\cD(4)^\bullet\bigr)\,$ is formed. The cohomology
that describes local data of $\,\cG,\ \Phi\,$ and $\,\varphi_n$,\ as
well as the usual (gauge) ambiguities in their definition is exactly
the cohomology of that double complex, defined as the cohomology of
the diagonal subcomplex
\qq\nn
A_\xcF^\bullet(\cO^\xcF)=\bigoplus_{r=0,1,\ldots}\,A_\xcF^r(\cO^\xcF)\,,
\qquad\qquad A_\xcF^r(\cO^\xcF)=\bigoplus_{p+q=r}\,\vC^p\bigl(
\cO^\xcF,\cD(4)^q\bigr)
\qqq
with respect to the Deligne differential $\,D_{r}:A_\xcF^r(
\cO^\xcF)\to A_\xcF^{r+1}(\cO^\xcF)$.\ We give explicit formul\ae
~for the latter on representative \v Cech-Deligne cochains for the
first few values of $\,r$.\ Thus, for cochains
\qq\nn
(p_i)&\in&
A_\xcF^0(\cO^\xcF)=\vC^0\bigl(\cO^\xcF,\unl{2\pi\bZ}_\xcF\bigr)
\,,\cr\cr (f_i,q_{ij})&\in& A_\xcF^1(\cO^\xcF)=\vC^0\bigl(\cO^\xcF,
\unl\Om^0(\xcF)\bigr)\oplus\vC^1\bigl(\cO^\xcF,\unl{2\pi\bZ}_\xcF\bigr)\,,
\cr\cr (P_i,k_{ij},r_{ijk})&\in& A_\xcF^2(\cO^\xcF)=\vC^0\bigl(
\cO^\xcF,\unl\Om^1(\xcF)\bigr)\oplus\vC^1\bigl(\cO^\xcF,\unl\Om^0(
\xcF)\bigr)\oplus\vC^2\bigl(\cO^\xcF,\unl{2\pi\bZ}_\xcF\bigr)\,,\cr\cr
(B_i,A_{ij},h_{ijk},s_{ijkl})&\in& A_\xcF^3(\cO^\xcF)=\vC^0\bigl(
\cO^\xcF,\unl\Om^2(\xcF)\bigr)\oplus\vC^1\bigl(\cO^\xcF,\unl\Om^1(
\xcF)\bigr)\oplus\vC^2\bigl(\cO^\xcF,\unl\Om^0(\xcF)\bigr)\cr\cr
&&\hspace{1.9cm}\oplus\,\vC^3\bigl(\cO^\xcF,\unl{2\pi\bZ}_\xcF\bigr)\,,
\qqq
we have
\qq\nn
D_{0}(p_i)&=&\bigl(p_i,(p_j-p_i)\vert_{\cO^\xcF_{ij}}\bigr)\,,\cr
\cr
D_{1}(f_i,q_{ij})&=&\bigl(\sfd f_i,(f_i-f_j)\vert_{\cO^\xcF_{ij}}+
q_{i j},(q_{jk}-q_{ik}+q_{ij})\vert_{\cO^\xcF_{ijk}}\bigr)\,,\cr\cr
D_{2}(P_i,k_{ij},r_{ijk})&=&\bigl(\sfd P_i,(P_j-P_i)
\vert_{\cO^\xcF_{ij}}+\sfd k_{ij},(-k_{jk}+k_{ik}-k_{ij})
\vert_{\cO^\xcF_{ijk}}+r_{ijk},\cr\cr
&&(r_{jkl}-r_{ikl}+r_{ijl}-r_{ijk})\vert_{\cO^\xcF_{ijkl}}\bigr)\,,
\cr\cr
D_{3}(B_i,A_{ij},h_{ijk},s_{ijkl})&=&\bigl(\sfd B_i,
(B_j-B_i)
\vert_{\cO^\xcF_{ij}}-\sfd A_{ij},(A_{jk}-A_{ik}+A_{ij})
\vert_{\cO^\xcF_{ijk}}+\sfd h_{ijk},\cr\cr
&&(-h_{jkl}+h_{ikl}-h_{ijl}+h_{ijk})\vert_{\cO^\xcF_{ijkl}}+s_{ijkl}
,\cr\cr
&&(s_{jklm}-s_{iklm}+s_{ijlm}-s_{ijkm}+s_{ijkl})\vert_{\cO^\xcF_{ij
klm}}\bigr)\,.
\qqq
We may now describe the gerbe $\,\cG\,$ with the curvature 3-form
$\,\txH\,$ in terms of its local data
$\,(B_i,A_{ij},h_{ijk},s_{ijkl})\in A_M^3(\cO^M)\,$ satisfying the
relation
\qq\label{eq:gerbe-loc}
D_{3}(B_i,A_{ij},h_{ijk},s_{ijkl})=(\txH\vert_{\cO^M_i},0,0,0,0)\,.
\qqq
Similarly, the 1-isomorphism $\,\Phi\,$ may be described by local data
$\,(P_i,k_{ij},r_{ij
k})\in A_Q^2(\cO^Q)\,$ verifying the identity
\qq
\iota_1^*(B_{\phi_1(i)},A_{\phi_1(i)\phi_1(j)},h_{\phi_1(i)\phi_1(j
)\phi_1(k)},s_{\phi_1(i)\phi_1(j)\phi_1(k)\phi_1(l)})+D_{2}(P_i,
k_{ij},r_{ijk})\cr\\\label{eq:Cext-bib-spell}
=\iota_2^*(B_{\phi_2(i)},A_{\phi_2(i)\phi_2(j)},h_{\phi_2(i)\phi_2(
j)\phi_2(k)},s_{\phi_2(i)\phi_2(j)\phi_2(k)\phi_2(l)})+(\om
\vert_{\cO^Q_i},0,0,0)\,,\nonumber
\qqq
written in terms of the index maps $\,\phi_\a:\xcI^Q\to\xcI^M\,$
that cover (or \v Cech-extend) the $\,\iota_\a\,$ for covers of
$\,M\,$ and $\,Q\,$ chosen in conformity with the condition
\qq\label{eq:Cext-bib}
\iota_\a(\cO^Q_i)\subset\cO^M_{\phi_\a(i)}\,,
\qqq
cf.\ \Rcite{Runkel:2008gr}. Finally, local data $\,(f_{n\,i},q_{n\,
ij})\in A_{T_n}^2(\cO^{T_n})\,$ of the 2-isomorphism $\,\varphi_n\,$
are readily seen to obey
\qq
\sum_{k=1}^n\,\vep_n^{k,k+1}\,\pi_n^{k,k+1\,*}(P_{\psi_n^{k,k+1}(i
)},k_{\psi_n^{k,k+1}(i)\psi_n^{k,k+1}(j)},r_{\psi_n^{k,k+1}(i)
\psi_n^{k,k+1}(j)\psi_n^{k,k+1}(k)})\ \cr
+\,D_{1}(f_{n\,i},q_{n\,ij})=0\,,\cr \label{eq:Cext-ibb-spell}
\qqq
where the index maps $\,\psi_n^{k,k+1}:\xcI^{T_n}\to\xcI^Q\,$
covering the inter-bi-brane maps $\,\pi_n^{k,k+1}\,$ have been used
for covers of $\,Q\,$ and $\,T_n\,$ chosen such that the relations
\qq\label{eq:Cext-ibb}
\pi_n^{k,k+1}(\cO^{T_n}_i)\subset\cO^Q_{\psi_n^{k,k+1}(i)}
\qqq
hold true.

\becon
Following \Rcite{Runkel:2008gr}, we shall use the symbols
$\,\check\iota_\a\,$ and $\,\check\pi_n^{k,k+1}\,$ for the \v
Cech-extended maps, e.g., given $\,b:=(B_i,A_{ij},h_{ijk},s_{ijkl})
\in A_M^3(\cO^M)\,$ and $\,p=(P_i,k_{ij},r_{ijk}) \in A^2_Q(\cO^Q)$
,\ we shall write
\qq\nn
\check\iota_\a^*b&\equiv&\iota_\a^*(B_{\phi_\a(i)},A_{\phi_\a(i)
\phi_\a(j)},h_{\phi_\a(i)\phi_\a(j)\phi_\a(k)},s_{\phi_\a(i)\phi_\a
(j)\phi_\a(k)\phi_\a(l)})\,,\cr\cr
\check\pi_n^{k,k+1\,*}p&\equiv&\pi_n^{k,k+1\,*}(P_{\psi_n^{k,k+1}(i
)},k_{\psi_n^{k,k+1}(i)\psi_n^{k,k+1}(j)},r_{\psi_n^{k,k+1}(i)
\psi_n^{k,k+1}(j)\psi_n^{k,k+1}(k)})\,.
\qqq
\econ

\subsection{$\ggt$-equivariant structures}

\noindent We shall now gradually descend in the diagonal subcomplex of the
triple complex obtained by extending $\,\vC^\bullet\bigl(\cO^\xcF,
\cD(4)^\bullet\bigr)\,$ in the direction of $\ggt$-cohomology, as
dictated by the constraints for a consistent gauging derived
earlier. Thus, we find
\berop\label{prop:HS-as-gequiv-str}
Let $\,M\,$ be a $\txG$-space with a gerbe $\,\cG\,$ of a
$\txG$-invariant curvature $\,\txH\,$ over it, and let $\,b:=(B_i,
A_{ij},h_{ijk},s_{ijkl})\in A^3_M(\cO^M)\,$ be local data of
$\,\cG\,$ with respect to a good open cover $\,\cO^M=\{\cO^M_i\}_{i
\in\xcI^M}\,$ of $\,M$,\ so that, in particular,
\qq\label{eq:gerbe-loc-bis}
D_{3}b=(\txH\vert_{\cO^M_i},0,0,0,0)\,.
\qqq
Given $\,\kappa_a\in\Om^1(M),\ a=1,2,\ldots,\dim\,\ggt$, \,that
satisfy \Reqref{eq:HS-exact}, write
\qq\label{eq:UpsA}
\Upsilon_a:=(-\kappa_a\vert_{\cO^M_i}+\ic_a B_i,\ic_a A_{ij},0)\in
A^2_M(\cO^M)
\qqq
and
\qq\label{eq:gammAB}
\g_{ab}:=\bigl(\txc_{ba}\vert_{\cO^M_i}+\ic_a\ic_b B_i,0\bigr)\in
A^1_M(\cO^M)
\qqq
for $\,\Mup\xcK_a\,$ the fundamental vector fields on $\,M\,$ from
Definition \ref{eq:Gact-M}. Relations \eqref{eq:HS1-ids-triv} and
\eqref{eq:HS2-ids-triv} are then satisfied iff there exists a
$\ggt$-equivariant structure on $\,\cG$,\ described locally by
relations
\qq
&&\pLie{a}b=D_{2}\Upsilon_a\,,\label{eq:Kab-Ups}\\\cr
&&\pLie{a}\Upsilon_b-\pLie{b}\Upsilon_a-f_{abc}\,
\Upsilon_c=D_{1}\g_{ab}\,,\label{eq:KaUps-gam}\\\cr
&&\pLie{a}\g_{bc}-\pLie{b}\g_{ac}+\pLie{c}\g_{ab}-
f_{abd}\,\g_{dc}+f_{acd}\,\g_{db}-f_{bcd}\,\g_{da}=0\,.\hspace*{0.7cm}
\label{eq:Kagam-0}
\qqq
\eerop
\noindent A proof of the proposition is given in Appendix
\ref{app:proof-gequiv-target}.\bigskip

Along similar lines, we may rephrase \Reqref{eq:bdry-exact} together
with the constraint \eqref{eq:FFM-ids-triv} in the presence of a
defect in terms of a $\cG$-twisted $\ggt$-equivariant structure on
the bi-brane.
\berop\label{prop:FFM-as-gequiv-str}
Let $\,M\,$ and $\,Q\,$ be a pair of $\txG$-spaces, equipped with a
pair of smooth $\txG$-equivariant maps $\,\iota_\a:Q\to M,\ \a=1
,2$,\ and let $\,\cO^M=\{\cO^M_i\}_{i\in\xcI^M}\,$ and $\,\cO^Q=
\{\cO^Q_i\}_{i\in\xcI^Q}\,$ be the respective good open covers,
chosen such that there exist \Cv ech extensions $\,\check\iota_\a=(
\iota_\a,\phi_\a)$. \,Furthermore, let $\,\cG\,$ be a gerbe over
$\,M\,$ with a $\txG$-invariant curvature $\,\txH\,$ and local data
$\,b\in A^3_M(\cO^M)$.\ Suppose that $\,\cG\,$ is endowed with a
$\ggt$-equivariant structure, understood in terms of its local data
given by the $\,(\Upsilon_a,\g_{ab})\,$ with the $\,\Upsilon_a\,$
and the $\,\g_{ab}\,$ as in Eqs.\,\eqref{eq:UpsA} and
\eqref{eq:gammAB} of Proposition \ref{prop:HS-as-gequiv-str},
respectively. Finally, let $\,p=(P_i,k_{ij},r_{ijk}) \in
A^2_Q(\cO^Q)\,$ be local data of a 1-isomorphism $\,\Phi:
\iota_1^*\cG\xrightarrow{\cong}\iota_2^*\cG\ox I_\om\,$ forming part
of the structure of a $\cG$-bi-brane
$\,\cB=(Q,\iota_1,\iota_2,\om,\Phi)\,$ with a $\txG$-invariant
curvature $\,\om$,\ so that, in particular,
\qq\nn
D_{2}p=\check\D_Q b+(\om\vert_{\cO^Q_i},0,0,0)\,,
\qqq
where we have introduced the \Cv ech-extended pullback operator
\qq\nn
\check\D_Q:=\check\iota_2^*-\check\iota_1^*\,.
\qqq
Given $\,k_a\in C^\infty(Q,\bR),\ a=1,2,\ldots,\dim\,\ggt\,$ that
satisfy \Reqref{eq:bdry-exact}, write
\qq\label{eq:XiA}
\Xi_a:=(-k_a \vert_{\cO^Q_i}+\ic_a P_i,0)\in A^1_Q(\cO^Q)\,.
\qqq
Relation \eqref{eq:FFM-ids-triv} is satisfied iff there exists a
$\cG$-twisted $\ggt$-equivariant structure on $\,\cB$,\ described
locally by relations
\qq
&&\pLie{a}p=\check\D_Q\Upsilon_a+D_{1}\Xi_a\,,
\label{eq:Kap-DupXi}\\\cr
&&\pLie{a}\Xi_B-\pLie{b}\Xi_a-f_{abc}\,\Xi_c=-
\check\D_Q\g_{ab}\,.\label{eq:KaXi-0}
\qqq
\eerop
\noindent A proof of the proposition is given in Appendix
\ref{app:proof-gequiv-bib}.\bigskip

We conclude the present section by incorporating the remaining
relation \eqref{eq:junct-exact} into the $\ggt$-equivariance scheme.
\berop\label{prop:djc-as-gequiv-str}
Let $\,(M,Q,T),\ T=\bigsqcup_{n\geq 3}\,T_n\,$ be a family of
$\txG$-spaces, equipped with smooth $\txG$-equivariant maps
$\,\iota_\a:Q\to M,\ \a=1,2\,$ and $\,\pi_n^{k,k+1}:T_n\to Q,\
k=1,2,\ldots,n$,\ and let $\,\cO^M=\{\cO^M_i\}_{i\in\xcI^M},\ \cO^Q=
\{\cO^Q_i\}_{i\in\xcI^Q}\,$ and $\,\cO^{T_n}=\{\cO^{T_n}_i
\}_{i\in\xcI^{T_n}}\,$ be the respective good open covers, chosen
such that there exist \Cv ech extensions $\,\check\iota_\a=(\iota_\a
,\phi_\a)\,$ and $\,\check\pi_n^{k,k+1}=(\pi_n^{k,k+1},\psi_n^{k,k+
1})$. \,Furthermore, let $\,\cG\,$ be a gerbe over $\,M\,$ with a
$\txG$-invariant curvature $\,\txH\,$ and local data $\,b\in A^3_M(
\cO^M)\,$ and let $\,\cB=(Q,\iota_1,\iota_2,\om,\Phi)\,$ be a
$\cG$-bi-brane with a $\txG$-invariant curvature $\,\om\,$ and a
1-isomorphism $\,\Phi:\iota_1^*\cG\xrightarrow{\cong}\iota_2^*\cG
\ox I_\om\,$ with local data $\,p\in A^2_Q(\cO^Q)$.\ Suppose that
both $\,\cG\,$ and $\,\cB\,$ are endowed with a $\ggt$-equivariant
structure, understood in terms of its local data given by the
$\,(\Upsilon_a,\g_{ab})\,$ as in Eqs.\,\eqref{eq:UpsA} and
\eqref{eq:gammAB} of Proposition \ref{prop:HS-as-gequiv-str}, and
the $\,\Xi_a\,$ as in \Reqref{eq:XiA} of Proposition
\ref{prop:FFM-as-gequiv-str}, respectively. Finally, let $\,h_n=
(f_{n\,i},q_{n\,ij})\in A^1_{T_n}(\cO^{T_n})\,$ be local data of
2-isomorphisms of diagram \eqref{diag:2iso} forming part of the
structure of an $(\cG,\cB)$-inter-bi-brane
$\,\cJ=\bigl(T_n,\bigl(\vep^{k,k+
1}_n,\pi^{k,k+1}_n\bigr);\varphi_n\bigr)$,\ so that, in particular,
\qq\nn
D_{1}h_n=-\check\D_{T_n}p\,,
\qqq
where we have introduced the \Cv ech-extended pullback operator
\qq\nn
\check\D_{T_n}:=\sum_{k=1}^n\,\vep_n^{k,k+1}\,\check\pi_n^{k,k+1\,
*}\,.
\qqq
Relation \eqref{eq:junct-exact} is satisfied iff there exists a
$\cB$-twisted $\ggt$-equivariant structure on $\,\cJ$,\ described
locally by the relation
\qq\nn
\pLie{a}h_n=-\check\D_{T_n}\Xi_a\,.
\qqq
\eerop
\noindent A proof of the proposition is given in Appendix
\ref{app:proof-gequiv-ibb}.

\section{$\txG$-equivariant string backgrounds}\label{sec:Gequiv}

\noindent Our discussion in Section \ref{sec:large} of the circumstances in
which the global gauge anomaly of the $\si$-model for arbitrary
network-field configurations in the background of topologically
trivial gauge fields vanishes revealed the necessity of endowing the
string background $\,\Bgt=(\cM,\cB,\cJ)\,$ with additional
structure, composed of a 1-isomorphism $\,\Upsilon\,$ over $\,\txG\x
M\,$ and a 2-isomorphism $\,\Xi\,$ over $\,\txG\x Q$,\ related as in
Theorem \ref{thm:equiv-12iso-constr}. This additional structure can
be viewed as the first component of a larger construct involving the
geometric structure $\,(\cG,\cB,\cJ)\,$ over the target space
$\,\xcF=M\sqcup Q\sqcup T\,$ equipped with the action of a group
$\,\txG\,$ of rigid symmetries of the $\,\si$-model.\ The subsequent
algebraic reinterpretation of conditions
\eqref{eq:HS1-ids-triv}-\eqref{eq:HS2-ids-triv} in terms of a
$\ggt$-equivariant structure on $\,(\cG,\cB,\cJ)$,\ provided in
Section \ref{sec:infinit-equiv}, suggests already an extension of
this first component to what we shall introduce below under the name
of a $\txG$-equivariant string background. Its existence is a
sufficient and rather natural condition for defining the gauged
$\si$-model coupled to gauge fields in an arbitrary principal
$\,\txG$-bundles over world-sheets, an issue discussed in Section
\ref{sec:nontriv}. It is also suggested by the previous studies of
orbifold and orientifold $\si$-models in
Refs.\,\cite{Gawedzki:2003pm,Gawedzki:2004tu,Schreiber:2005mi,Gawedzki:2007uz,Gawedzki:2008um} dealing with discrete symmetries of the string
background. Finally, it turns out to be a
necessary and sufficient condition for the gauged $\si$-model to
define a $\si$-model on the quotient $\,\xcF/\txG\,$ (at least
for smooth quotients $\,\xcF/\txG$), as elaborated in Section
\ref{sec:coset}.
\smallskip

The definitions that make up this section rest heavily on the
simplicial formalism and notation laid out in Section
\ref{sec:groupoid}. Their physical significance will be
substantiated in the remainder of the paper, in which the
fundamental argument will be the descent principle discussed in the
last part of the present section.

\subsection{$\txG$-equivariant structures}

\noindent Let us start by recalling
\bedef\cite[Def.\,5.1]{Gawedzki:2010rn}\label{def:equiv-gerbe}
Let $\,\xcM\,$ be a $\txG$-space, and let $\,\txG\xcM\,$ be the
nerve of the action groupoid $\,\txG\lx\xcM$,\ with the
corresponding face maps $\,\xcMup d^{(n+1)}_i$,\ as introduced in
Definition \ref{def:nerves}. Furthermore, let $\,\cG\,$ be a gerbe
over $\,\xcM\,$ with a $\txG$-invariant curvature $\,\txH\,$
admitting a $\ggt$-equivariantly closed $\txG$-equivariant
(Cartan-model) extension $\,\widehat\txH=\txH-\kappa,\
\kappa\in\Om^1(\xcM)\ox \ggt^*$.\ Finally, let $\,\rho\,$ be the
2-form on $\,\txG\x\xcM\,$ given in \Reqref{eq:rho-on-G}. A
\textbf{$\txG$-equivariant structure on gerbe $\,\cG\,$ relative to
2-form $\,\rho\,$} is a pair $\,(\Upsilon,\g)\,$ consisting of
\bit
\item a 1-isomorphism
\qq\label{eq:equiv-gerb-1}
\Upsilon\ :\ \xcMup d_1^{(1)\,*}\cG\xrightarrow{\cong}\xcMup d_0^{(1
)\,*}\cG\ox I_\rho
\qqq
of gerbes over $\,\txG\x\xcM$;
\item a 2-isomorphism
\qq\label{diag:2-iso-Gequiv}
\qquad\alxydim{@C=1cm@R=2cm}{ \bigl(\xcMup d^{(1)}_1\circ\xcMup d^{(2)}_1
\bigr)^*\cG \ar[r]^{\xcMup d^{(2)\,*}_2\Upsilon\hspace{1cm}}
\ar[d]_{\xcMup d^{(2)\,*}_1\Upsilon} & \bigl(\xcMup d^{(1)}_1\circ
\xcMup d^{(2)}_0\bigr)^*\cG\ox I_{\xcMup d^{(2)\,*}_2\rho}
\ar[d]^{\xcMup d^{(2)\,*}_0\Upsilon \ox\id} \ar@{=>}[dl]|{\,\g\ } \\
\bigl(\xcMup d^{(1)}_0\circ\xcMup d^{(2)}_1\bigr)^*\cG\ox I_{\xcMup
d^{(2)\,*}_1\rho} \ar@{=}[r] & \bigl(\xcMup d^{(1)}_0\circ\xcMup
d^{(2)}_0\bigr)^*\cG\ox I_{\xcMup d^{(2) *}_0\rho+\xcMup d^{(2)\,
*}_2\rho}}
\qqq
between the 1-isomorphisms over $\,\txG^2\x\xcM$,\ satisfying, over
$\,\txG^3\x\xcM$,\ the coherence condition
\qq\label{eq:Gerbe-1iso-coh}
\xcMup d_1^{(3)\,*}\g\bullet\bigl(\id\circ\xcMup d_3^{(3)\,*}\g
\bigr)=\xcMup d_2^{(3)\,*}\g\bullet\bigl(\bigl(\xcMup d_0^{(3)\,*}\g
\ox\id\bigr)\circ\id\bigr)\,,
\qqq
equivalently expressed by the commutative 2-diagram
\eqref{diag:Gerbe-1iso-coh} from Appendix \ref{app:Gequiv-diags}.
\eit
Henceforth, the quadruple $\,(\cG,\Upsilon,\g;\kappa)\,$ will be
called a \textbf{gerbe $\txG$-equivariant relative to 2-form
$\,\rho\,$}, or, simpler, a \textbf{$(\txG,\rho)$-equivariant
gerbe}, or, whenever there is no risk of confusion, just a
\textbf{$\txG$-equivariant gerbe}. \exdef \noindent Note that all
the arrows in diagram \eqref{diag:2-iso-Gequiv} make sense in virtue
of Eqs.\,\eqref{eq:simpl-rel-dddd} and \eqref{eq:delro}.\medskip

\noindent Another useful notion is introduced in
\bedef\cite[Sec.\,V.A]{Gawedzki:2010rn}\label{def:equiv-1iso}
Let $\,(\cG_\b,\Upsilon_\b,\g_\b;\kappa),\ \b=1,2\,$ be a pair of
$\txG$-equivariant gerbes of curvature $\,\txH\,$ over a
$\txG$-space $\,\xcM$,\ as characterised in Definition
\ref{def:equiv-gerbe}. A \textbf{$\txG$-equivariant 1-isomorphism}
between them is a pair $\,(\Psi_{1,2},\eta_{1,2})\,$ consisting of a
1-isomorphism
\qq\nn
\Psi_{1,2}\ :\ \cG_1\xrightarrow{\cong}\cG_2
\qqq
and of a 2-isomorphism
\qq\nn
\alxydim{@C=3cm@R=2cm}{ \xcMup d^{(1)\,*}_1\cG_1 \ar[r]^{\xcMup
d^{(1)\,*}_1\Psi_{1,2}} \ar[d]_{\Upsilon_1} & \xcMup d^{(1)\,
*}_1\cG_2 \ar[d]^{\Upsilon_2} \\ \xcMup d^{(1)\,*}_0\cG_1\ox I_\rho
\ar[r]_{\xcMup d^{(1)\,*}_0\Psi_{1,2}\ox\id} \ar@{=>}[ur]|{{\
\eta_{1,2}\ }} & \xcMup d^{(1)\,*}_0\cG_2\ox I_\rho}
\qqq
between the 1-isomorphisms over $\,\txG\x\xcM$,\ satisfying, over
$\,\txG^2\x\xcM$,\ the coherence condition
\qq\label{eq:coh-cond-equiv-1iso}\qquad\qquad
\xcMup d^{(2)\,*}_1\eta_{1,2}\bullet(\id\circ\g_1)=(\g_2\circ\id)
\bullet\bigl(\id\circ\xcMup d^{(2)\,*}_2\eta_{1,2}\bigr)\bullet
\bigl(\bigl(\xcMup d^{(2)\,*}_0\eta_{1,2}\ox\id\bigr)\circ\id\bigr)
\,,
\qqq
equivalently expressed by the commutative 2-diagram
\eqref{diag:Gequiv1iso-coh} from Appendix \ref{app:Gequiv-diags}.

The gerbes $\,(\cG_\b,\Upsilon_\b,\g_\b;\kappa)\,$ will be called
\textbf{$\txG$-equivariantly 1-isomorphic}. In the distinguished
case of $\,(\cG_1,\kappa_1)=(\cG_2,\kappa_2)=: (\cG,\kappa)$,\ we
shall speak of \textbf{equivalent $\txG$-equivariant structures on
$\,\cG\,$ relative to $\,\rho$}. \exdef \noindent Once more, it is
due to relations \eqref{eq:simpl-rel-dddd} and \eqref{eq:delro} that
the above diagram is well-defined.\medskip

\noindent We complete the basic description of $\txG$-equivariant
gerbes with
\bedef\cite[Sec.\,V.A]{Gawedzki:2010rn}
Let $\,(\Psi_{1,2}^\a,\eta_{1,2}^\a),\ \a=1,2\,$ be a pair of
$\txG$-equivariant 1-isomorphisms between two $\txG$-equivariant
gerbes $\,(\cG_\b,\Upsilon_\b,\g_\b;\kappa_\b),$\ $\b=1,2\,$ of
curvature $\,\txH\,$ over a $\txG$-space $\,\xcM$.\ A
\textbf{$\txG$-equivariant 2-isomorphism} between them is a
2-isomorphism
\qq\nn
\alxydim{}{\cG_1 \ar@/^1.6pc/[rrr]^{\Psi_{1,2}^1}="5"
\ar@/_1.6pc/[rrr]_{\Psi_{1,2}^2}="6"
\ar@{=>}"5"+(0,-4);"6"+(0,4)|{\psi_{1,2}} & & & \cG_2}\,,
\qqq
satisfying, over $\,\txG\x\xcM$,\ the coherence condition
\qq\label{eq:Gequiv-2iso-coh}\qquad\qquad
\bigl(\id\circ\xcMup d^{(1)\,*}_1\psi_{1,2}\bigr)\bullet\eta_{1,
2}^1=\eta_{1,2}^2\bullet\bigl(\bigl(\xcMup d^{(1)\,*}_0\psi_{1,2}\ox
\id\bigr)\circ\id\bigr)\,,
\qqq
equivalently expressed by the commutative 2-diagram
\eqref{diag:Gequiv2iso-coh} from Appendix \ref{app:Gequiv-diags}.
\exdef ~\medskip

In the next step, we extend the notion of $\txG$-equivariance to
bi-branes.
\bedef\label{def:Gequiv-bib}
Let $\,M,Q\,$ be a pair of $\txG$-spaces and let $\,\txG\xcM,\
\xcM=M,Q\,$ be the respective nerves of the action groupoids
$\,\txG\lx\xcM$,\ with face maps $\,\xcMup d^{(n+1)}_i$,\ as
introduced in Definition \ref{def:nerves}. Furthermore, let
$\,(\cG,\Upsilon,\g;\kappa)\,$ be a $(\txG,\rho)$-equivariant gerbe
over $\,M\,$ as described in Definition \ref{def:equiv-gerbe}.
Consider a $\cG$-bi-brane $\,\cB=(Q,\iota_\a,\om,\Phi)\,$ with
$\,Q\,$ as a world-volume, and a pair of smooth $\txG$-maps
$\,\iota_\a :Q\to M$.\ Suppose the curvature 2-form $\,\om\,$ is
$\txG$-invariant and admits a $\txG$-equivariant (Cartan-model)
extension $\,\widehat\om=\om-k,\ k\in\Om^0(Q)\ox\ggt^*\,$ satisfying
\Reqref{eq:hatdom-hatdelHa} and the relation
\qq\label{eq:dhatdro=hatHH}
\widehat\sfd\widehat\om=-\D_Q\widehat\txH
\qqq
in which $\,\D_Q=\iota_2^*-\iota_1^*\,$ and $\,\widehat\txH\,$ is
the $\txG$-equivariant extension of the curvature $\,\txH\,$ of
$\,\cG$.\ Finally, let $\,\la\,$ be the 1-form on $\,\txG\x Q\,$
given in \Reqref{eq:la-on-G}. A \textbf{$\txG$-equivariant structure
on $\cG$-bi-brane $\,\cB\,$ relative to 1-form $\,\la\,$} is a
2-isomorphism
\qq\label{diag:penta-first}
\xy (50,0)*{\bullet}="G12"+(2,5)*{\Qup d^{(1)\,*}_0\iota_2^*\cG\ox
I_{\Qup d^{(1)\,*}_0\om+\iota_1^{(1)\,*}\rho}};
(25,-20)*{\bullet}="G1r1"+(-15,0)*{\Qup d^{(1)\,*}_0\iota_1^*\cG\ox
I_{\iota_1^{(1)\,*}\rho}}; (75,-20)*{\bullet}="G2om"+(21,0)*{\Qup
d^{(1)\,*}_0\iota_2^*\cG\ox I_{\Qup d^{(1)\,*}_1\om+\iota_2^{(1)\,*}
\rho}}; (35,-40)*{\bullet}="G2or1"+(0,-5)*{\Qup d^{(1)\,*}_1
\iota_1^*\cG}; (65,-40)*{\bullet}="G2or2"+(8,-5)*{\Qup d^{(1)\,*}_1
\iota_2^*\cG\ox I_{\Qup d^{(1)\,*}_1\om}}; \ar@{->}|{\Qup d^{(1)\,
*}_0\Phi\ox\id} "G1r1";"G12" \ar@{->}|{\id\ox J_\la} "G12";"G2om"
\ar@{->}|{\iota_1^{(1)\,*}\Upsilon} "G2or1";"G1r1"
\ar@{->}|{\iota_2^{(1)\,*}\Upsilon^{-1}\ox\id} "G2om";"G2or2"
\ar@{->}|{\Qup d^{(1)\,*}_1\Phi} "G2or1";"G2or2" \ar@{=>}|{\Xi}
"G2or1"+(15,3);"G12"+(0,-3)
\endxy
\qqq
between the 1-isomorphisms over $\,\txG\x Q$,\ subject to the
coherence condition
\qq
\bigl(\bigl(\iota_2^{(1)\,*}\g^\sharp\ox\id\bigr)\circ\id\bigr)
\bullet\Qup d^{(2)\,*}_1\Xi=\bigl(\id\circ\iota_1^{(1)\,*}\g\bigr)
\bullet\bigl(\id\circ\bigl(\Qup d_0^{(2)\,*}\Xi\ox\id\bigr)\circ\id
\bigr)\bullet\Qup d_2^{(2)\,*}\Xi\,,\cr\label{eq:Gequiv-bimod-coh}
\qqq
equivalently expressed by the commutative 2-diagram
\eqref{diag:Gequiv-bimod-coh} from Appendix \ref{app:Gequiv-diags}.

Henceforth, the triple $\,(\cB,\Xi;k)\,$ will be called a
\textbf{$(\cG,\Upsilon,\g;\kappa)$-bi-brane $\txG$-equivariant
relative to 1-form $\,\la$},\ or, simpler, a
\textbf{$(\txG,\la)$-equivariant
$(\cG,\Upsilon,\g;\kappa)$-bi-brane}, or, whenever there is no risk
of confusion, just a \textbf{$\txG$-equivariant ($\cG$-)bi-brane}.
\exdef ~\medskip

\noindent In order to classify $\txG$-equivariant bi-branes, we
employ
\bedef\label{def:equiv-Gequiv-bi}
Let $\,(\cG,\Upsilon,\g;\kappa)\,$ be a $(\txG,\rho)$-equivariant
gerbe over a $\txG$-space $\,M$,\ as introduced in Definition
\ref{def:equiv-gerbe}, and let $\,(\cB_\b,\Xi_\b;k),\ \b=1,2\,$ be a
pair of $(\txG,\la)$-equivariant $(\cG,\Upsilon,\g;\kappa
)$-bi-branes, as described in Definition \ref{def:Gequiv-bib}, with
$\,\cB_\b=(Q,\iota_\a,\om,\Phi_\b\ |\ \a=1,2)\,$ (i.e.\ with the
same world-volume, the same bi-brane maps $\,\iota_\a\,$ and of
equal curvatures). We call the latter \textbf{$\txG$-equivariantly
equivalent} iff the respective $(\iota_1^*\cG,\iota_2^*\cG
)$-bi-modules $\,\Phi_\b\,$ are $\txG$-equivariantly 2-isomorphic,
that is iff there exists a 2-isomorphism
\qq\nn
\alxydim{}{\iota_1^*\cG \ar@/^1.6pc/[rrr]^{\Phi_1}="5"
\ar@/_1.6pc/[rrr]_{\Phi_2}="6" \ar@{=>}"5"+(0,-4);"6"+(0,4)|{\psi}
& & & \iota_2^*\cG\ox I_\om}\,,
\qqq
subject to the coherence condition
\qq\label{eq:Gequiv-bib-2iso-coh}
\Xi_2\bullet\Qup d^{(1)\,*}_1\psi=\bigl(\Qup d^{(1)\,*}_0\psi\ox\id
\bigr)\bullet\Xi_1\,,
\qqq
equivalently expressed by the commutative 2-diagram
\eqref{diag:Gequiv-bib-2iso-coh} from Appendix
\ref{app:Gequiv-diags}. Whenever $\,\Phi_1=\Phi_2=:\Phi$,\ we shall
speak of \textbf{equivalent $\txG$-equivariant structures on
$\cG$-bi-brane $\,\cB=(Q,\iota_1,\iota_2,\om,\Phi)\,$ relative to
$\,\la$}. \exdef ~\medskip

Finally, we pass to inter-bi-branes.
\bedef\label{def:Gequiv-ibb}
Let $\,M,Q\,$ and $\,T_n,\ n\geq 3$, \,be a collection of
$\txG$-spaces and let $\,\txG\xcM,\ \xcM=M,Q,T_n$, \,be the
respective nerves of the action groupoids $\,\txG\lx\xcM$,\ with
face maps $\,\xcMup d^{(n+1)}_i$,\ as in Definition
\ref{def:nerves}. Furthermore, let $\,(\cG,\Upsilon,\g;\kappa)\,$ be
a $(\txG,\rho)$-equivariant gerbe over $\,M$,\ and let $\,(\cB,\Xi;k
)\,$ be a $(\txG,\la)$-equivariant $(\cG,\Upsilon,\g;\kappa
)$-bi-brane with $\,\cB=(Q,\iota_1,\iota_2,\om,\Phi)\,$ and with a
\textbf{$(\txG,\la)$-equivariant $(\iota_1^*\cG,\iota_2^*\cG
)$-bi-module} $\,(\Phi,\Xi;k)$.\ Finally, let
$\,\cJ\hspace{-0.05cm}= \hspace{-0.05cm}\bigl(T_n,\hspace{-0.05cm}
\bigl(\vep^{k,k+1}_n\hspace{-0.05cm},\pi^{k,k +1}_n\bigr);
\varphi_n\bigr)$ be a $(\cG,\cB )$-inter-bi-brane with the $\,T_n\,$
as component world-volumes, and with smooth $\txG$-maps
$\,\pi_n^{k,k+1}:T_n\to Q\,$ and $\txG$-invariant orientation maps
$\,\vep^{k,k+1}_n:T_n\to\{-1,+1 \}$.\ Suppose also that the
2-isomorphisms $\,\varphi_n\,$ satisfy the conditions
\qq
\Tnup d^{(1)\,*}_1\varphi_n&=&\bigl(d_\Upsilon\bigr)_n^{1\,(1)}
\bullet\bigl(\id\circ\bigl(i_\Upsilon^{\vep_n^{1,2}}\bigr)_n^{1\,(1
)}\bigr)\bullet(\id\circ\la_{\Upsilon^{1\,(1)}_n})\bullet\bigl(\id
\circ\Tnup d^{(1)\,*}_0\varphi_n\circ\id\bigr)\cr\cr
&&\bullet\bigl(\id\circ\la_{(\Phi_n^{n,1\,(1)})_{2^*}\ox\id}\circ
\la_{(\Phi_n^{n-1,n})_{2^*}\ox\id}\circ\cdots\circ\la_{(\Phi_n^{2,
3})_{2^*}\ox\id}\circ\id\bigr)\cr\cr
&&\bullet\bigl(\id\circ\bigl(b_\Upsilon^{-1}\bigr)_n^{n\,(1)}\circ
\id\circ\bigl(b_\Upsilon^{-1}\bigr)_n^{n-1\,(1)}\circ\id\circ\cdots
\circ\id\circ\bigl(b_\Upsilon^{-1}\bigr)_n^{2\,(1)}\circ\id\bigr)
\label{eq:Gequiv-ibb-coh}\\\cr
&&\bullet\bigl(\id\circ\bigl(i_\Upsilon^{\vep_n^{n,1}}\bigr)_n^{n\,
(1)}\circ\id\circ\bigl(i_\Upsilon^{\vep_n^{n-1,n}}\bigr)_n^{n-1\,(1
)}\circ\id\circ\cdots\circ\id\circ\bigl(i_\Upsilon^{\vep_n^{2,3}}
\bigr)_n^{2\,(1)}\circ\id\bigr)\cr\cr
&&\bullet\bigl(\bigl(\Xi_n^{n,1\,(1)}\ox\id\bigr)\circ\bigl(
\Xi_n^{n-1,n\,(1)}\ox\id\bigr)\circ\cdots\circ\Xi_n^{1,2\,(1)}
\bigr)\,.\nonumber
\qqq
A $(\cG,\cB)$-inter-bi-brane with these properties will be called a
\textbf{$\txG$-equivariant $((\cG,\Upsilon,\g;\kappa),$ $(\cB,\Xi;k)
)$-inter-bi-brane}, or, whenever there is no risk of confusion, just
a \textbf{$\txG$-equivariant ($(\cG,\cB)$-)inter-bi-brane}. \exdef

\noindent Altogether, we are naturally led to the following
\bedef\label{def:equiv-backgrnd}
A \textbf{$(\txG,\rho,\la)$-equivariant string background} is a
string background whose target space carries the structure of a
$\txG$-space, and such that the corresponding gerbe is
$(\txG,\rho)$-equivariant in the sense of Definition
\ref{def:equiv-gerbe}, the corresponding bi-brane is
$(\txG,\la)$-equivariant in the sense of Definition
\ref{def:Gequiv-bib}, and the corresponding inter-bi-brane is
$\txG$-equivariant in the sense of Definition
\ref{def:Gequiv-ibb}.\exdef

$\txG$-equivariant string backgrounds are among those that permit to
consistently gauge the rigid $\txG$-symmetry of the $\si$-model on
general world-sheets in the presence of topologically trivial
world-sheet gauge fields in a manner that guarantees the absence of
local and global gauge anomalies. Indeed, the former are absent in
consequence of relations \eqref{eq:HS-exact}-\eqref{eq:junct-exact}
and \eqref{eq:HS1-ids-triv}-\eqref{eq:HS2-ids-triv}, cf.\ Corollary
\ref{cor:infinit-Ginv-act}, and the latter vanish due to the
existence of the 1-isomorphism $\,\Upsilon\,$ and of the
2-isomorphism $\,\Xi\,$ with properties listed in Theorem
\ref{thm:equiv-12iso-constr}. These conditions and properties are
among those imposed by $\txG$-equivariance of the string background.
In Section \ref{sec:nontriv} below, we shall show that the
full-blown $\txG$-equivariant structure carried by a string
background actually enables us to define the gauged $\si$-model also
in the topologically nontrivial sectors with the gauge fields that
represent a connection in a nontrivial principal $\,\txG$-bundle
over the world-sheet.

\subsection{The descent principle for $\txG$-equivariant string
backgrounds}\label{sub:descent}

\noindent The key feature of $(\txG,\rho,\la)$-equivariant string
backgrounds in the special case when $\,\rho=0=\la\,$ is that they
descend to consistent string backgrounds on the quotient of the
target space by the action of $\,\txG$, \,provided the latter
quotient is sufficiently regular. This fact will play a pivotal
r\^ole in defining the topologically non-trivial sectors of the
gauged $\si$-model on world-sheets containing a defect quiver.

We start by explaining the concept of 2-categorial descent. To this
end, we consider
\bedef\label{def:desc-cat}
Given a surjective submersion $\,\varpi_\xcM:\xcM\to\xcX$,\ and the
associated simplicial space composed of fibred products of $\,\xcM$,
\qq\nn
\alxydim{@C=1.cm@R=.05cm}{\xcM^{[\bullet]}(\varpi_\xcM)\ :
\hspace{-1cm} & \cdots \ar@<.75ex>[r]^{\hspace{-1.7cm}\pr_{i,j,k}}
\ar@<.25ex>[r] \ar@<-.25ex>[r] \ar@<-.75ex>[r] &
\xcM^{[3]}\equiv\xcM\x_\xcX\xcM \x_\xcX\xcM
\ar@<.5ex>[r]^{\hspace{.4cm}\pr_{i,j}} \ar@<0.ex>[r] \ar@<-.5ex>[r]
& \xcM^{[2]}\equiv\xcM\x_\xcX\xcM \ar@<.5ex>[r]^{\hspace{1cm}\pr_i}
\ar@<-.5ex>[r] & \xcM\,,}
\qqq
with face maps given by the canonical projections, the
\textbf{descent 2-category} $\,\gt{Des}(\varpi_\xcM)\,$ is composed
of
\bit
\item the object class with elements given by triples $\,(\cG,
\Upsilon,\g)$,\ consisting of a gerbe $\,\cG\,$ (with connection)
over $\,\xcM$,\ a 1-isomorphism $\,\Upsilon:\cG_{1^*}\xrightarrow{\
\cong\ }\cG_{2^*}\,$ over $\,\xcM^{[2]}$,\ and a 2-isomorphism
\qq\nn
\alxydim{@C=2cm@R=1cm}{\cG_{1^*} \ar[r]^{\Upsilon_{[1,2]^*}}
\ar[d]_{\Upsilon_{[1,3]^*}} & \cG_{2^*} \ar[d]^{\Upsilon_{[2,3]^*}}
\ar@{=>}[dl]|{\,\g\ } \\ \cG_{3^*} \ar@{=}[r] & \cG_{3^*}}
\qqq
subject to the coherence condition
\qq\label{eq:Desc-obj-coh}
\g_{[1,2,4]^*}\bullet\bigl(\g_{[2,3,4]^*}\circ\id\bigr)=\g_{[1,3,4
]^*}\bullet\bigl(\id\circ\g_{[1,2,3]^*}\bigr)\,,
\qqq
equivalently expressed by the commutative 2-diagram
\eqref{diag:Desc-obj-coh} from Appendix \ref{app:Gequiv-diags};
\item for any pair $\,\cO_\a=(\cG_\a,\Upsilon_\a,\g_\a),\ \a=1,2
\,$ of objects, a category $\,\gt{Hom}(\cO_1,\cO_2)$,\ with
\smallskip
\bit
\item objects (1-cells) given by pairs $\,(\Psi,\eta)$,\ consisting
of a 1-isomorphism $\,\Psi:\cG_1\xrightarrow{\ \cong\ }\cG_2\,$ and
a 2-isomorphism
\qq\nn
\qquad\qquad\xy (100,0)*{\bullet}="G12"+(0,5)*{\cG_{1\,[1]^*}};
(80,-15)*{\bullet}="G1r1"+(-8,0)*{\cG_{1\,[2]^*}};
(120,-15)*{\bullet}="G2om"+(8,0)*{\cG_{2\,[1]^*}};
(88,-30)*{\bullet}="G2orl"+(-5,-3)*{\cG_{2\,[2]^*}};
(112,-30)*{\bullet}="G2or2"+(5,-3)*{\cG_{2\,[2]^*}};
\ar@{->}|{\Upsilon_1} "G12";"G1r1" \ar@{->}|{\Psi_{1^*}}
"G12";"G2om" \ar@{->}|{\Psi_{2^*}} "G1r1";"G2orl"
\ar@{->}|{\Upsilon_2} "G2om";"G2or2" \ar@{=}
"G2orl"+(2,0);"G2or2"+(-2,0) \ar@{=>}|{\ \eta\ }
"G1r1"+(3,0);"G2om"+(-3,0)
\endxy
\qqq
subject to the coherence condition
\qq\label{eq:Desc-1cell-coh}
\eta_{[1,3]^*}\bullet(\id\circ\g_1)=(\g_2\circ\id)\bullet\bigl(\id
\circ\eta_{[1,2]^*}\bigr)\bullet\bigl(\eta_{[2,3]^*}\circ\id\bigr)
\,,
\qqq
equivalently expressed by the commutative 2-diagram
\eqref{diag:Desc-1cell-coh} from Appendix \ref{app:Gequiv-diags};
\item morphisms (2-cells) given by 2-isomorphisms
\qq\nn
\alxydim{}{\cG_1 \ar@/^1.6pc/[rrr]^{\Psi_1}="5"
\ar@/_1.6pc/[rrr]_{\Psi_2}="6" \ar@{=>}"5"+(0,-3);"6"+(0,+3)|{\psi}
& & & \cG_2}
\qqq
subject to the coherence condition
\qq\label{eq:Desc-2cell-coh}
\bigl(\id\circ\psi_{1^*}\bigr)\bullet\eta_1=\eta_2\bullet\bigl(
\psi_{2^*}\circ\id\bigr)\,,
\qqq
equivalently expressed by the commutative 2-diagram
\eqref{diag:Desc-2cell-coh} from Appendix \ref{app:Gequiv-diags}.
\eit
\eit
\exdef \noindent The significance of the descent category manifests
itself through the following
\bethe\cite[Prop.\,6.7]{Stevenson:2000wj}\cite[Thm.\,A.4.2]{Gawedzki:2010rn}\label{thm:desc}
Given a surjective submersion $\,\varpi_\xcM:\xcM\to\xcX$,\ the
induced pullback functor
\qq\nn
\widehat\varpi_\xcM^*\ :\ \bgrb^\nabla(\xcX)\to\gt{Des}(\varpi_\xcM)
\,,
\qqq
defined by the formul\ae
\qq\nn
\widehat\varpi_\xcM^*\cG=(\varpi_\xcM^*\cG,\id,\id),\quad
\widehat\varpi_\xcM^*\bigl(\Psi:\cG_1\xrightarrow{\ \cong\
}\cG_2\bigr)
=(\varpi_\xcM^*\Psi,\id),\quad\widehat\varpi_\xcM^*\bigl(\psi:
\Psi_1\xLongrightarrow{\cong}\Psi_2\bigr)=\varpi_\xcM^*\psi,
\qqq
is an equivalence of 2-categories. \ethe

For $\,\varpi_{\hat\xcM}:\hat\xcM\to\xcM\,$ a left principal
$\txG$-bundle, the simplicial space $\,\hat\xcM^{[\bullet]}(
\varpi_{\hat\xcM})\,$ can be $\txG$-equivariantly identified with
the nerve $\,\txG\hat\xcM\,$ of the action groupoid
$\,\txG\lx\hat\xcM$,\ and Definition \ref{def:desc-cat} is readily
seen to coincide with that of a sub-2-category
$\,\bgrb^\nabla(\hat\xcM)^\txG_0\,$ of $(\txG,0)$-equivariant bundle
gerbes with connection over $\,\hat\xcM$.\ As a simple corollary to
Theorem \ref{thm:desc}, we then obtain
\bethe\cite[Thm.\,5.3]{Gawedzki:2010rn}\label{thm:Gdesc}
Let $\,\varpi_{\hat\xcM}:\hat\xcM\to\xcM\,$ be a left principal
$\txG$-bundle. There exists a canonical equivalence between the
2-category $\,\bgrb^\nabla(\xcM)\,$ of bundle gerbes with connection
over $\,\xcM=\hat\xcM/\txG\,$ and the 2-category
$\,\bgrb^\nabla(\hat\xcM)^\txG_0\,$ of $(\txG,0)$-equivariant bundle
gerbes with connection over $\,\hat\xcM$. \ethe
\brem\label{rem:pull-stab-equiv} The equivalence from Theorem
\ref{thm:Gdesc} is preserved by pullback along $\txG$-equivariant
maps. \erem \noindent Theorem \ref{thm:Gdesc} was proven in
\Rcite{Gawedzki:2010rn}. Here, we want to study its consequences for
bi-modules for $\txG$-equivariant gerbes and for their
trivializations. We start with the former. We have the obvious
\becor
Let $\,\varpi_{\hat\xcM}:\hat\xcM\to\xcM\,$ be a left principal
$\txG$-bundle, and let $\,\hat\cO_\a=(\hat\cG_\a,
\hat\Upsilon_\a,\hat\g_\a),$ $\a=1,2,\,$ and the trivial gerbe
$\,\hat\cT=(I_{\hat\om},\hat\cE,\hat\vep)$,\ of an arbitrary curving
$\,\hat\om\in\Om^2(\hat\xcM)\,$ and $\txG$-equivariant structure
$\,(\hat\cE,\hat\vep)$,\ all belong to the object class of the
2-category $\,\bgrb^\nabla(\hat\xcM)^\txG_0$.\ Assume, furthermore,
that there exists a 1-cell $\,(\hat\Phi,
\hat\Xi)\in\gt{Hom}(\hat\cO_1,\hat\cO_2\ox\hat\cT)\,$ in
$\,\bgrb^\nabla(\hat\xcM)^\txG_0$.\ Then, there exist unique (up to
1-isomorphism) objects $\,\cG_1,\cG_2\,$ and $\,\cT\,$ and a unique
(up to 2-isomorphism) 1-cell $\,\Phi:\cG_1\to\cG_2\ox\cT\,$ in the
2-category $\,\bgrb^\nabla(\xcM)\,$ such that
$\,(\varpi_{\hat\xcM}^*\cG_\a,\id,\id)\,$ and $\,(\varpi_{\hat
\xcM}^*\cT,\id,\id)\,$ are $\txG$-equivariantly 1-isomorphic to
$\,\hat\cO_\a\,$ and $\,\hat\cT$,\ respectively, and such that
$\,(\varpi_{\hat\xcM}^*\Phi,\id)\,$ is $\txG$-equivariantly
2-isomorphic to $\,(\hat\Phi,\hat\Xi)$. \ecor \brem The above
corollary brings forward an issue that has to be resolved when
formulating the $\si$-model in the presence of defects. The latter
prerequires the structure of a bi-brane in the first place, cf.\
\Rcite{Runkel:2008gr}, and so we need a characterisation of the
circumstances in which the descent preserves the bi-brane structure,
taking into account that $\,\cT\,$ is \emph{not} trivial in general.
Thus, a bi-brane defined by a triple of $(\txG,0)$-equivariant
gerbes $\,\hat\cO_\a,\hat\cT\,$ over a $\txG$-space $\,\hat\xcM\,$
does \emph{not} always descend to a bi-brane over the quotient space
$\,\xcM$.\ Clearly, the problem here lies with the arbitrary choice
of the data $\,(\hat\om,\hat\cE,\hat\vep )\,$ (which appear to
provide us with a most natural definition of a trivial
$(\txG,0)$-equivariant gerbe on $\,\hat\xcM$). From the point of
view of the construction of a consistent field theory, these data
have to be further constrained.\erem

\noindent The solution is contained in the following
\berop\label{prop:triv-Gbas-triv-equiv}
The equivalence of Theorem \ref{thm:Gdesc} restricts to an
equivalence between, on the one hand, the sub-2-category
$\,\gt{TrivGrb}(\hat \xcM)^{\txG}_0\,$ of $(\txG,0 )$-equivariant
gerbes over the $\txG$-space $\,\hat\xcM\,$ $\txG$-equivariantly
1-isomorphic to $\txG$-equivariant gerbes of the form
$\,(I_{\hat\om},\id,\id;0)$,\ and, on the other hand, the
sub-2-category $\,\gt{TrivGrb}(\xcM)\,$ of gerbes over $\,\xcM=
\hat\xcM/\txG\,$ 1-isomorphic to trivial gerbes.
\eerop
\noindent
\beroof
In the light of Theorem \ref{thm:Gdesc}, the only thing that has to
be checked is that every object in $\,\gt{TrivGrb}(\hat\xcM
)^{\txG}_0\,$ descends to an object in $\,\gt{TrivGrb}( \xcM)$,\ and
that this mapping is essentially surjective, i.e.\ that each element
of the latter 2-category can be obtained from an object of the
former 2-category through descent. It is a straightforward exercise,
using an explicit construction of the inverse of the functor
$\,\widehat\varpi_\xcM^*\,$ given in \Rcite{Stevenson:2000wj}, to
verify that the gerbe over $\,\xcM\,$ associated to the trivial
$(\txG,0)$-equivariant gerbe $\,(I_{\hat \om},\id,\id;0)\,$ over
$\,\hat\xcM\,$ with a $\txG$-basic curving $\,\hat\om$ is
$\,I_\om$,\ where $\,\om\,$ is the unique 2-form on $\,\xcM\,$ such
that $\,\hat\om=\varpi_{\hat\xcM}^*\om$.\ We also have, by
construction, $\,(I_{\hat\om},\id,\id;0)=\widehat
\varpi_{\hat\xcM}^*I_\om$. \eroof \bigskip \noindent Put together,
Theorem \ref{thm:Gdesc}, Remark \ref{rem:pull-stab-equiv} and
Proposition \ref{prop:triv-Gbas-triv-equiv} lead to a conclusion
which is central to our subsequent considerations, and so we
formulate it as
\bethe\label{thm:Gequivbib-Gbas-bib-equiv}
Let $\,\varpi_{\hat M}:\hat M\to M\,$ and $\,\varpi_{\hat Q}: \hat
Q\to Q\,$ be left principal $\txG$-bundles with the respective
smooth bases $\,M=\hat M/\txG\,$ and $\,Q=\hat Q/\txG$.\ There
exists a canonical bijection
\qq\nn
\left\{\barr{c} \tx{$\txG$-equivariant isomorphism classes} \\
\tx{of $(\txG,0)$-equivariant bi-branes} \\ \tx{with world-volume
$\,\hat Q$}
\\ \tx{for $(\txG,0)$-equivariant gerbes over $\,\hat M$}
\earr\right\}\qquad\xlongleftrightarrow{1:1}\qquad\left\{\barr{c}
\tx{Equivalence classes} \\ \tx{of bi-branes} \\ \tx{with
world-volume $\,Q$} \\ \tx{for gerbes over $\,M$}\earr\right\}.
\qqq
\ethe \beroof Let us begin by identifying the bi-brane with
world-volume $\,Q\,$ associated to the given $(\txG,0)$-equivariant
bi-brane $\,(\hat\cB,\Xi;0)\,$ with $\,\hat\cB=( \hat
Q,\hat\iota_1,\hat\iota_2,\hat\om,\hat\Phi)\,$ for a
$(\txG,0)$-equivariant gerbe $\,\hat \cG\,$ over $\,\hat M$.\ First
of all, note that the $\txG$-equivariant bi-brane maps
$\,\hat\iota_\a:\hat Q \to\hat M\,$ induce smooth maps
$\,\ovl\iota_\a:Q\to M\,$ which render the following diagram
commutative:
\qq\label{diag:quot-bib-maps-from-Gequiv}
\alxydim{@C=1.cm@R=1.cm}{\hat Q \ar[r]^{\ \varpi_{\hat Q}}
\ar[d]_{\hat\iota_\a} & Q \ar[d]^{\ovl\iota_\a} \cr \hat M \ar[r]^{\
\varpi_{\hat M}} & M}\,.
\qqq
It is these quotient maps that enter the definition of the bi-brane
over $\,Q$.\ Second, the vanishing of $\,\rho\,$ and $\,\la\,$
implies $\,\kappa_a=0\,$ and $\,k_a=0$,\ and so, in virtue of
\Reqref{eq:bdry-exact}, we obtain
\qq\nn
\ic_a\hat\om=0\,.
\qqq
This, in conjunction with the assumed $\txG$-invariance of
$\,\hat\om$,\ shows that the latter 2-form is, in fact,
$\txG$-basic, hence, in particular, there exists a unique 2-form
$\,\om\in\Om^2(Q)\,$ such that
\qq\nn
\hat\om=\varpi_{\hat Q}^*\om\,.
\qqq
Next, we examine the bi-brane 1-isomorphism $\,\hat\Phi$.\ Let
$\,\cG\,$ be the gerbe over $\,M\,$ obtained from the $(\txG,0
)$-equivariant gerbe $\,(\hat\cG,\hat\Upsilon,\hat \g;0)\,$
according to Theorem \ref{thm:Gdesc}. Owing to the assumed
$\txG$-equivariance of the $\,\hat\iota_\a$,\ the pullback gerbes
$\,(\hat\iota_\a^*\hat\cG,\hat\iota_\a^{(1
)\,*}\Upsilon,\hat\iota_\a^{(2)\,*}\g;0)\,$ descend, in conformity
with the same theorem, to the respective gerbes $\,\ovl
\iota_\a^*\cG\,$ over $\,Q$.\ Finally, also $\,(I_{\hat\om},
\id,\id;0)\,$ with the $\txG$-basic curving $\,\hat\om= \varpi_{\hat
Q}^*\om\,$ descends to the trivial gerbe $\,I_\om$.\ Theorem
\ref{thm:Gdesc} now predicts the existence of a descendant
1-isomorphism
\qq\label{eq:desc-bimod}
\Phi\ :\ \ovl\iota_1^*\cG\xrightarrow{\ \cong\ }\ovl\iota_2^*\cG\ox
I_\om
\qqq
between the respective descendant gerbes, and thus yields the
desired descendant bi-module. Clearly, $\txG$-equivariantly
equivalent $(\txG,0)$-equivariant bi-branes over $\,\hat Q\,$ are
mapped to equivalent ones over $\,Q$.

For the reverse statement, note that the 2-isomorphism class of a
bi-module $\,\Phi:\ovl\iota_1^*\cG\xrightarrow{\ \cong\
}\ovl\iota_2^* \cG\ox I_\om\,$ over $\,Q\,$ for a gerbe $\,\cG\,$
over $\,M\,$ and with $\,\om\in\Om^2(Q)\,$ corresponds to the
$\txG$-equivariant 2-isomorphism class of the $(\txG,0)$-equivariant
bi-module $\,\hat\Phi=\varpi_{\hat Q}^*\Phi$,\ with
\qq\nn
\hat\Phi\ :\ \varpi_{\hat Q}^*\ovl\iota_1^*\cG \xrightarrow{\ \cong\
}\varpi_{\hat Q}^*\ovl\iota_2^*\cG\ox I_{\varpi_{\hat Q}^*\om}\,.
\qqq
The commutativity of diagram \eqref{diag:quot-bib-maps-from-Gequiv}
then enables us to rewrite the above as
\qq\nn
\hat\Phi\ :\ \hat\iota_1^*\varpi_{\hat M}^*\cG \xrightarrow{\ \cong\
}\hat\iota_2^*\varpi_{\hat M}^*\cG\ox I_{\varpi_{\hat Q}^*\om}\,,
\qqq
and the gerbe $\,\varpi_{\hat M}^*\cG\,$ over $\,\hat M\,$ is
manifestly $(\txG,0)$-equivariant as the functorial image (with
respect to the equivalence $\,\widehat\varpi_{\hat M}^*$) of the
gerbe $\,\cG\,$ entering the definition of $\,\Phi$.\eroof \brem The
one-to-one correspondence of Theorem
\ref{thm:Gequivbib-Gbas-bib-equiv} is preserved by pullback along
$\txG$-equivariant maps.\erem

It is equally straightforward to prove
\bethe\label{thm:Gequivibb-ibb-equiv}
Let $\,(\hat M,\varpi_{\hat M}),(\hat Q, \varpi_{\hat Q})\,$ and
$\,(\hat T_n,\varpi_{\hat T_n}),\ n\geq 3$, \,be principal
$\txG$-manifolds with the respective smooth bases $\,M=\hat
M/\txG,Q=\hat Q/ \txG\,$ and $\,T_n=\hat T_n/\txG$.\ There exists a
canonical bijection
\qq\nn
\quad\left\{\barr{c} \tx{$\txG$-equivariant inter-bi-branes} \\
\tx{with world-volume $\,\hat T=\bigsqcup_{n\geq 3}\,
\hat T_n$} \\ \tx{for $(\txG,0)$-equivariant bi-branes} \\
\tx{with world-volume $\,\hat Q$} \\ \tx{for gerbes over $\,\hat M$}
\earr\right\}\ \xlongleftrightarrow{1:1}\ \left\{\barr{c}
\tx{Inter-bi-branes} \\ \tx{with world-volume $\,T= \bigsqcup_{n\geq
3}\,T_n$} \\ \tx{for bi-branes with world-volume $\,Q$} \\ \tx{for
gerbes over $\,M$}\earr\right\}.
\qqq
\ethe \beroof Consider a $\txG$-equivariant inter-bi-brane
$\,\hat\cJ=\bigl(\hat T_n,\bigl(\hat\vep^{k,k+1}_n
,\hat\pi^{k,k+1}_n\bigr);\hat \varphi_n\bigr)\,$ for a $(\txG,0
)$-equivariant bi-brane $\,(\hat\cB,\hat\Xi;0)\,$ with
$\,\hat\cB=(\hat Q,\hat\iota_\a,\hat\om, \hat\Phi)$. \,The
$\txG$-equivariant maps $\,\hat\pi^{k,k+1}_n:\hat T_n\to\hat Q\,$
induce smooth maps $\,\ovl\pi^{k,k+1}_n:T_n\to Q\,$ which render the
following diagram commutative:
\qq\label{diag:quot-ibb-maps-from-Gequiv}
\alxydim{@C=1.cm@R=1.cm}{\hat T_n \ar[r]^{\ \varpi_{\hat T_n}}
\ar[d]_{\hat\pi^{k,k+1}_n} & T_n \ar[d]^{\ovl\pi^{k,k+ 1}_n}
\cr \hat Q \ar[r]^{\ \varpi_{\hat Q}} & Q}\,.
\qqq
We shall employ the induced maps to explicitly construct the
descendant inter-bi-brane $\,\cJ=\bigl(T_n,\bigl(\ovl\vep^{k,k+1}_n,
\ovl\pi^{k,k+1}_n\bigr),\varphi_n\bigr)\,$ for $\,\hat\cJ$.\ The
descendant inter-bi-brane has the orientation maps
$\,\ovl\vep^{k,k+1}_n\,$ trivially induced from the respective
orientation maps $\,\hat\vep^{k,k+1}_n$,\ which makes sense as the
$\txG$-action preserves each of the component world-volumes of
$\,\hat\cJ\,$ corresponding to fixed values of the
$\,\hat\vep^{k,k+1}_n$.\ It therefore remains to identify the
descendant 2-isomorphisms $\,\varphi_n$.\ Let $\,\Phi\,$ be the
gerbe bi-module over $\,Q\,$ obtained from $\,\hat\Phi\,$ through
Theorem \ref{thm:Gequivbib-Gbas-bib-equiv}. The maps
$\,\hat\pi_n^{k,k+1}\,$ being $\txG$-equivariant, the 1-isomorphisms
$\,\hat\Phi_n^{k,k+1}\,$ descend, in virtue of the same theorem, to
the respective 1-isomorphisms $\,\Phi_n^{k,k+1}
=\ovl\pi_n^{k,k+1\,*}\Phi^{\ovl\vep_n^{k,k+1}}\,$ over $\,T_n$,\ and
so also the composite 1-isomorphism $\,(\hat\Phi_n^{n,1}\ox\id
)\circ(\hat\Phi_n^{n-1,n}\ox\id)\circ\cdots\circ\hat \Phi_n^{1,2}\,$
descends to the 1-isomorphism $\,(\Phi_n^{n,1}\ox\id
)\circ(\Phi_n^{n-1,n}\ox\id)\circ\cdots\circ\Phi_n^{1,2}$.\
\,Invoking Theorem \ref{thm:Gdesc}, we then find descendant
2-isomorphisms
\qq\nn
\varphi_n\ :\ (\Phi_n^{n,1}\ox\id)\circ(\Phi_n^{n-1,n}\ox\id)\circ
\cdots\circ\Phi_n^{1,2}\xLongrightarrow{\cong}\id_{\cG_n^1}
\qqq
between the respective descendant 1-isomorphisms, the latter of them
being written for the descendant gerbe $\,\cG_n^1=(\ovl
\iota_1^{\ovl\vep_n^{1,2}}\circ\ovl\pi_n^{1,2})^*\cG$,\ \,as
dictated by the same theorem for the $\txG$-equivariant map $\,\hat
\iota_1^{\hat\vep_n^{1,2}}\circ\hat\pi_n^{1,2}$.

Conversely, 2-isomorphisms $\,\varphi_n:(\Phi_n^{n,1}\ox\id)\circ(
\Phi_n^{n-1,n}\ox\id)\circ\cdots\circ\Phi_n^{1,2}
\xLongrightarrow{\cong}\id_{\cG_n^1}\,$ trivialising the
concatenations of 1-isomorphisms $\,\Phi_n^{k,k+1}\,$ for the gerbe
$\,\cG\,$ over $\,M$,\ pulled back from $\,Q\,$ along the respective
maps $\,\ovl\pi_n^{k,k+1}\,$ as described above, yield the
manifestly $\txG$-equivariant 2-isomorphisms $\,\hat\varphi_n=
\varpi_{\hat T_n}^*\varphi_n$,\ with
\qq\nn
\hat\varphi_n\ :\ \bigl(\bigl(\varpi_{\hat T_n}^*
\Phi_n^{n,1}\ox\id\bigr)\circ\bigl(\varpi_{\hat T_n}^*
\Phi_n^{n-1,n}\ox\id\bigr)\circ\cdots\circ\varpi_{\hat T_n}^*
\Phi_n^{1,2}\xLongrightarrow{\cong}\id_{\varpi_{\hat T_n}^*
\cG_n^1}\,.
\qqq
Repeated application of the commutative diagram
\eqref{diag:quot-ibb-maps-from-Gequiv} enables to rewrite the above
in the sought-after form
\qq\nn
\hat\varphi_n\ :\ \bigl(\bigl(\varpi_{\hat Q}^*\Phi
\bigr)_n^{n,1}\ox\id\bigr)\circ\bigl(\bigl(\varpi_{\hat Q}^*
\Phi\bigr)_n^{n-1,n}\ox\id\bigr)\circ\cdots\circ\bigl( \varpi_{\hat
Q}^*\Phi\bigr)_n^{1,2}\xLongrightarrow{\cong} \id_{(\varpi_{\hat
M}^*\cG)_n^1}\,,
\qqq
with $\,(\varpi_{\hat M}^*\cG)_n^1=(\hat
\iota_1^{\hat\vep_n^{1,2}}\circ\hat\pi_n^{1,2})^* \varpi_{\hat
M}^*\cG\,$ and $\,\varpi_{\hat Q}^*\Phi\,$ manifestly
$(\txG,0)$-equivariant.\eroof\bigskip

We may summarise the findings of the present section in
\becor\label{cor:Gequiv-backgrnd-desc}
Let $\,\hat\xcF=\hat M\sqcup\hat Q\sqcup\hat T\,$ be the target
space of a string background, endowed with the structure of a
$\txG$-space with respect to a (left) action of the group $\,\txG$,\
the latter assumed such that the canonical projection
$\,\hat\xcF\to\hat\xcF/\txG=:\xcF\,$ gives a principal
$\txG$-bundle. There exists a canonical bijection
\qq\nn
\hspace*{0.2cm}\left\{\barr{c} \tx{Equivalence classes} \\ \tx{of
$(\txG,0,0 )$-equivariant string backgrounds} \\ \tx{with target
space $\,\hat\xcF$}
\earr\right\}\quad\xlongleftrightarrow{1:1}\quad\left\{\barr{c}
\tx{Equivalence classes} \\ \tx{of string backgrounds} \\ \tx{with
target space $\,\xcF$}\earr\right\}.
\qqq
\ecor

\section{Coset $\si$-models}\label{sec:coset}

\noindent One of the physical ideas underlying the concept of a
gauged $\si$-model, particularly amply illustrated on the example of
the (gauged) WZW models and the coset models of conformal field
theory, cf.\
Refs.\,\cite{Gawedzki:1988hq,Gawedzki:1988nj,Karabali:1988au}, is
that promoting global symmetries of the theory to the rank of local
ones may serve to reduce the number of field-theoretic degrees of
freedom. This systematic construction circumvents the usual problems
posed by attempts at defining the two-dimensional field theory
directly on the quotient $\,\xcF/\txG\,$ of the original target
space $\,\xcF$,\ \,such as metric singularities and the emergence of
another component of the string background, that is the dilaton. It
does so in a manner analogous to that in which $\txG$-equivariant
cohomology of a $\txG$-space $\,\xcM\,$ replaces the cohomology of
the orbit space of $\,\xcM\,$ in circumstances in which the orbit
space is no longer smooth, that is by giving us a field theory which
would, in the case of smooth $\,\xcF/\txG$,\ be a `pullback' to
$\,\xcF\,$ (in a sense that shall be clarified below) of a unique
field theory on the quotient. Below, we shall disregard - for the
sake of illustration - the possible problems of the latter field
theory mentioned above and explicitly carry out the descent from the
field space $\,\xcF\,$ to its quotient, which boils down to
integrating out the (non-dynamical) gauge field. This will make
apparent the r\^ole of the $\txG$-equivariant structure on the
string background.
\smallskip

We begin our discussion by remarking that the very structure of the
action functional of Corollary \ref{cor:triv-Ansatz}, at most
quadratic in the world-sheet gauge field $\,\txA^a_u$,\ indicates
that the effective two-dimensional field theory obtained by
integrating out the gauge field in the path integral of the gauged
$\si$-model can be read off, at least up to a dilaton term induced
from the gaussian path integral over the gauge fields, from
\Reqref{eq:gauges-sigmod} upon solving the classical equations for
$\,\txA^a_u\,$ and subsequently substituting the solution back into
the action functional, cf.\ \Rxcite{Sect.\,2.4}{Runkel:2008gr}.
\smallskip

In the conformal gauge of the world-sheet metric, the field
equations for the gauge field can be written as
\qq\nn
\bigg(\begin{smallmatrix} \txh_{ab} & \sfi\,\txc_{ab} \\ \\ -
\sfi\,\txc_{ab} & \txh_{ab}\end{smallmatrix}\bigg)\,
\bigg(\begin{smallmatrix} \txA^b_1 \\ \\ \txA^b_2
\end{smallmatrix}\bigg)=\bigg(\begin{smallmatrix} \txK_{a\,\mu} &
-\sfi\,\kappa_{a\,\mu} \\ \\ \sfi\,\kappa_{a\,\mu} & \txK_{a\,\mu}
\end{smallmatrix}\bigg)\,\bigg(\begin{smallmatrix} \p_1\varphi^\mu
\\ \\ \p_2\varphi^\mu \end{smallmatrix}\bigg)
\qqq
(in terms of the 1-forms $\,\txK_a\,$ of \Reqref{eq:KA-def} and of
the symmetric tensor $\,\txh_{ab}\,$ of \Reqref{eq:hAB-def}) and
\qq\label{eq:eff-bib-geom}
k_a=0\,.
\qqq
In order to be able to solve the former, we need to make certain
assumptions about the geometry of the (metric) target space
$\,(M,\txg)$.\ Namely, we shall assume that $\,M
\xrightarrow{\varpi_M}M/\txG\,$ carries the structure of a principal
$\txG$-bundle, so that, in particular, the $\txG$-action on $\,M\,$
is free and $\,\txg\,$ defines a positive-definite metric on the
fibres of $\,M$.\ From the last fact, it then follows that the
symmetric matrix $\,\txh\,$ with entries $\,\txh_{ab}\,$ is
invertible. It is now easy to see that the background matrix
\qq\nn
\txE:=\txh-\txc\,\txh^{-1}\,\txc
\qqq
is invertible as well. Indeed, the latter can be rewritten as
$\,\txh\,(\bd1-\txh^{-1}\,\txc)\,(\bd1+\txh^{-1}\,\txc )$,\ and so
its invertibility is ensured by the absence of eigenvectors of the
matrix $\,\txh^{-1}\,\txc\,$ associated with the eigenvalues $\,\pm
1$.\ Under the above assumptions, we obtain
\qq\nn
\txA^a_u=\txM^a_\mu(\varphi)\,\p_u\varphi^\mu-\sfi\,\vep_{uv}\,
\txN^a_\mu(\varphi)\,\p_v\varphi^\mu\,,
\qqq
with
\qq\nn
\txM^a_\mu:=\bigl(\txE^{-1}\bigr)^{ab}\,\txK_{b\,\mu}+\bigl(\txh^{-
1}\,\txc\,\txE^{-1}\bigr)^{ab}\,\kappa_{b\,\mu}\,,\quad
\txN^a_\mu:=\bigl(\txE^{-1}\bigr)^{ab}\,\kappa_{b\,\mu}+\bigl(
\txh^{-1}\,\txc\,\txE^{-1}\bigr)^{ab}\,\txK_{b\,\mu}\,.
\qqq
The effective field theory, defined by the Feynman amplitudes
\qq\nn
{\xcA}^{{\rm eff}}\bigl[(\varphi\,\vert\,\G)] :=\int\,\xcD\txA\,
\widehat{\xcA}\bigl[(\varphi\,\vert\,\G);\txA,\g\bigr]
\qqq
takes the form of a non-linear $\si$-model
\qq\nn
{\xcA}^{{\rm eff}}\bigl[(\varphi\,\vert\,\G)]=\ee^{-\tfrac{1}{2}\,
\int_\Si\,\unl\txg(\sfd\varphi
\overset{\wedge}{,}\star_\g\sfd\varphi)-W_{\rm dil}[\varphi]}\,\,
\Hol_{\unl\cG,\unl\Phi,\unl\varphi_n}(\varphi)\,,
\qqq
with $\,W_{\rm dil}\,$ the induced dilaton term (to be left
unspecified), and with a new background $\,\unl\Bgt\,$ composed of
\bit
\item a new target $\,\unl\cM=(M,\unl\txg,\unl\cG)\,$ with the metric
\qq\qquad\qquad\label{eq:new-metr}
\unl\txg:=\txg+\txE^{-1\,ab}\,(\kappa_a\ox\kappa_b-\txK_a\ox\txK_b)+
\bigl(\txh^{-1}\,\txc\,\txE^{-1}\bigr)^{ab}\,(\kappa_a\ox\txK_b+
\txK_b\ox\kappa_a)\,,
\qqq
and the gerbe
\qq\label{eq:new-gerb}
\unl\cG:=\cG\ox I_\b\,,
\qqq
where the curving $\,\b\in\Om^2(M)\,$ of the trivial factor reads
\qq\label{eq:beta}
\b=\tfrac{1}{2}\,\bigl(\txh^{-1}\,\txc\,\txE^{-1}\bigr)^{ab}\,
\bigl(\kappa_a\wedge\kappa_b-\txK_a\wedge\txK_b\bigr)+\txE^{-1\,
ab}\,\kappa_a\wedge\txK_b\,;
\qqq
\item a new bi-brane $\,\unl\cB=(\unl Q,\iota_\a,\unl\om,\unl\Phi\
\vert\ \a=1,2)\,$ of world-volume, assumed smooth,
\qq\label{eq:new-bib-wvol}
\unl Q:=\bigcap_{a=1}^{\dim\,\ggt}\,k_a^{-1}\bigl(\{0\}\bigr)\,,
\qqq
the curvature
\qq\label{eq:unlom}
\unl\om:=\om-\D_Q\b
\qqq
and with the bi-brane 1-isomorphism
\qq\label{eq:new-bib-iso}
\unl\Phi:=\Phi\ox\id_{I_{\iota_1^*\b}}\ :\ \iota_1^*\unl\cG
\xrightarrow{\ \cong\ }\iota_2^*\unl\cG\ox I_{\unl\om}\,,
\qqq
the latter two pulled back to $\,\unl Q$;
\item a new inter-bi-brane
$\,\unl\cJ=\bigl(\unl T_n,\bigl(\vep^{k,k+1}_n,\pi^{k,k+1}_n\bigr);
\unl\varphi_n\bigr)\,$ of the component world-volumes, assumed
smooth,
\qq\label{eq:new-ibb-wvol}
\unl T_n:=\bigcap_{k=1}^n\,\bigl(\pi_n^{k,k+1}\bigr)^{-1}(\unl Q)
\qqq
and with the inter-bi-brane 2-isomorphisms
\qq\label{eq:new-ibb-iso}
\unl\varphi_n:=\varphi_n\ox\id_{\id_{I_{\b^1_n}}}\,.
\qqq
\eit
We shall next study the symmetry properties of the new string
background with respect to the (induced) action of the group
$\,\txG$.\bigskip

First, we consider the target-space structures, i.e.\ the metric
$\,\unl\txg\,$ and the gerbe $\,\unl\cG$.\ We find
\berop\label{prop:g-basic}
Let $\,\varpi_M:M\to M/\txG\,$ be a left principal $\txG$-bundle
with a $\txG$-invariant metric $\,\txg$,\ and with a
$\txG$-invariant closed 3-form $\,\txH\,$ admitting a
$\ggt$-equivariantly closed extension $\,\widehat\txH=\txH-\kappa\,$
for $\,\kappa\in\Om^1(M) \ox\ggt^*$.\ The metric $\,\unl\txg\,$
introduced in \Reqref{eq:new-metr} is $\txG$-basic, and so there
exists a unique metric $\,\unl\txG\,$ on $\,M/\txG\,$ such that
\qq\nn
\unl\txg=\varpi_M^*\unl\txG\,.
\qqq
\eerop
\beroof
Using the identities
\qq\nn
\bigl(\txc\,\txE^{-1}\bigr)^{ab}=-\bigl(\txE^{-1}\,\txc\bigr)^{ba}
\,,\qquad\qquad\txh^{-1}\,\txc\,\txE^{-1}\,\txh=\txE^{-1}\,\txc\,,
\qqq
we readily verify the $\ggt$-horizontality of $\,\unl\txg$,
\qq\nn
\unl\txg\bigl(\Mup\xcK_a,\cdot\bigr)&=&\txK_a+\txE^{-1\,bc}\,(
\txc_{ab}\,\kappa_c-\txh_{ab}\,\txK_c)+\bigl(\txh^{-1}\,\txc\,
\txE^{-1}\bigr)^{bc}\,(\txc_{ab}\,\txK_c+\txh_{ac}\,\kappa_b)\cr\cr
&=&\bigl(\bd1-\txh\,\txE^{-1}+\txc\,\txh^{-1}\,\txc\,\txE^{-1}
\bigr)^{ab}\,\txK_b+\bigl(\txh^{-1}\,\txc\,\txE^{-1}\,\txh-\txE^{-
1}\,\txc\bigr)^{ba}\,\kappa_b=0\,.
\qqq
Its $\txG$-invariance is a consequence of the same property of
$\,\txg$,\ as well as of the $\txG$-equivariance of $\,\kappa\,$ and
that of the fundamental vector fields,
\qq\nn
\left(\Mup\ell_g\right)_*\Mup\xcK_a=(\Ad_g)_{ab}\,\Mup\xcK_b\,.
\qqq
The last identity readily follows from \Reqref{eq:K-shift}. \eroof
\bigskip
Similarly, we establish the symmetry properties of the 2-form
$\,\b$,\ summarised in
\berop
Under the assumptions and in the notation of Proposition
\ref{prop:g-basic}, the 2-form $\,\b\,$ introduced in
\Reqref{eq:beta} is $\txG$-invariant and satisfies the identity
\qq\label{eq:beta-con}
\ic_a\b&=&-\kappa_a\,.
\qqq
Consequently, the curvature
\qq\nn
\unl\txH:=\txH+\sfd\b
\qqq
of the gerbe $\,\unl\cG\,$ defined in \Reqref{eq:new-gerb} is
$\txG$-basic, and so there exists a unique 3-form $\,\unl\txh\,$ on
$\,M/\txG\,$ such that
\qq\nn
\unl\txH=\varpi_M^*\unl\txh\,.
\qqq
\eerop
\beroof
The $\txG$-invariance of $\,\b\,$ follows as before. Identity
\eqref{eq:beta-con} is now obtained as
\qq\nn
\ic_a\b&=&\bigl(\txh^{-1}\,\txc\,\txE^{-1}\bigr)^{bc}\,
\bigl(\txc_{ab}\,\kappa_c-\txh_{ab}\,\txK_c\bigr)+\txE^{-1\,bc}\,
(\txc_{ab}\,\txK_c-\txh_{ac}\,\kappa_b)\cr\cr
&=&\bigl(\bigl(\txc\,\txh^{-1}\,\txc-\txh\bigr)\,\txE^{-1}\bigr)^{a
b}\,\kappa_b=-\kappa_a\,,
\qqq
using the symmetry of the background matrix. Put together, the above
identities yield
\qq\nn
\ic_a\sfd\b=-\sfd\bigl(\ic_a\b\bigr)=\sfd\kappa_a \,,
\qqq
whence also
\qq\nn
\ic_a\unl\txH=-\sfd\kappa_a+\sfd\kappa_a=0\,.
\qqq
Since $\,\unl\txH\,$ is also $\txG$-invariant, it is $\txG$-basic.
\eroof
\bigskip
At this point, it seems pertinent to enquire as to the circumstances
under which the new gerbe $\,\unl\cG\,$ descends to the quotient
$\,M/\txG\,$ in the sense of Section \ref{sub:descent}. The answer
is provided by
\berop\label{prop:Gnew-equiv}
Let $\,\varpi_M:M\to M/\txG\,$ be a left principal $\txG$-bundle
with a $\txG$-invariant metric $\,\txg$,\ and let $\,\cG\,$ be a
gerbe over $\,M\,$ with a $\txG$-invariant curvature $\,\txH\,$
admitting a $\ggt$-equivariantly closed extension $\,\widehat\txH=
\txH-\kappa\,$ for $\,\kappa\in\Om^1(M)\ox\ggt^*$.\ The gerbe
$\,\unl\cG=\cG\ox I_\b\,$ defined in terms of the 2-form $\,\b\,$
given in \Reqref{eq:beta} admits a $(\txG,0)$-equivariant structure
iff $\,\cG\,$ is endowed with a $(\txG,\rho)$-equivariant structure,
with $\,\rho\,$ as in \Reqref{eq:rho-on-G}.
\eerop
\beroof
The claim of the proposition is an immediate consequence of the
following observation: The $\txG$-invariance of $\,\b$,\ expressed
by the identity
\qq\nn
\Mup\ell_g^*\b=\b\,,
\qqq
valid for any $\,g\in\txG$,\ implies, in virtue of
Eqs.\,\eqref{eq:pull-as-shift-act} and \eqref{eq:beta-con}, the
identity
\qq\nn
\Mup\ell^*\b(g,m)&=&\ee^{-\ovl{\theta_L(g)}}.\b(m)=\b(m)+\theta_L^a
(g)\wedge\kappa_a(m)+\tfrac{1}{2}\,\txc_{ab}(m)\,\theta_L^a(g)
\wedge\theta_L^b(g)\cr\cr
&\equiv&\b(m)-\rho(g,m)\,,
\qqq
and hence
\qq\nn
\Mup\ell^*\unl\cG=\Mup\ell^*\cG\ox I_{\Mup\ell^*\b}=\Mup\ell^*\cG
\ox I_{-\rho}\ox I_{\b_{2^*}}\,.
\qqq
Thus, the existence of a $(\txG,\rho)$-structure $\,(\Upsilon,\g;
\kappa)\,$ on $\,\cG\,$ is tantamount to the existence of the
desired $(\txG,0)$-structure $\,(\unl\Upsilon,\unl\g;0)\,$ on
$\,\unl\cG$,\ with $\,\unl\Upsilon:=\Upsilon\ox\id\,$ and $\,\unl\g
:=\g\ox\id$. \eroof ~\bigskip

Next, we take a closer look at the effective bi-brane structure
determined by the defect condition \eqref{eq:eff-bib-geom} and the
formerly established effective target-space geometry. The first
result is contained in
\berop
Under the assumptions and in the notation of Proposition
\ref{prop:Gnew-equiv}, let $\,\varpi_Q:Q\to Q/\txG\,$ be a left
principal $\txG$-bundle equipped with a pair of smooth $\txG$-maps
$\,\iota_\a:Q\to M,\ \a=1,2$,\ and with a 2-form $\,\om\,$ which
admits a $\txG$-equivariant extension $\,\widehat\om=\om-k,\
k\in\Om^0(Q)\ox \ggt^*\,$ satisfying \Reqref{eq:hatdom-hatdelHa}.
Then, the subset $\,\unl Q\subset Q\,$ defined in
\Reqref{eq:new-bib-wvol} is $\txG$-invariant and, assuming that
$\,\unl Q\,$ is a submanifold of $\,Q$,\ the pullback of the 2-form
$\,\unl\om\,$ of \Reqref{eq:unlom} to $\,\unl Q\,$ is $\txG$-basic.
\eerop
\beroof
The $\txG$-invariance of $\,\unl Q\subset Q\,$ follows from the
$\txG$-equivariance of $\,k$.\ The $\txG$-invariance of
$\,\unl\om\,$ on $\,Q\,$ (a consequence of the same property of
$\,\om\,$ and $\,\b$) then ensures the $\txG$-invariance of its
restriction (pullback) to $\,\unl Q$.\ Finally, we find
\qq\nn
\ic_a\unl\om\equiv\ic_a\om-\D_Q\bigl(\ic_a\b)=-\sfd k_a\,,
\qqq
and so we conclude that $\,\unl\om\vert_{\unl Q}\,$ is
$\ggt$-horizontal.\eroof
\bigskip
The last proposition enables us to formulate
\berop\label{prop:Bnew-equiv}
Let $\,\varpi_M:M\to M/\txG\,$ be a left principal $\txG$-bundle
with a $\txG$-invariant metric $\,\txg$,\ and let $\,(\cG,\Upsilon,
\g;\kappa)\,$ be a $(\txG,\rho)$-equivariant gerbe over $\,M$.\
Furthermore, let $\,(\unl\cG,\unl\Upsilon,\unl\g;0)\,$ be the
$(\txG,0)$-equivariant gerbe defined in (the proof of) Proposition
\ref{prop:Gnew-equiv} in terms of the 2-form $\,\b\,$ given in
\Reqref{eq:beta}. Finally, let
$\,\cB=(Q,\iota_1,\iota_2,\om,\Phi)\,$ be a $\cG$-bi-brane with a
world-volume given by a left principal $\txG$-bundle
$\,\varpi_Q:Q\to Q/\txG$,\ equipped with a pair of
$\txG$-equivariant bi-brane maps $\,\iota_\a$,\ and of a
$\txG$-invariant curvature $\,\om\,$ which admits a
$\txG$-equivariant extension $\,\widehat\om=\om-k,\ k\in\Om^0(Q)\ox
\ggt^*\,$ satisfying \Reqref{eq:hatdom-hatdelHa}. The $(\unl\cG,\unl
\Upsilon,\unl\gamma;0)$-bi-brane $\,\unl\cB=(\unl
Q,\iota_1,\iota_2,\unl \om,\unl\Phi)$,\ defined through
Eqs.\,\eqref{eq:new-bib-wvol}-\eqref{eq:new-bib-iso} and
\Reqref{eq:beta}, admits a $(\txG,0)$-equivariant structure iff the
restriction of $\,\cB\,$ to $\,\unl Q\,$ is endowed with a $(\txG,
\la)$-equivariant structure, with $\,\la\,$ as in
\Reqref{eq:la-on-G} and hence vanishing on $\,\unl Q$.
\eerop
\beroof
First of all, we verify, by recalling the claim of the previous
proposition and through direct calculation
\qq\nn
\sfd\unl\om\equiv\sfd\om-\D_Q\sfd\b=-\D_Q(\txH+\sfd\b)\equiv-\D_Q
\unl\txH\,,
\qqq
that the 2-form $\,\unl\om\,$ has all the properties of the
curvature of $\,\unl\cB$.\ We then readily check that the diagram
defining the 2-isomorphism $\,\unl \Xi\,$ of the $(\txG,0
)$-equivariant structure sought after can be obtained from the same
diagram for $\,\Xi\,$ by tensoring all gerbe entries by $\,I_{\Qup
d^{(1)\,*}_1\iota_1^*\b}\,$ and all 1-isomorphism entries by the
corresponding identity factor. Thus, altogether, the existence of a
$(\txG,\la=0)$-structure on $\,\cB\,$ over $\,\unl Q\,$ is
manifestly equivalent to the existence of the desired
$(\txG,0)$-structure $\,(\unl\Xi;0)\,$ on $\,\unl\cB$,\ with
$\,\unl\Xi:=\Xi\ox\id$. \eroof
\bigskip
Finally, we have
\berop
Under the assumptions and in the notation of Proposition
\ref{prop:Bnew-equiv}, and for $\,\unl\cB=( \unl
Q,\iota_\a,\unl\om,\unl\Phi)\,$ the
$(\unl\cG,\unl\Upsilon,\unl\gamma;0)$-bi-brane of that proposition,
with data as in
Eqs.\,\eqref{eq:new-bib-wvol}-\eqref{eq:new-bib-iso}, let $\,\cJ=
\bigl(T_n,\bigl(\vep^{k,k+1}_n,\pi^{k,k+1}_n\bigr);\varphi_n\bigr)\,$
be a $(\cG,\cB)$-inter-bi-brane with component world-volumes given
by left principal $\txG$-bundles $\,\varpi_{T_n}:T_n\to T_n/\txG$,\
equipped with the respective $\txG$-equivariant inter-bi-brane maps
$\,\pi_n^{k,k+1}: T_n\to Q\,$ and such that condition
\eqref{eq:hatdelom-nil} is satisfied on $\,T_n$.\ Then, the $(\unl
\cG,\unl\cB)$-inter-bi-brane $\,\unl\cJ=\bigl(\unl T_n,\bigl(
\vep^{k,k+ 1}_n,\pi^{k,k+1}_n\bigr);\unl \varphi_n\bigr)$,\ defined
through Eqs.\,\eqref{eq:new-ibb-wvol} and \eqref{eq:new-ibb-iso},
admits a $\txG$-equivariant structure iff the restriction of
$\,\cJ\,$ to $\,\bigsqcup_{n\geq 3}\,\unl T_n\,$ is endowed with a
$\txG$-equivariant structure.
\eerop
\beroof Obvious. \eroof
\bigskip
The findings of the present section, in conjunction with Corollary
\ref{cor:Gequiv-backgrnd-desc}, yield the important
\bethe
Let $\,\Bgt\,$ be a string background with a target space given by a
left principal $\txG$-bundle $\,\xcF\to\xcF/\txG$,\ and assume that
$\,\Bgt\,$ is endowed with a $(\txG,\rho,\la )$-equivariant
structure. It then canonically defines a \textbf{descendendant
background} $\,\bgt\,$ with the quotient $\,\unl\xcF/\txG\,$ as the
target space, where the disjoint components of
$\,\unl\xcF=M\sqcup\unl Q\sqcup\bigsqcup_{n\geq 3}\,\unl
T_n\,$ are, besides $\,M$,\ the spaces defined in
Eqs.\,\eqref{eq:new-bib-wvol} and \eqref{eq:new-ibb-wvol}, both
assumed smooth. Furthermore, it induces a dilaton field on
$\,\unl\xcF/\txG$.\ethe \noindent The theorem emphasises the r\^ole
played by a $\txG$-equivariant structure on the string background in
realizing a $\,\si$-model with the quotient target through the
coupling of the original $\,\si$-model to the fluctuating gauge
field, with the latter procedure well defined even if the quotient
target is singular. As we demonstrate in the sections that follow, a
$\txG$-equivariant structure on the string background is also
sufficient to couple the $\,\si$-model to topologically nontrivial
world-sheet gauge fields.

\section{The coupling to arbitrary gauge fields}\label{sec:nontriv}

\noindent The study (and partial resolution) in
Refs.\,\cite{Schellekens:1990sn,Schellekens:1990xy,Hori:1994nc,Fuchs:1995tq,Gawedzki:2010rn}
of the field-identification problem in coset models of CFT seems to
indicate that one should put topologically nontrivial world-sheet
gauge fields on the same footing as the topologically trivial ones.
The reasoning presented in Section \ref{sec:gauge-min} suggests an
extension of the gauging recipe worked out for gauge fields defined
as global Lie algebra $\ggt$-valued 1-forms on $\,\Si\,$ to the
situation in which the gauge field represents a connection on a
possibly nontrivial principal $\txG$-bundle
\qq\nn
\pi_\sfP\ :\ \sfP\to\Si\,.
\qqq
The bundle comes with the structure of a $\txG$-space with the left
$\txG$-action given by
\qq\nn
\sfPup\ell\ :\ \txG\x\sfP\to\sfP\ :\ p\mapsto g.p\equiv r\left(p,
g^{-1}\right)\,,
\qqq
where $\,r\,$ is the defining (right) $\txG$-action on $\,\sfP$.\
The connections on $\,\sfP\,$ are described by $\ggt$-valued 1-forms
$\,\cA\,$ on $\,\sfP\,$ such that
\qq\label{eq:conn-trans}
\sfPup\ell_g^*\cA=\Ad_g\cA\,,
\qqq
and such that the identity
\qq\label{eq:Kill-dual-conn}
\ic_a\cA=t_a\qquad\text{for}\qquad\ic_a=\ic_{_{\sfPup\xcK_a}}
\qqq
holds for the fundamental vector fields $\,\sfPup\xcK_a\,$
generating the $\txG$-action $\,\sfPup\ell$. \,It may then seem
natural to replace the world-sheet $\,\Si\,$ in
the definition of the extended target $\,\Sigma\times\xcF\,$ by
$\,\sfP\,$ (and its restrictions to subsets of $\,\Sigma$). We shall
set
\qq\label{eq:tilxcF-def}
\tilde\xcF:=\bigl(\sfP\x M\bigr)\sqcup\bigl(\sfP\vert_{\G
\setminus\Vgt_\G}\x Q\bigr)\sqcup\bigsqcup_{n\geq 3}\,
\bigl(\sfP\vert_{\Vgt_\G^{(n)}}\x T_n\bigr)\equiv\tilde M\sqcup
\tilde Q\sqcup\bigsqcup_{n\geq 3}\,\tilde T_n\,.
\qqq
It is for such extended target space that we can define a natural
counterpart of the extended string background $\,\Bgt_\cA\,$ from
Corollary \ref{cor:triv-Ansatz}. \,In close analogy with the
discussion of Section \ref{sec:gauge-min}, and - in particular -
with Corollary \ref{cor:triv-Ansatz}, we give
\bedef\label{def:Pext-backgnd}
Let $\,\Si\,$ be a world-sheet with an embedded defect quiver
$\,\G$,\ and let $\,\pi_\sfP:\sfP\to\Si\,$ be a principal
$\txG$-bundle over $\,\Si\,$ with connection
$\,\cA\in\Om^1(\sfP)\ox\ggt$.\ Consider a string background
$\,\Bgt=(\cM,\cB,\cJ)$,\ with target space $\,\xcF=M\sqcup Q\sqcup
T$,\ and suppose that $\,\xcF\,$ carries the structure of a
$\txG$-space, and that the string background $\,\Bgt\,$ is
$\txG$-equivariant. A \textbf{$\sfP$-extension of string background
$\,\Bgt$},\ or a \textbf{$\sfP$-extended string background}, is the
string background $\,\tilde\Bgt_\cA:=(\tilde\cM_\cA,\tilde\cB_\cA,
\tilde\cJ_\cA)\,$ with the following components
\bit
\item the \textbf{$\sfP$-extended target} $\,\tilde\cM_\cA\,$
composed of the target space $\,\tilde M=\sfP\x M\,$ with the metric
\qq\label{eq:Pext-metric}
\tilde\txg_\cA=\txg_{2^*}-\txK_{a\,2^*}\ox\cA^a_{1^*}-\cA^a_{1^*}
\ox\txK_{a\,2^*}+\txh_{ab\,2^*}\,\left(\cA^a\ox\cA^b\right)_{1^*}
\qqq
(defined in terms of the 1-forms $\,\txK_a\,$ of \Reqref{eq:KA-def}
and of the symmetric tensor $\,\txh_{ab}\,$ of \Reqref{eq:hAB-def})
and the gerbe $\,\tilde\cG_\cA=\cG_{2^*}\ox I_{\tilde\rho_\cA}$,\
with
\qq\label{eq:trhoA-def}
\tilde\rho_\cA=\kappa_{a\,2^*}\wedge\cA_{1^*}^a-\tfrac{1}{2}\,
\txc_{ab\,2^*}\,\cA_{1^*}^a\wedge\cA_{1^*}^b\,,
\qqq
cf.\ \Reqref{eq:rhoA}, of the curvature
\qq\label{eq:tHA-def}
\tilde\txH_\cA=\txH_{2^*}+\sfd\tilde\rho_\cA\,,
\qqq
cf.\ \Reqref{eq:txHA};
\item the \textbf{$\sfP$-extended $\tilde\cG_\cA$-bi-brane}
$\,\tilde\cB_\cA\,$ with the world-volume $\,\tilde Q=\sfP\vert_\G\x
Q$,\ the bi-brane maps $\,\tilde\iota_\a=\id_\sfP\x\iota_\a,\
\a=1,2$,\ the curvature
\qq\label{eq:tilomA-def}
\tilde\om_\cA=\om_{2^*}-\D_{\tilde Q}\tilde\rho_\cA+\sfd\tilde
\la_\cA\,,
\qqq
cf.\ \Reqref{eq:omA}, written in terms of
\qq\label{eq:tlaA-def}
\tilde\la_\cA=-k_{a\,2^*}\,\cA^a_{1^*}\,,
\qqq
cf.\ \Reqref{eq:laA}, and the 1-isomorphism $\,\tilde\Phi_\cA=
\Phi_{2^*}\ox J_{\tilde\la_\cA}$;
\item the \textbf{$\sfP$-extended $(\tilde\cG_\cA,\tilde\cB_\cA
)$-inter-bi-brane} $\,\tilde\cJ_\cA\,$ with component world-volumes
$\,\tilde T_n=\sfP \vert_{\Vgt_\G^{(n)}}\x T_n,\ n\geq 3$,\ with
inter-bi-brane maps $\,\tilde\pi_n^{k,k+1}=\id_\sfP\x\pi_n^{k,
k+1},\ k=1,2,\ldots, n$,\ and 2-isomorphisms $\,\tilde\varphi_{n\,
\cA}=\varphi_{n\,2^*}$.\eit \exdef~\medskip

\noindent The rationale behind the introduction of the
$\sfP$-extended string background is that it permits to rephrase the
gauge coupling in terms of purely geometric constructs in a manner
that generalises the treatment of the topologically trivial case. On
the other hand, the target space $\,\tilde\xcF\,$ of the
$\sfP$-extended string background is not the physical space of the
corresponding gauged $\si$-model that we strive to define. In order
to keep the original field content, we have to pass to the (smooth)
quotient $\,\tilde\xcF/\txG\cong\xcF\,$ with respect to the combined
(left) action
\qq\label{eq:Gact-on-Fext}
\txcFup\ell\ :\ \txG\x\tilde\xcF\to\tilde\xcF\ :\ \bigl(g,(p,x)
\bigr)\mapsto\bigl(\sfPup\ell(g,p),\xcFup\ell(g,x)\bigr)\,,
\qqq
thereby arriving at associated bundles. This is straightforward on
the level of the target space, and the true challenge is to ensure
that also the geometric structure supported by $\,\widehat \xcF\,$
descends to the quotient space. Here, the results of Section \ref{sub:descent}
will prove instrumental.

\subsection{Equivariance properties of $\sfP$-extended
string backgrounds}\label{sec:Gequiv-Pext}

\noindent A convenient language in which to discuss the behaviour of
various tensorial and cohomological structures on the
$\sfP$-extended target space $\,\tilde\xcF\,$ was introduced in
Section \ref{sub:simplicial}. Here, we merely specialise it to the
case of interest. Thus, we consider $\,\tilde\xcF\,$ as a
$\txG$-space equipped with the left action $\,\txcFup\ell\,$ defined
in \Reqref{eq:Gact-on-Fext} and giving rise to the fundamental
vector fields
\qq\nn
\txcFup\xcK_a(p,x)=\sfPup\xcK_a(p)+\xcFup\xcK_a(x)\,,\qquad
a=1,2,\ldots,\dim\,\ggt\,,
\qqq
and subsequently construct the nerve $\,\txG\tilde\xcF\,$ of the
action groupoid $\,\txG\lx\tilde\xcF$,\ with its face maps
$\,\txcFup d^{(n+1)}_i$.\ Furthermore, the $\txG$-equivariant maps
$\,\iota_\a^{(m)}\,$ and $\,\pi_n^{k,k+1\,( m)}$,\ introduced in
\Reqref{eq:funct-bibb-maps}, admit $\txG$-equivariant extensions
\qq\nn
\tilde\iota^{(m)}_\a&:=&\id_{\txG^m}\x(\id_\sfP\x\iota_\a)\ :\
\txG^m\x\tilde Q\to\txG^m\x\tilde M\,,\cr\cr
\tilde\pi_n^{k,k+1\,(m)}&:=&\id_{\txG^m}\x\bigl(\id_\sfP,
\pi_n^{k,k+1}\bigr)\ :\ \txG^m\x\tilde T_n\to\txG^m\x \tilde Q\,.
\qqq
They provide us with the standard pullback operators $\,\D_{\tilde
Q}^{(m)}\,$ and $\,\D_{\tilde T_n}^{(m)}$,\ respectively. Together
with the pullback-cohomology operators $\,\txcMup\d^{(m )}_\txG\,$
produced from the face maps of $\,\txG\tilde\xcF$,\ they satisfy the
familiar algebra
\qq
&\hspace*{-1.5cm}\txcMup\d_\txG^{(m)}\circ\txcMup\d_\txG^{(m-1)}=0\,,\quad
\tilde\xcM=\tilde M,\tilde Q,\tilde T_n\,, \label{eq:tildel-nil}&\\
\cr
&\hspace*{-0.3cm}\tQup\d^{(m)}_\txG\circ\D_{\tilde Q}^{(m)}\hspace{-0.1cm}
=\hspace{-0.07cm} \D_{\tilde
Q}^{(m+1)}\circ\tMup\d^{(m)}_\txG\,,\quad
\tTnup\d^{(m)}_\txG\circ\D_{\tilde
T_n}^{(m)}\hspace{-0.1cm}=\hspace{-0.05cm} \D_{\tilde
T_n}^{(m+1)}\circ\tQup\d^{(m)}_\txG\,.\label{eq:tildel-tilDel}&
\qqq
Finally, upon equipping the constituent subspaces of $\,\txG\tilde
\xcF\,$ with the structure of a $\txG$-space through
\Reqref{eq:Gact-on-nerve}, we obtain the corresponding fundamental
vector fields
\qq
\txcFup\xcK_a^{(m)}(g_m,g_{m-1},\ldots,g_1,p,x)=\sfPup\xcK_a(p)+
\xcFup\xcK_a^{(m)}(g_m,g_{m-1},\ldots,g_1,x)\,,\label{eq:til-Killing}
\qqq
in conformity with \Reqref{eq:fund-vect-on-nerve}.

In the remainder of this section, we shall study the behaviour of
various tensors and the associated geometric objects (gerbes, gerbe
bi-modules etc.), involved in the description of the $\si$-model
coupled to topologically nontrivial gauge fields, under the
transport within $\,\txG\tilde\xcF\,$ effected by operators
$\,\txcMup\d_\txG^{(n)}\,$ and intertwined by $\,\D_{\tilde
Q}^{(m)}\,$ and $\,\D_{\tilde T_n}^{(m)}$.\ (We shall systematically
add the tilde over all objects that live on the $\sfP$-extended
spaces.) This will pave the way to a subsequent formalisation of our
description of the gauged $\si$-model in the presence of the defect
$\,\G$.\medskip

We begin with
\berop\label{prop:Ginv-gA}
Let $\,\tilde\cM_\cA\,$ be a $\sfP$-extended target of Definition
\ref{def:Pext-backgnd}, with a (left) $\txG$-action on $\,\tilde
M\,$ given by \Reqref{eq:Gact-on-Fext}. The metric $\,\tilde
\txg_\cA\,$ on the target space $\,\tilde M\,$ is $\txG$-basic.
\eerop
\beroof
The $\txG$-invariance of $\,\tilde\txg_\cA\,$ is a consequence of
the same property of $\,\txg$,\ as well as of the
$\txG$-equivariance of the fundamental vector fields
$\,\Mup\xcK_a$.\ That the extended metric is also $\ggt$-horizontal
readily follows from \Reqref{eq:Kill-dual-conn}. \eroof
\bigskip
Passing to the differential forms, we establish
\berop\cite[Lemma 5.4 \& Eq.\,(5.13)]{Gawedzki:2010rn}\label{prop:Gequiv-props-tilomA}
Let $\,\tilde\cM_\cA=(\tilde M,\tilde\txg_\cA,\tilde\cG_\cA)\,$ be a
$\sfP$-extended target of Definition \ref{def:Pext-backgnd}, with a
(left) $\txG$-action on $\,\tilde M\,$ given by
\Reqref{eq:Gact-on-Fext} and the canonical projections $\,\pr_1:
\tilde M\to\sfP\,$ and $\,\pr_2:\tilde M\to M$.\ Moreover, let
$\,\txG\tilde M\,$ be the nerve, with face maps $\,\tMup d^{(m
)}_i\,$ and the corresponding coboundary operators
\eqref{eq:pullback-cob}, of the action groupoid $\,\txG\lx\tilde
M\,$ over $\,\tilde M$.\ Then, the following holds true for
$\,\tilde\txH_\cA\,$ and $\,\tilde\rho_\cA\,$ defined in
\Reqref{eq:tHA-def} and \Reqref{eq:trhoA-def}, respectively.
\bit
\item[i)] $\tilde\txH_\cA\,$ and $\,\tilde\rho_\cA\,$
satisfy the relations
\qq
\tMup\d^{(0)}_\txG\tilde\rho_\cA+\rho_{[1,3]^*}&=&0\,,
\label{eq:tildel-rhoA}\\\cr
\tMup\d^{(0)}_\txG\tilde\txH_\cA&=&0\,,\label{eq:tildelHA}\\\cr
\tMup\ell^{*}_h\tilde\rho_\cA&=&\tilde\rho_\cA\,,
\label{eq:G-inv-tilrhoA}
\qqq
for $\,\pr_{1,3}:\txG\x\tilde M\to\txG\x M\,$ the canonical
projection, and $\,\rho\,$ as in \Reqref{eq:rho-on-G};
\item[ii)] $\tilde\txH_\cA\,$ is $\txG$-basic and
$\ggt$-equivariantly closed.
\eit
\eerop
\beroof
\bit
\item[Ad i)] Identities \eqref{eq:tildel-rhoA} and
\eqref{eq:tildelHA} were proven in \Rcite{Gawedzki:2010rn} as Lemma
5.4 and Eq.\,(5.13), respectively. The former implies identity
\eqref{eq:G-inv-tilrhoA} upon pullback along the map $\,\Mup\tilde
\iota_h:\tilde M\to\txG\x\tilde M:\tilde m\mapsto(h,\tilde m)$.
\item[Ad ii)] For the sake of the proof, it will be convenient to
view $\,\tilde\txH_\cA\,$ as the Weil transform
\qq\label{eq:HA-as-Weil}
\tilde\txH_\cA=\ee^{-\ovl{\cA_{1^*}}}.\widehat\txH_{2^*}\bigl(
\bigl(\pi_\sfP^*\txF\bigr)_{1^*}\bigr)
\qqq
of $\,\widehat\txH\,$ evaluated on the pullback
\qq\nn
\pi_\sfP^*\txF:=\sfd\cA+\cA\wedge\cA
\qqq
of the curvature $\,\txF\,$ of $\,\cA\,$ along (the composition of
the canonical projection $\,\pr_1\,$ with) the bundle projection
$\,\pi_\sfP$,\ cf.\ Proposition \ref{prop:Weil-ext-H}. Indeed, upon
recalling Eqs.\,\eqref{eq:impl-HS}, we obtain
\qq\nn
&&\hspace{-0.6cm}\ee^{-\ovl{\cA_{1^*}}}.\widehat\txH_{2^*}\bigl(\bigl(\pi_\sfP^*\txF
\bigr)_{1^*}\bigr)\cr\cr
&=&\txH_{2^*}-\kappa_{a\,2^*}\wedge\bigl(\sfd
\cA^a+\tfrac{1}{2}\,f_{abc}\,\cA^b\wedge\cA^c\bigr)_{1^*}+
\cA^a_{1^*}\wedge\sfd\kappa_{a\,2^*}\cr\cr
&&+\tfrac{1}{2}\,\bigl(\cA^a\wedge\cA^b)_{1^*}\,\bigl(f_{abc}\,
\kappa_c-\sfd\txc_{ab}\bigr)_{2^*}\cr\cr
&&-\tfrac{1}{3!}\,\bigl(\cA^a\wedge\cA^b\wedge\cA^c\bigr)_{1^*}\,
\bigl(\bigl(f_{bcd}\,\ic_a+f_{abd}\,\ic_c-f_{acd}\,\ic_b\bigr)
\kappa_d\bigr)_{2^*}\cr\cr
&&+\txc_{ab\,2^*}\,\bigl(\cA^a\wedge\bigl(\sfd\cA^b+\tfrac{1}{2}\,
f_{bcd}\,\cA^c\wedge\cA^d\bigr)\bigr)_{1^*}\cr\cr
&=&\txH_{2^*}-\kappa_{a\,2^*}\wedge\sfd\cA^a_{1^*}+\cA^a_{1^*}
\wedge\sfd\kappa_{a\,2^*}-\tfrac{1}{2}\,\sfd\bigl(\ic_a
\kappa_b\bigr)_{2^*}\wedge\bigl(\cA^a\wedge\cA^b)_{1^*}\cr\cr
&&+\txc_{ab\,2^*}\,\bigl(\cA^a\wedge\sfd\cA^b\bigr)_{1^*}\,,
\qqq
where the two terms trilinear in $\,\cA\,$ are readily seen to
cancel out. That the above gives the desired result is verified
through a direct computation using identity \eqref{eq:HS2-ids-triv}.

With the help of Eqs.\,\eqref{eq:HS-exact}, \eqref{eq:HS1-ids-triv}
and \eqref{eq:Kill-dual-conn}, as well as of the identity
\qq\nn
\ic_a\sfd\cA^b=f_{abc}\,\cA^c
\qqq
that follows from the definition of $\,\txF\,$ and
\Reqref{eq:Kill-dual-conn}, we now establish
\qq\nn
&&\hspace{-0.6cm}\ic_a\tilde\txH_\cA\cr\cr
&=&\bigl(\ic_a\txH\bigr)_{2^*}-\txc_{ab\,2^*}
\wedge\sfd\cA^b_{1^*}+f_{abc}\,\kappa_{b\,2^*}\wedge\cA^c_{1^*}+
\sfd\kappa_{a\,2^*}+(\ic_a\sfd\kappa_b)_{2^*}\wedge\cA^b_{1^*}
\cr\cr
&&-\tfrac{1}{2}\,\bigl(\ic_a\sfd\bigl(\ic_b\kappa_c\bigr)
\bigr)_{2^*}\,\bigl(\cA^b\wedge\cA^c\bigr)_{1^*}+\sfd\txc_{ab\,2^*}
\wedge\cA^b_{1^*}+\txc_{ab\,2^*}\,\sfd\cA^b_{1^*}\cr\cr
&&-f_{acd}\,\bigl(\ic_b\kappa_c\bigr)_{2^*}\,\bigl(\cA^b\wedge\cA^d
\bigr)_{1^*}\cr\cr
&=&\bigl(\tfrac{1}{2}\,\bigl(f_{acd}\,\txc_{bd}-f_{abd}\,\txc_{dc}+
\ic_b\pLie{a}\kappa_c\bigr)\bigr)_{2^*}\,\bigl(\cA^b\wedge\cA^c
\bigr)_{1^*}=0\,,
\qqq
which shows that $\,\tilde\txH_\cA\,$ is $\txG$-horizontal. Since it
is also $\txG$-invariant, which can be seen by pulling back
\Reqref{eq:tildelHA} along $\,\Mup\tilde\iota_h$,\ it is
$\txG$-basic, as claimed. Being closed, it is then automatically
$\ggt$-equivariantly closed.
\eit
\eroof\bigskip

Passing to the defect data, we find
\berop
Under the assumptions and in the notation of Proposition
\ref{prop:Gequiv-props-tilomA}, let $\,\tilde\cB_\cA=(\tilde
Q,\tilde\iota_\a,\tilde\om_\cA,\tilde \Phi_\cA\ |\ \a=1,2)\,$ be a
$\sfP$-extended $\tilde \cG_\cA$-bi-brane of Definition
\ref{def:Pext-backgnd}, with a (left) $\txG$-action on $\,\tilde
Q\,$ given by \Reqref{eq:Gact-on-Fext} and the canonical projections
$\,\pr_1: \tilde Q\to\sfP\vert_\G\,$ and $\,\pr_2:\tilde Q\to Q$.\
Moreover, let $\,\txG\tilde Q\,$ be the nerve, with face maps
$\,\tQup d^{(m)}_i\,$ and the corresponding coboundary operators
\eqref{eq:pullback-cob}, of the action groupoid $\,\txG\lx\tilde
Q\,$ over $\,\tilde Q$.\ Then, the following holds true for
$\,\tilde\om_\cA\,$ and $\,\tilde\la_\cA\,$ defined in
\Reqref{eq:tilomA-def} and \Reqref{eq:tlaA-def}, respectively.
\bit
\item[i)] $\tilde\om_\cA\,$ and $\,\tilde\la_\cA\,$
satisfy the relations
\qq
\tQup\d^{(0)}_\txG\tilde\la_\cA+\la_{[1,3]^*}&=&0\,,
\label{eq:tildel-lA}\\\cr
\tQup\d_\txG^{(0)}\tilde\om_\cA&=&0\,,\label{eq:tildel-omA}\\
\cr
\tQup\ell^*_g\tilde\la_\cA&=&\tilde\la_\cA\,,\label{eq:G-inv-tillA}\,,\\\cr
\qqq
for $\,\pr_{1,3}:\txG\x\tilde Q\to\txG\x Q\,$ the canonical
projection, and $\,\la\,$ as in \Reqref{eq:la-on-G};
\item[ii)] $\tilde\om_\cA\,$ is $\txG$-basic and it verifies
\qq\label{eq:domA-DelHA}
\widehat\sfd\tilde\om_\cA=-\D_{\tilde Q}\tilde\txH_\cA\,.
\qqq
\eit
\eerop
\beroof
\bit
\item[Ad i)] Identity \eqref{eq:tildel-lA} follows from
Eqs.\,\eqref{eq:pull-as-shift-act}, \eqref{eq:conn-trans} and
\eqref{eq:Kill-dual-conn}, upon taking into account the
$\txG$-equivariance of $\,k$.\ It yields identity
\eqref{eq:G-inv-tillA} through pullback along the map $\,\Qup\tilde
\iota_h:\tilde Q\to\txG\x\tilde Q:\tilde q\mapsto(h,\tilde q)$.\
Finally, in conjunction with Eqs.\,\eqref{eq:dla} and
\eqref{eq:tildel-rhoA}, it gives identity \eqref{eq:tildel-omA}.
\item[Ad ii)] Pulling back \Reqref{eq:tildel-omA} along $\,\Qup\tilde
\iota_h$,\ we establish the $\txG$-invariance of $\,\tilde\om_\cA$.\
From the simple identities
\qq\nn
\ic_a\tilde\rho_\cA=-\kappa_{a\,2^*}\,,\qquad\qquad
\ic_a\sfd\tilde\la_\cA=\sfd k_{a\,2^*}\,,
\qqq
taken together with \Reqref{eq:bdry-exact}, we infer that
$\,\tilde\om_\cA\,$ is also $\ggt$-horizontal,
\qq\nn
\ic_a\tilde\om_\cA=0\,,
\qqq
so that, altogether, it is $\txG$-basic as claimed. Using the last
identity, we then show
\qq\nn
\widehat\sfd\tilde\om_\cA(t_a)\equiv\sfd\tilde\om_\cA-\ic_a\tilde
\om_\cA=(\sfd\om)_{2^*}-\D_{\tilde Q}\sfd\tilde\rho_\cA=-\D_{\tilde
Q}\tilde\txH_\cA(t_a)\,,
\qqq
whence also \Reqref{eq:domA-DelHA}.
\eit
\eroof~\medskip \brem We may interpret $\,\tilde\om_\cA\,$ as the
Weil transform
\qq\label{eq:tilom-as-trans}
\tilde\om_\cA=\ee^{-\ovl\cA_{1^*}}.\widehat\om_{2^*}\bigl(\bigl(
\pi_\sfP^*\txF\bigr)_{1^*}\bigr)\,,
\qqq
cf.\ Proposition \ref{prop:Weil-ext-om}. Indeed, proceeding along
the same lines as in the proof of \Reqref{eq:HA-as-Weil} and
invoking Eqs.\,\eqref{eq:bdry-exact} and \eqref{eq:FFM-ids-triv}
along the way, we find
\qq\nn
&&\ee^{-\ovl\cA_{1^*}}.\widehat\om_{2^*}\bigl(\bigl(\pi_\sfP^*\txF
\bigr)_{1^*}\bigr)\ =\ \om_{2^*}-k_{a\,2^*}\,\bigl(\sfd\cA^a+
\tfrac{1}{2}\,f_{abc}\,\cA^b\wedge\cA^c\bigr)_{1^*}+\cA^a_{1^*}
\wedge(\D_Q\kappa_a+\sfd k_a)_{2^*}\cr\cr
&&\hspace*{7cm}+\tfrac{1}{2}\,\bigl(\cA^a\wedge\cA^b\bigr)_{1^*}\,\bigl(\ic_a
(\D_Q\kappa_b+\sfd k_b)\bigr)_{2^*}\cr\cr
&&=\ \om_{2^*}-k_{a\,2^*}\,\sfd\cA^a_{1^*}-\sfd k_{a\,2^*}\wedge
\cA^a_{1^*}-\D_{\tilde Q}\bigl(\kappa_{a\,2^*}\wedge
\cA^a_{1^*}\bigr) +\tfrac{1}{2}\,\D_{\tilde Q}\bigl(\bigl(\ic_a
\kappa_b\bigr)_{2^*}\,\bigl(\cA^a\wedge\cA^b\bigr)_{1^*}\bigr)\,.
\qqq
\erem

We are now ready to discuss the global geometric structures
associated with the forms considered above. Thus, we have
\berop\cite[Prop.\,5.5]{Gawedzki:2010rn}
\label{prop:Gequiv-gerbe-ext} Let $\,(\cG,\Upsilon,\g;\kappa)\,$ be
a $(\txG,\rho)$-equivariant gerbe over a $\txG$-space $\,M$,\ and
let $\,\pi_\sfP:\sfP\to\Si\,$ be a principal $\txG$-bundle with
connection $\,\cA\in\Om^1(\sfP)\ox\ggt\,$ over an oriented
two-dimensional surface $\,\Si$.\ Denote by $\,\tilde M\equiv\sfP\x
M\,$ the $\sfP$-extension of $\,M$.\ The quadruple
$\,(\tilde\cG_\cA,\tilde\Upsilon_\cA,\tilde\g_\cA;0):=(\cG_{2^*}\ox
I_{\tilde\rho_\cA},\Upsilon_{[1,3]^*}\ox\id,\g_{[1,
2,4]^*}\ox\id;0)$,\ written in the notation of Proposition
\ref{prop:Gequiv-props-tilomA} and in terms of the canonical
projections $\,\pr_1:\tilde M\to \sfP$,\ $\,\pr_2:\tilde M\to M$,\
$\,\pr_{1,3}:\txG\x\tilde M\to\txG\x M\,$ and
$\,\pr_{1,2,4}:\txG^2\x\tilde M\to\txG^2\x M$,\ carries a canonical
structure of a $(\txG,0)$-equivariant gerbe over $\,\tilde M$.
\eerop

Proposition \ref{prop:Gequiv-gerbe-ext} points towards a relation between
a general $\txG$-equivariant structure over a given manifold and a
distinguished one over its principal $\txG$-extension. This
observation leads us immediately to the counterpart of the above
that pertains to bi-branes.
\berop\label{prop:Gequiv-bibrane-ext}
Under the assumptions and in the notation of Proposition
\ref{prop:Gequiv-gerbe-ext}, let $\,Q\,$ be a $\txG$-space equipped
with a pair of smooth $\txG$-equivariant maps $\,\iota_\a:Q\to M,\
\a=1,2$,\ and denote by $\,\tilde Q\equiv\sfP \vert_\G\x Q\,$ its
$\sfP$-extension, for $\,\G\,$ a defect quiver without junctions
embedded in $\,\Si$.\ Moreover, let
$\,(\tilde\cG_\cA,\tilde\Upsilon_\cA,\tilde\g_\cA; 0)\,$ be the
$(\txG,0)$-equivariant gerbe over $\,\tilde M\,$ associated to a
$(\txG,\rho)$-equivariant gerbe $\,(\cG,\Upsilon,\g;\kappa)\,$ over
$\,M$.\ Finally, let $\,(\cB, \Xi;k)$,\ with
$\,\cB=(Q,\iota_1,\iota_2,\om,\Phi)$,\ be a $(\txG,\la)$-equivariant
$(\cG,\Upsilon,\g;\kappa)$-bi-brane and denote by
$\,\tilde\Phi_\cA\,$ the 1-isomorphism
\qq\label{eq:tilPhiA}
\tilde\Phi_\cA=\Phi_{2^*}\ox J_{\tilde\la_\cA}\ :\
\tilde\iota_1^*\tilde\cG_\cA\xrightarrow{\ \cong\ }\tilde
\iota_2^*\tilde\cG_\cA\ox I_{\tilde\om_\cA}
\qqq
of gerbes over $\,\tilde Q$,\ written in terms of the smooth maps
$\,\tilde\iota_\a\equiv\tilde\iota_\a^{(0)}\,$ and the canonical
projection $\,\pr_2:\tilde Q\to Q$,\ as well as of the trivial
1-isomorphism $\,J_{\tilde\la_\cA}\,$ defined by the 1-form
$\,\tilde\la_\cA\,$ on $\,\tilde Q\,$ given in
\Reqref{eq:tlaA-def},\ and of the 2-form $\,\tilde\om_\cA\,$ on
$\,\tilde Q\,$ defined in \Reqref{eq:tilomA-def}.\ Then, there
exists a canonical structure of a $(\txG,0)$-equivariant $(
\tilde\cG_\cA,\tilde\Upsilon_\cA,\tilde\g_\cA;0 )$-bi-brane on the
triple $\,(\tilde\cB_\cA,\tilde\Xi_\cA;0 )\,$ consisting of the
bi-brane $\,\tilde\cB_\cA=(\tilde Q,
\tilde\iota_\a,\tilde\om_\cA,\tilde\Phi_\cA)\,$ over $\,\tilde Q\,$
and of the 2-isomorphism $\,\tilde\Xi_\cA= \Xi_{[1,3]^*}\ox\id$.
\eerop
\beroof
First of all, we have to demonstrate that $\,\tilde\Phi_\cA\,$ does,
indeed, define a $(\tilde\iota_1^*\tilde\cG_\cA,
\tilde\iota_2^*\tilde\cG_\cA)$-bi-module of curvature
$\,\tilde\om_\cA$.\ That it yields the desired 1-isomorphism follows
trivially from the very definition of $\,\tilde \om_\cA\,$ as
\qq\nn
\tilde\iota_1^*\tilde\cG_\cA\equiv\pr_2^*\iota_1^*\cG\ox
I_{\tilde\iota_1^*\tilde\rho_\cA}&\xrightarrow{\pr_2^*\Phi \ox
J_{\tilde\la_\cA}}&\pr_2^*(\iota_2^*\cG\ox I_\om)\ox
I_{\tilde\iota_1^*\tilde\rho_\cA+\sfd\tilde\la_\cA}\cr\cr
&&=\pr_2^*\iota_2^*\cG\ox I_{\tilde\iota_2^*\tilde\rho_\cA}\ox
I_{\om_{2^*}-\D_{\tilde Q}^*\tilde\rho_\cA+\sfd\tilde \la_\cA}\cr\cr
&&\equiv\tilde\iota_2^*\tilde\cG_\cA\ox I_{\tilde
\om_\cA}\,.
\qqq
\Reqref{eq:domA-DelHA} and the $\txG$-equivariance of
$\,\tilde\om_\cA\,$ ensure that the trivial $\txG$-equivariant
(Cartan-model) extension of $\,\tilde \om_\cA\,$ has the properties
required of a curvature of the
$(\tilde\iota_1^*\tilde\cG_\cA,\tilde\iota_2^*\tilde
\cG_\cA)$-bi-module which is $\txG$-equivariant relative to the
vanishing 1-form, as stated in Definition \ref{def:Gequiv-bib}.

In the next step, we check that $\,\tilde\Xi_\cA\,$ renders the
appropriate 2-diagram \eqref{diag:penta-first} commutative. To this
end, we write out the objects appearing in the diagram in the case
in hand using the respective definitions given in the proposition,
and subsequently apply the (commutation) relations for various
projections, bi-brane and face maps involved, and use
\Reqref{eq:tildel-lA},
\qq\label{diag:penta-last}
\xy (50,0)*{\bullet}="G12"+(2,5)*{\tx{\scriptsize $\pr_{1,3}^*\Qup
d^{(1)\,*}_0\iota_2^*\cG\ox I_{\tQup d^{(1)\,*}_0\tilde\om_\cA+
\tilde\iota_2^{(1)\,*}\tMup d^{(1)\,*}_0\tilde\rho_\cA}$}};
(20,-25)*{\bullet}="G1r1"+(-21,0)*{\tx{\scriptsize ${\pr_{1,3}^*\Qup
d^{(1)\,*}_0\iota_1^*\cG\atop\ox I_{\tilde\iota_1^{(1)\,*}\tMup
d^{(1 )\,*}_0\tilde\rho_\cA}}\hspace*{-1cm}$}};
(80,-25)*{\bullet}="G2om"+(28,0)*{\tx{\scriptsize $\hspace*{-1cm}
{\pr_{1,3}^*\Qup d^{(1)\,*}_0 \iota_2^*\cG\atop\ox I_{\tQup
d^{(1)\,*}_1\tilde\om_\cA+ \tilde\iota_2^{(1)\,*}\tMup
d^{(1)\,*}_0\tilde\rho_\cA}}$}};
(30,-55)*{\bullet}="G2or1"+(-10,-5)*{\tx{\scriptsize$\pr_{1,3}^*
\Qup d^{(1)\,*}_1\iota_1^*\cG\atop\ox I_{\tilde\iota_1^{(1)\,*}\tMup
d^{(1)\,*}_1\tilde\rho_\cA}$}};
(70,-55)*{\bullet}="G2or2"+(18,-5)*{\tx{\scriptsize $\hspace*{-1cm}
{\pr_{1,3}^*\Qup d^{(1)\,*}_1\iota_2^*\cG\atop\ox I_{\tQup
d^{(1)\,*}_1\tilde\om_\cA+ \tilde\iota_2^{(1)\,*}\tMup
d^{(1)\,*}_1\tilde\rho_\cA}}$}}; \ar@{->}|{\pr_{1,3}^*\Qup
d^{(1)\,*}_0\Phi\ox J_{\tQup d^{(1)\,*}_0\tilde\la_\cA}\qquad}
"G1r1";"G12" \ar@{=}|{\id} "G12";"G2om"
\ar@{->}|{\pr_{1,3}^*\iota_1^{(1)\,*}\Upsilon\ox\id} "G2or1";"G1r1"
\ar@{->}|{\pr_{1,3}^*\iota_2^{(1)\,*}\Upsilon^{-1} \ox\id}
"G2om";"G2or2" \ar@{->}|{\pr_{1,3}^*\Qup d^{(1)\,*}_1\Phi\ox
J_{\tQup d^{(1)\,*}_1\tilde\la_\cA}} "G2or1";"G2or2"
\ar@{=>}|{\tilde\Xi_\cA} "G2or1"+(20,3);"G12"+(0,-3)
\endxy
\qqq
whereupon it becomes clear, in virtue of the assumed properties of
$\,(\cB,\Xi;k)$,\ that the 2-isomorphism $\,\tilde\Xi_\cA=
\pr_{1,3}^*\Xi\ox\id\,$ renders the diagram meaningful. Having
carried out an analogous exercise for diagram
\eqref{diag:Gequiv-bimod-coh} in the present setting, we conclude
that all the 2-isomorphisms on the faces of the trigonal prism
factor as tensor products of an identity 2-isomorphism with
pullbacks, along the canonical projection
$\,\pr_{1,2,4}:\txG^2\x\tilde Q\to\txG^2\x Q$,\ of the
2-isomorphisms from the respective faces of the trigonal prism of
the similar diagram for the assumed $\txG$-equivariant
$\cG$-bi-brane structure on $\,(\cB,\Xi;k)$,\ and so the
2-isomorphism $\,\tilde\Xi_\cA\,$ satisfies the necessary coherence
condition of Definition \ref{def:Gequiv-bib}. This completes the
proof of the proposition. \eroof\medskip

We complete our discussion with a statement concerning an
inter-bi-brane.
\berop\label{prop:Gequiv-ibbrane-ext}
Under the assumptions and in the notation of Propositions
\ref{prop:Gequiv-gerbe-ext} and \ref{prop:Gequiv-bibrane-ext}, let
$\,T_n,\ n\geq 3$, \,be a collection of $\txG$-spaces, equipped with
smooth $\txG$-equivariant maps $\,\pi_n^{k,k+1}:T_n\to Q,\
k=1,2,\ldots,n$,\ and denote by $\,\tilde T_n\equiv
\sfP\vert_{\Vgt_\G^{(n)}}\x T_n\,$ the respective $\sfP$-extensions,
for $\,\Vgt_\G^{(n)}\,$ the sets of junctions of valence $n$ of the
defect quiver $\,\G\subset\Si$.\ Furthermore, let $\,(\tilde\cG_\cA,
\,\tilde\Upsilon_\cA,\tilde\g_\cA;0)\,$ be $(\txG,0)$-equivariant
gerbe over $\,\tilde M\,$ associated to a $(\txG,\rho)$-equivariant
gerbe $\,(\cG ,\Upsilon,\g;\kappa)\,$ over $\,M$,\ and let
$\,(\tilde\cB_\cA,\tilde \Xi_\cA;0)$ be the $(\txG,0 )$-equivariant
$(\tilde\cG_\cA,\tilde\Upsilon_\cA, \tilde\g_\cA;0)$-bi-brane over
$\,\tilde Q\,$ associated to a $(\txG,\la)$-equivariant
$(\cG,\Upsilon,\g;\kappa)$-bi-brane $\,(\cB,\Xi; k)$.\ Finally, let
$\,\cJ=\bigl(T_n,\bigl( \vep^{k,k+1}_n,\pi^{k,k
+1}_n\bigr);\varphi_n \bigr)\,$ be a $\txG$-equivariant $((\cG,
\Upsilon,\g;\kappa),(\cB,\Xi;k))$-inter-bi-brane, and denote by
$\,\varphi_{n\,\cA}\,$ the 2-isomorphisms
\qq\nn
\tilde\varphi_{n\,\cA}:=\varphi_{n\,2^*}
\qqq
between 1-isomorphisms of gerbes over the respective $\,\tilde
T_n$,\ with $\,\pr_2:\tilde T_n\to T_n\,$ the canonical projections.
Then, there exists a canonical structure of a $\txG$-equivariant
$((\tilde\cG_\cA,\tilde\Upsilon_\cA,
\tilde\g_\cA;0),(\tilde\cB_\cA,$ $\tilde\Xi_\cA;0) )$-inter-bi-brane
on $\,\tilde\cJ_\cA=\bigl(\tilde T_n,
\bigl(\tilde\vep^{k,k+1}_n,\tilde\pi^{k,k+1}_n\bigr);\tilde\varphi_{n\,\cA}\bigr)$,
\,the latter being defined in terms of the maps
$\,\tilde\vep^{k,k+1}_n:=\vep^{k,k+1}_n\circ\pr_2\,$ and
$\,\tilde\pi^{k,k+1}_n\equiv\tilde\pi^{k,k+1\,(0)}_n$.
\eerop
\beroof
The claim of the proposition is a direct consequence of the chain of
equalities
\qq\nn
\bigl(\tilde\pi_n^{n,1\,*}\Phi^{\tilde\vep_n^{n,1}}\ox\id
\bigr)\circ\bigl(\tilde\pi_n^{n-1,n\,*}\Phi^{\tilde
\vep_n^{n-1,n}}\ox\id\bigr)\circ\cdots\circ\tilde\pi_n^{1,2\,*}
\Phi^{\tilde\vep_n^{1,2}}\cr\cr
=\pr_2^*\bigl(\bigl(\Phi_n^{n,1}\ox\id\bigr)\circ\bigl(\Phi_n^{n-1,
n}\ox\id\bigr)\circ\cdots\circ\Phi_n^{1,2}\bigr)\ox J_{\D_{\tilde
T_n}\tilde\la_\cA}\cr\cr
=\pr_2^*\bigl(\bigl(\Phi_n^{n,1}\ox\id\bigr)\circ\bigl(\Phi_n^{n-1,
n}\ox\id\bigr)\circ\cdots\circ\Phi_n^{1,2}\bigr)\,,
\qqq
inferred by \Reqref{eq:junct-exact}. \eroof~\medskip

We shall now apply the propositions collected in the present
section, in conjunction with Theorems \ref{thm:Gdesc},
\ref{thm:Gequivbib-Gbas-bib-equiv} and \ref{thm:Gequivibb-ibb-equiv}
(as summarised in Corollary \ref{cor:Gequiv-backgrnd-desc}) to the
$\sfP$-extended target space $\,\tilde\xcF\,$ of
\Reqref{eq:tilxcF-def}, equipped with the $\txG$-action of
\Reqref{eq:Gact-on-Fext}. The latter turns $\,\varpi_{\tilde
M}:\tilde M\to\tilde M/\txG,\ \varpi_{\tilde Q}:\tilde Q\to\tilde
Q/\txG\,$ and $\,\varpi_{\tilde T_n}:\tilde T_n\to\tilde T_n/\txG\,$
into left principal $\txG$-bundles. For these, we obtain the
fundamental
\becor\label{cor:desc-gerbib}
Under the assumptions and in the notation of Definition
\ref{def:Pext-backgnd} and Propositions \ref{prop:Gequiv-gerbe-ext},
\ref{prop:Gequiv-bibrane-ext} and \ref{prop:Gequiv-ibbrane-ext}, the
$\sfP$-extension $\,\tilde\Bgt_\cA\,$ of a $(\txG,\rho,\la
)$-equivariant string background with the $\sfP$-extended target
space $\,\tilde\xcF=\tilde M\sqcup\tilde Q\sqcup\bigsqcup_{n\geq 3}
\,\tilde T_n$,\ descends to a unique string background
$\,\Bgt_\cA\,$ with target space $\,\tilde\xcF/\txG$,\ and the
pullbacks of the gerbe $\,\cG_\cA$, of the bi-brane 1-isomorphism
$\,\Phi_\cA\,$ and of the inter-bi-brane 2-isomorphisms
$\,\varphi_{n\,\cA}\,$ of $\,\Bgt_\cA\,$ along the respective smooth
maps $\,\varpi_{\tilde\xcM}\,$ are $\txG$-equivariantly equivalent
to the $(\txG,0)$-equivariant gerbe, the $(\txG,0)$-equivariant
bi-brane 1-isomorphism and the $\txG$-equivariant inter-bi-brane
2-isomorphism of $\,\tilde\Bgt_\cA$,\ respectively, in the sense of
Definitions \ref{def:equiv-1iso}, \ref{def:Gequiv-bib} and
\ref{def:Gequiv-ibb}. The quotient target space $\,\tilde M/\txG\,$
is equipped with the unique metric $\,\txg_\cA\,$ determined by the
condition that its pullback to $\,\tilde M\,$ along
$\,\varpi_{\tilde M}\,$ give $\,\tilde\txg_\cA$.\ecor

\subsection{Fully gauged $\si$-models and their gauge invariance}

\noindent Network-field configurations $\,(\psi\,|\,\G)\,$ over an oriented
Riemann surface $\,\Si\,$ with an embedded defect quiver $\,\G\,$ in
the $\si$-model coupled to a connection $\,\cA\,$ on a principal
$\txG$-bundle $\,\sfP\to\Si\,$ are, by definition (and in accord
with the discussion from the opening paragraphs), sections of the
disjoint union
\qq\label{eq:assoc-field-space}
\tilde\xcF/\txG=(\tilde M/\txG)\sqcup(\tilde Q/\txG)\sqcup
\bigsqcup_{n\geq 3}\,(\tilde T_n/\txG)\cong\xcF
\qqq
of the associated bundles $\,\tilde M/\txG\equiv\sfP\x_\txG M\to
\Si$,\ $\,\tilde Q/\txG\equiv\sfP\vert_\G\x_\txG Q\to\G\,$ and
$\,\tilde T_n/\txG\equiv\sfP\vert_{\Vgt_\G^{(n)}}\x_\txG T_n\to
\Vgt_\G^{(n)}$.\ The foregoing analyses motivate
\bedef\label{def:sigma-gauged}
Under the assumptions and in the notation of Definition
\ref{def:Pext-backgnd} and Corollary \ref{cor:desc-gerbib}, let
$\,(\psi\,|\,\G)$,\ with $\,\psi\,$ a section of the associated
bundles $\,\tilde\xcF/\txG\,$ of \Reqref{eq:assoc-field-space}, be a
network-field configuration over a closed oriented (euclidean)
world-sheet $\,(\Si,\g)\,$ with an embedded defect quiver $\,\G$,\
in the string background $\,\Bgt_\cA\,$ descended from a
$(\txG,0,0)$-equivariant $\sfP$-extended string background
$\,\tilde\Bgt_\cA\,$ and consisting, in particular, of a metric
target space $\,(\tilde M/\txG,\txg_\cA)\,$ with gerbe $\,\cG_\cA\,$
over it, of a $\cG_\cA$-bi-brane $\,\cB_\cA\,$ with a gerbe
1-isomorphism $\,\Phi_\cA$,\ and of a $(\cG_\cA,\cB_\cA
)$-inter-bi-brane $\,\cJ_\cA\,$ with component gerbe 2-isomorphisms
$\,\varphi_{n\,\cA}$.\ The euclidean Feynman amplitude of $\,(\psi\,
|\,\G)\,$ is
\qq\nn
\xcA[(\psi\,|\,\G);\g,\cA]\ =\ \exp\Big[-\tfrac{1}{2}\,\int_\Si\,
\txg_\cA(\sfd\psi\overset{\wedge}{,}\star_\g\sfd\psi)\Big]\
\Hol_{\cG_\cA,\Phi_\cA,\varphi_{n\,\cA}}(\psi\,\vert\,\G)\,.
\qqq
\exdef

In order to introduce gauge transformations in the present context,
we should recall the notion of the adjoint bundle $\,\sfP
\x_{\Ad\txG}\txG\to\Si$.\ Points in the total space $\,\sfP\x_{\Ad
\txG}\txG\,$ are classes $\,[(p,g)]\,$ of the equivalence relation
\qq\nn
\bigl(p.h^{-1},\Ad_h(g)\bigr)\sim(p,g)\,,\qquad h\in\txG
\qqq
on $\,\sfP\x\txG$,\ and so they admit point-wise multiplication
\qq\label{eq:mult-adj-bdle}
[(p,g_1)]\cdot[(p,g_2)]:=[(p,g_1\cdot g_2)]
\qqq
which endows $\,\sfP\x_{\Ad\txG}\txG\,$ with the structure of a
bundle of groups. The fibre-wise (left) action of $\,\sfP
\x_{\Ad\txG}\txG\,$ on $\,\sfP\,$
\qq\nn
(\sfP\x_{\Ad\txG}\txG)\x\sfP\to\sfP\ :\ \bigl(\bigl[(p,g_1)\bigr],p.
g_2\bigr)\mapsto p.(g_1\cdot g_2)
\qqq
induces an action
\qq\label{eq:lachi}
\la\ :\ \G(\sfP\x_{\Ad\txG}\txG)\x\sfP\to\sfP\ :\ (\chi,p)\mapsto
\la_\chi(p)
\qqq
of a section $\,\chi\,$ of the adjoint bundle $\,\sfP\x_{\Ad\txG}
\txG\,$ by an automorphism of $\,\sfP$.\ Note that
\qq\nn
\la_\chi\circ\la_{\chi'}=\la_{\chi\cdot\chi'}\,.
\qqq
The automorphisms $\,\la_\chi\,$ induce fibre maps $\,\xcMup L_\chi:
\sfP\x_\txG\xcM\to\sfP\x_\txG\xcM\,$ that descend from the
$\txG$-equivariant (with respect to the action
\eqref{eq:Gact-on-Fext}) maps
\qq\label{eq:tiLchi}
\xcMup\tilde L_\chi=(\la_\chi,\id_\xcM)\ :\ \sfP\x\xcM\to\sfP\x
\xcM\,.
\qqq
We have
\bedef
In a field theory on a space-time $\,\xcS\,$ with fields given by
sections of a bundle $\,\xcM\emb\sfP\x_\txG\xcM\to\xcS\,$ associated
to the principal $\txG$-bundle $\,\sfP\to\xcS$,\ and for
$\,\xcM\,$ a $\txG$-space with a (left) $\txG$-action given by
\Reqref{eq:Gact-M}, the \textbf{group $\,\txG_\sfP\,$ of gauge
transformations} is the set of sections
$\,\G(\sfP\x_{\Ad\txG}\txG)\,$ of the adjoint bundle
$\,\sfP\x_{\Ad\txG}\txG\to\xcS$,\ with the group operation induced
by the point-wise multiplication of \Reqref{eq:mult-adj-bdle}. A
\textbf{gauge transformation of the field} $\,\psi\in\G(\sfP\x_\txG
\xcM)\,$ by $\,\chi\in\txG_\sfP\,$ is given by
\qq\nn
(\chi,\psi)\mapsto\xcMup L_\chi\circ\psi=:{}^\chi\hspace{-2pt}
\psi\,,
\qqq
where $\,\xcMup L_\chi\,$ is the unique mapping induced by the
$\txG$-equivariant map $\,\xcMup\tilde L_\chi\,$ of
\Reqref{eq:tiLchi}. The associated \textbf{gauge transformation of
the connection} $\,\cA\,$ on $\,\sfP\,$ reads
\qq\label{eq:lachiA}
\cA\mapsto\la_{\chi^{-1}}^*\cA=:{}^\chi\hspace{-2pt}\cA\,,
\qqq
where $\,\chi^{-1}\,$ is to be understood as the point-wise group
inverse of $\,\chi$. \exdef

We are finally ready to present the fundamental result of the paper.
\bethe
The euclidean Feynman amplitude of the gauged $\si$-model from
Definition \ref{def:sigma-gauged} is invariant under arbitrary gauge
transformations, that is, for all $\,\chi\in\txG_\sfP$,\ the
identity
\qq\nn
\xcA[({}^\chi\hspace{-2pt}\psi\,|\,\G);\g,{}^\chi\hspace{-2pt}\cA]=
\xcA[(\psi\,|\,\G);\g,\cA]
\qqq
holds true. \ethe
\beroof
The gauge invariance of the `metric' term follows immediately from
the relation
\qq\nn
\Mup L_\chi^*\tilde\txg_{{}^\chi\hspace{-2pt}\cA}=\tilde\txg_\cA\,,
\qqq
cf.\ \Reqref{eq:Pext-metric}. Thus, we are left with the task of
proving the same property of the `topological' term, which is
tantamount to verifying the identity
\qq
\Hol_{\Mup L_\chi^*\cG_{{}^\chi\hspace{-2pt}\cA},\Qup L_\chi^*
\Phi_{{}^\chi\hspace{-2pt}\cA},\Tnup L_\chi^*\varphi_{n\,{}^\chi
\hspace{-2pt}\cA}}(\psi\,\vert\,\G)\equiv\Hol_{\cG_{{}^\chi
\hspace{-2pt}\cA},\Phi_{{}^\chi\hspace{-2pt}\cA},\varphi_{n\,
{}^\chi\hspace{-2pt}\cA}}\bigl({}^\chi\hspace{-2pt}\psi\,\vert\,\G
\bigr)=\Hol_{\cG_\cA,\Phi_\cA,\varphi_{n\,\cA}}(\psi\,\vert\,\G)\cr
\label{eq:ext-hol-ginv}
\qqq
for arbitrary $\,\chi,\psi\,$ and $\,\cA\,$ from the respective
domains of definition. The proof extends that of the gauge
invariance of the `topological' term for $\,\G=\emptyset$,\ given in
\Rcite{Gawedzki:2010rn}. Thus, we start by showing the existence of
a 1-isomorphism
\qq\nn
\Psi_{\cA,\chi}\ :\ \Mup L_\chi^*\cG_{{}^\chi\hspace{-2pt}\cA}
\xrightarrow{\ \cong\ }\cG_\cA\,,
\qqq
inferred, via Theorem \ref{thm:Gdesc}, and owing to the
$\txG$-equivariance of $\,\Mup\tilde L_\chi$, \,from the existence
of the corresponding 1-cell in $\,\bgrb^\nabla(\tilde M)^\txG_0$,
\qq\nn
(\tilde\Psi_{\cA,\chi},\tilde{\rm Z}_{\cA,\chi})\ :\
\bigl(\Mup\tilde L_\chi^*\tilde\cG_{{}^\chi\hspace{-2pt}
\cA},(\id_\txG\times\Mup\tilde L_\chi)^*\tilde\Upsilon_{{}^\chi
\hspace{-2pt}\cA},(\id_{\txG^2}\times\Mup\tilde L_\chi)^*\tilde
\g_{{}^\chi\hspace{-2pt}\cA}\bigr)\xrightarrow{\ \cong\ }(\tilde
\cG_\cA,\tilde\Upsilon_\cA,\tilde\g_\cA)\,.
\qqq
Upon writing out the components of the transformed gerbe
\qq\nn
&&\Mup\tilde L_\chi^*\tilde\cG_{{}^\chi\hspace{-2pt}\cA}\
\equiv\ \Mup\tilde L_\chi^*(\cG_{2^*}\ox I_{\tilde\rho_\cA})
=\cG_{2^*}\ox I_{\Mup\tilde L_\chi^*\tilde\rho_{{}^\chi
\hspace{-2pt}\cA}}=\cG_{2^*}\ox I_{\tilde\rho_{\la_\chi^*
{}^\chi\hspace{-2pt}\cA}}=\cG_{2^*}\ox I_{\tilde\rho_\cA}\equiv
\tilde\cG_\cA\,,\cr\cr
 &&\bigl(\id_\txG\times\Mup\tilde L_\chi
\bigr)^*\tilde\Upsilon_{{}^\chi\hspace{-2pt}\cA}\ \equiv\ \bigl(
\id_\txG\times\Mup\tilde L_\chi\bigr)^*(\Upsilon_{[1,3]^*}\ox\id)=
\Upsilon_{[1,3]^*}\ox\id\equiv\tilde\Upsilon_\cA\,,\cr\cr
&&\bigl(\id_{\txG^2}\times\Mup\tilde L_\chi\bigr)^*\tilde
\g_{{}^\chi\hspace{-2pt}\cA}\ \equiv\ \bigl(\id_{\txG^2}\times\Mup
\tilde L_\chi\bigr)^*(\g_{[1,2,4]^*}\ox\id)=\g_{[1,2,4]^*}\ox\id
\equiv\tilde\g_\cA\,,
\qqq
the 1-cell is found to be
\qq\label{eq:tilPsi-and-Zet-triv}
(\tilde\Psi_{\cA,\chi},\tilde{\rm Z}_{\cA,\chi})=\bigl(\id,
\la_{\Upsilon_{[1,3]^*}\ox\id}^{-1}\bullet\rho_{\Upsilon_{[1,3]^*}
\ox\id}\bigr)\,,
\qqq
and the coherence condition \eqref{eq:coh-cond-equiv-1iso} readily
follows from the naturality of the 2-isomorphisms
$\,\la_{\Upsilon_{[1,3]^*}\ox\id}\,$ and $\,\rho_{\Upsilon_{[1,3]^*}
\ox\id}\,$ in $\,\Upsilon_{[1,3]^*}\ox\id$,\ as demonstrated in
\Rxcite{Lemma 2.3.1}{Waldorf:2007phd}. The manifest
$\txG$-equivariance of the map $\,\Mup\tilde L_\chi\,$ further
implies that the gerbe $\,\Mup L_\chi^*\cG_{{}^\chi\hspace{-2pt}
\cA}\,$ is actually the descendant of $\,\Mup\tilde L_\chi^*
\tilde\cG_{{}^\chi\hspace{-2pt}\cA}$,\ and so, upon invoking Theorem
\ref{thm:Gdesc}, we infer from \Reqref{eq:tilPsi-and-Zet-triv}
\qq\nn
\Psi_{\cA,\chi}=\id_{\cG_\cA}\,,
\qqq
or, equivalently, the \emph{equality}
\qq\nn
\Mup L_\chi^*\cG_{{}^\chi\hspace{-2pt}\cA}=\cG_\cA\,.
\qqq

In the next step, we compare the 1-isomorphisms $\,\Qup L_\chi^*
\Phi_{{}^\chi\hspace{-2pt}\cA}\,$ and $\,\Phi_\cA$,\ a task which -
once more by virtue of Theorem \ref{thm:Gdesc} - reduces to
examining the respective parent 1-isomorphisms $\,(\Qup\tilde
L_\chi^*\tilde\Phi_{{}^\chi\hspace{-2pt}\cA},(\id_\txG\times\Qup
\tilde L_\chi)^*\tilde\Xi_{{}^\chi\hspace{-2pt}\cA};0)\,$ and
$\,(\tilde\Phi_\cA,\tilde\Xi_\cA;0)\,$ from $\,\bgrb^\nabla(\tilde
Q)^\txG_0$.\ An explicit calculation:
\qq\nn
&&\Qup\tilde L_\chi^*\tilde\Phi_{{}^\chi\hspace{-2pt}\cA}
\ \equiv\ \Qup\tilde L_\chi^*(\Phi_{2^*}\ox J_{\tilde
\la_{{}^\chi\hspace{-2pt}\cA}})=\Phi_{2^*}\ox J_{\Qup\tilde
L_\chi^*\tilde\la_{{}^\chi\hspace{-2pt}\cA}}=\Phi_{2^*}\ox
J_{\tilde\la_{\la_\chi^*{}^\chi\hspace{-2pt}\cA}}=\Phi_{2^*}\ox
J_{\tilde\la_\cA}\equiv\tilde\Phi_\cA\,,\cr\cr
&&\bigl(\id_\txG\times\Qup\tilde L_\chi\bigr)^*\tilde\Xi_{{}^\chi
\hspace{-2pt}\cA}\ \equiv\ \bigl(\id_\txG\times\Qup\tilde L_\chi
\bigr)^*(\Xi_{[1,3]^*}\ox\id)=\Xi_{[1,3]^*}\ox\id\equiv\tilde
\Xi_\cA
\qqq
yields the \emph{equality}
\qq\nn
\Qup L_\chi^*\Phi_{{}^\chi\hspace{-2pt}\cA}=\Phi_\cA
\qqq
in consequence of the same Theorem \ref{thm:Gdesc}.

At this stage, it remains to establish a relation between the
2-isomorphisms $\,\Tnup L_\chi^*\varphi_{n\,{}^\chi\hspace{-2pt}
\cA}\,$ and $\,\varphi_{n\,\cA}\,$ by examining the parent
2-isomorphisms $\,\Tnup\tilde L_\chi^*\tilde\varphi_{n\,
{}^\chi\hspace{-2pt}\cA}$.\ The latter rewrite as
\qq\nn
\Tnup\tilde L_\chi^*\tilde\varphi_{n\,{}^\chi\hspace{-2pt}
\cA}\equiv\Tnup\tilde L_\chi^*\varphi_{n\,2^*}=\varphi_{n\,2^*}
\equiv\tilde\varphi_{n\,\cA}\,,
\qqq
which, again by virtue of Theorem \ref{thm:Gdesc}, implies the
equality
\qq\nn
\Tnup L_\chi^*\varphi_{n\,{}^\chi\hspace{-2pt}\cA}=\varphi_{n\,\cA}
\,.
\qqq

All in all, we have
\qq\nn
\bigl(\Mup L_\chi^*\cG_{{}^\chi\hspace{-2pt}\cA},\Qup L_\chi^*
\Phi_{{}^\chi\hspace{-2pt}\cA},\Tnup L_\chi^*\varphi_{n\,{}^\chi
\hspace{-2pt}\cA}\bigr)=(\cG_\cA,\Phi_\cA,\varphi_{n\,\cA})\,,
\qqq
which immediately implies \Reqref{eq:ext-hol-ginv} and thus
concludes the proof of the theorem. \eroof

\section{The cohomological classification of $\txG$-equivariant string
backgrounds}\label{sec:class-equiv-back}

\noindent The prime r\^ole played by a $\txG$-equivariant structure on the
string background of the $\si$-model in gauging a rigid
$\txG$-symmetry and in descending the ensuing model to the quotient
of the target space by the action of $\,\txG\,$ motivate further
study of obstructions to the existence of such structures and their
classification. A particularly convenient approach to these problems
bases on a cohomological presentation of the various components of
the background (the gerbe, the bi-brane 1-isomorphism and the
inter-bi-brane 2-isomorphism) alongside the 1- and 2-morphisms of
the $\txG$-equivariant structure. Such a presentation uses local
data of the geometric structures associated with a judiciously
chosen open cover of the simplicial $\txG$-space $\,\txG\xcF$.\ \,A
proper treatment of the issue prerequires further elaboration of the
basic simplicial framework laid out in Section \ref{sub:simplicial},
which we carry out in Appendix \ref{app:nat-simpl-cov} and put to
work in the opening part of the present section. Once the formalism
is established, we proceed with a systematic reconstruction of the
cohomological description of a $\txG$-equivariant background.

\subsection{Setting up the local description}\label{sub:Deligne}

\noindent The point of departure in a local description of the geometric
objects of interest, extending the one of Section \ref{subsec:Locdesc},
is a choice of an open cover of the manifolds
that support them. In the case in hand, in which the manifolds
combine into simplicial $\txG$-spaces and definitions of the
geometric objects use the attendant face maps, it is natural to
render the choice of a cover consistent with the group action. The
proper notion is that of aligned simplicial sequences of
$\txG$-invariant refinements of open covers, as defined in Appendix
\ref{app:nat-simpl-cov}. Below, we specialise the general
construction presented therein to the setting of interest, thus
preparing the tools for later developments.
\smallskip

We shall work with sequences $\,\{\xcMup\cO^m\}_{m=0,1,\ldots}\,$ of
open covers $\,\xcMup\cO^m=\{\xcMup\cO^m_i\}_{i\in\xcMup \xcI^m}\,$
of the manifolds $\,\xcM^m\equiv\txG\xcM([m])=\txG^m\x \xcM,\
\xcM=M,Q,T_n$,\ \,chosen so that there exist index maps
\qq\nn
\phi^{(m)}_\a\ :\ \Qup\xcI^m\to\Mup\xcI^m\,,\qquad\qquad\psi_n^{k,k
+1\,(m)}\ :\ \Tnup\xcI^m\to\Qup\xcI^m
\qqq
which cover (or \v Cech-extend) the corresponding manifold maps in
the sense made precise by the conditions
\qq\nn
\iota_\a^{(m)}\bigl(\Qup\cO^m_i\bigr)\subset\Mup\cO^m_{\phi^{(m
)}_\a(i)}\,,\qquad\qquad\pi_n^{k,k+1\,(m)}\bigl(\Tnup\cO^m_i\bigr)
\subset\Qup\cO^m_{\psi_n^{k,k +1\,(m)}(i)}\,.
\qqq
A straightforward application of Proposition \ref{prop:nat-trans-J}
now yields simplicial sequences of refinements
$\,\xcMup\cU^m=\{\xcMup\cU^m_\jmath\}_{\jmath\in\xcMup \xcJ^m}\,$ of
\,the $\,\xcMup\cO^m\,$ engendered by the $\,\txG\xcM\,$ and indexed
by the corresponding incomplete simplicial sets
$\,\{\xcMup\xcJ^m\}_{m=0,1,\ldots},\ \xcMup\xcJ^m:=\xcMup J([m])$,\
where $\,\xcMup J:\D\to\Set\,$ are the contravariant functors from
Proposition \ref{prop:J-funct}. The functors are related by the
natural transformations of Proposition \ref{prop:nat-trans-J},
\qq\nn
j_\a\ :\ \Qup J\to\Mup J\,,\qquad\qquad j^{k,k+1}_n\ :\ \Tnup J\to
\Mup J\,,
\qqq
with $\,j_\a^{(m)}:=j_\a([m])\,$ and $\,j^{k,k+1\,(m)}_n:=j^{k,k+
1}_n([m])\,$ such that the covering relations
\qq\nn
\iota_\a^{(m)}\bigl(\Qup\cU^m_\jmath\bigr)\subset\Mup\cU^m_{j_\a^{(
m)}(\jmath)}\,,\qquad\qquad\pi_n^{k,k+1\,(m)}\bigl(\Tnup\cU^m_\jmath
\bigr)\subset\Qup\cU^m_{j_n^{k,k+1\,(m)}(\jmath)}
\qqq
hold true, and the diagram
\qq\nn
\alxydim{@C=2cm@R=2cm}{\Tnup\xcJ^m \ar[r]^{\Tnup J(\theta)}
\ar[d]_{j_n^{k,k+1\,(m)}} & \Tnup\xcJ^{m'} \ar[d]^{j_n^{k,k+1\,(m')}} \\
\Qup\xcJ^m \ar[r]^{\Qup J(\theta)} \ar[d]_{j_\a^{(m)}} &
\Qup\xcJ^{m'} \ar[d]^{j_\a^{(m')}} \\ \Mup\xcJ^m \ar[r]^{\Mup
J(\theta)} & \Mup \xcJ^{m'}}
\qqq
is commutative for any $\,\theta\in\D(m',m)$.
\smallskip

The foregoing discussion affords a particularly compact description
of the \v Cech cohomology of differential sheaves over the
simplicial $\txG$-manifolds $\,\txG\xcM$.\ Indeed, we may associate
with the simplicial covers $\,\xcMup\cU^m\,$ a number of \v
Cech-extended maps (in the sense of \Rcite{Runkel:2008gr}), namely,
the `horizontal' ones: $\,\xcMup\check{d}^{(m)}_k:=\bigl(\xcMup d^{(
m)}_k,\xcMup\d^{(m)}_k\bigr),\ \xcMup\d^{(m)}_k:=\xcMup J\bigl(
\theta^{(m)}_k\bigr)$,\ and the `vertical' ones:
$\,\check{\iota}_\a^{(m)}:=\bigl(\iota_\a^{(m)},j_\a^{(m)}\bigr)\,$
and $\,\check{\pi}_n^{k,k+1\,(m)}:=\bigl(\pi_n^{k,k+1\,(m)},j^{k,k+
1\,(m)}_n\bigr)$.\ These may subsequently be used to pull back
sections of differential sheaves. Let $\,\cS_{\xcM^m}^0=\unl{
2\pi\bZ}_{\xcM^m}\,$ be the sheaf of locally constant $2\pi\bZ$-valued
functions on $\,\xcM^m$,\ and $\,\cS_{\xcM^m}^{p+1}=\unl\Om^p(\xcM^m
),\ p\geq 0\,$ the sheaves of locally smooth (real) $p$-forms on the
same space. Take an arbitrary section $\,s\in\G(\cS_{\xcM^m}^p)\,$
given as a collection $\,s=(s_\jmath)_{\jmath\in\xcMup\xcJ^m}$.\ We
define its pullback along $\,\xcMup\check{d}^{(m)}_k\,$ by the
formula
\qq\nn
\xcMup\check{d}^{(m)\,*}_k s:=\bigl(\xcMup d^{(m)\, *}_k
s_{\xcMup\d^{(m)}_k(\jmath)}\bigr)_{\jmath\in\xcMup\xcJ^{m}}\,.
\qqq
Extending this prescription to arbitrary $\cS^p_{\xcM^m}$-valued \v
Cech $q$-cochains, we obtain maps
\qq\nn
\xcMup\check{d}^{(m)\,*}_k\ :\ \vC^q\bigl(\xcMup\cU^{m-1},
\cS_{\xcM^{m-1}}^p\bigr)\to\vC^q\bigl(\xcMup\cU^{m},\cS_{\xcM^m}^p
\bigr)
\qqq
satisfying the cosimplicial identities
\qq\nn
\xcMup\check{d}^{(m+1)\,*}_k\circ\xcMup\check{d}^{(m)\,*}_l=\xcMup
\check{d}^{(m+1)\,*}_l\circ\xcMup\check{d}^{(m)\,*}_{k-1}
\qqq
for all $\,0\leq l<k\,$ and $\,k=0,1,2,\ldots,m+1$. These imply the
cohomological identity
\qq\nn
\xcMup\check{\d}_{\txG,r}^{(m)}\circ\xcMup\check{\d}_{\txG,r}^{(m-1)}=0
\qqq
for the coboundary operators
\qq\nn
\xcMup\check{\d}_{\txG,r}^{(m)}:=\sum_{k=0}^{m+1}\,(-1)^{m+1-k}\,\xcMup
\check{d}^{(m+1)\,*}_k\ :\ A^r\bigl(\xcMup\cU^{m}\bigr)\to A^r\bigl(
\xcMup\cU^{m+1}\bigr)
\qqq
defined on the cochain groups
\qq\nn
A^r\bigl(\xcMup\cU^m\bigr):=\oplus_{p+q=r}\,\vC^q\bigl(\xcMup\cU^m,
\cD(4)_{\xcM^m}^p\bigr)\,,\qquad\cD(4)_{\xcM^m}^p:=
\cS^p_{\xcM^m}
\qqq
of the \v Cech-Deligne bicomplex obtained as an extension of the
Deligne complex
\qq\nn
\cD(4)_{\xcM^m}^\bullet\,:\ 0\to\cS^0_{\xcM^m}\hspace{-0.05cm}
\xrightarrow{\ \sfd^{( 0)}:=\id_{\cS^0_{\xcM^m}}\
}\cS^1_{\xcM^m}\hspace{-0.05cm} \xrightarrow{\ \sfd^{(1)}: =\sfd\
}\cS^2_{\xcM^m}\hspace{-0.05cm}\xrightarrow{\ \sfd^{(2)}:=\sfd\ }
\cS^3_{\xcM^m}\hspace{-0.05cm}\xrightarrow{\ \sfd^{(3)}:=\sfd\
}\cS^4_{\xcM^m}\,,
\qqq
cf.\ \Reqref{Delcom}, through \v Cech cohomology of the form
\qq\nn
0\to\vC^0(\xcMup\cU^m,\cS_{\xcM^m}^p)\xrightarrow{\ \vd^{(0)}\ }
\vC^1(\xcMup\cU^m,\cS_{\xcM^m}^p)\xrightarrow{\ \vd^{(1)}\ }\vC^2(
\xcMup\cU^m,\cS_{\xcM^m}^p)\xrightarrow{\ \vd^{(2)}\ }\cdots\,.
\qqq
Here, the $\,\vd^{(q)}\,$ are the standard \v Cech coboundary
operators
\qq\nn
\vd^{(q)}\ &:&\ \vC^q(\xcMup\cU^m,\cS_{\xcM^m}^p)\to\vC^{q+1}\bigl(
\xcMup\cU^m,\cS_{\xcM^m}^p\bigr)\cr\cr
&:&\ (s_{i_0 i_1 \ldots i_q})\mapsto\bigl((\vd^{(q)}s)_{i_0 i_1
\ldots i_{q+1}}):=\left(\sum_{k=0}^{q+1}\,(-1)^k\,s_{i_0 i_1
\underset{\widehat{i_k}}{\ldots} i_{q+1}}\vert_{\xcMup\cO^m_{i_0 i_1
\ldots i_{q+1}}}\right)\,,
\qqq
which we use to define the Deligne differentials $\,\xcMup D^{(m
)}_r\,$ with restrictions
\qq\nn
\xcMup D^{(m)}_r\ :\ A^r\bigl(\xcMup\cU^m\bigr)\to A^{r+1}\bigl(
\xcMup\cU^m\bigr)\,,\qquad\xcMup D^{(m)}_r\vert_{\vC^q(\xcMup
\cU^m,\cS^p_{\xcM^m})}=\sfd^{(p)}+(-1)^{p+1}\,\vd^{(q)}\,.
\qqq
The corresponding Deligne (hyper-)cohomology groups are denoted as
\qq\nn
\bH^r\bigl(\xcM^m,\cD(4)_{\xcM^m}^\bullet\bigr):=\frac{\ker\,\xcMup
D^{(m)}_r}{\im\,\xcMup D^{(m)}_{r-1}}\,.
\qqq
We have the important isomorphisms
\qq\nn
&\bH^3\bigl(\xcM^m,\cD(4)_{\xcM^m}^\bullet\bigr)\cong H^2\left(
\xcM^m,\uj\right)\,,\qquad
\bH^2\bigl(\xcM^m,\cD(4
)_{\xcM^m}^\bullet\bigr)\cong H^1\left(\xcM^m,\uj\right)\,,&\cr\cr
&\bH^1\bigl(\xcM^m,\cD(4 )_{\xcM^m}^\bullet\bigr)\cong H^0\left(
\xcM^m,\uj\right)\cong \uj^{\pi_0(\xcM_m)}\,.&
\qqq
The coboundary operators $\,\xcMup\check{\d}_{\txG,r}^{(m)}\,$ commute
with Deligne differentials,
\qq\nn
\xcMup\check{\d}_{\txG,r+1}^{(m)}\circ\xcMup D^{(m)}_r=\xcMup D^{(m+1)}_r
\circ\xcMup\check{\d}_{\txG,r}^{(m)}\,,
\qqq
and so they induce cohomology maps (denoted by the same symbols)
\qq\nn
\xcMup\check{\d}_{\txG,r}^{(m)}\ :\ \bH^r\left(\xcM^{m},\cD(4
)_{\xcM^{m}}^\bullet\right)\to\bH^r\left(\xcM^{m+1},\cD(4
)_{\xcM^{m+1}}^\bullet\right)\ :\ [x]\mapsto[\xcMup\check{\d}_{\txG,r}^{(m
)}x]
\qqq
that will be used amply in what follows.

Analogously, we define maps
\qq\nn
\check{\iota}_\a^{(m)\,*}\ :\ \vC^q\bigl(\Mup\cU^m,\cS\bigr)\to
\vC^q\bigl(\Qup\cU^m,\cS\bigr)\,,\qquad\check{\pi}_n^{k,k
+1\,(m)\,*}\ :\ \vC^q\bigl(\Qup\cU^m,\cS\bigr)\to\vC^q\bigl(\Tnup
\cU^m,\cS\bigr)\,,
\qqq
and their distinguished linear combinations
\qq\nn
\check{\D}_{Q,r}^{(m)}&:=&\check{\iota}_2^{(m)\,*}-\check{\iota}_1^{(m)
\,*} \ :\ A^r\bigl(\Mup\cU^m\bigr)\to A^r\bigl(\Qup\cU^m\bigr)\,,
\cr\cr \check{\D}_{T_n,r}^{(m)}&:=&\sum_{k=1}^n\,\vep^{k,k+1}_n\,
\check{\pi}_n^{k,k+1\,(m)\,*}\ :\ A^r\bigl(\Qup\cU^m\bigr)\to A^r
\bigl(\Tnup\cU^m\bigr)\,.
\qqq
These satisfy the identities
\qq\nn
\check{\D}_{T_n,r}^{(m)}\circ\check{\D}_{Q,r}^{(m)}=0\,,
\qqq
cf.\ \Reqref{eq:DelTnoDelQ}.

Clearly, all the above pullback operators intertwine the respective
Deligne differentials,
\qq\nn
\check{\D}_{Q,r+1}^{(m)}\circ\Mup D^{(m)}_r=\Qup D^{(m)}_r\circ
\check{\D}_{Q,r}^{(m)}\,,\qquad\check{\D}_{T_n,r+1}^{(m)}\circ\Qup
D^{(m)}_r=\Tnup D^{(m)}_r\circ\check{\D}_{T_n,r}^{(m)}\,,
\qqq
and so they give rise to cohomology maps (denoted by the same
symbols)
\qq\nn
\check{\D}_{Q,r}^{(m)}\ &:&\ \bH^r\left(\txG^m\x M,\cD(4
)_{\txG^m\x M}^\bullet\right)\to\bH^r\left(\txG^m\x Q,\cD(4
)_{\txG^m\x Q}^\bullet\right)\ :\ [x]\mapsto[\check{\D}_{Q.r}^{(m)}x]\,,
\cr\cr
\check{\D}_{T_n,r}^{(m)}\ &:&\ \bH^r\left(\txG^m\x Q,
\cD(4)_{\txG^m\x Q}^\bullet\right)\to\bH^r\left(\txG^m\x T_n,\cD(4
)_{\txG^m\x T_n}^\bullet\right)\ :\ [x]\mapsto[\check{\D}_{T_n,r}^{(m
)}x]\,.
\qqq
The mutual commutation relations between the `horizontal' and the
`vertical' pullback operators can be succinctly expressed by the
commutative diagram
\qq\nn
\alxydim{@C=2cm@R=2cm}{A^r\bigl(\Mup\cU^{m}\bigr)
\ar[r]^{\Mup\check{\d}^{(m)}_{\txG, r}} \ar[d]_{\check{\D}^{(m)}_{Q,r}} &
A^r\bigl(\Mup\cU^{m+1}\bigr) \ar[d]^{\check{\D}^{(m+1)}_{Q,r}} \\
A^r\bigl(\Qup\cU^{m}\bigr) \ar[r]^{\Qup\check{\d}^{(m )}_{\txG,r}}
\ar[d]_{\check{\D}^{(m)}_{T_n,r}} & A^r\bigl(\Qup\cU^{m+1}\bigr)
\ar[d]^{\check{\D}^{(m+1)}_{T_n,r}} \\ A^r\bigl(\Tnup\cU^{m}\bigr)
\ar[r]^{\Tnup\check{\d}^{( m)}_{\txG,r}} &
A^r\bigl(\Tnup\cU^{m+1}\bigr)}\,.
\qqq
We are now fully equipped to reconstruct a local description of a
$\txG$-equivariant string background.

As a first step, we give a local presentation of the background
$\,\Bgt=(\cM,\cB,\cJ)\,$ from Definition \ref{def:bckgrnd}. Thus,
the gerbe $\,\cG\,$ is presented by a 3-cochain
\qq\nn
\cG\xrightarrow{\rm loc.}b\in A^3 \bigl(\Mup\cU^0\bigr)
\qqq
satisfying the cohomological identity
\qq\label{eq:DG-is-H}
\Mup D^{(0)}_3 b=\ovl\txH\,,\qquad\qquad\ovl\txH:=(\txH\vert_{\Mup
\cU^0_i},0,0,0,0)\,.
\qqq
1-isomorphism classes of local data of $\,\cG\,$ with arbitrary
curvature (collecting local data of $\,\cG\,$ and of gerbes
1-isomorphic to $\,\cG$) \,are enumerated by elements of the
hyper-cohomology group $\,\bH^3\bigl(M,\cD(3)^\bullet_M\bigr)$,\
with the Deligne complex $\,\cD(3)^\bullet_M\,$ defined similarly as
$\,\cD(4)^\bullet_M\,$ but without the terminal node $\,\cD(4)^4_M$.
\ To the $\cG$-bi-brane 1-isomorphism $\,\Phi$,\ we associate a
2-cochain
\qq\nn
\Phi\xrightarrow{\rm loc.}p\in A^2\bigl(\Qup\cU^0\bigr)
\qqq
subject to the identity
\qq\label{eq:DPhi-is}
\Qup D^{(0)}_2 p=\check{\D}^{(0)}_{Q,3}b+\ovl\om\,,\qquad\ovl\om=(\om
\vert_{\Qup\cU^0_i},0,0,0)\,,
\qqq
cf.\ \Reqref{eq:Cext-bib-spell}. Finally, the inter-bi-brane
2-isomorphisms $\,\varphi_n\,$ define the respective 1-cochains
\qq\nn
\varphi_n\xrightarrow{\rm loc.}h_n\in A^1\bigl(\Tnup\cU^0\bigr)
\qqq
obeying the identities
\qq\label{eq:DFN-is}
\Tnup D^{(0)}_1 h_n=-\check{\D}^{(0)}_{T_n,2}\hspace{0.03cm}p\,,
\qqq
cf.\ \Reqref{eq:Cext-ibb-spell}.
\smallskip

Next, we write out local data of a $(\txG,\rho)$-equivariant
structure $\,(\Upsilon,\g;\kappa)\,$ on $\,\cG$.\ These consist of a
2-cochain describing the 1-isomorphism $\,\Upsilon$,
\qq\nn
\Upsilon\xrightarrow{\rm loc.}a\in A^2\bigl(\Mup\cU^1\bigr)\,,
\qqq
a 1-cochain induced by the 2-isomorphism $\,\g$,
\qq\nn
\g\xrightarrow{\rm loc.}e\in A^1\bigl(\Mup\cU^2\bigr)\,,
\qqq
and a 0-cochain $\,\nu\in A^0\left(\Mup\cU^3\right)\,$ that encodes
the coherence constraint \eqref{eq:coh-cond-equiv-1iso}. They are
required to satisfy the identities
\qq
\Mup D^{(1)}_2 a&=&-\Mup\check{\d}^{(0)}_{\txG,3}b+\ovl\rho\,,\qquad\ovl
\rho=(\rho\vert_{\Mup\cU^1_i},0,0,0)\,,\label{eq:a-def}\\
\nonumber\\ \Mup D^{(2)}_1 e&=&-\Mup\check{\d}^{(1)}_{\txG,2}\hspace{0.02cm}a\,,
\label{eq:d-vs-a} \\\nonumber\\ \Mup D^{(3)}_0\nu&=&\Mup\check{\d}^{(
2)}_{\txG,1}e\,,\label{eq:nu-vs-e}\\ \cr \Mup\check{\d}^{(3)}_{\txG,0}\nu&=&0\,.
\label{eq:d3nu}
\qqq
\smallskip

Passing to local data of a $(\txG,\la)$-equivariant structure $\,(
\Xi;k)\,$ on $\,\cB$,\ we find a 1-cochain representing the
2-isomorphism $\,\Xi\,$ as
\qq\nn
\Xi\xrightarrow{\rm loc.}s\in A^1\bigl(\Qup\cU^1\bigr)\,,
\qqq
alongside a 0-cochain $\,\theta\in A^0\left(\Qup\cU^2\right)\,$ that
accounts for the coherence condition \eqref{eq:Gequiv-bimod-coh}.
They verify the identities\footnote{We are representing various
canonical 2-isomorphisms entering the definition of $\,\g^\sharp\,$
by trivial local data.}
\qq
\Qup D^{(1)}_1 s&=&-\Qup\check{\d}^{(0)}_{\txG,2}p-\check{\D}^{(1)}_{Q,2}a+
\ovl\la\,,\qquad\ovl\la=(\la\vert_{\Qup\cU^1_i},0,0)\,,
\label{eq:s-def} \\ \cr
\Qup D^{(2)}_0\theta&=&\Qup\check{\d}^{(1)}_{\txG,1}s-\check{\D}^{(2)}_{Q,1}
e\,,\label{eq:th-vs-se}\\ \cr
\Qup\check{\d}^{(2)}_{\txG,0}\theta&=&-\check{\D}^{(3)}_{Q,0}\hspace{0.02cm}\nu\,.
\label{eq:d2theta}
\qqq

Finally, we can rephrase the defining relation
\eqref{eq:Gequiv-ibb-coh} of a $\txG$-equivariant structure on
$\,\cJ\,$ in terms of a 0-cochain $\,\varsigma\in A^0\left(\Tnup
\cU^1\right)\,$ subject to the cohomological identities
\qq
\Tnup D^{(1)}_0\varsigma&=&\Tnup\check{\d}^{(0)}_{\txG,1}h_n-\check{\D}^{(
1)}_{T_n,1}s\,,\label{eq:vas1}\\ \cr
\Tnup\check{\d}^{(1)}_{\txG,0}\hspace{0.02cm}\varsigma&=&
-\check{\D}^{(2)}_{T_n,0}\theta\,.
\label{eq:vas2}
\qqq

Since $\,\xcMup D_0^{(\bullet)}\,$ is injective, relations \eqref{eq:d3nu},
\eqref{eq:d2theta} and \eqref{eq:vas2} follow from the preceding
ones, so they are spurious.
We may now examine cohomological conditions for the existence of a
$\txG$-equivariant structure on a fixed string background.

\subsection{$(\txG,\rho)$-equivariant gerbes}

\noindent We shall first consider circumstances in which a gerbe $\,\cG\,$ can
be endowed with a $\txG$-equivariant structure.

\subsubsection{Obstructions to the existence of $\,\Upsilon\,$ and
$\,\gamma$}

\,Given a 3-cochain $\,b\in A^3 \bigl(\Mup\cU^0\bigr)\,$ on the target
space $\,M$,\ representing, as in \Reqref{eq:DG-is-H}, a gerbe
$\,\cG\,$ with a $\txG$-invariant curvature $\,\txH\,$ satisfying
the identity
\qq\nn
\Mup\check{\d}^{(0)}_{\txG,4}\ovl\txH=\Mup D^{(1)}_3\ovl\rho\,,
\qqq
cf.\ \Reqref{eq:delH-rho}, we find through direct calculation
\qq\nn
\Mup D^{(1)}_3\Mup\check{\d}^{(0)}_{\txG,3}b=\Mup\check{\d}^{(0)}_{\txG,4}\Mup
D^{(0)}_3 b=\Mup\check{\d}^{(0)}_{\txG,4}\ovl\txH=\Mup D^{(1)}_3\ovl\rho\,,
\qqq
that the 3-cochain $\,\Mup\check{\d}^{(0)}_{\txG,3}b-\ovl\rho\in A^3\big(
\Mup\cU^{(1)}\big)\,$ defines a class
\qq\nn
[\Mup\check{\d}^{(0)}_{\txG,3}b-\ovl\rho]\in H^2\big(\txG\x M,\uj\big)
\,.
\qqq
This is the class of the flat gerbe $\,\cD\cong\Mup\ell^*\cG\ox
\cG_{2*}^\vee\ox I_{-\rho}\,$ of Proposition
\ref{prop:flat-gerbe-D}, obstructing the existence of a
1-isomorphism $\,\Upsilon\ :\ \Mup \ell^*\cG\to\cG_{2*}\ox I_\rho$.\
Decomposing the relevant cohomology group with the help of the
Universal Coefficient Theorem and the K\"unneth Theorem as
\qq\nn
H^2\big(\txG\x M,\uj\big)\cong H^2
\big(\txG,\uj\big)^{\pi_0(M
)}\hspace{-0.07cm}\oplus\mor_\bZ\left(H_1(\txG)\ox H_1(M),\uj\right)\oplus
H^2\big(M,\uj\big)^{\pi_0(\txG)}\hspace{-0.07cm},
\qqq
we arrive at
\berop\cite[Prop.\,6.1]{Gawedzki:2010rn}
The class $\,[\cD]\in H^2\big(\txG\x M,\uj\big)\,$ that obstructs
the existence of a 1-isomorphism $\,\Upsilon\,$ of a $(\txG,\rho
)$-equivariant structure on $\,\cG\,$ decomposes as a direct sum
\qq\nn
[\cD]=[\cD]_{2,0}\oplus[\cD]_{1,1}\oplus[\cD]_{0,2}
\qqq
of terms from the subspaces $\,H^2\big(\txG,\uj\big)^{\pi_0(M)}
\ni[\cD]_{2,0},\ \mor_\bZ\big(H_1(\txG)\ox H_1(M),\uj\big)\ni[
\cD]_{1,1}\,$ and $\,H^2\big(M,\uj\big)^{\pi_0(\txG)}\ni[\cD]_{0
,2}$.
\eerop
\noindent As a particular case, we obtain
\becor\cite[Cor.\,6.2]{Gawedzki:2010rn}
If the connected components of $\,\txG\,$ and $\,M\,$ are
2-connected, there is no obstruction to the existence of a
1-isomorphism $\,\Upsilon$. \ecor

Assume, next, that the first obstruction has been lifted, i.e.\ that
there exists, for $\,\{\Mup\cU^n\}\,$ sufficiently fine, a 2-cochain
$\,a\in A^2\big(\Mup\cU^1\big)\,$ such that \Reqref{eq:a-def}
holds. The 2-cochain provides local data of a 1-isomorphism
$\,\Upsilon$.\ Taking into account the identity
\qq\nn
\Mup\check{\d}^{(1)}_{\txG,3}\hspace{0.03cm}\ovl\rho=0\,,
\qqq
tantamount to \Reqref{eq:delro}, we readily establish that
\qq\nn
\Mup D^{(2)}_2\Mup\check{\d}^{(1)}_{\txG,2}\hspace{0.02cm}a
=\Mup\check{\d}^{(1)}_{\txG,3}\Mup
D^{(1)}_2 a=-\Mup\check{\d}^{(1)}_{\txG,3}\Mup\check{\d}^{(0)}_{\txG,3}
\hspace{0.02cm}b+\Mup\check{\d}^{(1)}_{\txG,3}\hspace{0.03cm}\ovl\rho=0\,.
\qqq
We conclude that $\,\Mup\check{\d}^{(1)}_{\txG,2}\hspace{0.02cm}a\,$ defines
a class
\qq\nn
[\Mup\check{\d}^{(1)}_{\txG,2}\hspace{0.02cm}a]\in H^1\big(\txG^2\x M,\uj\big)
\qqq
that obstructs the existence of a 2-isomorphism $\,\g\ :\ \big(
\Mup d^{(2)\,*}_0\Upsilon\ox\id\big)\circ\Mup d^{(2)\,*}_2
\Upsilon\Rightarrow\Mup d^{(2)\,*}_1\Upsilon$.\ Given that $\,a\,$
is determined by \Reqref{eq:a-def} only up to shifts
\qq\nn
a\mapsto a+a'\,,\qquad a'\in\ker\,\Mup D^{(1)}_2\,,
\qqq
capturing the freedom of choice of (local data for) $\,\Upsilon$,
\,we infer that the above class is defined modulo elements of the
group
\qq\nn
\Mup\ceH^{1,2}:=\Mup\check{\d}^{(1)}_{\txG,2}\big(H^1\big(\txG
\x M,\uj\big)\big)\,.
\qqq
In this manner, we establish
\berop\cite[Prop.\,6.3]{Gawedzki:2010rn}
\,Let $\,a\in A^2\big(\Mup\cU^1\big)\,$ be local data of a
1-isomorphism $\,\Upsilon\ :\ \Mup \ell^*\cG\to\cG_{2*}\ox I_\rho$.\
Then, there exists a 2-isomorphism $\,\g\ :\ \big( \Mup
d^{(2)\,*}_0\Upsilon\ox\id\big)\circ\Mup d^{(2)\,*}_2
\Upsilon\Rightarrow\Mup d^{(2)\,*}_1\Upsilon\,$ iff the obstruction
class
\qq\nn
[\Mup\check{\d}^{(1)}_{\txG,2}\hspace{0.02cm}a]
+\Mup\ceH^{1,2}\in H^1\big(\txG^2\x M,\uj
\big)/\Mup\ceH^{1,2}
\qqq
is trivial.
\eerop
We have, in particular,
\becor\cite[Cor.\,6.4]{Gawedzki:2010rn}
If the connected components of $\,\txG\,$ and $\,M\,$ are
1-connected, there is no obstruction to the existence of a
2-isomorphism $\,\g$. \ecor One may get a more precise understanding
of the structure of the obstruction by decomposing the cohomology
groups,
\qq\nn
&&H^1\big(\txG\x M,\uj\big)\,\cong\,H^1\big(\txG,\uj\big)^{\pi_0(M
)}\oplus H^1\big(M,\uj\big)^{\pi_0(\txG)}\,,\cr\cr
&&H^1\big(\txG^2\x M,\uj\big)\cr
&&\quad\cong\,H^1\big(\txG,\uj
\big)^{\pi_0(\txG)\x\pi_0(M)}\oplus H^1\big(\txG,\uj
\big)^{\pi_0(\txG)\x\pi_0(M)}\oplus H^1\big(M,\uj\big)^{\pi_0(
\txG)^2}\,,
\qqq
and examining in detail the image of the former in the latter under
$\,\Mup\check{\d}^{(1)}_{\txG,2}$.\ This way, one finds,
among other things,
\becor\cite[Cor.\,6.5]{Gawedzki:2010rn}
On the $\txG$-space $\,M:=\xcG\,$ in the WZW-model context of
Definition \ref{def:WZW-target}, a suitable choice of a 1-isomorphism
$\,\Upsilon\,$ ensures the existence of a 2-isomorphism $\,\g$.
\ecor

In the last step, we presuppose that obstructions to the existence
of $\,(\Upsilon,\g)\,$ have both been lifted, so that there exists,
besides the 2-cochain $\,a\,$ introduced formerly, a 1-cochain
$\,e\in A^1\big(\Mup\cU^2\big)\,$ that satisfies relation
\Reqref{eq:d-vs-a}. We now find
\qq\nn
\Mup D^{(3)}_1\Mup\check{\d}^{(2)}_{\txG,1}e=\Mup\check{\d}^{(2)}_{\txG,2}\Mup
D^{(2)}_1 e=-\Mup\check{\d}^{(2)}_{\txG,2}\Mup\check{\d}^{(1)}_{\txG,2}
\hspace{0.02cm}a=0\,,
\qqq
whence we conclude that to $\,\Mup\check{\d}^{(2)}_{\txG,1}e\,$ there is
associated a class
\qq\nn
[\Mup\check{\d}^{(2)}_{\txG,1}e]\in H^0\big(\txG^3\x M,\uj\big)\,.
\qqq
This is the class of a locally constant map $\,d=\xcMup d_2^{(3)\,
*}\g\bullet\bigl(\bigl(\xcMup d_0^{(3)\,*}\g\ox\id\bigr)\circ\id
\bigr)\bullet\bigl(\id\circ\xcMup d_3^{(3)\,*}\g^{-1}\bigr)\bullet
\big(\id\circ\xcMup d_1^{(3)\,*}\g^{-1}\big)\,$ that captures the
last obstruction to the existence of a full-blown $(\txG,\rho
)$-equivariant structure on $\,\cG$,\ namely the obstruction to the
commutativity of diagram \eqref{diag:Gerbe-1iso-coh}. Under changes
of $\,a\,$ and $\,e\,$ admitted by their respective definitions (and
chosen such as to preserve the 2-isomorphism class of $\,\Upsilon$),
\qq\nn
e&\mapsto&e+\Mup\check{\d}^{(1)}_{\txG,1}e'
+e''\,,\qquad e'\in A^2\hspace{-0.05cm}
\big(
\Mup\cU^1\big)\,,\qquad e''\in\ker\,\Mup D^{(1)}_1\,,\cr\cr
a&\mapsto&a-\Mup D^{(1)}_1 e'\,,
\qqq
this class gets shifted as
\qq\nn
[\Mup\check{\d}^{(2)}_{\txG,1}e]\,\mapsto\,[\Mup\check{\d}^{(2)}_{\txG,1}e]
+[\Mup\check{\d}^{(2)}_{\txG,1}e'']=[\Mup\check{\d}^{(2)}_{\txG,1}e]+
\Mup\check{\d}^{(2)}_{\txG,1}[e'']\,.
\qqq
Write
\qq\nn
\Mup\ceH^{0,3}:=\Mup\check{\d}^{(2)}_{\txG,1}\big(H^0\big(
\txG^2\x M,\uj\big)\big)\subset H^0\big(\txG^3\x M,\uj\big)\,.
\qqq
The upshot of the above analysis can now be phrased as
\berop\cite[Prop.\,6.6]{Gawedzki:2010rn}
\,Let $\,e\in A^1\big(\Mup\cU^2\big)\,$ be local data of a
2-isomorphism $\,\g\ :\ \big(\Mup d^{(2)\,*}_0\Upsilon\ox\id\big)
\circ\Mup d^{(2)\,*}_2\Upsilon\Rightarrow\Mup d^{(2)\,*}_1
\Upsilon\,$ for an appropriate choice of a \emph{fixed}
1-isomorphism $\,\Upsilon\ :\ \Mup\ell^*\cG\to\cG_{2^*}\ox I_\rho$.\
The 2-isomorphism can be chosen so that it satisfies the coherence
condition of \Reqref{eq:Gerbe-1iso-coh} iff the obstruction class
\qq\nn
[\Mup\check{\d}^{(2)}_{\txG,1}e]+\Mup\ceH^{0,3}\in H^0\big(\txG^3\x M,\uj
\big)/\Mup\ceH^{0,3}
\qqq
is trivial.
\eerop

It is convenient to rephrase the statement of the last proposition,
along the lines of \Rcite{Gawedzki:2010rn}, in terms of the
cohomology theory of the group $\,\pi_0(\txG)\,$ with values in the
$\pi_0(\txG)$-module $\,\uj^{\pi_0(M)}\cong H^0\big(M,\uj\big)$,\
the module structure being induced from the action of $\,\txG\,$ on
$\,M$,
\qq\nn
\pi_0(\txG)\x\uj^{\pi_0(M)}\to\uj^{\pi_0(M)}\ :\ \left([g],u\right)
\mapsto[g].u\,,\qquad[g].u\left([m]\right):=u\left([g^{-1}.m]
\right)\,.
\qqq
The rephrasing is possible owing to the existence of a natural
identification between classes in $\,H^0\big(\txG^p\x M,\uj \big)$,\
viewed as locally constant $\,\uj$-valued functions on $\,\txG^p\x
M$,\ and $p$-chains from $\,C^p\big(\pi_0(\txG),\uj^{\pi_0(M
)}\big)$.\ The identification is defined by the formula
\qq\nn
&H^0\big(\txG^p\x M,\uj\big)\ \ni\ f^p
\ \ \mathop{\longmapsto}\limits^{\Mup\z_p}\ \ u^p\ \in\
C^p\big(\pi_0(\txG
),\uj^{\pi_0(M)}\big)&\\ \label{eq:zetap-id}\\
&u^p_{[g_1],[g_2],\ldots,[g_p]}([m]):=f^p\big(g_p^{-1},g_{p-
1}^{-1},\ldots,g_1^{-1},m\big)\,.&\nonumber
\qqq
The cohomological maps $\,\Mup\check{\d}^{(p)}_{\txG,1}\,$ are
transformed into the coboundary operators of the group cohomology
under this identification,
\qq\nn
\Mup\z_{p+1}\circ\Mup\check{\d}^{(p)}_{\txG,1}=\Mup\d^{(p
)}_{\pi_0(\txG)}\circ\Mup\z_p\,.
\qqq
Since
\qq\nn
\Mup\check{\d}^{(3)}_{\txG,1}[\Mup\check{\d}^{(2)}_{\txG,1}e]=[
\Mup\check{\d}^{(3)}_{\txG,1}\Mup\check{\d}^{(2)}_{\txG,1}e]=0\,,
\qqq
we obtain, as a result of the above identification,
\becor\cite[Cor.\,6.7]{Gawedzki:2010rn}
The image $\,\Mup\z_3\big([\Mup\check{\d}^{(2)}_{\txG,1}e]\big)=:\Mup
u^3\,$ of the obstruction class $\,[\Mup\check{\d}^{(2)}_{\txG,1}e]\,$ for
a \emph{coherent}\footnote{That is one for which the coherence
condition of \Reqref{eq:Gerbe-1iso-coh} is satisfied.} 2-isomorphism
$\,\g\,$ under the identification of \Reqref{eq:zetap-id} is a
$\uj^{\pi_0(M)}$-valued 3-cocycle of the group $\,\pi_0(\txG)$,\ and
that of the obstruction coset $\,[\Mup\check{\d}^{(2)}_{\txG,1}e]+\Mup
\ceH^{0,3}\,$ is the cohomology class $\,[\Mup u^3]\in H^3\big(
\pi_0(\txG),\uj^{\pi_0(M)}\big)$. \ecor The advantage of working
with group cohomology is that it is amenable to explicit
computations. There are several general results about it that
can be employed in the context in hand. E.g., for the
trivial group $\,\bd1$,\ we have
\qq\label{eq:GHp-triv}
H^p\big(\bd1,\uj^{\pi_0(M)}\big)=\bd1\,,\qquad p\geq 1\,.
\qqq
This gives us
\becor\cite[Cor.\,6.8]{Gawedzki:2010rn}
If the symmetry group $\,\txG\,$ is connected and the 2-isomorphism
$\,\g\,$ exists, then it can always be chosen in a coherent form.
\ecor

In the previous paragraph, we restricted our choice of redefinitions
of local data of $\,\Upsilon\,$ and $\,\g\,$ to those that leave the
2-isomorphism class of $\,\Upsilon\,$ unaltered. From the point of
view of obstruction analysis, it is well justified to allow more
general redefinitions with view to further reducing the obstruction
cohomology group and thus rendering it possible to put \emph{some}
$(\txG,\rho)$-equivariant structure on a given gerbe. Accordingly,
let us consider (in addition to the above) transformations of the
type
\qq\nn
a\mapsto a+a'\,,\qquad a'\in\ker\,\Mup D^{(1)}_2\,,\qquad\qquad e
\mapsto e+e'\,,\qquad e'\in A^1\big(\Mup\cU^2\big)\,,
\qqq
the two cochains being related through
\qq\nn
\Mup\check{\d}^{(1)}_{\txG,2}\hspace{0.02cm}a'=-\Mup D^{(2)}_1 e'\,.
\qqq
The last constraint ensures that Eqs.\,\eqref{eq:a-def} and
\eqref{eq:d-vs-a} still hold true. It fixes $\,e'\,$ only up to
shifts
\qq\nn
e'\mapsto e'+e''\,,\qquad\qquad e''\in\ker\,\Mup D^{(2)}_1\,,
\qqq
and so we recover a connecting (Bockstein-type) homomorphism
\qq\label{eq:Bokh1}
[a']\,\mathop{\longmapsto}\limits^{\Mup\xcB^{(1)}_2}
\,[\Mup\check{\d}^{(2)}_{\txG,1}e']+\Mup\ceH^{0,3}\,.
\qqq
from the kernel of $\,\Mup\check{\d}^{(1)}_{\txG,2}\,$
acting on $\,H^1\big(\txG\x M,\uj\big)\,\,$ to
$\,\,H^0\big(\txG^3\x M,\uj\big)/\Mup\ceH^{0,3}$.
\berop
The component of the class
$\,[\Mup\check{\d}^{(2)}_{\txG,1}e]+\Mup\ceH^{0 ,3}\in
H^0\big(\txG^3\x M,\uj\big)/\Mup\ceH^{0,3}$,\ obstructing the
existence of a \emph{coherent} 2-isomorphism $\,\g$,\ that lies in
the image of the connecting homomorphism $\,\Mup\xcB^{(1)}_2\,$ of
\Reqref{eq:Bokh1} can be removed by a suitable choice of a
1-isomorphism $\,\Mup\ell^*\cG\to \cG_{2*}\ox I_\rho$.
\eerop
\vskip 0.1cm

We may now proceed to classify inequivalent $(\txG,\rho)$-structures
on a given gerbe, in the sense of Definition \ref{def:equiv-1iso}.

\subsubsection{Classification}

\,We are interested here in enumerating all inequivalent $(\txG,\rho
)$-equi\-vari\-ant structures on a gerbe $\,\cG\,$ over $\,M\,$ with
local data given by some $\,b\in A^3\big(\Mup\cU^0\big)$.\ To this
end, we should first write out cohomological relations satisfied by
local data of a $\txG$-equivariant isomorphism $\,(\Psi ,\eta)\,$
between two such structures $\,(\cG,\Upsilon_\b,\g_\b;\rho ),\
\b=1,2$.\ These data consist of a 2-cochain $\,f\in A^2\big(
\Mup\cU^0\big)\,$ for a 1-isomorphism $\,\Psi\ :\ \cG\to\cG$, \,a
1-cochain $\,l\in A^1\big(\Mup\cU^1\big)\,$ for a 2-isomorphism
$\,\eta\ :\ \big(\Mup d^{(1)\,*}_0\Psi\ox\id\big)\circ\Upsilon_1
\Rightarrow\Upsilon_2\circ\Mup d^{(1)\,*}_1\Psi$, \,and a 0-cochain
$\,\z\in A^0\big(\Mup\cU^2\big)\,$ that expresses the coherence of
the latter. Denote local data of $\,\Upsilon_\b\,$ by $\,a_\b\,$ and
those of $\,\g_\b\,$ by $\,e_\b\,$ and $\,\nu_\b\,$ (the coherence
0-cochain). They obey
\qq\nn
\Mup D^{(1)}_2 a_\b=-\Mup\check{\d}^{(0)}_{\txG,3}\hspace{0.02cm}b
+\ovl\rho\,,\ \quad
\Mup D^{(2)}_1 e_\b=-\Mup\check{\d}^{(1)}_{\txG,2}\hspace{0.02cm}a_\b\,,
\ \quad\Mup D^{(3)}_0\nu_\b=\Mup\check{\d}^{(2)}_{\txG,1}e_\b\,,
\quad\ \Mup\check{\d}^{(3)}_{\txG,0}\hspace{0.02cm}\nu_\b=0\,.
\qqq
Therefore, the respective differences
\qq\nn
a_{2,1}:=a_2-a_1\,,\qquad\qquad e_{2,1}:=e_2-e_1\,,\qquad
\nu_{2,1}:=\nu_2-\nu_1
\qqq
automatically satisfy the identities
\qq\nn
\Mup D^{(1)}_2 a_{2,1}=0\,,\ \quad\Mup D^{(2)}_1 e_{2,1}=-
\Mup\check{\d}^{(1)}_{\txG,2}\hspace{0.02cm}a_{2,1}\,,\ \quad\Mup D^{(3)}_0\nu_{2,1}
=\Mup\check{\d}^{(2)}_{\txG,1}e_{2,1}\,,\ \quad\Mup\check{\d}^{(3)}_{\txG,0}
\hspace{0.02cm}\nu_{2,1}=0\,.
\qqq
From Definition \ref{def:equiv-1iso}, we infer that the relations
\qq\nn
&a_{2,1}=\Mup\check{\d}^{(0)}_{\txG,2}f+\Mup D^{(1)}_1 l\,,\qquad\Mup
D^{(0)}_2 f=0\,,&\cr\cr
&e_{2,1}=-\Mup\check{\d}^{(1)}_{\txG,1}l+\Mup D^{(2)}_0\z\,,\qquad
\nu_{2,1}=\Mup\check{\d}^{(2)}_{\txG,0}\z&
\qqq
are tantamount to the existence of a $\txG$-equivariant isomorphism
between the two $(\txG,\rho)$-equivariant structures. Define a
bicomplex $\,\Mup\cK\,$:
\qq\label{diag:J-bicompl}
\alxydim{@C=1.5cm@R=1.cm}{A^0\big(\Mup\cU^0\big)
\ar[d]^{\Mup\check{\d}^{(0)}_{\txG,0}} \ar[r]^-{\Mup D^{(0)}_0} &
A^1\big(\Mup\cU^0\big) \ar[d]^{\Mup\check{\d}^{(0)}_{\txG,1}}
\ar[r]^-{\Mup D^{(0)}_1} &
\ker\,\Mup D^{(0)}_2 \ar[d]^{\Mup\check{\d}^{(0)}_{\txG,2}} \\
A^0\big(\Mup\cU^1\big) \ar[d]^{\Mup\check{\d}^{(1)}_{\txG,0}}
\ar[r]^-{\Mup D^{(1)}_0} & A^1\big(\Mup\cU^1\big)
\ar[d]^{\Mup\check{\d}^{(1)}_{\txG,1}} \ar[r]^-{\Mup D^{(1)}_1} &
*+++[o][F]{\ker\,\Mup D^{(1)}_2\ni a_{2,1}}
\ar[d]^{\Mup\check{\d}^{(1)}_{\txG,2}} \\ A^0\big(\Mup\cU^2\big)
\ar[d]^{\Mup\check{\d}^{(2)}_{\txG,0}} \ar[r]^-{\Mup D^{(2)}_0} &
*+++[o][F]{A^1\big(\Mup\cU^2\big)\ni e_{2,1}}
\ar[d]^{\Mup\check{\d}^{(2)}_{\txG,1}} \ar[r]^-{\Mup D^{(2)}_1} &
\ker\,\Mup D^{(2)}_2 \ar[d]^{\Mup\check{\d}^{(2)}_{\txG,2}} \\
*+++[o][F]{A^0\big(\Mup\cU^3\big)\ni\nu_{2,1}}
\ar[d]^{\Mup\check{\d}^{(3)}_{\txG,0}} \ar[r]^-{\Mup D^{(3)}_0} &
A^1\big(\Mup\cU^3\big) \ar[d]^{\Mup\check{\d}^{(3)}_{\txG,1}}
\ar[r]^-{\Mup D^{(3)}_1} & \ker\,\Mup D^{(3)}_2
\ar[d]^{\Mup\check{\d}^{(3)}_{\txG,2}} \\ \vdots & \vdots & \vdots\,.}
\qqq
Clearly, its cohomology can be used to classify inequivalent
$(\txG,\rho)$-structures on a given gerbe $\,\cG$.\ We readily
establish
\berop\cite[Prop.\,6.10]{Gawedzki:2010rn}
The set of equivalence classes of $(\txG,\rho)$-equivariant
structures on gerbe $\,\cG\,$ over $\,M\,$ is a torsor of the third
hypercohomology group $\,\bH^3\big(\Mup\cK\big)\,$ of the bicomplex
$\,\Mup \cK\,$ of \eqref{diag:J-bicompl}.
\eerop
In order to obtain a presentation of the classifying
cohomology for $(\txG,\rho)$-equivariant structures, we may
further map $\,\bH^3\big(\Mup\cK\big)\,$ to the coset
$\,H^1\big(\txG\x M,\uj\big)/\Mup\ceH^{1,1}\,$ with respect to the
subgroup
\qq\nn
\Mup\ceH^{1,1}:=\Mup\check{\d}^{(0)}_{\txG,2}\left(H^1\big(M,\uj
\big)\right)\,.
\qqq
Properties of this map
\qq\nn
\kappa\ :\ \bH^3\big(\Mup\cK\big)\to H^1\big(\txG\x M,\uj\big)
/\Mup\ceH^{1,1}\ :\ [(a,e,\nu)]\mapsto[a]+\ceH^{1,1}\,,
\qqq
whose definition is based on the implication
\qq\nn
a\mapsto a+\Mup\check{\d}^{(0)}_{\txG,2}f+\Mup D^{(1)}_1 l\qquad\Rightarrow
\qquad[a]\mapsto[a]+\Mup\check{\d}^{(0)}_{\txG,2}[f]\,,
\qqq
can be obtained through a standard diagram-chasing analysis,
performed in \Rcite{Gawedzki:2010rn}, to the following effect.
\berop\cite[Lemmas 6.11 \& 6.12]{Gawedzki:2010rn}
\,The coset $\,[a]+\ceH^{1,1}\,$ belongs to the image of $\,\kappa\,$
iff
\bit
\item[(i)] $\Mup\check{\d}^{(1)}_{\txG,2}[a]=0$;
\item[(ii)] the image in $\,H^3\big(\pi_0(\txG),\uj^{\pi_0(M)}
\big)\,$ under identification $\,\Mup\z_3\,$ of \Reqref{eq:zetap-id}
of the class $\,\Mup\xcB^{(1)}_2([a])$,\ assigned to $\,[a]\,$ by
the connecting homomorphism $\,\Mup\xcB^{(1 )}_2\,$ of
\Reqref{eq:Bokh1}, is trivial.
\eit
Furthermore, the kernel of $\,\kappa\,$ may be naturally identified
with the group $\,H^2\big(\pi_0(\txG),\uj^{\pi_0(M)}\big)$.
\eerop

As a simple consequence of the above, we obtain
\becor\cite[Corollaries 6.12 \& 6.13]{Gawedzki:2010rn}
If the connected components of $\,\txG\,$ and $\,M\,$ are simply
connected, then
\qq\nn
\bH^3\big(\Mup\cK\big)\cong H^2\big(\pi_0(\txG),\uj^{\pi_0(M)}
\big)\,.
\qqq
If $\,\txG\,$ and $\,M\,$ are connected, then
\qq\nn
\bH^3\big(\Mup\cK\big)\cong H^1\big(\txG,\uj\big)/(r_m^*)_*
H^1\big(M,\uj\big)\cong Z_M^*\,,
\qqq
where $\,Z_M^*\,$ is the group of characters of the kernel $\,Z_M\,$
of the homomorphism $\,H_1(\txG)\to H_1(M)\,$ induced by the map
$\,r_m\ :\ \txG\to M\ :\ g\mapsto g.m$. \ecor Finally, in the WZW
setting, we find
\becor\cite[Cor.\,6.14]{Gawedzki:2010rn}
On the $\txG$-space $\,M:=\xcG\,$ in the WZW-model
context of Definition \ref{def:WZW-target},
\qq\nn
\bH^3\big({}^{\tx{\tiny $\xcG$}}\hspace{-2pt}\cK\big)\cong
H^1\big(\txG,\uj\big)\,.
\qqq
\ecor
\vskip 0.1cm

\subsection{$(\txG,\la)$-equivariant bi-branes}

\noindent Having recalled the cohomological description of $(\txG,\rho
)$-equivariant gerbes, as formulated in \Rcite{Gawedzki:2010rn}, we
shall discuss bi-branes in a similar vein.

\subsubsection{Obstructions to the existence of $\,\Xi$}

\,Take a 2-cochain $\,p\in A^2\left(\Qup\cU^0\right)\,$ on the
world-volume $\,Q\,$ of a $\cG$-bi-brane $\,\cB=(Q,\iota_1,\iota_2,
\om,\Phi)$,\ giving local data of a 1-isomorphism $\,\Phi\,$ and
hence obeying \Reqref{eq:DPhi-is}. By virtue of Eqs.\,\eqref{eq:dla}
and \eqref{eq:della}, the curvature of the bi-brane is subject to
the identity
\qq\label{eq:preced1}
\Qup\check{\d}^{(0)}_{\txG,3}\hspace{0.03cm}\ovl\om
=-\check{\D}^{(1)}_{Q,3}\hspace{0.03cm}
\ovl\rho+\Qup D^{(1)}_2\ovl\la\,,
\qqq
where
\qq\label{eq:preced2}
\Qup\check{\d}^{(1)}_{\txG,2}\hspace{0.03cm}\ovl\la=0\,.
\qqq
We now obtain, using relations \eqref{eq:DPhi-is} and
\eqref{eq:a-def},
\qq\nn
\Qup D^{(1)}_2\,\Qup\check{\d}^{(0)}_{\txG,2}\hspace{0.03cm}p
=\Qup\check{\d}^{(0)}_{\txG,3}\,\Qup
D^{(0)}_2 p=\check{\D}^{(1)}_{Q,3}\,\Mup\check{\d}^{(0)}_{\txG,3}\hspace{0.02cm}b
+\Qup\check{\d}^{(0)}_{\txG,3}\hspace{0.02cm}\ovl\om
=-\Qup D^{(1)}_2\check{\D}^{(1)}_{Q,2}
\hspace{0.02cm}a+\Qup D^{(1)}_2
\ovl\la\,,
\qqq
from which we extract the first obstruction to the existence of a
$(\txG,\la)$-equivariant structure, namely the class
\qq\label{eq:obcl}
[D]:=[\Qup\check{\d}^{(0)}_{\txG,2}\hspace{0.02cm}p
+\check{\D}^{(1)}_{Q,2}\hspace{0.02cm}a-\ovl\la]\in H^1
\big(\txG\x Q,\uj\big)
\qqq
of the flat line bundle $\,D\to\txG\x Q\,$ of
\Reqref{eq:triv-bund-corr}, with $$D\cong\left(\Qup \ell^*\Phi\ox
J_{-\la}\right)\circ\left(\iota_1^{(1)\,*}\Upsilon^{-
1}\ox\id\right)\circ\left(\Phi_{2^*}^{-1}\ox\id\right)\circ
\iota_2^{(1)\,*}\Upsilon\,.$$ This class obstructs the existence of
a 1-isomorphism $\,\Xi\,$ of \Reqref{Xi} with local data $\,s\in A^1
\big(\Qup\cU^1\big)\,$ such that relation \eqref{eq:s-def} holds.
Note that due to Eqs.\,\eqref{eq:preced2} and \eqref{eq:d-vs-a},
\qq
\Qup\check{\d}^{(1)}_{\txG,2}\big(\Qup\check{\d}^{(0)}_{\txG,2}\hspace{0.02cm}p
+\check{\D}^{(1)}_{Q,2}\hspace{0.02cm}a-\ovl\la\big)
=\check{\D}^{(2)}_{Q,2}\,\Mup\check{\d}^{(1)}_{\txG,2}\hspace{0.02cm}a
=-\Qup D^{(2)}_1\,\check{\D}^{(2)}_{Q,1}\hspace{0.01cm}e\,,
\qqq
hence the isomorphism class $\,[D]\,$ is constrained by the relation
\qq\label{eq:constrD}
H^1\big(\txG^2\x Q,\uj\big)\ \ni\ \Qup\check{\d}^{(1)}_{\txG,2}[D]\,=\,0\,.
\qqq
Let us abbreviate
\qq
{\rm
ker}\left(\xcMup\check{\d}^{(1)}_{\txG,2}:H^1\big(\txG\x\xcM,\uj\big)
\rightarrow H^1\big(\txG^2\x\xcM,\uj\big)\right)\
=:\,\xcMup\cP^{1,1}\,.
\qqq
\berop
Let $\,a\in A^2\big(\Mup\cU^1\big)\,$ be local data of a
1-isomorphism $\,\Upsilon\ :\ \Mup \ell^*\cG\to\cG_{2*}\ox I_\rho\,$
of a $\txG$-equivariant gerbe $\,(\cG,\Upsilon,\g;\rho)$\,\ and let
$\,p\in A^2\big(\Qup\cU^0 \big)\,$ be local data of a 1-isomorphism
$\,\Phi\ :\ \iota_1^*\cG \to\iota_2^*\cG\ox I_\om\,$ of
$\cG$-bi-brane $\,(Q,\iota_1, \iota_2,\om,\Phi)$.\ Then, there
exists a 2-isomorphism $\,\Xi\ :\
\Qup\ell^*\Phi\xLongrightarrow{\cong}\bigl(\bigl(\iota_2^{(1)\,*}
\Upsilon^{-1}\ox\id\bigr)\circ\bigl(\Phi_{2^*}\ox\id\bigr)\circ
\iota_1^{(1)\,*}\Upsilon\bigr)\ox J_\la\,$ iff the obstruction class
\qq\nn
[\Qup\check{\d}^{(0)}_{\txG,2}\hspace{0.02cm}p+\check{\D}^{(1)}_{Q,2}
\hspace{0.02cm}a-\ovl\la]\ \in\ \Qup\CP^{1,1}
\qqq
is trivial.
\eerop
The cohomology group $\,H^1\big(\txG\x\xcM,\uj\big)\,$ decomposes naturally as
\qq\label{eq:Kun-H1}
H^1\big(\txG\x\xcM,\uj\big)\cong H^1\big(\txG,\uj\big)^{\pi_0(\xcM
)}\oplus H^1\big(\xcM,\uj\big)^{\pi_0(\txG)}\,.
\qqq
Taking $\,\xcM=Q$, \, we obtain
\becor
If the connected components of $\,\txG\,$ and $\,Q\,$ are
1-connected, there is no obstruction to the existence of a
2-isomorphism $\,\Xi$. \ecor \noindent If $\,\txG\,$ is connected
then it is easy to see (cf.\ \cite[Eq.\,(6.37)]{Gawedzki:2010rn})
that
\qq
\xcMup\cP^{1,1}\ =\ H^1\big(\txG,\uj\big)^{\pi_0(\xcM)}
\qqq
in terms of the decomposition \eqref{eq:Kun-H1}. Taking $\,\xcM=Q\,$
again, \, we infer that
\becor\label{cor:obstrXi}
If $\,\txG\,$ is connected then the obstruction class $\,[D]\,$ to the
existence of 2-isomorphism $\,\Xi$ lies
in $\,\,H^1\big(\txG,\uj\big)^{\pi_0(Q)}\subset\,H^1\big(\txG\x Q,\uj\big)$.
\ecor

It is natural to ask to what extent the obstruction just described
can be removed by a change of the equivalence class of the original
$(\txG,\rho)$-equivariant structure on $\,\cG$,\ resp.\ by a change
of a representative within this class. In order to answer this
question, we study the dependence of $\,[D]\,$ on the choice of the
class from $\,\bH^3\big(\Mup\cK\big)$ of (local data of) a
$(\txG,\rho)$-equivariant structure.\ Note that under a transformation
\qq\nonumber
&a\mapsto a+a'\,,\qquad\qquad e\mapsto e+e'\,,\qquad\qquad\nu
\mapsto\nu+\nu'&\\\label{eq:aenu-trafo}\\
&\Mup D^{(1)}_2 a'=0\,,\ \quad\Mup D^{(2)}_1 e'=-\Mup\check{\d}^{(
1)}_{\txG,2}\hspace{0.02cm}a'\,,\
\quad\Mup D^{(3)}_0\nu'=\Mup\check{\d}^{(2)}_{\txG,1}
e'\,,\ \quad\Mup\check{\d}^{(3)}_{\txG,0}\hspace{0.03cm}\nu'=0&\nonumber
\qqq
the obstruction class shifts by
\qq\nn
[D]\mapsto[D]+\check{\D}^{(1)}_{Q,2}[a']\,.
\qqq
The constraints imposed upon the triple $\,(a',e',\nu')\,$ identify
it as a 3-cocycle in the hypercohomology of the bicomplex
$\,\Mup\cK\,$ of \Reqref{diag:J-bicompl}. Denote by $\,\cZ^3\left(
\Mup\cK\right)\,$  the abelian group of all such 3-cocycles, and by
$\,\Mup\pi^{1,1}\,$ the canonical projection
\qq\label{eq:obv-cohom-proj}
\cZ^3\big(\Mup\cK\big)\,\ni\,(a',e',\nu')
\ \,\mathop{\longmapsto}\limits^{\Mup\pi^{1,1}}\,\ [a']\,\in\,
H^1\big(\txG\x M,\uj\big)
\qqq
with the image
\qq\label{eq:P11-def}
\Mup\cK^{1,1}:=\Mup\pi^{1,1}\left(\cZ^3\big(\Mup\cK\big)\right)\,\subset\,
\Mup\cP^{1,1}\,.
\qqq
We thus arrive at
\becor
The component of the class $\,[D]\in\Qup\cP^{1,1}$,\ obstructing the
existence of 2-isomorphism $\,\Xi$,\ that lies in the image of the
group $\,\Mup\cK^{1,1}\,$ of \Reqref{eq:P11-def} under the induced
cohomology map $\ \check{\D}^{(1)}_{Q,2}:H^1\big(\txG\x
M,\uj\big)\to H^1\big(\txG\x Q,\uj\big)\ $ can be removed by a
suitable choice of a $(\txG,\rho)$-equivariant structure on the
gerbe $\,\cG$.\ecor \noindent If $\,\txG\,$ is connected and $\,M\,$
is connected, then it is easy to see that
$\,\check{\D}_{Q,2}^{(1)}\big(\Mup\cP^{1,1}\big)=\{0\}\,$ (since
both $\,\check{\iota}_1^{(1)\,*}\,$ and
$\,\check{\iota}_1^{(2)\,*}\,$ when restricted to
$\,H^1\big(\txG,\uj\big)\subset\,H^1\big(\txG\x M,\uj\big) \,$ map
as identities into each of the factors in
$\,H^1\big(\txG,\uj\big)^{\pi_0(Q)}\subset H^1\big(\txG\x
Q,\uj\big)$.
\becor
\noindent If $\,\txG\,$ and $\,M\,$ are connected, then
$\,\check{\D}_{Q,2}^{(1)}\big(\Mup\cK^{1,1}\big)=\{0\}$. \ecor
\noindent This applies, in particular, to the case of maximally
symmetric non-boundary WZW $\cGk$-bi-brane from Definition
\ref{def:WZW-non-bdry-bib} with the (connected) world-volume given
by the $\txG$-space $\,Q_\la=\xcG\x\xcC_\la\,$ where
$\,\Mup\CK^{1,1} =\Mup\CP^{1,1}=H^1\big(\txG,\uj\big)$. \,On the
other hand, for the maximally symmetric boundary WZW $\cGk$-bi-brane
from Definition \ref{def:WZW-bdry-bib} with the world-volume
$\,\xcC_\la\,$ and $\,M=\xcG\sqcup\{ \bullet\}$,
$\,\Mup\CK^{1,1}=\Mup\CP^{1,1}=H^1\big(\txG, \uj\big)^2\,$ and it is
mapped by $\,\check{\D}^{(1)}_{Q,2}\,$ onto $\,\Qup\cP^{1,1}=
H^1\big(\txG,\uj\big)$, \,permitting to remove the obstruction to
the existence of a 2-isomorphism $\,\Xi\,$ in this case. If,
however, we take a disjoint union of two maximally symmetric
boundary WZW $\cGk$-bi-branes with
$\,Q=\xcC_{\la_1}\sqcup\,\xcC_{\la_2}\,$ then
$\,\check{\D}^{(1)}_{Q,2} \big(\Mup\cK^{1,1}\big)\,$ becomes the
diagonal subgroup in $\,\Qup\cP^{1,1}=H^1\big(\txG,\uj\big)^2\,$ and
one can remove only the diagonal component of the anomaly
obstructing the existence of a 2-isomorphism $\,\Xi$. \vskip 0.2cm

Whenever the first obstruction vanishes, that is whenever a
1-cochain $\,s\in A^1\big(\Qup\cU^1\big)\,$ can be found (for
$\,\{\Qup\cU^1\}\,$ sufficiently fine) for which identity
\eqref{eq:s-def} is satisfied, we may enquire whether the
2-isomorphism $\,\Xi\,$ with local data given by $\,s\,$ obeys
constraints \eqref{eq:Gequiv-bimod-coh}. In general, we only find
\qq\nn
\ \,\Qup D^{(2)}_1\Qup\check{\d}^{(1)}_{\txG,1}s=\Qup\check{\d}^{(1)}_{\txG,2}\Qup
D^{(1)}_1 s=-\Qup\check{\d}^{(1)}_{\txG,2}\,\check{\D}^{(1)}_{Q,2}a=-
\check{\D}^{(2)}_{Q,2}\Mup\check{\d}^{(1)}_{\txG,2}a=\check{\D}^{(2)}_{Q,2}\Mup
D^{(2)}_1 e=\Qup D^{(2)}_1\check{\D}^{(2)}_{Q,1}e\,,
\qqq
and so we end up with a class
\qq\label{eq:S}
[\si]:=[\Qup\check{\d}^{(1)}_{\txG,1}s-\check{\D}^{(2)}_{Q,1}e]\in H^0\big(
\txG^2\x Q,\uj\big)
\qqq
that may be nontrivial. Note that
\qq\label{eq:QdS}
\Qup\check{\d}^{(2)}_{\txG,1}[\si]=[-\Qup\check{\d}^{(2)}_{\txG,1}
\check{\D}^{(2)}_{Q,1}e]=-[\check{\D}^{(3)}_{Q,1}\Mup\check{\d}^{(2)}_{\txG,1}e]
=-[\Qup D_0^{(3)}\check{\D}^{(3)}_{Q,0}\hspace{0.02cm}\nu]=0\,,
\qqq
whence
\qq
[\si]\ \in\ {\rm ker}\left(\Qup\check{\d}^{(2)}_{\txG,1}:H^0\big(
\txG^2\x Q,\uj\big)\rightarrow H^0\big(\txG^3\x Q,\uj\big)\right)\ =:\
\Qup\cP^{0,2}\,.
\qqq
The actual obstruction to the commutativity of diagram
\eqref{diag:Gequiv-bimod-coh} belongs to a coset of the above group
that can be identified by exploiting, similarly as for the gerbe,
the intrinsic ambiguity in the definition of the 1-cochain $\,s$.\
First, the latter may be shifted as
\qq\label{eq:ss-trafo}
s\mapsto s+s''\,,\qquad\qquad s''\in\ker\,\Qup D^{(1)}_1\,,
\qqq
whereby $\,[\si]\,$ undergoes a transformation
$\,[\si]\mapsto[\si]+\Qup\check{\d}^{(1)}_{\txG,1}[s'']$. \,Write
\qq\nn
\Qup\ceH^{0,2}:=\Qup\check{\d}^{(1)}_{\txG,1}\left(H^0\big(\txG
\x Q,\uj\big)\right).
\qqq
The upshot of our analysis is the following
\berop
Let $\,e\in A^1\big(\Mup\cU^2\big)\,$ be local data of a
2-isomorphism $\,\g\,$ of a $(\txG,\rho)$-equi\-vari\-ant structure
on a gerbe $\,\cG\,$ over $\,M$,\ and let $\,s\in A^1\big(\Qup\cU^1
\big)\,$ be local data of a 2-isomorphism $\,\Xi\ :\ \Qup\ell^*
\Phi\xLongrightarrow{\cong}\bigl(\bigl(\iota_2^{(1)\,*}\Upsilon^{-
1}\ox\id\bigr)\circ\bigl(\Phi_{2^*}\ox\id\bigr)\circ\iota_1^{(1)\,
*}\Upsilon\bigr)\ox J_\la$.\ The 2-isomorphism $\,\Xi\,$ may be
chosen so that it satisfies the coherence condition of
\Reqref{eq:Gequiv-bimod-coh} iff the obstruction class
\qq\nn
[\Qup\check{\d}^{(1)}_{\txG,1}s-\check{\D}^{(2)}_{Q,1}e]+\Qup\ceH^{0,2}\,\in\,
\Qup\cP^{0,2}/\Qup\ceH^{0,2}
\qqq
is trivial.
\eerop
It is to be noted that the obstruction identified in the previous
proposition can be further reduced by exploiting the residual
freedom of choice of a $(\txG,\rho)$-equivariant structure on
$\,\cG$.\ Indeed, consider an admissible transformation
\eqref{eq:aenu-trafo} of local data of a $(\txG,\rho)$-equivariant
structure on $\,\cG$,\ accompanied by a shift
\qq\label{eq:s-trafo}
s\mapsto s+s'\,,\qquad\qquad\Qup D^{(1)}_1 s'=-\check{\D}^{(1)}_{Q,2} a'
\,,\qquad\qquad s'\in A^1\big(\Qup\cU^1\big)\,,
\qqq
i.e.\ altogether constrained so that they do not spoil the anomaly
cancellations ensured previously. The resultant shift of $\,[\si]\,$
is given by
\qq\nn
[\si]\mapsto[\si]+[\Qup\check{\d}^{(1)}_{\txG,1}s'-\check{\D}^{(2)}_{Q,1}
e']\,,
\qqq
and the intrinsic ambiguity of the 1-cochain $\,s'\,$ captured by
\Reqref{eq:ss-trafo} leads us, once more, to consider the coset
$\,[\Qup\check{\d}^{(1)}_{\txG,1}s'-\check{\D}^{(2)}_{Q,1} e']+\Qup
\ceH^{0,2}\,$ rather than the class itself. In order to properly
account for the observed freedom of readjustment of the structure
under construction, we consider the tricomplex $\,{}^{\tx{\tiny
$(M,Q)$}}\hspace{-2pt}\cK\,$ formed by the pair of bicomplexes
$\,\Mup\cK\,$ and $\,\Qup\cK$, cf.\ \Reqref{diag:J-bicompl},
{\scriptsize
\qq\nn \hspace{-0.4cm}
\alxydim{@C=.55cm@R=.55cm}{ & & & & A^0\big(\Mup\cU^0\big)
\ar[dl]_{\Mup\check{\d}^{(0)}_{\txG,0}} \ar[r]^{\Mup D^{(0)}_0} &
A^1\big(\Mup\cU^0\big) \ar[dl]_{\Mup\check{\d}^{(0)}_{\txG,1}}
\ar[r]^{\Mup D^{(0)}_1} &
\ker\,\Mup D^{(0)}_2 \ar[dl]^{\Mup\check{\d}^{(0)}_{\txG,2}} \\
& & & A^0\big(\Mup\cU^1\big) \ar[dl]_{\Mup\check{\d}^{(1)}_{\txG,0}}
\ar[r]^-{\Mup D^{(1)}_0} & A^1\big(\Mup\cU^1\big)
\ar[dl]_{\Mup\check{\d}^{(1)}_{\txG,1}} \ar[r]^-{\Mup D^{(1)}_1} &
*++[o][F]{\ker\,\Mup D^{(1)}_2\ni a'} \ar[dddd]^{\check{\D}^{(1)}_{Q,2}}
\ar[dl]^{\Mup\check{\d}^{(1)}_{\txG,2}} & \\ & & A^0\big(\Mup\cU^2\big)
\ar[dl]_{\Mup\check{\d}^{(2)}_{\txG,0}} \ar[r]^-{\Mup D^{(2)}_0} &
*++[o][F]{A^1\big(\Mup\cU^2\big)\ni e'}
\ar[dl]_{\Mup\check{\d}^{(2)}_{\txG,1}} \ar[r]^-{\Mup D^{(2)}_1}
\ar'[d][dddd]^{\check{\D}^{(2)}_{Q,1}} & \ker\,\Mup D^{(2)}_2
\ar[dl]^{\Mup\check{\d}^{(2)}_{\txG,2}} & & \\
& *++[o][F]{A^0\big(\Mup\cU^3\big)\ni\nu'} \ar[r]^-{\Mup D^{(3)}_0}
\ar[dl]_{\Mup\check{\d}^{(3)}_{\txG,0}} & A^1\big(\Mup\cU^3\big)
\ar[r]^-{\Mup D^{(3)}_1} \ar[dl]_{\Mup\check{\d}^{(3)}_{\txG,1}} &
\ker\,\Mup D^{(3)}_2 \ar[dl]^{\Mup\check{\d}^{(3)}_{\txG,2}} & & & \\
\udots & \udots & \udots & & A^0\big(\Qup\cU^0\big)
\ar[dl]_{\Qup\check{\d}^{(0)}_{\txG,0}} \ar[r]^{\Qup D^{(0)}_0} &
A^1\big(\Qup\cU^0\big) \ar[dl]_{\Qup\check{\d}^{(0)}_{\txG,1}}
\ar[r]^{\Qup D^{(0)}_1} &
\ker\,\Qup D^{(0)}_2 \ar[dl]^{\Qup\check{\d}^{(0)}_{\txG,2}} \\
& \ar[d]^{\check{\D}^{(\bullet)}_{Q,\bullet}} & & A^0\big(\Qup\cU^1
\big) \ar[dl]_{\Qup\check{\d}^{(1)}_{\txG,0}} \ar[r]^-{\Qup D^{(1)}_0}
& *++[o][F]{A^1\big(\Qup\cU^1\big)\ni s'}
\ar[dl]_{\Qup\check{\d}^{(1)}_{\txG,1}} \ar[r]^-{\Qup D^{(1)}_1} &
\ker\,\Qup D^{(1)}_2 \ar[dl]^{\Qup\check{\d}^{(1)}_{\txG,2}} & \\
& & A^0\big(\Qup\cU^2\big) \ar[dl]_{\Qup\check{\d}^{(2)}_{\txG,0}}
\ar[r]^-{\Qup D^{(2)}_0} & *++[F-,]{A^1\big(\Qup\cU^2\big)}
\ar[dl]_{\Qup\check{\d}^{(2)}_{\txG,1}} \ar[r]^-{\Qup D^{(2)}_1} &
\ker\,\Qup D^{(2)}_2 \ar[dl]^{\Qup\check{\d}^{(2)}_{\txG,2}} & & \\
& A^0\big(\Qup\cU^3\big) \ar[r]^-{\Qup D^{(3)}_0}
\ar[dl]_{\Qup\check{\d}^{(3)}_{\txG,0}} & A^1 \big(\Qup\cU^3\big)
\ar[r]^-{\Qup D^{(3)}_1} \ar[dl]_{\Qup\check{\d}^{(3)}_{\txG,1}} &
\ker\,\Qup D^{(3)}_2 \ar[dl]^{\Qup\check{\d}^{(3)}_{\txG,2}} & & & \\
\udots &\udots& \udots & & & & }
\qqq}
\vskip 0.2cm
\noindent using the family $\,\{\check{\D}^{(q)}_{Q,p}\}\,$ of chain
maps, and with the following assignment of the total degree
\qq\nn
\deg\,A^p\big(\Mup\cU^q\big)=p+q\,,\qquad\qquad\deg\,A^p\big(
\Qup\cU^q\big)=p+q+1\,.
\qqq
Our discussion points to the r\^ole of a group homomorphism
\qq\label{eq:H-map}
\ker\big(\check{\D}^{(1)}_{Q,2}\circ\Mup
\pi^{1,1}\big)\ \ni\ (a',e',\nu')\ \ \mathop{\longmapsto}\limits^\xcH\ \
[\Qup\check{\d}^{(1)}_{\txG,1}s'- \check{\D}^{(2
)}_{Q,1} e']+\Qup\ceH^{0,2}\ \in\ \Qup\cP^{0,2}/\Qup\ceH^{0,2}.\ \quad
\qqq
We are thus led to the following
\becor\label{cor:terrible-coset}
The component from the image of map $\,\xcH\,$ of \Reqref{eq:H-map}
of the class $\,[\Qup\check{\d}^{(1)}_{\txG,1}s-\check{\D}^{(
2)}_{Q,1} e]\in \Qup\cP^{0,2}/\Qup\ceH^{0,2}\,$ obstructing the
existence of a \emph{coherent} 2-isomorphism $\,\Xi\,$ of
\Reqref{Xi} (and defined in terms of local data $\,s\in A^1\big(\Qup
\cU^1\big)\,$ of $\,\Xi\,$ and local data $\,e\in A^1\big(\Mup\cU^2
\big)\,$ of a 2-isomorphism $\,\g\,$ of \Reqref{diag:2-iso-Gequiv})
can be removed by a suitable choice of a $(\txG,\rho)$-equivariant
structure on gerbe $\,\cG$,\ and a choice of a 2-isomorphism
$\,\Xi\,$ compatible with the former in the sense of
\Reqref{eq:s-trafo}.\ecor

At this stage, we may employ identification \eqref{eq:zetap-id} (for
$\,\txG Q$) to reformulate the description of the latter obstruction
to the existence of a coherent $(\txG,\la)$-equivariant structure on
$\,\cB\,$ in more tractable terms. Noting that identity
\eqref{eq:QdS} translates into the statement
\qq\nn
\Qup\z_2([\si])\in\ker\,\d_{\pi_0(\txG)}^{(2)}\,,
\qqq
we arrive at
\becor
The image
$\,\Qup\z_2\big([\Qup\check{\d}^{(1)}_{\txG,1}s-\check{\D}^{(2
)}_{Q,1}e]\big)=:\Qup u^2\,$ of the obstruction class $\,[
\Qup\check{\d}^{(1)}_{\txG,1}s-\check{\D}^{(2)}_{Q,1}e]\,$ for a
\emph{coherent} 2-isomorphism $\,\Xi\,$ under the identification of
\Reqref{eq:zetap-id} is a $\uj^{\pi_0(Q)}$-valued 2-cocycle of the
group $\,\pi_0(\txG)$,\ and that of the obstruction coset
$\,[\Qup\check{\d}^{(1)}_{\txG,1}s-\check{\D}^{(2)}_{Q,1}e]+\Qup\ceH^{0,2}\,$
is the cohomology class $\,[\Qup u^2]\in H^2\big(\pi_0(\txG),
\uj^{\pi_0(Q)}\big)$. \ecor \noindent \Reqref{eq:GHp-triv} now
yields
\becor
If the symmetry group $\,\txG\,$ is connected and a 2-isomorphism
$\,\Xi\,$ exists, then it can always be chosen in a coherent
form.\ecor

\noindent Clearly, this applies to the maximally symmetric WZW
$\cGk$-bi-branes from Definitions \ref{def:WZW-bdry-bib} and
\ref{def:WZW-non-bdry-bib} in Sections \ref{sub:maxym-bound} and
\ref{sub:maxym-nonbound}.

\subsubsection{Classification}

\,Let us now examine equivalences among $(\txG,\la)$-equivariant
structures on a given $\cG$-bi-brane $\,\cB=(Q,\iota_1,\iota_2,\om,
\Phi)\,$ in detail. Suppose that there exist two such structures,
$\,(\cB,\Xi_\b;\kappa),\ \b=1,2\,$ on $\,\cB$.\ Denote local data of
the respective 2-isomorphisms as $\,s_\b\,$ and $\,\theta_\b\,$ (the
coherence 0-cochain), and their differences as
\qq\nn
s_{2,1}:=s_2-s_1\,,\qquad\qquad\theta_{2,1}:=\theta_2-\theta_1\,.
\qqq
For a fixed $(\txG,\rho )$-equivariant structure on $\,\cG$,\ we
have
\qq\nn
\Qup D^{(1)}_1 s_\b=-\Qup\check{\d}^{(0)}_{\txG,2}\hspace{0.02cm}p
-\check{\D}^{(1)}_{Q,2}\hspace{0.02cm}a+
\ovl\la\,,\qquad\qquad\Qup D^{(2)}_0\theta_\b=\Qup\check{\d}^{(1)}_{\txG,1}
s_\b-\check{\D}^{(2)}_{Q,1} e\,,
\qqq
and so the difference 1-cochain $\,s_{2,1}\,$ automatically
satisfies the identities
\qq\label{eq:s21etc}
\Qup D^{(1)}_1 s_{2,1}=0\,,\qquad\qquad\Qup\check{\d}^{(1)}_{\txG,1}s_{2,1}
=\Qup D_0^{(2)}\theta_{2,1}\,,
\qqq
from which it follows that it defines a class
\qq\nn
[s_{2,1}]\in H^0\big(\txG\x Q,\uj\big)
\qqq
with the property
\qq\nn
\Qup\check{\d}^{(1)}_{\txG,1}[s_{2,1}]=0\,.
\qqq
According to Definition \ref{def:equiv-Gequiv-bi}, an equivalence
between the two structures, whenever it exists, can be expressed in
purely cohomological terms with the help of local data $\,t\in
A^1\left(\Qup\cU^0 \right)\,$ and $\,\tau\in
A^0\left(\Qup\cU^1\right)\,$ of a 2-isomorphism $\,\psi\ :\
\Phi\Rightarrow\Phi\,$ subject to the additional coherence
constraint tantamount to \Reqref{eq:Gequiv-bib-2iso-coh},
\qq\label{eq:sttau}
s_{2,1}&=&\Qup\check{\d}^{(0)}_{\txG,1}t+\Qup D^{(1)}_0\tau\,,\qquad\qquad
\Qup D^{(0)}_1 t=0\,,\cr\cr \theta_{2,1}&=&\Qup\check{\d}^{(1)}_{\txG,0}
\tau\,,\label{eq:Dtau}
\qqq
where the last relation follows from Eqs.\,\eqref{eq:s21etc} and
\eqref{eq:sttau}. The above equations hold iff
\qq\nn
[s_{2,1}]=\Qup\check{\d}^{(0)}_{\txG,1}[t]\,.
\qqq
As a result, we obtain, using identification \eqref{eq:zetap-id},
\berop
The set of equivalence classes of $(\txG,\la)$-equivariant
$\cG$-bi-branes with world-volume $\,Q\,$ is, for a fixed $(\txG,
\rho)$-equivariant structure on a gerbe $\,\cG$,\ a torsor of the
first cohomology group $\,H^1\left(\pi_0(\txG),\uj^{\pi_0(Q)}
\right)\,$ of the group $\,\pi_0(\txG)\,$ with values in the $\pi_0(
\txG)$-module $\,\uj^{\pi_0(Q)}$.
\eerop
\noindent In particular,
\becor
If the symmetry group $\,\txG\,$ is connected, a $(\txG,\la
)$-equivariant structure on a $\cG$-bi-brane, whenever it exists, is
unique up to an equivalence.\ecor

\noindent This is the case for the maximally symmetric WZW
$\cGk$-bi-branes from Sections \ref{sub:maxym-bound} and
\ref{sub:maxym-nonbound}. \vskip 0.3cm

\subsection{$\txG$-equivariant inter-bi-branes}

\noindent The completion of the construction of a full-blown $(\txG,\rho,\la
)$-equivariant string background requires one further step that
consists in checking the compatibility of the structures, introduced
hitherto as an extension of gerbe and bi-brane data, with the
structure of an inter-bi-brane. We address this final issue below.

\subsubsection{Obstructions}

\,Given local data $\,h_n\in A^1\big(\Tnup\cU^0\big)$,\ constrained
by identity \eqref{eq:DFN-is}, of 2-isomorphism $\,\varphi_n\,$ of a
$(\cG,\cB)$-inter-bi-brane
$\,\left(T_n\left(\vep_n^{k,k+1},\pi_n^{k,k+1}\right);\varphi_n
\right)\,$ with component world-volumes $\,T_n$,\ we readily compute
\qq\nn
\Tnup D^{(1)}_1\Tnup\check{\d}^{(0)}_{\txG,1}h_n&=&\Tnup\check{\d}^{(0)}_{\txG,2}
\Tnup D^{(0)}_1 h_n=-\Tnup\check{\d}^{(0)}_{\txG,2}\check{\D}^{(0)}_{T_n,2}
\hspace{0.02cm}p=-\check{\D}^{(1)}_{T_n,2}\Qup\check{\d}^{(0)}_{\txG,2}
\hspace{0.02cm}p\cr\cr
&=&\check{\D}^{(1)}_{T_n,2}\Qup D^{(1)}_1 s+\check{\D}^{(1)}_{T_n,2}
\check{\D}^{(1)}_{Q,2}\hspace{0.02cm}a-\check{\D}^{(1)}_{T_n,2}
\hspace{0.02cm}\ovl\la=\Tnup D^{(1)}_1\check{\D}^{(1)}_{T_n,1}s\,.
\qqq
This gives us a class
\qq\label{eq:cF}
d_n\,:=\,[\Tnup\check{\d}^{(0)}_{\txG,1}h_n-\check{\D}^{(1)}_{T_n,1}s]\,\in\, H^0
\big(\txG\x T_n,\uj\big)
\qqq
that obstructs a coherent extension of a $(\txG,\rho,\la
)$-equivariant structure on the gerbe $\,\cG\,$ and the bi-brane
$\,\cB\,$ (whose existence we are assuming) to the complete
background. Viewed as a locally constant $\,\uj$-valued function on
$\,\txG\x T_n$, $\,d_n\,$ coincides with the function defined by
equality \eqref{eq:locfctn}. In this manner, we establish
\berop
Let $\,s\in A^1\big(\Qup\cU^1\big)\,$ be local data of a
2-isomorphism $\,\Xi\,$ of the $(\txG,\la)$-equivariant
$\cG$-bi-brane $\,(\cB,\Xi;\kappa)$,\ and let $\,h_n\in
A^1\big(\Tnup\cU^0 \big)\,$ be local data of the 2-isomorphism
$\,\varphi_n\,$ of \eqref{diag:2iso} of the
$(\cG,\cB)$-inter-bi-brane $\,\cJ_n\,$. \,Then, a
$(\txG,\rho)$-equivariant structure on $\,\cG\,$ together with a
$(\txG,\la)$-equivariant structure on $\,\cB\,$ compatible with it
extend coherently to a $(\txG,\rho,\la )$-equivariant structure on
$\,(\cG,\cB,\cJ_n)$,\ as expressed by Definition
\ref{def:Gequiv-ibb}, iff the obstruction class
\qq\nn
d_n\,\in\, H^0\big(
\txG\x T_n,\uj\big)
\qqq
of \eqref{eq:cF} is trivial.
\eerop
\noindent We have
\qq\nn
\Tnup\check{\d}^{(1)}_{\txG,1}d_n=[-\check{\D}^{(2)}_{T_n,
1}\Qup\check{\d}^{(1)}_{\txG,1}s]=[-\check{\D}^{(2)}_{T_n,1}\check{\D}^{(2
)}_{Q,1} e-\Tnup D^{(2)}_0\check{\D}^{(2)}_{T_n,0}\theta]=0\,,
\qqq
i.e.
\qq\nn
d_n\,\in\,\ker\left(\Tnup\check{\d}^{(1)}_{\txG,1}:H^0\left(\txG\x
T_n,\uj\right)\rightarrow H^0\left(\txG^2\x T_n,\uj\right) \right)\
=:\ \Tnup\cP^{0,1}\,.
\qqq
The identification \eqref{eq:zetap-id} permits, as before, to
reformulate the description of the obstruction in the language of
the cohomology of the group $\,\pi_0(\txG)$, \,this time with
values in $\,\uj^{\pi_0(T_n)}$.\ We have
\becor
The image $\,\Tnup\z_1\big([d_n]\big)=:\Tnup u^1$,\ under the
identification of \Reqref{eq:zetap-id}, of the obstruction class
$\,d_n\,$ of \eqref{eq:cF} for a \emph{coherent} extension of a
$(\txG,\rho)$-equivariant structure on the gerbe $\,\cG\,$ together
with a $(\txG,\la)$-equivariant structure on the $\cG$-bi-brane
$\,\cB\,$ compatible with it to a $(\txG,\rho,\la)$-equivariant
structure on the background $\,(\cG,\cB,\cJ_n)\,$ is a
$\uj^{\pi_0(T_n)}$-valued 1-cocycle of the group $\,\pi_0(\txG)$.
\ecor

\noindent Since 1-cocycles on the trivial group $\,\bd1\,$ are
trivial, we obtain
\becor\label{cor:glancF}
For a connected group $\,\txG$, \,the obstruction classes $\,d_n\,$
are trivial. \ecor

\noindent This applies, in particular, to the maximally symmetric
WZW $\cGk$-bi-branes and inter-bi-branes from Sections
\ref{sub:maxym-nonbound} and \ref{sub:maxsyminter}. \vskip 0.2cm

In the general case, pursuing the logic of the preceding sections,
we admit the
possibility of using the freedom of choice, consistent with previous
stages of the construction, of a $(\txG,\rho )$-equivariant
structure on $\,\cG\,$ and of a $(\txG,\la)$-equivariant structure
on $\,\cB\,$ compatible with it to rid the string
background\footnote{We implicitly assume that the metric on $\,M\,$
has the desired invariance properties.} $\,(\cG,\cB,\cJ_n )\,$ of at
least a part of the obstruction to the existence of a full-blown
coherent $(\txG,\rho,\la)$-equivariant structure. Thus, we assume
Eqs.\,\eqref{eq:a-def}-\eqref{eq:th-vs-se} to be obeyed by a
collection of data $\,(a,e,\nu;s,\theta)\,$ introduced previously,
and examine the effect on $\,d_n\,$ of the most general transformation of
the initial data,
\qq\nn
(a,e,\nu;s,\theta)\,\longmapsto\,(a,e,\nu;s,\theta)+(a',e',\nu';s',\theta')
\,,
\qqq
that preserves the structure of the cohomological relations
\eqref{eq:a-def}-\eqref{eq:th-vs-se}. Components of such a
transformation are constrained to satisfy a `homogeneous' variant of
those relations, i.e.
\qq\nn
&\Mup D^{(1)}_2 a'=0\,,\qquad\qquad\Mup D^{(2)}_1 e'=-\Mup\check{\d}^{(1
)}_{\txG,2}\hspace{0.02cm}a'\,,\qquad\qquad
\Mup D^{(3)}_0\nu'=\Mup\check{\d}^{(2)}_{\txG,1}e'
\,,&\cr\cr
&\Qup D^{(1)}_1 s'=-\check{\D}^{(1)}_{Q,2} a'\,,\qquad\qquad\Qup D^{(2
)}_0\theta'=\Qup\check{\d}^{(1)}_{\txG,1}s'-\check{\D}^{(2)}_{Q,1} e'\,,\qquad
\qquad\Qup\check{\d}^{(2)}_{\txG,0}\theta'=-\check{\D}^{(3)}_{Q,0}\nu'\,.&
\qqq
This identifies $\,(a',e',\nu';0,s',\theta')\,$ as a 3-cocycle of
the hypercohomology of the tricomplex $\,{}^{\tx{\tiny $(M,Q)$}}
\hspace{-2pt}\cK\,$ from the kernel of the canonical projection
\qq\nn
\cZ^3\left( {}^{\tx{\tiny $(M,Q)$}}\hspace{-2pt}\cK\right)\ \ni\
(a',e',\nu';x,s',\theta')\ \
\mathop{\longmapsto}\limits^{{}^{\tx{\tiny $(M,Q)$}}
\hspace{-2pt}\pi^{1,0}}\ \ [x]\ \in\ H^1\left(Q,\uj \right)\,,
\qqq
defined on the abelian group $\,\cZ^3\left({}^{\tx{\tiny $(M,Q)$}}
\hspace{-2pt}\cK\right)\,$ of 3-cocycles of the hypercohomology of
the tricomplex $\,{}^{\tx{\tiny $(M,Q)$}}\hspace{-2pt}\cK$.\ On this
subspace, there exists a well-defined group homomorphism
\qq
\ker\big({}^{\tx{\tiny $(M,Q)$}}\hspace{-2pt}\pi^{1,0}\big)\,\ni\,
(a',e',\nu';0,s', \theta')\ \ \mathop{\longmapsto}\limits^\xcK\ \
[\check{\D}^{(1)}_{T_n,1}s']\, \in\,\Tnup\cP^{0,1}\label{eq:Khom}
\qqq
and we have
\becor
The component of the obstruction class $\,d_n\,$ from the image of
map $\,\xcK\,$ of \Reqref{eq:Khom} may be removed by a suitable
choice of a $(\txG,\rho )$-equivariant structure on $\,\cG$,\ and
that of a $(\txG,\la )$-equivariant structure on $\,\cB$. \ecor

Let us close this Section by summarizing the situation for the
WZW $\,\si$-model.
\becor
In the WZW-model context of Definition \ref{def:WZW-target} with the
maximally symmetric $\,\cG_\sfk$-bi-branes and inter-bi-branes
discussed in Section \ref{sec:eg-backgrnd}, the only obstructions to
the existence of a $\txG$-equivariant structure on the corresponding
string background come from the global gauge anomalies for
topologically trivial gauge fields that were discussed in Section
\ref{sec:eg-backgrnd}. In their absence, the set of equivalence
classes of $\,\txG$-equivariant structures forms a torsor under the
group $\,H^1(\txG,\uj)$. \ecor

\brem The freedom of choice of a $\,\txG$-equivariant structure
appears in the WZW-model gauged Feynman amplitudes in topologically
nontrivial sectors with gauge-fields given by connections in
nontrivial principal $\,\txG$-bundles classified by
$\,\pi_1(\txG)\,$. \,A change of the equivariant structure by an
element of $\,H^1(\txG,\uj)\,$ that may be viewed as a character of
$\,\pi_1(\txG)\,$ multiplies the amplitude by the corresponding
value of the character. This was discussed in detail in
\Rcite{Gawedzki:2010rn} for the defect-free amplitudes and the
discussion extends to the case with defects. \erem

\section{Conclusions}

\noindent We have presented a detailed study of the gauging of rigid
symmetries of the two-dimensional bosonic $\si$-model on
world-sheets $\,\Si\,$ with defects forming a graph $\,\G\,$
embedded in the world-sheet, including the case of boundary defects.
Classical fields mapping $\,\Si\setminus\G\,$ into the target space
of the model may jump across the defect lines and at defect
junctions. The jumps across the lines of the graph are described by
maps from the lines into bi-brane world-volumes and the jumps at the
line junctions by maps from the junction points to inter-bi-brane
world-volumes. The Wess--Zumino contributions to the action of the
$\,\si$-model are given in terms of a holonomy of a gerbe on the
target manifold supplemented with bi-branes and inter-bi-branes
equipped with a gerbish structure formulated in terms of the
2-category of gerbes with connections. The rigid symmetries of the
model have been identified with group actions on the $\,\si$-model
target space (including the bi-brane and inter-bi-brane
world-volumes) that respect the target-space metric and the gerbish
structure in a way that ensures the invariance of the $\,\si$-model
Feynman amplitudes. Promoting the rigid symmetries to the rank of
local ones required coupling the model to the world-sheet gauge
fields of the rigid-symmetry group $\,\txG$.\ For a gauge field
given by a connection in a trivial principal $\txG$-bundle over
$\,\Si$, \, the appropriate gauging procedure that renders the
gauged amplitudes invariant under infinitesimal gauge transformation
has been defined as a slight extension of that introduced in
Refs.\,\cite{Jack:1989ne,Hull:1990ms} for bulk amplitudes and in
\Rcite{Figueroa:2005} for the boundary ones. Various interpretations
of the conditions required by such a construction have been
presented. We have analysed the invariance of the corresponding
gauged amplitudes under large gauge transformations non-homotopic to
the identity, identifying the defect- and boundary-contributions to
the global gauge anomalies. These results have been illustrated on
the example of the WZW amplitudes, extending the work of
\Rcite{Gawedzki:2010rn} where this was done for the defect-free
models. We have subsequently extended the notion of equivariant
gerbes introduced in \Rcite{Gawedzki:2010rn} to include equivariant
structures on bi-branes and inter-bi-branes, and showed that such
structures permit a natural coupling of the $\,\si$-model amplitudes
with defects to world-sheet gauge fields given by connections in
arbitrary principal $\txG$-bundles in a way that assures the
invariance of the resulting amplitudes under all gauge
transformations, including the large ones. Obstructions to the
existence and the classification of such equivariant structures have
been fully analysed. In particular, we have showed that in the
context of WZW models with maximally symmetric defects, all
obstructions manifest themselves already in the presence of
topologically trivial gauge fields, similarly as in the defect-free
case. In summary, the present paper provides a realisation of a
major step in the study of global gauge anomalies in two-dimensional
bosonic $\,\si$-models with Wess--Zumino terms in the action,
initiated in \Rcite{Gawedzki:2010rn}.

\newpage

\appendix

\section{A proof of the ``only if'' part of Proposition \ref{prop:rigid-inf-inv}}
\label{app:rigid-inf-inv}

\noindent That conditions \eqref{eq:Kill} and \eqref{eq:HS-exact}
are necessary for the infinitesimal rigid invariance of Feynman
amplitudes was noticed already in \Rcite{Gawedzki:2010rn} where
field configurations without defects were considered. It is
straightforward to see in that case that the vanishing of the
right-hand side of \Reqref{eq:var-act-gen} for arbitrary world-sheet
metrics $\,\gamma\,$ and arbitrary $\,\varphi:\Sigma\rightarrow M\,$
requires that condition \eqref{eq:Kill} be satisfied. \,Using
subsequently the fact that the homology group $\,H_2(M)\,$ is
generated by singular 2-cycles $\,c_2^M(\varphi)\,$ obtained from
fields $\,\varphi\,$ and triangulations of $\,\Sigma$, \,one infers
condition \eqref{eq:HS-exact}.
\smallskip

Next, let us consider world-sheets $\,\Sigma\,$ with defect quivers
$\,\G\,$ without defect junctions, i.e. composed of closed loops
only. Given a triangulation of $\,\Sigma\,$ compatible with
$\,\Gamma$, \,each network-field configuration
$\,(\varphi\,|\,\G)\,$ gives rise to a singular 1-cycle
$\,c_1^Q(\varphi)\,$ in $\,Q\,$ and a singular 2-chain
$\,c_2^M(\varphi)\,$ in $\,M$. \, The 1-cycle $\,c_1^Q(\varphi)\,$
is obtained from $\,\varphi|_\G\,$ and the induced triangulation of
$\,\G$, and the 2-chain $\,c_2^M(\varphi)\,$ from
$\,\varphi|_{\Sigma\setminus\G}\,$ extended by the boundary values
and the triangulation of $\,\Sigma$. \,Let
\qq\nn
\Delta^{\hspace{-0.05cm}Q}\,=(\iota_2)_*-(\iota_1)_*
\qqq
be the combination of maps on singular chains induced by the
$\,\iota_\alpha:Q\rightarrow M$. \,The defining requirements for a
network-field configuration imply that
\qq\label{eq:DQ-partial}
\,\DQ c_1^Q(\varphi)\,=\,-\partial c_2^M(\varphi)\,.
\qqq
Note that this means that the homology class $\,[c_1^Q(\varphi)]\in
H_1(Q)\,$ belongs to the kernel $\,\ker\,\DQ\,$ of the induced map
$\,\DQ:H_1(Q)\rightarrow H_1(M)$. \,It is not difficult to see that
$\,\ker\,\DQ\,$ may be generated this way. \,Now, assuming that
\Reqref{eq:HS-exact} holds, we have
\qq\label{eq:intred}
\int\limits_\Sigma\varphi^*(\iota_a\txH)
+\int\limits_\G\hspace{-0.1cm}\varphi^*(\iota_a\om)\ &=&
\int\limits_{c^M_2(\varphi)}\hspace{-0.2cm}\iota_a\txH
\,+\hspace{-0.1cm}\int\limits_{c_1^Q(\varphi)}\hspace{-0.1cm}\iota_a\om\
=\ -\hspace{-0.2cm}\int\limits_{\partial c_2^M(\varphi)}
\hspace{-0.2cm}\kappa_a\
+\hspace{-0.1cm}\int\limits_{c_1^Q(\varphi)} \hspace{-0.2cm}
\iota_a\om\\
&&=\hspace{-0.2cm}\int\limits_{\DQ c_1^Q(\varphi)}\hspace{-0.4cm}\kappa_a
\ +\hspace{-0.1cm}\int\limits_{c_1^Q(\varphi)}\hspace{-0.2cm}\iota_a\om
\ =\int\limits_{c_1^Q(\varphi)}\hspace{-0.1cm}(\iota_a\om+\D_Q\kappa_a)\,.\nn
\qqq
The 1-forms $\,\iota_a\om+\D_Q\kappa_a\,$ are closed due to
relations \eqref{eq:triv-restr} and \eqref{eq:HS-exact}. We want to
show that the vanishing of the right-hand side of \Reqref{eq:intred}
for all $\,c_1^Q(\varphi)\,$ implies that for an appropriate choice
of 1-forms $\,\kappa_a\,$ on $\,M$, \,fixed by \Reqref{eq:HS-exact}
up to addition of closed 1-forms $\,\kappa_a'$, \,the induced
cohomology classes $\,[\iota_a\om+\D_Q\kappa_a]\in H^1(Q,\bR)\,$
vanish, whence relation \eqref{eq:bdry-exact} ensues. This follows
from the cohomology exact sequence
\qq\label{coh-ES}
H^1(M,\bR)\ \xrightarrow{\ \D_Q\ }\ H^1(Q,\bR)\ \longrightarrow\
\mor_{\bZ}\left(\ker\,\DQ,\bR\right)
\qqq
obtained by applying the exact functor $\,\mor_\bZ(\cdot\,,\bR)\,$
to the homology exact sequence
\qq\nn
\ker\,\Delta^{\hspace{-0.05cm}Q}\ \hookrightarrow\ H_1(Q)\
\xrightarrow{\ \DQ\ }\ H_1(M)\,.
\qqq

Finally, consider closed world-sheets $\,\Si\,$ with general defect
quivers $\,\G$. \,A network-field configuration
$\,(\varphi\,|\,\G)\,$ together with a triangulation of $\,\Sigma\,$
compatible with $\,\G\,$ gives rise to a singular 0-cycle
$\,c_0^T(\varphi)\,$ in $\,T\,$, \,to a singular 1-chain
$\,c_1^Q(\varphi)\,$ in $\,Q\,$ and to a singular 2-chain
$\,c_2^M(\varphi)\,$ in $\,M$,
\,$$c_0^T(\varphi)\,=\,\sum\limits_{j\in\Vgt_\G}\pm\varphi(j)\,,$$
where the $+$ sign is taken for the positive defect junctions (i.e.
the ones with the counter-clockwise ordering of defect lines) and
the $-$ sign for the negative junctions (that have defect lines
ordered clockwise). \,The 1-chain $\,c_1^Q(\varphi)\,$ results from
extending $\,\varphi|_{\Gamma\setminus\Vgt_\G}\,$ by the boundary
values, and the 2-chain $\,c_2^M(\varphi)\,$ is obtained as before.
\,Let
\qq
\DTn\,=\,\sum_{k=1}^n
\vep_n^{k,k+1}\,(\pi_n^{k,k+1})_*
\qqq
be the combination of maps on singular chains induced by the
$\,\pi_n^{k,k+1}:T_n\rightarrow Q$. \,Note that $\,\DQ\circ\DTn=0$.
\,Let $\,\DT=\sum\limits_{n\geq3}\DTn$. \,One has
\qq\label{eq:c0c1c2}
\DT c_0^T(\varphi)\,=\,\partial c_1^Q(\varphi)\,,\qquad
\DQ c_1^Q(\varphi)\,=\,-\partial c_2^M(\varphi)\,.
\qqq
The first of these relations implies that the homology class
$\,[c_0^T(\varphi)]\in H_0(T)\,$ belongs to the kernel
$\,\ker\,\DT\,$ of the induced map $\,\DT:H_0(T)\rightarrow H_0(Q)$.
\,Consider the connecting homomorphism
\qq\nn
B\ :\ \ker\,\DT\longrightarrow H_1(M)/\DQ(H_1(Q))
\qqq
defined by associating to the homology class of the 0-cycle
$\,c_0^T\,$ such that $\,\DT\hspace{-0.06cm}c_0^T=\partial c_1^Q\,$
the coset $\,[\DQ c_1^Q]+\DQ(H_1(Q))$. \,The second of relations
\eqref{eq:c0c1c2} implies that $\,B\left([c_0^T(\varphi)]\right)=0$.
\,One may see again that the homology classes $\,[c_0^T(\varphi)]\,$
generate $\,\ker\,B\subset H_0(T)$. \,Assuming relations
\eqref{eq:HS-exact} and \eqref{eq:bdry-exact}, \,the chain of
equations \eqref{eq:intred} now implies that
\qq\label{eq:intredit}
\int\limits_\Sigma\varphi^*(\iota_a\txH)
+\int\limits_\G\varphi^*(\iota_a\om)\ =
\ -\hspace{-0.2cm}\int\limits_{\partial c_1^Q(\varphi)}\hspace{-0.2cm}k_a\
=\ -\hspace{-0.3cm}\int\limits_{\DT c_0^T(\varphi)}\hspace{-0.3cm}k_a\
=\ -\hspace{-0.1cm}\int\limits_{c_0^T(\varphi)}\hspace{-0.2cm}\D_Tk_a\,.
\qqq
Note that functions $\,k_a\,$ are defined modulo addition of
$\,k_a'\,$ such that $\,dk_a'=\Delta_Q\kappa'_a\,$ for some closed
1-forms $\,\kappa'_a\,$. \,We have to show that the vanishing of the
right-hand side of \Reqref{eq:intredit} for all $\,c_1^Q(\varphi)\,$
implies that functions $\,\Delta_Tk_a\,$ on $\,T\,$ may be
annihilated by such a modification. Not much seems to be needed
since functions $\,\D_Tk_a\,$ are locally constant,
\qq\label{eq:const}
\sfd\D_{T_n}k_a=-\D_{T_n}(\iota_a\om+\Delta_Q\kappa_a)=
-\iota_a\D_{T_n}\om=0
\qqq
due to \Reqref{eq:triv-restr2}. \,There is a cohomology exact sequence
\qq\label{cohoESB}
\hspace*{0.9cm}\mor_{\bZ}\left(H_1(M)/\DQ(H_1(Q)),\bR\right)\
\xrightarrow{\ B^*\ }\ \mor_{\bZ}\left(\ker\,\DT,\bR\right)\
\longrightarrow\mor_{\bZ}\left(\ker\,B,\bR\right)
\qqq
obtained by applying the exact functor $\,\mor_{\bZ}(\cdot\,,\bR)\,$ to the
homology exact sequence
\qq
\ker\,B\ \hookrightarrow\ \ker\,\DT\ \xrightarrow{\ B\ }\
H_1(M)/\DQ(H_1(Q))\,.
\qqq
For 0-cycles $\,c_0^T\,$ such that $\,\DT\hspace{-0.06cm}
c_0^T=\partial c_1^Q\,$ for some 1-chain $\,c_1^Q$,  \,the pairing
\qq\label{eq:pairing}
\ker\,\DT\,\ni\,[c_0^T]\ \mapsto\
\int\limits_{c_0^T}\Delta_Tk_a\,\in\,\bR
\qqq
defines a $\,\bZ$-homomorphism on $\,\ker\,\DT\,$ that, according to
our assumption, vanishes when restricted to $\,\ker\,B$. It must
then lie in the image of
$\,\mor_{\bZ}\left(H_1(M)/\DQ(H_1(Q)),\bR\right)\,$ under the linear
map $\,B^*\,$ dual to $\,B$. \,But
$\,\mor_{\bZ}\left(H_1(M)/\DQ(H_1(Q)),\bR\right)\cong\ker\,\D_Q\subset
H^1(M,\bR)\,$ and elements of $\,\ker\,\D_Q\,$ are the cohomology
classes of closed 1-forms $\,\kappa'\,$ on $\,M\,$ such that there
exists a function $\,k'\,$ on $\,Q\,$ for which $\,\D_Q\kappa'=dk'$.
\,One has
\qq\nn
\langle[c_0^T],B^*[\kappa']\rangle\,=\,\langle B[c_0^T],[\kappa']\rangle\,=
\,\langle[\DQ c_1^Q],[\kappa']\rangle\,=\int\limits_{c_1^Q}\D_Q\kappa'\,=
\int\limits_{\partial c_1^Q}k'\,=\int\limits_{c_0^T}\Delta_Tk'\,.
\qqq
It follows that through allowed modifications of functions $\,k_a$,\
adding to them appropriate functions $\,k_a'\,$ with
$\,dk_a'=\D_Q\kappa_a'\,$ for some closed 1-forms $\,\kappa_a'$,
\,one may achieve that the pairing of \Reqref{eq:pairing} vanishes.
But then the cohomology exact sequence
\qq\label{coh-ES1}
H^0(Q,\bR)\ \xrightarrow{\ \D_T\ }\ H^0(T,\bR)\ \longrightarrow\
\mor_{\bZ}\left(\ker\,\DT,\bR\right)
\qqq
shows that by an eventual further modifications of functions
$\,k_a\,$ that adds to them locally constant functions $\,k'_a$,
\,we may obtain relation \eqref{eq:junct-exact}.

\section{\mbox{Proofs of the ``only if'' parts of Proposition \ref{prop:equiv-2iso}
and Theorem \ref{thm:equiv-12iso-constr}}} \label{app:equiv-2isom}

\noindent We shall argue similarly as in Appendix
\ref{app:rigid-inf-inv}. Consider a closed world-sheet $\,\Si\,$
with a defect quiver $\,\G\,$ without defect junctions and a
triangulation of $\,\Si\,$ compatible with $\,\G$. \,A gauge
transformation $\,\chi:\Sigma\rightarrow\txG\,$ and a network-field
configuration $\,(\varphi\,|\,\G)$, \,both restricted to $\,\G$,
\,determine a singular 1-cycle $\,c_1^{Q^{(1)}}\hspace{-0.1cm}
(\chi,\varphi)\,$ in $\,Q^{(1)}=\txG\times Q$.\ When both are
restricted to $\,\Sigma\setminus\G\,$ (and extended by the boundary
values), they also define a singular 2-chain
$\,c_2^{M^{(1)}}\hspace{-0.1cm}(\chi,\varphi)\,$ in
$\,M^{(1)}=\txG\times M$. \,Denoting by $\,\Delta^{Q^{(1)}}\,$ the
combination $\,(\iota^{(1)}_2)_*-(\iota^{(1)}_1)_*\,$ of maps on
chains originating from the
$\,\iota_\alpha^{(1)}=\id_\txG\times\iota_\alpha$, \,we have
\qq\nn
\Delta^{Q^{(1)}} c_1^{Q^{(1)}}\hspace{-0.1cm}(\chi,\varphi)\,=\,-
\partial\hspace{0.03cm}c_2^{M^{(1)}}\hspace{-0.1cm}(\chi,\varphi)\,,
\qqq
cf.\ \Reqref{eq:DQ-partial}. Hence, the homology class
$\,[c_1^{Q^{(1)}}\hspace{-0.1cm}(\chi,\varphi)]\,$ lies in the
kernel $\,\ker\,\Delta^{Q^{(1)}}\,$ of the induced map
$\,\Delta^{Q^{(1)}}:H_1(Q^{(1)}) \rightarrow H_1(M^{(1)})$.
\,Besides, $\,\ker\,\Delta^{Q^{(1)}} \subset H_1(Q^{(1)})\,$ may be
generated by homology classes obtained that way. The triviality of
the holonomy $\,\Hol_D\bigl((\chi,\varphi) \vert_\G\bigr)\,$ then
implies that the isomorphism class $\,[D]\in
H^1\left(Q^{(1)},\uj\right)\,$ of the flat line bundle $\,D\,$ is in
the kernel of the right morphism of the the cohomology exact
sequence
\qq\label{cohoES1}
\quad H^1\left(M^{(1)},\uj\right)\xrightarrow{\ \D_Q^{(1)}\ } H^1
\left(Q^{(1)},\uj\right)\xrightarrow{\quad r\quad}\mor_{\bZ}
\left(\ker\,\Delta^{Q^{(1)}},\uj\right)\,,
\qqq
obtained by applying the exact functor $\,\mor_{\bZ}\left(\cdot\,,\uj
\right)\,$ to the exact sequence of $\bZ$-modules
\qq\nn
\ker\,\Delta^{Q^{(1)}}\,\hookrightarrow\,H_1(Q^{(1)})
\,\xrightarrow{\ \Delta^{Q^{(1)}}\ }\,H_1(M^{(1)})\,.
\qqq
Thus, $\,[D]\,$ lies in the image of the left morphism of the exact
sequence of \Reqref{cohoES1}. This means that $\,[D]\,$ is
trivializable by a redefinition of the 1-isomorphism $\,\Upsilon\,$
of \Reqref{Upsilon} by a flat bundle, cf.\ Proposition
\ref{prop:torsors}, completing the proof of the ``only if'' part of
Proposition \ref{prop:equiv-2iso}. \,Note that such a redefinition
of $\,\Upsilon\,$  is fixed only modulo a flat bundle $\,D'\,$ over
$\,M^{(1)}\,$ with isomorphism class $\,[D']\,$ in the kernel of the
left homomorphism in the exact sequence of \Reqref{cohoES1}, i.e.\
such that there exists a trivialisation $\,d'\,$ of flat bundles
$\,\D_Q^{(1)}D'\,$ over $\,Q^{(1)}$. \,Since the choice of a
2-isomorphism $\,\Xi\,$ of \Reqref{Xi} is given by a choice of a
trivialisation of the bundle $\,D$, \,it is defined modulo
trivializations $\,d'$. \,Pulling back $\,d'\,$ to $\,T^{(1)}\,$
with the help of $\,\D_T^{(1)}$, \,one obtains a trivialisation of
the flat bundle $\,\D_T^{(1)}\D_Q^{(1)}D'\,$ which is naturally
trivial since each fibre of $\,D'\,$ that appears in it is
accompanied by the dual fibre. The two trivializations differ on
$\,T_n^{(1)}\,$ by multiplication by a locally constant $\uj$-valued
function $\,d'_n\,$ that describes possible multiplicative
ambiguities of the function $\,d_n\,$ defined by
\Reqref{eq:loc-const-corr}. We shall use such ambiguities below.
\smallskip

In order to prove the ``only if'' part of Theorem
\ref{thm:equiv-12iso-constr}, we shall employ Proposition
\ref{prop:loc-const-junct}, proceeding similarly as in Appendix
\ref{app:rigid-inf-inv}. For a closed surface with an arbitrary
defect quiver $\,\G\,$ and a triangulation compatible with $\,\G$,
\,a gauge transformation $\,\chi:\Sigma\to\txG\,$ and a
network-field configuration $\,(\varphi\,|\,\G)\,$ determine a
0-cycle $\,c_0^{T^{(1)}}\hspace{-0.1cm}(\chi,\varphi)\,$ in
$\,T^{(1)}=\txG\times T$, a 1-chain
$\,c_1^{Q^{(1)}}\hspace{-0.1cm}(\chi,\varphi)\,$ in $\,Q^{(1)}\,$
and a 2-chain $\,c_2^{M^{(1)}}\hspace{-0.1cm}(\chi,\varphi)\,$ in
$\,M^{(1)}\,$ such that
\qq\nn
\Delta^{T^{(1)}}c_0^{T^{(1)}}\hspace{-0.1cm}(\chi,\varphi)\,=\,
\partial\hspace{0.03cm}c_1^{Q^{(1)}}\hspace{-0.1cm}(\chi,\varphi)\,,
\qquad\Delta^{Q^{(1)}} c_1^{Q^{(1)}}\hspace{-0.1cm}(\chi,\varphi)\,=\,-
\partial\hspace{0.03cm}c_2^{M^{(1)}}\hspace{-0.1cm}(\chi,\varphi)\,,
\qqq
with $\,\Delta^{T^{(1)}}\,$ defined similarly as $\,\DT$,\ cf.\
\Reqref{eq:c0c1c2}. This means that the homology class
$\,[c_0^{T^{(1)}}\hspace{-0.1cm}(\chi,\varphi)]\,$ is in the kernel
$\,\ker\,B^{(1)}\,$ of the homomorphism
\qq\nn
B^{(1)}\ :\ \ker\,\Delta^{T^{(1)}}\longrightarrow
H_1(M^{(1)})/\Delta^{Q^{(1)}}\left(H_1(Q^{(1)})\right)
\qqq
defined as before, i.e.\ assigning to a homology class of a 0-cycle
$\,c_0^{T^{(1)}}\,$ such \,that $\,\Delta^{T^{(1)}}c_0^{T^{(1)}}
=\partial\hspace{0.02cm}c_1^{Q^{(1)}}\,$ the coset
$\,[\Delta^{Q^{(1)}}c_1^{Q^{(1)}}]+\Delta^{Q^{(1)}}\left(H_1(Q^{(1)})\right)$.
\,One may see that $\,\ker\,B^{(1)}\,$ is generated by classes
$\,[c_0^{T^{(1)}}\hspace{-0.1cm}(\chi,\varphi)]\,$ (this follows
from the fact that $\,\ker\,B\,$ was generated by classes
$\,[c_0^T(\varphi)]$). \,The collection $\,(d_n)\,$ of locally
constant $\uj$-valued functions may be viewed as an element of
$\,H^0\left(T^{(1)},\uj\right)\cong
\mor_\bZ\left(H_0(T^{(1)}),\uj\right)$,\ and one has the identity
\qq\nn
\prod_{\jmath\in\Vgt_\G}\,(\chi,\varphi)^*d_{n_\jmath}(\jmath)^{\pm 1}\,=\,
\big\langle [c_0^{T^{(1)}}\hspace{-0.1cm}(\chi,\varphi)],(d_n)\big\rangle\,,
\qqq
where the factor on the right-hand side of \Reqref{eq:viol-inv}
describes an eventual violation of the gauge invariance. The
triviality of that factor for all classes
$\,[c_0^{T^{(1)}}\hspace{-0.1cm}(\chi,\varphi)]\,$ implies that
$\,(d_n)\,$ defines an element in the kernel of the restriction
homomorphism on the right of the cohomology exact sequence
\qq\label{cohoESB(1)}
\hspace*{0.7cm}\mor_{\bZ}\left(H_1(M^{(1)})/\Delta^{Q^{(1)}}\left(H_1(Q^{(1)})\right),\uj\right)\
\xrightarrow{\ (B^{(1)})^*\ }\,
\mor_{\bZ}\left(\ker\,\Delta^{T^{(1)}},\uj\right)\qquad
\\ \cr\nn \longrightarrow\
\mor_{\bZ}\left(\ker\,B^{(1)},\uj\right)\,,
\qqq
and hence an element in the image of the left homomorphism. The
group on the left of \Reqref{cohoESB(1)} is naturally identified
with the kernel of the homomorphism
$\,\D_Q^{(1)}:H^1\left(M^{(1)},\uj\right)\rightarrow
H^1\left(Q^{(1)},\uj\right)$. \,For $\,[D']\in
H^1\left(M^{(1)},\uj\right)\,$ in that kernel and for a 1-cycle
$\,c_0^{T^{(1)}}\,$ such that $\,\Delta^{T^{(1)}}c_0^{T^{(1)}}
=\partial\hspace{0.02cm}c_1^{Q^{(1)}}$,
\qq\nn
\big\langle c_0^{T^{(1)}},(B^{(1)})^*[D']\big\rangle
\,=\,\big\langle B^{(1)}[c_0^{T^{(1)}}],[D']\big\rangle\,=\,
\big\langle\Delta^{Q^{(1)}}c_1^{Q^{(1)}},D'\big\rangle\,=\,
\big\langle\partial c_1^{Q^{(1)}},d'\big\rangle\,,
\qqq
where the expression on the right-hand side is identified with an
element of $\,\uj\,$ using the fact that $\,\Delta^{Q^{(1)}}\partial
c_1^{Q^{(1)}}=0$. \,But then
\qq
\big\langle\partial c_1^{Q^{(1)}},d'\big\rangle\,=\,
\big\langle c_0^{T^{(1)}},\D_T^{(1)}d'\big\rangle\,=\,
\big\langle c_0^{T^{(1)}},(d_n')\big\rangle\,,
\qqq
where the right-hand side provides an identification of the middle
term with an element of $\,\uj$. \,It then follows that by an
allowed modification of functions $\,(d_n)$,\ we may achieve that
\qq\nn
\big\langle[c_0^{T^{(1)}}],(d_n)\big\rangle\,=\,1
\qqq
for all classes $\,[c_0^{T^{(1)}}]\in\ker\,\Delta^{T^{(1)}}$. \,The
exact cohomology sequence
\qq\nn
H^0\left(Q^{(1)},\uj\right)\ \xrightarrow{\ \D_T^{(1)}\ }\
H^0\left(T^{(1)},\uj\right)\ \longrightarrow\
\mor_{\bZ}\left(\ker\,\Delta^{T^{(1)}},\uj\right)
\qqq
implies, in turn, that a further modification of the 2-isomorphism
$\,\Xi\,$ by locally constant $\,\uj$-valued functions on
$\,Q^{(1)}\,$ allows to get rid of functions $\,(d_n)\,$ altogether.
This completes the proof of the ``only if'' part of Theorem
\ref{thm:equiv-12iso-constr}.

\section{Groupoids and algebroids}\label{app:oid}

\noindent In this appendix, we recall several basic concepts from the theory
of (Lie) groupoids and algebroids, of relevance to the discussion
presented in the main text. The interested reader is urged to
consult the literature on the subject, e.g.,
Refs.\,\cite{MacKenzie:1987,Moerdijk:2003mm}.

We begin with
\bedef\label{def:grpd}
A \textbf{groupoid} is the septuple $\,\Gr=(\obj\,\Gr,\morf\,\Gr
,s,t,\Id,\Inv,\circ)\,$ composed of a pair of sets: the
\textbf{object set} $\,\obj\,\Gr\,$ and the \textbf{arrow set}
$\,\morf\,\Gr$,\ and a quintuple of \textbf{structure maps}: the
\textbf{source map} $\,s:\morf\,\Gr\to\obj\,\Gr\,$ and the
\textbf{target map} $\,t:\morf\,\Gr\to\obj\,\Gr$,\ the \textbf{unit
map} $\,\Id:\obj\,\Gr\to \morf\,\Gr:m\mapsto\Id_m$,\ the
\textbf{inverse map} $\,\Inv:\morf\,\Gr\to \morf\,\Gr:
\overrightarrow g\mapsto\overrightarrow g^{-1}\equiv\Inv(
\overrightarrow g)$,\ and the \textbf{multiplication map} $\,\circ:
\morf\,\Gr{}_s\hspace{-3pt}\x_t\hspace{-1pt}\morf\,\Gr\to\morf\,\Gr
:(\overrightarrow g,\overrightarrow h)\mapsto\overrightarrow g\circ
\overrightarrow h$.\ Here,
$\,\morf\,\Gr{}_s\hspace{-3pt}\x_t\hspace{-1pt}\morf\,\Gr\,$ is the
subset of composable morphisms,
\qq\nn
\morf\,\Gr{}_s\hspace{-3pt}\x_t\hspace{-1pt}\morf\,\Gr=\{\
(\overrightarrow g, \overrightarrow h)\in\morf\,\Gr\x\morf\,\Gr
\quad\vert\quad s(\overrightarrow g)=t(\overrightarrow h) \ \}\,.
\qqq
The structure maps satisfy the consistency conditions (whenever the
expressions are well-defined):
\bit
\item[(i)] $s(\overrightarrow g\circ\overrightarrow h)=s(
\overrightarrow h),\ t(\overrightarrow g\circ\overrightarrow h)=t(
\overrightarrow g)$;
\item[(ii)] $(\overrightarrow g\circ\overrightarrow h)\circ
\overrightarrow k=\overrightarrow g\circ(\overrightarrow h\circ
\overrightarrow k)$;
\item[(iii)] $\Id_{t(\overrightarrow g)}\circ\overrightarrow g=
\overrightarrow g=\overrightarrow g\circ\Id_{s(\overrightarrow g)}$
;
\item[(iv)] $s(\overrightarrow g^{-1})=t(\overrightarrow g),\ t(
\overrightarrow g^{-1})=s(\overrightarrow g),\ \overrightarrow g
\circ\overrightarrow g^{-1}=\Id_{t(\overrightarrow g)},\
\overrightarrow g^{-1}\circ\overrightarrow g=\Id_{s(\overrightarrow
g)}$.
\eit
Thus, a groupoid is a (small) category with all morphisms
invertible.

A \textbf{morphism} between two groupoids $\,\Gr_i,\ i=1,2\,$ is a
functor $\,\Phi:\Gr_1\to\Gr_2$.

A \textbf{Lie groupoid} is a groupoid whose object and arrow sets
are smooth manifolds, whose structure maps are smooth, and whose
source and target maps are surjective submersions. A morphism
between two Lie groupoids is a functor between them with smooth
object and morphism components. \exdef We also have
\bedef
Let $\,\xcM\,$ be a smooth manifold. A \textbf{Lie algebroid} over
the \textbf{base} $\,\xcM\,$ is a triple $\,\gtGr=\bigl(V,[\cdot,
\cdot],\a_{\sfT\xcM}\bigr)\,$ composed of a vector bundle $\,\pi_V:V
\to\xcM$,\ a Lie bracket $\,[\cdot,\cdot]\,$ on the vector space
$\,\G(V)\,$ of its sections, and a bundle map $\,\a_{\sfT\xcM}
:V\to\sfT\xcM$,\ termed the \textbf{anchor (map)}. These are
required to have the following properties:
\bit
\item[(i)] the induced map $\,\G(\a_{\sfT\xcM}):\G(V)\to\G(\sfT\xcM
)$, to be denoted by the same symbol $\,\a_{\sfT\xcM}\,$ below, is a
Lie-algebra homomorphism;
\item[(ii)] $[\,\cdot\,,\,\cdot\,]\,$ obeys the \textbf{Leibniz
identity}
\qq\nn
[X,f\,Y]=f\,[X,Y]+\G(\a_{\sfT\xcM})(X)(f)\,Y
\qqq
for all $\,X,Y\in\G(V)\,$ and any $\,f\in C^\infty(\xcM,\bR)$.
\eit

A \textbf{morphism} between two Lie algebroids $\,\gtGr_i=\bigl(V_i,
[\cdot,\cdot]_i,\a_{\sfT\xcM\,i}\bigr),\ i=1,2\,$ (over the same
base $\,\xcM$) is a bundle map $\,\phi:V_1\to V_2\,$ that satisfies
the relations
\qq\nn
\a_{\sfT\xcM\,1}=\a_{\sfT\xcM\,2}\circ\phi\,,\qquad\qquad [\cdot,
\cdot]_1=[\cdot,\cdot]_2\circ(\phi\x\phi)\,.
\qqq
\exdef Furthermore, we shall need
\bedef\label{def:tan-alg}
Let $\,\Gr=(\obj\,\Gr,\morf\,\Gr,s,t,\Id,\Inv,\circ)\,$ be a Lie
groupoid. Denote by $\,\sfd s\,$ and $\,\sfd t\,$ the tangents of
the source map and the target map, respectively, and by $\,\sfd
R_{\overrightarrow g},\ \overrightarrow g\in\morf\,\Gr\,$ the
tangent of the smooth map
\qq\nn
R_{\overrightarrow g}\ :\ s^{-1}(\{t(\overrightarrow g)\})\to s^{-1}
(\{s(\overrightarrow g)\})\ :\ \overrightarrow h\mapsto
R_{\overrightarrow g}(\overrightarrow h):=\overrightarrow h\circ
\overrightarrow g\,.
\qqq
Furthermore, let $\,\Xgt^s_{\rm inv}(\morf\,\Gr)\,$ denote the
vector space of \textbf{right $\Gr$-invariant vector fields} on
$\,\morf\,\Gr$,\ given by
\qq\nn
\Xgt^s_{\rm inv}(\morf\,\Gr)=\{\ \xcV\in\G\bigl(\ker\,\sfd s\bigr)
\quad\vert\quad \sfd R_\cdot(\xcV)=\xcV \ \}\,.
\qqq
The \textbf{tangent algebroid} of $\,\Gr\,$ is the Lie algebroid
$\,\gtgr=\bigl(\Id^*\ker\,\sfd s,[\,\cdot\,,\,\cdot\,],\a_{\sfT
(\obj\,\Gr)}\bigr)\,$ over $\,\obj\,\Gr\,$ with the anchor
$\,\a_{\sfT(\obj\,\Gr)}\,$ inducing the map $\,\sfd t\circ i\,$
between spaces of sections, defined in terms of the canonical
vector-space isomorphism
\qq\nn
i\ :\ \G(\Id^*\ker\,\sfd s)\xrightarrow{\ \cong\ }\Xgt^s_{\rm inv}(
\morf\,\Gr)\,,
\qqq
and with the Lie bracket given by the unique bracket on $\,\G(\Id^*
\ker\,\sfd s)\,$ for which $\,i\,$ is an isomorphism of Lie
algebras.\exdef

\section{A proof of Theorem \ref{thm:gtAlgebroid}}
\label{app:proof-algebroid}

\noindent By way of proof, we present a detailed reconstruction of the tangent
algebroid $\,\ggt\lx\xcF\,$ along the lines of the abstract
definition \ref{def:tan-alg}, specialising the general concepts to
the case of interest, $\,\Gr=\txS_\Bgt\equiv\txG\lx\xcF$.\ We
commence by extracting the vector sub-bundle
\qq\nn
\sfT^s(\txG\x\xcF):=\ker\,\sfd s\subset\sfT(\txG\x\xcF)
\qqq
of the tangent bundle $\,\pi\ :\ \sfT(\txG\x\xcF)\to\txG\x\xcF$,\
defined in terms of the tangent map $\,\sfd s\,$ and hence spanned
by vector fields tangent to the $s$-fibres in $\,\txG\x\xcF$.\ In
the case in hand, these vector fields are given by
$C^\infty(\txG\x\xcF,\bR )$-linear combinations
$\,f^i(g,m)\,\xcV_i(g)\,$ of vector fields on $\,\txG$.\ On sections
$\,\Xgt^s(\txG\x\xcF):=\G\bigl(\sfT^s(\txG\x \xcF)\bigr)$,\ we have
the natural (right) fibre-wise $\txG\lx \xcF$-action
\qq\nn
\rho\ :\ \Xgt^s(\txG\x\xcF){}_{\mu_\pi}\hspace{-3pt}\x_t
\hspace{-1pt}(\txG\x\xcF)\to\Xgt^s(\txG\x\xcF)\ :\ (\xcV,
\overrightarrow g)\mapsto\sfd R_{\overrightarrow g}(\xcV)=:
\rho_{\overrightarrow g}(\xcV)\equiv\xcV.\overrightarrow g
\qqq
with momentum
\qq\nn
\mu_\pi:=s\circ\pi\,,
\qqq
defined in terms of the tangent $\,\sfd R\,$ of the right action
\qq\nn
R\ :\ (\txG\x\xcF){}_{\mu_R}\hspace{-3pt}\x_t\hspace{-1pt}(\txG\x
\xcF)\to\txG\x\xcF\ :\ \bigl(\overrightarrow g,\overrightarrow h
\bigr)\mapsto\overrightarrow g\circ\overrightarrow h=:
R_{\overrightarrow h}(\overrightarrow g)
\qqq
with momentum
\qq\nn
\mu_R:=s\,.
\qqq
Above, and in what follows, we employ
\becon
Given a pair $\,f_i: \xcM_i\mapsto\xcN,\ i=1,2\,$ of maps from the
respective domains $\,\xcM_i\,$ to the common codomain $\,\xcN$,\
the symbol $\,\xcM{}_{f_1}\hspace{-3pt}\x_{f_2}
\hspace{-1pt}\xcM_2\,$ denotes the fibred product
\qq\nn
\xcM_1{}_{f_1}\hspace{-3pt}\x_{f_2}\hspace{-1pt}\xcM_2=\{\ (m_1,m_2)
\in\xcM_1\x\xcM_2 \quad \vert \quad f_1(m_1)=f_2(m_2)\ \}\,.
\qqq
\econ \noindent The restriction to the sub-bundle $\,\sfT^s(\txG\x
\xcF)\,$ ensures that $\,\rho\,$ covers the right action $\,R\,$ on
the base,
\qq\nn
\alxydim{@C=1.5cm@R=1.cm}{\sfT^s(\txG\x\xcF){}_{\pi^*\mu_R}
\hspace{-3pt}\x_t\hspace{-1pt}(\txG\x\xcF) \ar[r]^{\hspace{1cm}\rho}
\ar[d]_{(\pi,\Id)} & \sfT^s(\txG\x\xcF) \ar[d]^{\pi}\cr
(\txG\x\xcF){}_{\mu_R} \hspace{-3pt}\x_t\hspace{-1pt}(\txG\x\xcF)
\ar[r]^{\hspace{.8cm}R} & \txG\x\xcF}\,.
\qqq
Among sections $\,\Xgt^s(\txG\x\xcF)$,\ we subsequently distinguish
those, denoted jointly by $\,\Xgt^s_{\rm inv}(\txG\x\xcF)\equiv\G
\bigl(\sfT^s_{\rm inv}(\txG\x\xcF)\bigr)$,\ which are invariant
under $\,\rho$,\ i.e.\ such that
\qq\nn
\sfd R_{\overrightarrow h}(\xcV)(\overrightarrow
g\circ\overrightarrow h)\equiv\xcV(\overrightarrow
g).\overrightarrow h=\xcV(\overrightarrow g\circ\overrightarrow
h)\,.
\qqq
They close to a Lie subalgebra of the Lie algebra $\,\Xgt(\txG\x\xcF
)\,$ of vector fields on $\,\txG\x\xcF\,$ since
\qq
\sfd R_{\overrightarrow h}\left([\xcV,\xcW]\right)=[\sfd
R_{\overrightarrow h}(\xcV),\sfd R_{\overrightarrow h}(\xcW)
]=[\xcV,\xcW]\,, \label{eq:Rinv-subalg}
\qqq
and the subalgebra homomorphically projects, along $\,\sfd t$,\ to
the Lie algebra $\,\Xgt(\xcF)\,$ of vector fields on $\,\xcF$,\ a
straightforward consequence of the previous property and of the
identity
\qq\label{eq:dt-proj}\qquad\qquad
\sfd t\bigl(\xcV(\overrightarrow g)\bigr)=\sfd t\bigl(\sfd
R_{\overrightarrow g}\bigl(\xcV (\Id_{t(\overrightarrow g)})\bigr)
\bigr)=\sfd(t\circ R_{\overrightarrow g})\bigl(\xcV(\Id_{t(
\overrightarrow g)})\bigr)=\sfd t\bigl(\xcV(\Id_{t(\overrightarrow
g)})\bigr)\,.
\qqq
The subalgebra may be reconstructed as follows: Represent an
arbitrary right-invariant vector field as a
$C^\infty(\txG\x\xcF,\bR)$-linear combination of the basic
right-invariant vector fields $\,R_a,\ a=1,2,\ldots,\dim\,\ggt\,$ on
$\,\txG\,$ as discussed above. We readily establish the defining
property of the functional coefficients in this decomposition of a
\emph{right-invariant} vector field:
\qq\nn
f^i(g\cdot h,h^{-1}.m)=f^i(g,m)\qquad\Longrightarrow\qquad f^i(g,m)
=f^i(e,g.m)\,,
\qqq
which shows that each $\,f^i\,$ factors through a function $\,\ovl
f^i=f^i\circ\Id\in C^\infty(\xcF,\bR)\,$ as
\qq\nn
f^i=\ovl f^i\circ t\,.
\qqq
Thus, altogether, a right-invariant vector field on $\,\txG\x\xcF\,$
can be written in the form
\qq\label{eq:ffactt}
\xcV=(\upsilon^a\circ t)\,(R_a\circ\pr_1)\,,\qquad\qquad\upsilon^a
\in C^\infty(\xcF,\bR)\,,\qquad\pr_1\ :\ \txG\x\xcF\to\txG\,.
\qqq
This explicitly demonstrates the isomorphism of vector spaces
\qq\nn
\Xgt^s_{\rm inv}(\txG\x\xcF)\cong\G\bigl(\Id^*\sfT^s(\txG\x\xcF)
\bigr)\,.
\qqq
Indeed, in general, every right-invariant vector field on $\,\txG\x
\xcF\,$ defines a vector field on $\,\Id(\xcF)\,$ via restriction,
and, conversely, every vector field $\,\xcV\in\G\bigl(\Id^*\sfT^s(
\txG\x\xcF)\bigr)\,$ extends to a unique right-invariant vector
field $\,\overrightarrow\xcV\,$ on the arrow manifold as
\qq\nn
\overrightarrow\xcV(\overrightarrow g):=\xcV(\Id_{t(\overrightarrow
g)}).\overrightarrow g\,.
\qqq
The bijectivity of this correspondence follows from the fact that a
right-invariant vector field on $\,\txG\x\xcF\,$ is completely
determined by its restriction to $\,\Id(\xcF)$.\ Note also that the
correspondence is compatible with the $C^\infty(\xcF,\bR)$-module
structure on both vector spaces as
\qq\nn
\overrightarrow{f\,\xcV}=(f\circ t)\,\overrightarrow\xcV\,,\qquad
f\in C^\infty(\xcF,\bR)\,,
\qqq
cf.\ \Reqref{eq:ffactt}. The vector bundle
\qq\nn
\Id^*\sfT^s_{\rm inv}(\txG\x\xcF)\to\xcF
\qqq
is the first ingredient of the Lie algebroid under reconstruction.
By virtue of \Reqref{eq:ffactt}, we obtain, in the case of interest,
\qq\nn
\Id^*\sfT^s_{\rm inv}(\txG\x\xcF)\cong\ggt\x\xcF\,.
\qqq

The next step consists in defining the anchor map
\qq\nn
\a_{\sfT\xcF}\ :\ \Id^*\sfT^s_{\rm inv}(\txG\x\xcF)\to\sfT\xcF
\qqq
that induces a Lie-algebra homomorphism between the Lie-bracket
algebras of sections. Comparison with
Eqs.\,\eqref{eq:Rinv-subalg}-\eqref{eq:dt-proj} immediately suggests
the choice
\qq\nn
\a_{\sfT\xcF}(\xcV)=\sfd t(\overrightarrow\xcV)\,.
\qqq
We find, for the restriction $\,\xcR_a\,$ of the right-invariant
vector field $\,R_a\circ\pr_1$,
\qq\nn
\bigl(\a_{\sfT\xcF}(\xcR_a)\,(f)\bigr)(m)&\equiv&t^*\bigl(\a_{\sfT\xcF}(
\xcR_a)(f)\bigr)(e,m)=t^*\bigl(\ic_{t_*R_a}\sfd
f\bigr)(e,m)=\bigl(\ic_{R_a}\sfd(t^*f)\bigr)(e,m)\cr\cr
&=&\tfrac{\sfd\ }{\sfd s}\big\vert_{s=0}(t^*f)\bigl(\ee^{-s\,t_a}
\cdot e,m\bigr)\equiv\tfrac{\sfd\ }{\sfd s}\big\vert_{s=0}f\bigl(
\ee^{-s\,t_a}.m\bigr)=\bigl(\xcFup\xcK_a\,(f)\bigr)(m)\,,
\qqq
or simply
\qq\label{eq:Ka-on-M}
\sfd t(R_a\circ\pr_1)=\a_{\sfT\xcF}(\xcR_a)=\xcFup\xcK_a\,,
\qqq
with $\,\xcFup\xcK_a,\ a=1,2,\ldots,\dim\,\ggt\,$ generating the
infinitesimal action of $\,\txG\,$ on $\,C^\infty(\xcF,\bR)$,\ as in
Definition \ref{eq:Gact-M}. This fixes the anchor map
completely.

At this stage, it remains to identify the unique Lie bracket on
$\,\G(\ggt\x\xcF)\equiv C^\infty(M,\ggt)\,$ with the Leibniz
property for which $\,\Xgt^s_{\rm
inv}(\txG\x\xcF)\cong\G(\ggt\x\xcF)\,$ is a Lie-algebra isomorphism,
that is the unique bracket such that
\qq\nn
\overrightarrow{[\xcV,f\,\xcW]}_{\ggt\lx\xcF}&=&[\overrightarrow
\xcV,\overrightarrow{f\,\xcW}]=[\overrightarrow\xcV,(f\circ t)\,
\overrightarrow\xcW]=(f\circ t)\,[\overrightarrow\xcV,
\overrightarrow\xcW]+\overrightarrow\xcV(f\circ t)\,\overrightarrow
\xcW\cr\cr
&=&(f\circ t)\,\overrightarrow{[\xcV,\xcW]}_{\ggt\lx\xcF}+\bigl(\sfd
t(\overrightarrow\xcV)(f)\circ t\bigr)\,\overrightarrow\xcW=
\overrightarrow{f\,[\xcV,\xcW]_{\ggt\lx\xcF}+\sfd t(
\overrightarrow{\xcV})(f)\,\xcW}\,,
\qqq
whence
\qq\nn
[\xcV,f\,\xcW]_{\ggt\lx\xcF}=f\,[\xcV,\xcW]_{\ggt\lx\xcF}+\a_{\sfT
\xcF}(\xcV)(f)\,\xcW\,.
\qqq
Using the last result, in conjunction with the obvious implication
\qq\nn
\overrightarrow{[\xcR_a,\xcR_b]}_{\ggt\lx\xcF}=[R_a,R_b]\circ\pr_1=
f_{abc}\,\overrightarrow\xcR_c\qquad\Longrightarrow\qquad[\xcR_a,
\xcR_b]_{\ggt\lx\xcF}=f_{abc}\,\xcR_c\,,
\qqq
we ultimately derive the general expression
\qq\nn
[\la^a\,\xcR_a,\mu^b\,\xcR_b]_{\ggt\lx\xcF}=f_{abc}\,\la^a\,\mu^b\,
\xcR_c+\bigl(\pLie{\la^a\,\xcK_a}\mu^B-\pLie{\mu^b\,\xcK_a}\la^b
\bigr)\,\xcR_b\,.
\qqq

Thus, after the dust has cleared, we find the full-fledged structure
of the tangent algebroid
\qq\nn
\ggt\lx\xcF=\bigl(\oplus_{a=1}^{\dim\,\ggt}\,C^\infty(\xcF,\bR)\,
\xcR_a,[\cdot,\cdot]_{\ggt\lx\xcF},\a_{\sfT\xcF}\bigr)\,.
\qqq
of the action groupoid $\,\txG\lx\xcF$,\ also termed the
\textbf{action algebroid}. It is now a matter of a straightforward
calculation to check that the map
\qq\label{eq:subcour-lie}
\iota\ :\ \Sgt_\Bgt\to\ggt\lx\xcF\ :\ \la^a\,\Kgt_a\mapsto\la^a\,
\xcR_a
\qqq
is an isomorphism of Lie algebroids over the target space.

\section{A proof of Proposition \ref{prop:HS-as-gequiv-str}}
\label{app:proof-gequiv-target}

We first compute
\qq\nn
D_{3}\pLie{a}b=\pLie{a}(\txH\vert_{\cO^M_i},0,0 ,0,0)=0\,,
\qqq
whereby it becomes clear that condition \eqref{eq:HS-exact} lifts
the obstruction to trivialising the 3-cocycle $\,\pLie{a} b\,$ in
the Deligne hypercohomology. Indeed, we have
\qq\nn
\pLie{a}B_i&=&\ic_a\txH\vert_{\cO^M_i}+\sfd(\ic_a B_i)=
\sfd(-\kappa_a\vert_{\cO^M_i}+\ic_a B_i)\,,\cr\cr
\pLie{a}A_{ij}&=&\ic_a(B_j-B_i)\vert_{\cO^M_{ij}} +\sfd(\ic_a
A_{ij})\,,\cr\cr
\pLie{a}h_{ijk}&=&\ic_a(-A_{jk}+A_{ik}-A_{ij})\vert_{\cO^M_{ijk}}\,,\cr\cr
\pLie{a}s_{ijkl}&=&0\,,
\qqq
and so
\qq\nn
\pLie{a}b=D_{2}\Upsilon_a\,.
\qqq
The last identity, in conjunction with \Reqref{eq:gerbe-loc-bis},
implies, in turn,
\qq\nn
D_{2}\bigl(\pLie{a}\Upsilon_b-\pLie{b}
\Upsilon_a\bigr)=[\,\pLie{a}\,,\,\pLie{b}\,]b=
f_{abc}\,\pLie{c}b=D_{2}(f_{abc}\,\Upsilon_c)\,,
\qqq
or, simply,
\qq\nn
\pLie{a}\Upsilon_b-\pLie{b}\Upsilon_a-f_{abc}\,
\Upsilon_c\in\ker\,D_{2}\,.
\qqq
This time, it is condition \eqref{eq:HS1-ids-triv} that turns the
latter 2-cocycle into a 2-coboundary. We compute
\qq\nn
\pLie{a}(-\kappa_b\vert_{\cO^M_i}+\ic_b B_i)=f_{abc}\,(-\kappa_c
\vert_{\cO^M_i}+\ic_c B_i)+\ic_b\pLie{a}B_i
\qqq
and
\qq\nn
\pLie{b}(-\kappa_a\vert_{\cO^M_i}+\ic_a B_i)&=&f_{abc}\,\kappa_c
\vert_{\cO^M_i}+\sfd(\ic_b\ic_a B_i)+\ic_b\sfd(\ic_a B_i)\cr\cr
&=&f_{abc}\,\kappa_c\vert_{\cO^M_i}+\sfd(\ic_b\ic_a B_i)+\ic_b
\pLie{a}B_i+\ic_b\sfd\kappa_a\vert_{\cO^M_i}\cr\cr
&=&\ic_b\pLie{a}B_i+\sfd\bigl(-(\ic_b\kappa_a)\vert_{\cO^M_i}+
\ic_b\ic_a B_i\bigr)\,,
\qqq
whence
\qq\nn
\pLie{a}(-\kappa_b\vert_{\cO^M_i}+\ic_b B_i)-
\pLie{b}(-\kappa_a\vert_{\cO^M_i}+\ic_a B_i)\cr\cr
=f_{abc}\,(-\kappa_c\vert_{\cO^M_i}+\ic_c B_i)+\sfd(\txc_{ba}
\vert_{\cO^M_i}+\ic_a\ic_b B_i)\,.
\qqq
Similarly, we find
\qq\nn
\pLie{a}(\ic_b A_{ij})-\pLie{b}(\ic_a A_{ij})=f_{abc}\,\ic_c A_{ij}
+\ic_b\ic_a(B_j-B_i)\vert_{\cO^M_{ij}}\,,
\qqq
so that, altogether,
\qq\nn
\pLie{a}\Upsilon_B-\pLie{b}\Upsilon_a=f_{abc}\,
\Upsilon_c+D_{1}\g_{ab}\,.
\qqq
In the last step of this descent, we establish, with the help of the
Jacobi identity for the structure constants $\,f_{abc}$,
\qq\nn
&&D_{1}(\pLie{a}\g_{bc}-\pLie{b}\g_{ac}+
\pLie{c}\g_{ab})\cr\cr
&=&\pLie{a}(\pLie{b}\Upsilon_c-\pLie{c}
\Upsilon_b)-\pLie{b}(\pLie{a}\Upsilon_c-
\pLie{c}\Upsilon_a)+\pLie{c}(\pLie{a}
\Upsilon_b-\pLie{b}\Upsilon_a)\cr\cr
&&-f_{bcd}\,\pLie{a}\Upsilon_d+f_{acd}\,\pLie{b}
\Upsilon_d-f_{abd}\,\pLie{c}\Upsilon_d\cr\cr
&=&f_{abd}\,(\pLie{d}\Upsilon_c-\pLie{c}
\Upsilon_d)-f_{acd}\,(\pLie{d}\Upsilon_b-\pLie{b}
\Upsilon_d)+f_{bcd}\,(\pLie{d}\Upsilon_a-\pLie{a} \Upsilon_d)\cr\cr
&=&D_{1}(f_{abd}\,\g_{dc}-f_{acd}\,\g_{db}+f_{bcd}\,\g_{da})\,,
\qqq
and so we may seek to trivialise the 1-cocycle
\qq\nn
\pLie{a}\g_{bc}-\pLie{b}\g_{ac}+\pLie{c}
\g_{ab}-f_{abd}\,\g_{dc}+f_{acd}\,\g_{db}-f_{bcd}\,\g_{da}\in\ker\,
D_{1}\,.
\qqq
A direct computation, using the shorthand notation $\,Y_{a,i}:=-
\kappa_a\vert_{\cO^M_i}+\ic_a B_i\,$ alongside the relations
\qq\nn
g_{ab,i}=-\ic_b Y_{a,i}\,,\qquad\qquad g_{ab,i}+g_{ba,i}=2
c_{(ab),i}
\qqq
satisfied by $\,g_{ab,i}:=\txc_{ba}\vert_{\cO^M_i}+\ic_a\ic_b B_i\,$
with
\qq\nn
c_{(ab),i}=\tfrac{1}{2}\,(\txc_{ab}+\txc_{ba})\vert_{\cO^M_i}\,,
\qquad\qquad\sfd c_{(ab),i}=0\,,
\qqq
yields
\qq\label{eq:g-equiv-gerbe-last}
\pLie{a}g_{bc,i}-\pLie{b}g_{ac,i}+\pLie{c}g_{ab,i}\\\nonumber\cr
=\ic_a\sfd g_{bc,i}-\ic_b\sfd
g_{ac,i}+\ic_c(\pLie{a}Y_{b,i}-\pLie{b}Y_{a,i}-f_{abd}\,Y_{d,i})
\cr\cr =\ic_a\sfd g_{bc,i}-\ic_b\sfd g_{ac,i}+\ic_c\pLie{a}
Y_{b,i}-\ic_b\sfd(\ic_c Y_{a,i})\cr\cr +f_{bcd}\,\ic_d
Y_{a,i}-f_{abd}\,\ic_c Y_{d,i}
\cr\cr =\ic_a\sfd g_{bc,i}-\ic_b(\pLie{a}Y_{c,i}-f_{acd}
\,Y_{d,i})+\ic_c\pLie{a} Y_{b,i}+\ic_b\ic_c\sfd
Y_{a,i}\cr\cr +f_{bcd}\,\ic_d Y_{a,i}-f_{abd}\,\ic_c Y_{d,i}
\cr\cr =\ic_a\sfd(\ic_b Y_{c,i})-\pLie{a}(\ic_b Y_{c,i})
+\ic_c\pLie{a}Y_{b ,i}+\ic_b\ic_c\sfd Y_{a,i}\cr\cr
+f_{abd}\,\ic_d Y_{c,i}+f_{acd}\,\ic_b Y_{d,i}+f_{bcd}\,\ic_d
Y_{a,i}-f_{abd}\,\ic_c Y_{d,i}\cr\cr
=f_{abd}\,g_{dc,i}-f_{acd}\,g_{db,i}+f_{bcd}\,g_{da,i}-2f_{abd}\,
c_{(dc),i}-2f_{bcd}\,c_{(da),i}\cr\cr
+\ic_c\pLie{a}Y_{b,i}+\ic_b\ic_c\sfd Y_{a,i}-f_{abd}\,\ic_c
Y_{d,i}\,.\nonumber
\qqq
Taking a closer look at the expression in the last line, we find
\qq\nn
\ic_c\pLie{a}Y_{b,i}+\ic_b\ic_c \sfd Y_{a,i}-f_{abd}\,\ic_c
Y_{d,i}\cr\cr =-f_{abd}\,\txc_{cd}\vert_{\cO^M_i}+\ic_c\pLie{a}
(\ic_b B_i)+\ic_b\ic_c\ic_a\txH \vert_{\cO^M_i}\cr\cr
+\ic_b\ic_c\sfd(\ic_a
B_i)-f_{abd}\,\ic_c(-\kappa_d\vert_{\cO^M_i}+\ic_d B_i)\cr\cr
=\ic_c\ic_b\pLie{a}B_i+\ic_b
\ic_c\ic_a\txH\vert_{\cO^M_i}+\ic_b\ic_c\sfd(\ic_a B_i)=0\,.
\qqq
Next, upon using the identity
\qq\nn
f_{bcd}\,c_{(da),i}+f_{bad}\,c_{(dc),i}=\pLie{b}c_{(ac),i}
\equiv\ic_b\sfd c_{(ac),i}=0\,,
\qqq
we ultimately reduce \Reqref{eq:g-equiv-gerbe-last} to the relation
\qq\nn
\pLie{a}g_{bc,i}-\pLie{b}g_{ac,i}+\pLie{c}g_{ab,i}\cr\cr
=f_{abd}\,g_{dc,i}-f_{acd}\,g_{db,i}+f_{bcd}\,g_{da,i}-4f_{abd}\,
c_{(dc),i}\,,
\qqq
which explicitly identifies the (local) constants $\,c_{(ab),i}\,$
as the last obstruction to the $\ggt$-equivariance of the gerbe
$\,\cG\,$ and thus concludes the proof.

\section{A proof of Proposition \ref{prop:FFM-as-gequiv-str}}
\label{app:proof-gequiv-bib}

\noindent First of all, we have the identity
\qq\nn
D_{2}\pLie{a}p=\check\D_Q\pLie{a}b+\pLie{a}\ovl\om
=D_{2}\check\D_Q\Upsilon_a\,,
\qqq
and the obstruction to trivialising the 2-cocycle
\qq\nn
\pLie{a}p-\check\D_Q\Upsilon_a\in\ker\,D_{2}
\qqq
is lifted by imposing \Reqref{eq:bdry-exact}. Indeed, we have
\qq\nn
\pLie{a}P_i&=&\sfd(\ic_a P_i)+\check\D_Q(\ic_a
B_i)+\ic_a\om\vert_{\cO^Q_i}=\sfd(-k_a+\ic_a P_i)+\check\D_Q
Y_{a,i}\cr\cr \pLie{a}k_{ij}&=&\ic_a\sfd k_{ij}=\ic_a
(P_i-P_j)\vert_{\cO^Q_{ij}}+\check\D_Q(\ic_a A_{ij})\cr\cr
\pLie{a}r_{ijk}&=&0\,,
\qqq
and so we obtain
\qq\nn
\pLie{a}p=\check\D_Q\Upsilon_a+D_{1}\Xi_a\,.
\qqq
Similarly, given
\qq\nn
D_{1}(\pLie{a}\Xi_b-\pLie{b}\Xi_a)&=&[\,
\pLie{a}\,,\,\pLie{b}\,]p-\check\D_Q(\pLie{a}\Upsilon_b-\pLie{b}\Upsilon_a)
\cr\cr
&=&f_{abc}\,(\pLie{c}p-\check\D_Q\Upsilon_c)-D_{1}\check
\D_Q\g_{ab}\cr\cr
&=&D_{1}(f_{abc}\,\Xi_c-\check\D_Q\g_{ab})\,,
\qqq
we can employ \Reqref{eq:FFM-ids-triv} to trivialise the 1-cocycle
\qq\nn
\pLie{a}\Xi_b-\pLie{b}\Xi_a-f_{abc}\,\Xi_c+\check
\D_Q\g_{ab}\in\ker\,D_{1}\,,
\qqq
to the effect
\qq\nn
\pLie{a}(-k_b\vert_{\cO^Q_i}+\ic_b
P_i)-\pLie{b}(-k_a\vert_{\cO^Q_i}+\ic_a P_i)\cr\cr
=f_{abc}\,(-k_c\vert_{\cO^Q_i}+\ic_c P_i)+\ic_b
\pLie{a}P_i+\pLie{b}k_a\vert_{\cO^Q_i}-\ic_b\sfd(\ic_a P_i)\cr\cr
=f_{abc}\,(-k_c\vert_{\cO^Q_i}+\ic_c P_i)+\ic_b( \ic_a\sfd P_i+\sfd
k_a\vert_{\cO^Q_i})\cr\cr =f_{abc}\,(-k_c\vert_{\cO^Q_i}+\ic_c
P_i)-\check\D_Q g_{(ab),i}\,.
\qqq
This completes the proof of the proposition.

\section{A proof of Proposition \ref{prop:djc-as-gequiv-str}}
\label{app:proof-gequiv-ibb}

\noindent We have the identity
\qq\nn
D_{1}\pLie{a}h_n=-\check\D_{T_n}\pLie{a}p=-
\check\D_{T_n}\bigl(\check\D_Q\Upsilon_a+D_{1}\Xi_a\bigr)=-D_{1}
\check\D_{T_n}\Xi_a\,,
\qqq
whence
\qq\nn
\pLie{a}h_n+\check\D_{T_n}\Xi_a\in\ker\,D_{1}\,.
\qqq
Relation \eqref{eq:junct-exact} ensures the triviality of the latter
cocycle as it gives
\qq\nn
\pLie{a}f_{n\,i}&=&-\ic_a\check\D_{T_n}P_i=-
\check\D_{T_n}\bigl(\ic_a P_i\bigr)=-\check\D_{T_n}\Xi_a\,,
\cr\cr
\pLie{a}q_{n\,ij}&=&0\,,
\qqq
and so, indeed,
\qq\nn
\pLie{a}h_n=-\check\D_{T_n}\Xi_a\,,
\qqq
as claimed.

\section{Commuting diagrams  for $\txG$-equivariant string
backgrounds}\label{app:Gequiv-diags}

\noindent In the appendix, we have gathered the (rather heavily decorated)
2-diagrams whose commutativity expresses the various coherence
conditions imposed on elements of a $\txG$-equivariant string
background.
\smallskip

Thus, we have a diagrammatic representation of
\bit
\item \Reqref{eq:Gerbe-1iso-coh}
\qq
\xy (40,0)*{\bullet}="34"+(2,3)*{\tx{\tiny $(\xcMup d^{(1)}_1\circ
\xcMup d^{(2)}_0\circ\xcMup d^{(3)}_1)^*\cG\ox I_{(\xcMup d^{(2)}_2
\circ\xcMup d^{(3)}_1)^*\rho}$}};
(0,-50)*{\bullet}="1234"+(2,-5)*{\tx{\tiny $(\xcMup d^{(1)}_1\circ
\xcMup d^{(2)}_1\circ\xcMup d^{(3)}_1)^*\cG$}};
(65,-50)*{\bullet}="4"+(10,-5)*{\tx{\tiny $(\xcMup d^{(1)}_0\circ
\xcMup d^{(2)}_1\circ\xcMup d^{(3)}_1)^*\cG\ox I_{(\xcMup d^{(2)}_1
\circ\xcMup d^{(3)}_1)^*\rho}$}};
(90,-35)*{\bullet}="234"+(15,5)*{\tx{\tiny $(\xcMup d^{(1)}_1\circ
\xcMup d^{(2)}_0\circ\xcMup d^{(3)}_3 )^*\cG\ox I_{(\xcMup d^{(2)}_2
\circ\xcMup d^{(3)}_3)^*\rho}$}}; (77,-41)*{}="234-4";
(38,-49)*{}="1234-4"; (35,-48)*{}="1234-4-bis";
(22,-25)*{}="1234-34"; (49,-38)*{\tx{\tiny $(\xcMup d^{(2)}_2\circ
\xcMup d^{(3)}_3)^*\Upsilon$}}; (48,-13)*{\tx{\tiny $(\xcMup d^{(2
)}_0\circ\xcMup d^{(3)}_1)^*\Upsilon\ox\id$}}; \ar@{->}^{\tx{\tiny
$(\xcMup d^{(2)}_2\circ\xcMup d^{(3)}_1)^*\Upsilon$}} "1234";"34"
\ar@{->}_{\tx{\tiny $(\xcMup d^{(2)}_1\circ\xcMup d^{(3)}_1)^*
\Upsilon$}} "1234";"4" \ar@{->} "34";"4" \ar@{->}^{\tx{\tiny
$(\xcMup d^{(2)}_1\circ\xcMup d^{(3)}_0)^*\Upsilon\ox\id$}}
"234";"4" \ar@{->}_(.4){\tx{\tiny $(\xcMup d^{(2)}_0\circ\xcMup
d^{(3 )}_3)^*\Upsilon\ox\id$}} "234";"34" \ar@{->} "1234";"234"
\ar@{=>}|{\tx{\tiny $\xcMup d^{(3)\,*}_0\g\ox\id$}}
"34"+(3,-3);"234-4" \ar@{==>}|{\tx{\tiny $\xcMup d^{(3)\,*}_2\g$}}
"234"+(-4,-1.5);"1234-4" \ar@{=>}|{\tx{\tiny $\xcMup d^{(3)\,*}_1
\g$}} "34"+(-.5,-3);"1234-4-bis" \ar@{==>}|{\tx{\tiny $\xcMup d^{(3
)\,*}_3\g$}} "234"+(-3,1);"1234-34"
\endxy\,.\cr\cr
\label{diag:Gerbe-1iso-coh}
\qqq
\item \Reqref{eq:coh-cond-equiv-1iso}
\qq
\xy (-25,-40)*{\bullet}="A"+(-25,-2)*{(\xcMup d^{(1)}_0\circ\xcMup
d^{(2)}_1)^*\cG_1\ox I_{\xcMup d^{(2)\,*}_1\rho_1}};
(25,-40)*{\bullet}="B"+(15,-2)*{(\xcMup d^{(1)}_1\circ\xcMup
d^{(2)}_1)^*\cG_1}; (10,-5)*{\bullet}="C"+(-5,5)*{(\xcMup
d^{(1)}_0\circ\xcMup d^{(2)}_2)^*\cG_1\ox I_{\xcMup d^{(2)\,*}_2
\rho_1}}; (-25,-100)*{\bullet}="X"+(-25,-2)*{(\xcMup d^{(1)}_0
\circ\xcMup d^{(2)}_1)^*\cG_2\ox I_{\xcMup d^{(2)\,*}_1\rho_2}};
(25,-100)*{\bullet}="Y"+(15,-2)*{(\xcMup d^{(1)}_1\circ\xcMup
d^{(2)}_1)^*\cG_2}; (10,-65)*{\bullet}="Z"+(25,3)*{(\xcMup
d^{(1)}_0\circ\xcMup d^{(2)}_2)^*\cG_2\ox I_{\xcMup
d^{(2)\,*}_2\rho_2}}; \ar@{->}|{\xcMup d^{(2)\,*}_1\Upsilon_1}
"B";"A" \ar@{->}|{\qquad\xcMup d^{(2)\,*}_2 \Upsilon_1} "B";"C"
\ar@{->}|{\xcMup d^{(2)\,*}_0\Upsilon_1\ox\id\qquad} "C";"A"
\ar@{->}|{\qquad\qquad\quad(\xcMup d^{(1)}_0\circ\xcMup
d^{(2)}_2)^*\Psi_{1,2}\ox\id} "C";"Z" \ar@{->}|{(\xcMup
d^{(1)}_1\circ \xcMup d^{(2)}_1)^*\Psi_{1,2}} "B";"Y"
\ar@{->}|{(\xcMup d^{(1)}_0\circ \xcMup
d^{(2)}_1)^*\Psi_{1,2}\ox\id} "A";"X" \ar@{->}|{\xcMup d^{(2)\,*}_1
\Upsilon_2} "Y";"X" \ar@{->}|{\xcMup d^{(2)\,*}_2\Upsilon_2} "Y";"Z"
\ar@{->}|{\xcMup d^{(2)\,*}_0\Upsilon_2\ox\id\qquad} "Z";"X"
\ar@{=>}|{\g_1} "C"+(-.5,-3);"A"+(23.5,3) \ar@{==>}|{\g_2}
"Z"+(-.5,-3);"X"+(23.5,3) \ar@{==>}|{\xcMup d^{(2)\,*}_2\eta_{1,2}}
"C"+(1,-5);"Y"+(-1,5) \ar@{=>}|{\xcMup d^{(2)\,*}_1\eta_{1,2}}
"A"+(3,-3);"Y"+(-3,3) \ar@{==>}|{\xcMup
d^{(2)\,*}_0\eta_{1,2}\ox\id} "A"+(3,-1.5);"Z"+(-3,1.5)
\endxy\,.\cr\cr
\label{diag:Gequiv1iso-coh}
\qqq
\item \Reqref{eq:Gequiv-2iso-coh}
\qq
\xy (30,-10)*{\bullet}="A"+(-9,0)*{\xcMup d_1^{(1)\,*}\cG_1};
(70,-10)*{\bullet}="B"+(9,0)*{\xcMup d_1^{(1)\,*}\cG_2};
(30,-50)*{\bullet}="C"+(-13,0)*{\xcMup d_0^{(1)\,*}\cG_1\ox I_\rho};
(70,-50)*{\bullet}="D"+(13,0)*{\xcMup d_0^{(1)\,*}\cG_2\ox I_\rho};
\ar@{->}@/^2pc/^{\xcMup d_1^{(1)\,*}\Psi_{1,2}^1} "A";"B"
\ar@{->}@/_2pc/_{\xcMup d_1^{(1)\,*}\Psi_{1,2}^2} "A";"B"
\ar@{=>}|{\xcMup d_1^{(1)\,*}\psi_{1,2}} "A"+(20,7);"A"+(20,-7)
\ar@{->}@/^2pc/^{\xcMup d_0^{(1)\,*}\Psi_{1,2}^1\ox\id} "C";"D"
\ar@{->}@/_2pc/_{\xcMup d_0^{(1)\,*}\Psi_{1,2}^2\ox\id} "C";"D"
\ar@{=>}|{\xcMup d_0^{(1)\,*}\psi_{1,2}\ox\id}
"C"+(20,7);"C"+(20,-7) \ar@{->}_{\Upsilon_1} "A";"C"
\ar@{->}^{\Upsilon_2} "B";"D" \ar@{=>}@/_2pc/_(.6){\eta^2_{1,2}}
"C";"B" \ar@{==>}@/^2pc/^(.4){\eta^1_{1,2}} "C";"B"
\endxy\,.\cr\cr
\label{diag:Gequiv2iso-coh}
\qqq
\item \Reqref{eq:Gequiv-bimod-coh}
\qq
\xy (-25,-40)*{\bullet}="A"+(-25,-2)*{(\Qup d^{(1)}_0\circ\Qup d^{(2
)}_1)^*\iota_1^*\cG\ox I_{\Qup d^{(2)\,*}_1\iota_1^{(1)\,*}\rho}};
(25,-40)*{\bullet}="B"+(15,-2)*{(\Qup d^{(1)}_1\circ\Qup d^{(2)}_1
)^*\iota_1^*\cG}; (10,-5)*{\bullet}="C"+(-5,5)*{(\Qup d^{(1)}_0\circ
\Qup d^{(2)}_2)^*\iota_1^*\cG\ox I_{\Qup
d^{(2)\,*}_2\iota_1^{(1)\,*} \rho}};
(-25,-100)*{\bullet}="X"+(-10,-5)*{(\Qup d^{(1)}_0\circ\Qup
d^{(2)}_1)^*\iota_2^*\cG\ox I_{\Qup
d^{(2)\,*}_1\iota_2^{(1)\,*}\rho+ (\Qup d^{(1)}_1\circ\Qup
d^{(2)}_1)^*\om}}; (25,-100)*{\bullet}="Y"+(10,-5)*{(\Qup
d^{(1)}_1\circ\Qup d^{(2)}_1 )^*\iota_2^*\cG\ox I_{(\Qup
d^{(1)}_1\circ\Qup d^{(2)}_1)^*\om}};
(10,-65)*{\bullet}="Z"+(37,0)*{(\Qup d^{(1)}_0\circ\Qup d^{(2)}_2)^*
\iota_2^*\cG\ox I_{\Qup d^{(2)\,*}_2\iota_2^{(1)\,*}\rho+(\Qup
d^{(1)}_1\circ\Qup d^{(2)}_1)^*\om}}; \ar@{->}|{\Qup d^{(2)\,*}_1
\iota_1^{(1)\,*}\Upsilon} "B";"A" \ar@{->}|{\qquad\Qup d^{(2)\,*}_2
\iota_1^{(1)\,*}\Upsilon} "B";"C" \ar@{->}|{\Qup d^{(2)\,*}_0
\iota_1^{(1)\,*}\Upsilon\ox\id\qquad} "C";"A"
\ar@{->}|{\qquad\qquad\quad(\Qup d^{(1)}_0\circ\Qup d^{(2)}_2)^*
\Phi\ox J_{\Qup d^{(2)\,*}_2\la}} "C";"Z" \ar@{->}|(.55){(\Qup d^{(
1)}_1\circ\Qup d^{(2)}_1)^*\Phi} "B";"Y" \ar@{->}|(.55){(\Qup d^{(1
)}_0\circ\Qup d^{(2)}_1)^*\Phi\ox J_{\Qup d^{(2)\,*}_1\la}} "A";"X"
\ar@{->}|{\Qup d^{(2)\,*}_1\iota_2^{(1)\,*}\Upsilon^{-1}\ox\id}
"X";"Y" \ar@{->}|{\qquad\Qup
d^{(2)\,*}_2\iota_2^{(1)\,*}\Upsilon^{-1} \ox\id} "Z";"Y"
\ar@{->}|{\Qup d^{(2)\,*}_0\iota_2^{(1)\,*}\Upsilon^{-
1}\ox\id\qquad} "X";"Z" \ar@{=>}|{\iota_1^{(1)\,*}\g}
"C"+(-.5,-3);"A"+(23.5,3)
\ar@{==>}|(.3){\iota_2^{(1)\,*}\g^\sharp\ox\id}
"X"+(23.5,3);"Z"+(-.5,-3) \ar@{==>}|{\Qup d^{(2)\,*}_2\Xi}
"B"+(-3,-29);"C"+(2,-38) \ar@{=>}|{\Qup d^{(2)\,*}_1\Xi}
"B"+(-11,-33);"A"+(17,-33) \ar@{==>}|{\Qup d^{(2)\,*}_0\Xi\ox\id}
"C"+(-3,-32);"A"+(3,-29)
\endxy\,.\label{diag:Gequiv-bimod-coh}
\qqq
\brem Note that the first diagram makes sense in virtue of relation
\eqref{eq:dhatdro=hatHH}. The only element of the second one that
calls for a word of explanation is the left rear face, which
represent a slightly more complex structure (we have dropped the
obvious gerbe labels of the inner vertices)
\qq\nn
\xy (30,-5)*{\bullet}="A"+(-15,5)*{(\Qup d^{(1)}_0\circ\Qup d^{(2
)}_2)^*\iota_1^*\cG\ox I_{\Qup d^{(2)\,*}_2\iota_1^{(1)\,*}\rho}};
(70,-5)*{\bullet}="B"+(15,5)*{(\Qup d^{(1)}_0\circ\Qup d^{(2)}_1)^*
\iota_1^*\cG\ox I_{\Qup d^{(2)\,*}_1\iota_1^{(1)\,*}\rho}};
(30,-40)*{\bullet}="C"; (70,-40)*{\bullet}="D";
(0,-60)*{\bullet}="E"+(10,-5)*{(\Qup d^{(1)}_0\circ\Qup d^{(2)}_2)^*
\iota_2^*\cG\ox I_{\Qup d^{(2)\,*}_2\iota_2^{(1)\,*}\rho+(\Qup d^{(1
)}_1\circ\Qup d^{(2)}_1)^*\om}};
(100,-60)*{\bullet}="F"+(-10,-5)*{(\Qup d^{(1)}_0\circ\Qup
d^{(2)}_1)^*\iota_2^*\cG\ox I_{\Qup
d^{(2)\,*}_1\iota_2^{(1)\,*}\rho+( \Qup d^{(1)}_1\circ\Qup
d^{(2)}_1)^*\om}}; \ar@{->}|{\Qup
d^{(2)\,*}_0\iota_1^{(1)\,*}\Upsilon\ox\id} "A";"B"
\ar@{->}|{\qquad\qquad\quad (\Qup d^{(1)}_0\circ\Qup
d^{(2)}_0)^*\Phi\ox J_{\Qup d^{(2)\,*}_0\la}} "B";"D"
\ar@{->}|{(\Qup d^{(1)}_1\circ\Qup
d^{(2)}_0)^*\Phi\ox\id\qquad\qquad} "A";"C" \ar@{->}|{\Qup
d^{(2)\,*}_0\iota_2^{(1)\,*}\Upsilon^{-1}\ox\id} "D";"C"
\ar@{->}|{(\Qup d^{(1)}_0\circ\Qup d^{(2)}_2)^*\Phi\ox J_{\Qup
d^{(2)\,*}_2\la}\qquad\qquad\qquad} "A";"E"
\ar@{->}|{\qquad\qquad\qquad(\Qup d^{(1)}_0\circ\Qup
d^{(2)}_1)^*\Phi\ox J_{\Qup d^{(2)\,*}_1\la}} "B";"F"
\ar@{->}|{\id\ox J_{\Qup d^{(2)\,*}_2\la}} "C";"E" \ar@{->}|{\id\ox
J_{\Qup d^{(2)\,*}_2\la}} "D";"F" \ar@{->}|{\Qup
d^{(2)\,*}_0\iota_2^{(1)\,*}\Upsilon^{-1}\ox\id} "F";"E"
\ar@{=>}|{\Qup d^{(2)\,*}_0\Xi\ox\id} "C"+(7,18);"D"+(-9,18) \ar@{=}
"F"+(-5,1.75);"C"+(5,-1.75) \ar@{=} "C"+(-2,1.5);(17,-32.5)
\ar@{=}_{(*)} "D"+(2,1.5);(83,-32.5)
\endxy
\qqq~\\[10pt]
We recover the simplified form of the face upon employing (the
pullback of) \Reqref{eq:dla} in conjunction with
Eqs.\,\eqref{eq:delro} and \eqref{eq:della}, the latter giving rise
to an identity 2-isomorphism for the right triangle in the above
subdiagram (indicated by $(*)$). \erem
\item \Reqref{eq:Gequiv-bib-2iso-coh}
\qq
\xy (30,-10)*{\bullet}="A"+(-10,0)*{\Qup d^{(1)\,*}_1\iota_1^*\cG};
(70,-10)*{\bullet}="B"+(17,0)*{\Qup d^{(1)\,*}_1\iota_2^*\cG\ox
I_{\Qup d^{(1)\,*}_1\om}}; (30,-60)*{\bullet}="C"+(-16,0)*{\Qup
d^{(1)\,*}_0\iota_1^*\cG\ox I_{\iota_1^{(1)\,*}\rho}};
(70,-60)*{\bullet}="D"+(22,0)*{\Qup d^{(1)\,*}_0\iota_2^*\cG\ox
I_{\Qup d^{(1)\,*}_1\om+\iota_2^{(1)\,*}\rho}};
\ar@{->}@/^2pc/^{\Qup d^{(1)\,*}_1\Phi_1} "A";"B"
\ar@{->}@/_2pc/_{\Qup d^{(1)\,*}_1\Phi_2} "A";"B" \ar@{=>}|{\Qup
d^{(1)\,*}_1\psi} "A"+(20,7);"A"+(20,-7) \ar@{->}@/^2pc/^{\Qup
d^{(1)\,*}_0\Phi_1\ox J_\la} "C";"D" \ar@{->}@/_2pc/_{\Qup
d^{(1)\,*}_0\Phi_2\ox J_\la} "C";"D" \ar@{=>}|{\Qup
d^{(1)\,*}_0\psi\ox\id} "C"+(20,7);"C"+(20,-7)
\ar@{->}_{\iota_1^{(1)\,*}\Upsilon} "A";"C"
\ar@{->}_{\iota_2^{(1)\,*}\Upsilon^{-1}\ox\id} "D";"B"
\ar@{=>}|{\Xi_2} "B"+(-10,-9);"D"+(-10,-4) \ar@{==>}|{\Xi_1}
"A"+(10,4);"C"+(10,9)
\endxy\,.\cr\cr\label{diag:Gequiv-bib-2iso-coh}
\qqq
\item \Reqref{eq:Desc-obj-coh}
\qq
\qquad\qquad\xy (40,0)*{\bullet}="34"+(2,3)*{\cG_{3^*}};
(0,-50)*{\bullet}="1234"+(2,-3)*{\cG_{1^*}};
(65,-50)*{\bullet}="4"+(2,-3)*{\cG_{4^*}};
(90,-35)*{\bullet}="234"+(5,0)*{\cG_{2^*}}; (77,-41)*{}="234-4";
(38,-49)*{}="1234-4"; (35,-48)*{}="1234-4-bis";
(22,-25)*{}="1234-34"; (49,-38)*{\Upsilon_{[1,2]^*}};
(48,-13)*{\Upsilon_{[3,4]^*}}; \ar@{->}^{\Upsilon_{[1,3]^*}}
"1234";"34" \ar@{->}_{\Upsilon_{[1,4]^*}} "1234";"4" \ar@{->}
"34";"4" \ar@{->}^{\Upsilon_{[2,4]^*}} "234";"4"
\ar@{->}_{\Upsilon_{[2,3]^*}} "234";"34" \ar@{->} "1234";"234"
\ar@{=>}|{\g_{[2,3,4]^*}} "34"+(3,-3);"234-4"
\ar@{==>}|{\g_{[1,2,4]^*}} "234"+(-4,-1.5);"1234-4"
\ar@{=>}|{\g_{[1,3,4]^*}} "34"+(-.5,-3);"1234-4-bis"
\ar@{==>}|{\g_{[1,2,3]^*}} "234"+(-3,1);"1234-34"
\endxy\,.\cr\cr\label{diag:Desc-obj-coh}
\qqq
\item \Reqref{eq:Desc-1cell-coh}
\qq
\xy (-25,-10)*{\bullet}="A"+(-7,0)*{\cG_{1\,[3]^*}};
(25,-10)*{\bullet}="B"+(7,0)*{\cG_{1\,[1]^*}};
(10,25)*{\bullet}="C"+(0,4)*{\cG_{1\,[2]^*}};
(-25,-70)*{\bullet}="X"+(-7,0)*{\cG_{2\,[3]^*}};
(25,-70)*{\bullet}="Y"+(7,0)*{\cG_{2\,[1]^*}};
(10,-35)*{\bullet}="Z"+(0,4)*{\cG_{2\,[2]^*}};
\ar@{->}|{\Upsilon_{1\,[1,3]^*}} "B";"A"
\ar@{->}|{\Upsilon_{1\,[1,2]^*}} "B";"C"
\ar@{->}|{\Upsilon_{1\,[2,3]^*}} "C";"A" \ar@{->}|{\Psi_{2^*}}
"C";"Z" \ar@{->}|{\Psi_{1^*}} "B";"Y" \ar@{->}|{\Psi_{3^*}} "A";"X"
\ar@{->}|{\Upsilon_{2\,[1,3]^*}} "Y";"X"
\ar@{->}|{\Upsilon_{2\,[1,2]^*}} "Y";"Z"
\ar@{->}|{\Upsilon_{2\,[2,3]^*}} "Z";"X" \ar@{=>}|{\g_1}
"C"+(-.5,-3);"A"+(23.5,3) \ar@{==>}|{\g_2} "Z"+(-.5,-3);"X"+(23.5,3)
\ar@{==>}|{\eta_{[1,2]^*}} "C"+(1,-5);"Y"+(-1,5)
\ar@{=>}|{\eta_{[1,3]^*}} "A"+(3,-3);"Y"+(-3,3)
\ar@{==>}|{\eta_{[2,3]^*}} "A"+(3,-1.5);"Z"+(-3,1.5)
\endxy\,.\cr\cr\label{diag:Desc-1cell-coh}
\qqq
\item \Reqref{eq:Desc-2cell-coh}
\qq
\xy (35,-10)*{\bullet}="A"+(-7,0)*{\cG_{1\,[1]^*}};
(65,-10)*{\bullet}="B"+(7,0)*{\cG_{2\,[1]^*}};
(35,-40)*{\bullet}="C"+(-7,0)*{\cG_{1\,[2]^*}};
(65,-40)*{\bullet}="D"+(7,0)*{\cG_{2\,[2]^*}};
\ar@{->}@/^2pc/^{\Psi_{1\,[1]^*}} "A";"B"
\ar@{->}@/_2pc/_{\Psi_{2\,[1]^*}} "A";"B" \ar@{=>}|{\psi_{1^*}}
"A"+(15,7);"A"+(15,-7) \ar@{->}@/^2pc/^{\Psi_{1\,[2]^*}} "C";"D"
\ar@{->}@/_2pc/_{\Psi_{2\,[2]^*}} "C";"D" \ar@{=>}|{\psi_{2^*}}
"C"+(15,7);"C"+(15,-7) \ar@{->}_{\Upsilon_1} "A";"C"
\ar@{->}^{\Upsilon_2} "B";"D" \ar@{=>}@/^2pc/^{\eta_2} "C";"B"
\ar@{==>}@/_2pc/_{\eta_1} "C";"B"
\endxy\,.\cr\cr
\label{diag:Desc-2cell-coh}
\qqq
\eit

\section{\hspace*{0.12cm}Natural simplicial refinements of open covers}
\label{app:nat-simpl-cov}

\noindent In this appendix, we extend the basic simplicial framework
introduced in Section \ref{sub:simplicial} with view to applying it
in a cohomological classification, carried out in Section
\ref{sec:class-equiv-back}, of $\txG$-equivariant string
backgrounds, the latter being regarded of as sheaf-theoretic
structures supported by the simplicial $\txG$-space $\,\txG\xcF$.\
In what follows, we use the notation of Section
\ref{sub:simplicial}.

Let $\,M:\D\to\Man\,$ be a contravariant functor, so that the family
$\,\{M^m\}_{m=0,1,\ldots}\,$ of manifolds $\,M^m:=M([m])$,\ together
with the collection of maps
$\,\mu^{(m+1)}_k:=M\bigl(\theta^{(m+1)}_k \bigr):M^{m+1}\to M^m$,\
forms an incomplete simplicial manifold. Furthermore, let
$\,\{\cO^m\}_{m=0,1,\ldots}\,$ be a sequence of open covers, each
$\,\cO^m=\{\cO^m_i\}_{i\in\xcI^m}\,$ covering the corresponding
smooth manifold $\,M^m\,$ (here, the $\,\xcI^m\,$ are index sets).
Denote
\qq\nn
\D^m:=\bigcup_{k=0}^m\,\D(k,m)\,,\qquad\qquad\cI^m:=\bigcup_{k=0}^m\,
\xcI^k\,.
\qqq
Define an index set
\qq\nn
\xcJ^m:=\{\ \jmath\ :\ \D^m\to\cI^m \quad\vert\quad \theta\in\D(k,m
)\ \Rightarrow\ \jmath(\theta)\in\xcI^k  \ \}
\qqq
and use its elements to label open\footnote{The sets are open as
finite intersections of continuous preimages of open sets.} sets
\qq\nn
\cU^m_\jmath:=\bigcap_{k=0}^m\,\bigcap_{\theta\in\D(k,m)}\,M(\theta
)^{-1}\bigl(\cO^k_{\jmath(\theta)}\bigr)\,.
\qqq
There exists a simple construction, advanced in \Rcite{Tu:2006}, of
a simplicial sequence of open covers refining a given open cover of
a simplicial manifold, which we now review. In so doing, we follow
its transcription given in \Rcite{Gawedzki:2009jj}. We have
\berop
The collection $\,\cU^m:=\{\cU^m_\jmath\}_{\jmath\in\xcJ^m}\,$ of
open sets is a refinement of the open cover $\,\cO^m\,$ of
$\,M^m$.\eerop
\beroof
First, we show that $\,\cU^m\,$ is a cover of $\,M^m$.\ Indeed, for
any $\,x\in M^m\,$ and each $\,\theta\in\D(k,m)$,\ there exists an
index $\,j_\theta\in\xcI^k\,$ such that $\,M(\theta)(x)\in
\cO^k_{j_\theta}$,\ and the ensuing assignment $\,\jmath:\theta
\mapsto j_\theta,\ \theta\in\D(k,m)\,$ defines an index $\,\jmath
\in\xcJ^m\,$ such that, clearly, $\,x\in\cU^m_\jmath$.

Next, we convince ourselves by exploiting the functoriality of
$\,M$,\ that the new cover is a refinement of the original one:
Indeed, $\,\cU^m_\jmath\subset M(\id_{[m]})^{-1}\bigl(\cO^m_{\jmath(
\id_{[m]})}\bigr)\equiv\id_{M^m}^{-1}\bigl(\cO^m_{\jmath(\id_{[m]})}
\bigr)=\cO^m_{\jmath(\id_{[m]})}$,\ and so each $\,\cU^m_\jmath\,$
is contained in the corresponding open set $\,\cO^m_{r(\jmath)}\,$
with $\,r:\xcJ^m\to\xcI^m:\jmath \mapsto\jmath(\id_{[m]})$.\eroof
\bigskip
In the next step, we verify
\berop\label{prop:J-funct}
Let $\,J:\D\to\Set\,$ be a map between the object and morphism
classes of the two categories defined, for arbitrary $\,\theta\in
\D(l,m)\,$ and $\,\xi\in\D(k,l)$,\ by the respective formul\ae:
\qq\nn
J([m]):=\xcJ^m\,,\qquad\qquad J(\theta)\ :\ \xcJ^m\to\xcJ^l\ :\
\jmath\mapsto J(\theta)(\jmath)\,,\qquad\bigl(J(\theta)(\jmath)
\bigr)(\xi):=\jmath(\theta\circ\xi)\in\xcI^k\,.
\qqq
Then, $\,J\,$ is a contravariant functor which makes the family
$\,\{\xcJ^m\}_{m=0,1,\ldots}\,$ into an incomplete simplicial set.
\eerop
\beroof
Clearly, $\,J(\theta)\in\mor_{{\rm Set}}\bigl(J([m]),J([l])
\bigr)\,$ by the very definition of the morphism component of the
map $\,J$.\ Besides, for any $\,\eta\in\D(m,p)\,$ and $\,\jmath\in
\xcJ^p$,\ we find
\qq\nn
\bigl(J(\eta\circ\theta)(\jmath)\bigr)(\xi)=\jmath(\eta\circ\theta
\circ\xi)=\bigl(J(\eta)(\jmath)\bigr)(\theta\circ\xi)=\bigl(J(
\theta)\bigl(J(\eta)(\jmath)\bigr)\bigr)(\xi)\equiv\bigl(\bigl(J(
\theta)\circ J(\eta)\bigr)(\jmath)\bigr)(\xi)\,,
\qqq
whence
\qq\nn
J(\eta\circ\theta)=J(\theta)\circ J(\eta)\,,
\qqq
as desired. The remaining part of the statement of the proposition
now follows from the one-to-one correspondence between incomplete
simplicial sets and contravariant functors $\,S:\D\to\Set$. \eroof
\bigskip
We have a natural
\bedef
Let $\,\{M^m\}_{m=0,1,\ldots}\,$ be an incomplete simplicial
manifold with face maps $\,\mu^{(m+1)}_k:M^{(m+1)}\to M^m$.\ A
sequence $\,\{ \cO^m\}_{m=0,1,\ldots}\,$ of open covers
$\,\cO^m=\{\cO^m_i\}_{i\in \xcI^m}$,\ each $\,\cO^m\,$ covering the
corresponding $\,M^m$,\ is termed \textbf{simplicial} iff the
sequence $\,\{\xcI^m\}_{m=0,1,\ldots}\,$ forms an incomplete
simplicial set with respect to the collection of face maps
$\,\iota^{(m+1)}_k:\xcI^{m+1}\to\xcI^m,\ k=0,1,\ldots,m+1\,$ and the
condition
\qq\label{eq:simp-cov-id}
\mu^{(m+1)}_k(\cO^{m+1}_i)\subset\cO^m_{\iota^{(m+1)}_k(i)}
\qqq
is satisfied. \exdef \noindent The basic feature of the construction
just described is stated in the following
\berop
The sequence $\,\{\cU^m\}_{m=0,1,\ldots}\,$ of open covers
$\,\cU^m=\{ \cU^m_\jmath\}_{\jmath\in\xcJ^m}\,$ of the incomplete
simplicial manifold engendered by a contravariant functor
$\,M:\D\to\Man\,$ is simplicial.
\eerop
\beroof
The sole thing that has to be checked is identity
\eqref{eq:simp-cov-id}. It follows from the more general one:
\qq\nn
M(\theta)(\cU^m_\jmath)\subset\cU^l_{J(\theta)(\jmath)}\,,
\qqq
valid for all $\,\jmath\in\xcJ^m\,$ and an arbitrary map $\,\theta
\in\D(l,m)$,\ upon restricting the latter to the universal coface
maps $\,\theta^{(m)}_k$.\ It therefore remains to prove the above
identity, which is tantamount to $\,M(\theta)(\cU^m_\jmath)\,$ being
contained in all the open sets $\,M(\eta)^{-1}\bigl(\cO^k_{\bigl(J(
\theta)(\jmath)\bigr)(\eta)}\bigr),\ k=0,1,\ldots,m,\ \eta\in\D(k,l
)\,$ obtained, with the help of the functor $\,M$,\ from some
sequence $\,\{\cO^m\}_{m=0,1,\ldots}\,$ of open covers $\,\cU^m=\{
\cO^m_i\}_{i\in\xcI^m}\,$ of the $\,M^m=M([m])\,$ in the manner
detailed. The desired inclusions are readily inferred from
\qq\nn
&&M(\eta)\bigl(M(\theta)\bigl(\cU^m_\jmath\bigr)\bigr)\ =\ M(\theta
\circ\eta)\bigl(\cU^m_\jmath\bigr)\equiv M(\theta\circ\eta)\bigl(
\bigcap_{k=0}^m\,\bigcap_{\psi\in\D(k,m)}\,M(\psi)^{-1}\bigl(
\cO^k_{\jmath(\psi)}\bigr)\bigr)\cr\cr
&&\subset\ \bigcap_{k=0}^m\,\bigcap_{\psi\in\D(k,m)}\,M(\theta\circ
\eta)\bigl(M(\psi)^{-1}\bigl(\cO^k_{\jmath(\psi)}\bigr)\bigr)
\subset M(\theta\circ\eta)\bigl(M(\theta\circ\eta)^{-1}\bigl(
\cO^k_{\jmath(\theta\circ\eta)}\bigr)\bigr)\cr\cr
&&=\ \cO^k_{\jmath(\theta\circ\eta)}\equiv\cO^k_{\bigl(J(\theta)(
\jmath)\bigr)(\eta)}\,.
\qqq
\eroof
\bigskip
An obvious question arises at this stage as to the naturality of the
above construction of simplicial covers of incomplete simplicial
manifolds. The answer to this question is given below.
\berop\label{prop:nat-trans-J}
Let $\,M_N:\D\to\Man,\ N=1,2\,$ be a pair of contravariant functors,
so that the respective families $\,\{M^m_N\}_{m=0,1,\ldots}\,$ of
manifolds $\,M^m_N:=M_N([m])$,\ together with the corresponding
collections of maps $\,\mu^{(m+1)}_{N,k}:=M_N(\theta^{(m+1)}_k):
M_N^{m+1}\to M_N^m$,\ form incomplete simplicial manifolds.
Furthermore, let $\,\{\cO_N^m\}_{m=0,1,\ldots},\ N=1,2\,$ be
sequences of open covers $\,\cO_N^m=\{\cO^m_{N,i}\}_{i\in
\xcI_N^m}\,$ of the respective $\,M_N^m$,\ with the corresponding
index sets $\,\xcI_N^m$.\ Finally, let
$\,\{\cU_N^m\}_{m=0,1,\ldots},\ N=1,2\,$ be the simplicial sequences
of refinements $\,\cU_N^m=
\{\cU^m_{N,\jmath}\}_{\jmath\in\xcJ_N^m}\,$ of the $\,\cO_N^m\,$
engendered by the $\,M_N\,$ and indexed by the corresponding
incomplete simplicial sets $\,\{\xcJ_N^m\}_{m=0,1,\ldots},\
\xcJ_N^m:= J_N([m])$,\ the latter being induced by the respective
contravariant functors $\,J_N:\D \to\Set$.\ Every natural
transformation $\,\mu:M_1 \to M_2$,\ with
$\,\mu^{(m)}:=\mu([m])\in\mor_{\Man}(M_1^m,M_2^m)$,\ and a sequence
$\,\{i^{(m)}\}_{m=0,1,\ldots}\,$ of maps $\,i^{(m)}:\xcI_1^n
\to\xcI_2^m\,$ satisfying the \textbf{covering relations}
\qq\label{eq:cov-rels-ass}
\mu^{(m)}\bigl(\cO^m_{1,i}\bigr)\subset\cO^m_{2,i^{(m)}(i)}\,,
\qqq
canonically induces a natural transformation $\,j:J_1\to J_2$,\ with
\qq
j^{(m)}:=j([m])\in\mor_{\Set}(J_1^m,J_2^m)\,,
\qqq
satisfying the covering relations
\qq\label{eq:cov-rels}
\mu^{(m)}\bigl(\cU^m_{1,\jmath}\bigr)\subset\cU^m_{2,j^{(m)}(\jmath)}
\,.
\qqq
\eerop
\beroof
Consider a map $\,j^{(m)}:\xcJ_1^m\to\xcJ_2^m\,$ assigning to an
arbitrary index $\,\jmath\in\xcJ_1^m\,$ a map $\,\jmath^{(m)}\equiv
j^{(m)}(\jmath)\,$ defined subset-wise over $\,\D^m\,$ by the
formul\ae
\qq\nn
\jmath^{(m)}\vert_{\D(k,m)}=i^{(k)}\circ\jmath\,.
\qqq
Clearly, this definition makes sense as
\qq\nn
\theta\in\D(k,m)\ \Rightarrow\ \jmath^{(m)}(\theta)=i^{(k)}\bigl(
\jmath(\theta)\bigr)\in i^{(k)}(\xcI_1^k)\subset\xcI_2^k\,.
\qqq
Take the universal coface map $\,\theta^{(m+1)}_k\in\D(m,m+1)\,$ and
an arbitrary map $\,\xi\in\D(k,m)$.\ We obtain, for any index
$\,\jmath\in\xcJ_1^m$,
\qq\nn
\bigl(J_2\bigl(\theta^{(m+1)}_k\bigr)\bigl(j^{(m+1)}(\jmath)\bigr)
\bigr)(\xi)&=&\bigl(j^{(m+1)}(\jmath)\bigr)\bigl(\theta^{(m+1)}_k
\circ\xi\bigr)=i^{(k)}\bigl(\jmath\bigl(\theta^{(m+1)}_k\circ\xi
\bigr)\bigr)\cr\cr
&=&\bigl(i^{(k)}\circ\bigl(J_1\bigl(\theta^{(m+1)}_k\bigr)(\jmath)
\bigr)\bigr)(\xi)=\bigl(j^{(m)}\bigl(J_1(\theta^{(m+1)}_k)(\jmath)
\bigr)\bigr)(\xi)\,,
\qqq
which assures the commutativity of the diagram
\qq\nn
\alxydim{@C=2.cm}{\xcJ_1^{m+1} \ar[r]^{J_1\bigl(\theta^{(m+1)}_k
\bigr)} \ar[d]_{j^{(m+1)}} & \xcJ_1^m \ar[d]^{j^{(m)}} \\
\xcJ_2^{m+1} \ar[r]_{J_2\bigl(\theta^{(m+1)}_k\bigr)} & \xcJ_2^m }
\qqq
of index maps. Upon expressing an arbitrary map $\,\theta\in\D(l,m
)\,$ in terms of the universal coface maps, we readily infer from
the above the naturality of $\,j$.

At this stage, it remains to check the covering relations
\eqref{eq:cov-rels} for the natural transformation constructed.
These follow immediately from
\qq\nn
M_2(\theta)\bigl(\mu^{(m)}\bigl(\cU^m_{1,\jmath}\bigr)\bigr)=\mu^{(k)}
\bigl(M_1(\theta)\bigl(\cU^m_{1,\jmath}\bigr)\bigr)\subset\mu^{(k)}
\bigl(\cO^k_{1,\jmath(\theta)}\bigr)\subset\cO^k_{2,(i^{(k)}\circ
\jmath)(\theta)}=\cO^k_{2,\bigl(j^{(m)}(\jmath)\bigr)(\theta)}\,,
\qqq
valid for any $\,\jmath\in\xcJ^m_1\,$ and an arbitrary $\,\theta\in
\D(k,m)\,$ in consequence of the assumed naturality of $\,\mu\,$
(ensuring the first equality) and of the very definition of the
$\,\cU^m_{1,\jmath}\,$ (explaining the first inclusion), as well as
the covering relations \eqref{eq:cov-rels-ass} (the second
inclusion).\eroof
\bigskip
In the context of $\txG$-equivariant structures on gerbes and
related geometric objects, we have
\bedef
Whenever the functors $\,M_N,\ N=1,2\,$ and the natural
transformation $\,\mu\,$ between them, appearing in Proposition
\ref{prop:nat-trans-J}, are the nerve functors $\,M_N:=\txG\xcM_N\,$
for the action groupoids $\,\txG\lx\xcM_N\,$ of $\txG$-spaces
$\,\xcM_N\,$ and the associated family of $\txG$-maps, respectively,
we shall call the $\,\cU^m_N\,$ \textbf{aligned simplicial sequences
of $\txG$-invariant refinements of open covers}. \exdef

\bibliographystyle{amsalpha}

\end{document}